# CAPMAP
# A New Instrument to Measure the E-mode CMB Polarization on Angular Scales of $4'$ to $40'$

Denis Barkats

A dissertation
presented to the faculty
of Princeton University
in candidacy for the degree
of Doctor of Philosophy

Recommended for acceptance
by the department of Physics

June 2004

# Abstract

The CMB polarization is the Everest in the quest to characterize the earliest photons from the Universe. After a long list of ever-decreasing upper limits, a detection of polarization was made in 2002 by the DASI team at $\ell \simeq 500$. The experiment described in this thesis is designed to make a more detailed measurements at higher angular resolution.

The E-mode polarization power spectrum not only provides a more direct link to the properties of the last scattering surface than the temperature anisotropy but also offers complementary information which can be used to break various degeneracies in the determination of cosmological parameters. Most importantly, the existence of polarization is a robust prediction of the standard cosmological picture so a precise measurement of the CMB polarization should come as a confirmation of the standard model. However, polarization measurements represent an experimental challenge. The weakness of the polarization signal requires both a demanding instrumental sensitivity and focused attention to all sources of systematic error.

This thesis describes the design, construction, and testing of a 90 GHz four-element array of correlation polarimeters to probe the E-mode polarization power spectrum at multipoles ($\ell$) ranging from 500 to 1500. The array was fielded in Jan 2003 on the 7-meter Crawford Hill antenna, in Holmdel, New Jersey and observed for two months. The receiver calibration is described in detail, as well as the characterization of the pointing and beams. Preliminary analysis indicates that the instrument is sufficiently sensitive to detect the few-$\mu$K signal of the CMB polarization.

# Acknowledgments

In these 200-or-so pages, I describe the CAPMAP instrument. Although it is the experiment which provides the content of my dissertation, it is the people behind the whole experiment, immersed between the lines, present behind each anecdote, which make the thesis worth the long effort. All the people I will thank are an integral part of my 6-year experience at Princeton. So I apologize for the lengthy Oscar-like list of acknowledgments but I don't often get a chance to express my appreciation to these people.

My first thanks goes to Suzanne Staggs, my advisor. She has been a truly wonderful advisor to work with, specifically because she makes it clear that you work with her, as her equal, not for her. I have to admit; I am truly grateful for her philosophy of giving graduate students so much responsibility. That has given me a lot of confidence, especially as I had little hands-on lab experience when I started. And yet, Suzanne always pushes herself to be available to answer questions from her students. And that despite her busy schedule with other experiments, teaching, proposal writing, undergraduate advising, and taking care of her two kiddos, Sarah and Zoe. Suzanne is one of those people who has kept the life at Princeton enjoyable; from chatting at 2AM in lab about anything to inviting me over to her house for hey spicy vegetarian dinners, she has made the time at Princeton into a human experience rather than a mere student-advisor relationship.

The CAPMAP experiment is a the work of team, small enough that everyone has played a critical role. I'd like to thank Matt Hedman, with whom I worked on the PIQUE experiment during my first two years at Princeton. He has answered a lot my "idiot's" questions about radiometers and the CMB. He is also the only person I know who can instantly come up with an explanation for the strangest behavior in any device or observation. Jeff McMahon has also been instrumental in the design of the CAPMAP optics. But above all, thanks for the comic relief: "Your awesomeness is only magnified by your ability to rule! Dude". He'll understand. Josh Gundersen was a postdoc and Bruce Winstein was on sabbatical in Princeton during my first years at Princeton. I've learned a ton about the CMB field in their presence through day to day discussion. Todd Gaier provided precious advice during the receiver integration and helped us make critical decisions about the LO configuration. I'd like to also thank Phil Farese, Dorothea Samtleben, Keith Vanderlinde, Lewis Hyatt, Colin Bishoff, and Eugenia Stefanescu for putting in so much of their time on various aspects of this experiment. Those shifts with each one of you are memorable.

Although not formally part of the CAPMAP collaboration, many other people deserve credit for a specific part of the experiment: Michelle Yeh for the complete overhaul of the telescope control software and writing the Labview monitoring software, Jennifer Hou for much of the IDL analysis programs, Ashish Gupta for the DAQ to monitoring communication software. Big thanks also to all the undergraduate students who worked over the summers but whose names escape right now. They have taught me patience.

I'd like to thank Bob Wilson and Greg Wright, not only for introducing me to the intricate working of the 7-meter telescope but also for being willing to help us out when the telescope displayed some new unpredictable inexplicable (to us), unforgettable behavior. Thanks also to all the Crawford Hill security guards for keeping us safe and especially to Floyd, the gate keeper.

The existence of the instrument is also due to the skills of many people. Norm Jarosik was always there for enlightening discussion about radiometers and electronics. The guys in the Princeton machine shop are amazing: Bill Dix, Lazlo Varga, Glenn Atkinson, Ted Lewis, Ted Griffith, and Mike Peloso. Not just because they can pretty much build anything you draw for them, but also because they will do it even when you tell them you need it yesterday. Thanks especially to Lazlo who was willing to go out of his way to teach Dan Angelescu and myself how to weld. I also thank all the folks who guided me through the daily life in the Physics department: John Washington, Claude Champagne, Kathy Warren, Joe Horvath, Helen Ju, Angela Glenn, Ray Lau, Eva Zeisky, and Aric Davala.

Thanks also to all grad students and postdocs who were part of the active and enthusiastic gravity group during my stay in Princeton: Huan, Randy, Yeong, Juan, Asad, Mike (and Mike the Katsu), Glen, Dan, Toby, Judy, Jeff, Phil, Lewis, Joe. You made the days in Jadwin so much brighter and also reminding me that no matter how much work there is to do, there is always time to goof off.

My friends in Princeton and in France are responsible for most of the fun stuff I did during these years. I can't name them all but they will recognize themselves in these few words. They were always there to play soccer, go swimming, go hiking, built a go-cart, dream about aviation, stuff ourselves on crepes, or simply watch a movie together. Merci mes amis, je vous salue tous bien bas.

To my parents finally who have constantly found the right words encourage me.

I also wish to thank my two reader Joe Fowler, and Suzanne Staggs for providing so many comments and improvements to my thesis. Finally I thank my defense committee, Suzanne, Lyman Page, and David Huse for helping me through this final step at Princeton.

To the memory of Jeff Willick,
for his inspiration and enthusiasm.

On ne voit bien qu'avec le coeur, l'essentiel est invisible pour les yeux.

–Antoine de St-Exupery,
*Le Petit Prince*

# Acronyms

This is a list of acronyms, abbreviations, and radio-astronomy jargon used throughout this dissertation. Conversion factors are also provided.

| | | | Section |
|---|---|---|---|
| JPL | - | Jet Propulsion Laboratory | 2.1 |
| FWHM | - | Full Width at Half Maximum | 2.1 |
| HDPE | - | High Density Polyethylene | 2.1 |
| NCP | - | North Celestial Pole | 2.1 |
| FP | - | Focal Point | 2.4 |
| RF | - | Radio Frequency (84-100 GHz at W-band) | 3.1 |
| IF | - | Intermediate Frequency, (2-18 GHz) | 3.1.4 |
| LO | - | Local Oscillator (82 GHz) | 3.1.3 |
| OMT | - | Ortho-Mode Transducer | 3.1.1 |
| VNA | - | Vector Network Analyzer | 3.1.1 |
| MMIC | - | Monolithic Microwave Integrated Circuit | 3.1.1 |
| HEMT | - | High Electron Mobility Transistor | 3.1.1 |
| GPS | - | Global Positioning System | 4 |
| DAQ | - | Data Acquisition | 5.2 |
| A/D | - | Analog-to-Digital Converter | 5.2.1 |
| NI | - | National Instruments | 5.2.3 |
| ABOB | - | Analog Breakout Box | 5.1.1 |
| ICS | - | Interactive Circuits and Systems | 5.2 |
| BCD | - | Binary Coded Decimal | 5.2.3 |

| | | |
|---|---|---|
| 1 inch | - | 2.54 cm |
| 39.3 mil | - | 1 mm |
| 1 pound | - | 0.543 kg |

# Contents

# List of Tables

# List of Figures

# Chapter 1

# Introduction

## 1.1 The Tripod of Cosmology

Cosmology is the study of the universe as a whole. In its details, the universe is rather complicated with a multitude of amazing objects ranging from minuscule dust grains, to massive black holes. Despite this complexity, cosmologists view the universe as a simple place characterized by:

- a homogenous and isotropic distribution of the constituents of the universe.
- a hotter and denser past and an expansion of the distances between galaxies as a function of time.
- a total energy density consistent with a spatially flat geometry.
- an interesting energy density composition (Table 1.1) comprised predominantly of a repulsive dark energy field whose general characteristics match that of Einstein's cosmological constant, and partially of weakly interacting cold dark matter, and a very small amount of the ordinary matter as we know it (baryonic matter).
- a hierarchy of gravitationally bound structures from planets and stars, to galaxies, to clusters and superclusters of galaxies.

This description is accepted today as the standard "Big-Bang" model of cosmology. Modern cosmology is backed up by an imposing backbone of observational evidence. During the twentieth century, three observations in particular have shaped a consistent model: the Hubble expansion of galaxies, the measurement of the light elements' abundances, and the Cosmic Microwave Background radiation.

| Description | $n_i$ ($cm^{-3}$) | $\Omega_i = \rho_i/\rho_c$ |
|---|---|---|
| Total density | - | $\Omega_{tot} = 1.02 \pm 0.02$ |
| Dark "Vacuum" density | - | $\Omega_\Lambda = 0.73 \pm 0.04$ |
| Matter density | ? | $\Omega_m = 0.27 \pm 0.04$ |
| Baryons density | $n_b = (2.55 \pm 0.1) \times 10^{-7}$ | $\Omega_b = 0.044 \pm 0.004$ |
| Neutrino density | (3/11) $n_\gamma$ per species | $\Omega_\nu < 0.015$ (95% CL) |
| CMB photons density | $n_\gamma = 410.4 \pm 0.9$ | $\Omega_\gamma = 5.06 \times 10^{-5}$ |

Table 1.1: Known and suspected contents of the universe. For each constituent, the table lists the number density, $n_i$, and the energy density, $\Omega_i = \rho_i/\rho_c$, relative to the critical density, $\rho_c = 0.92 * 10^{-29}$ g m$^{-3}$. All values assume a Hubble constant equal to $H_0 = 71 \pm 4$ km s$^{-1}$ Mpc. All values are gathered from the WMAP "best" cosmological parameters [16]. The matter number density remains undetermined because the particles which make up cold dark matter are still unknown.

### 1.1.1 Hubble Diagram

The first leap was the discovery by Edwin Hubble in 1929 of the recession of galaxies. Hubble measured the redshift and the distance of various galaxies in our close galactic neighborhood (z < 0.003) and found that almost all the galaxies he observed were moving away from us at a velocity proportional to their distance. This is exactly what is expected from an expanding universe. At earlier times, the distance between any two galaxies was smaller than it is today and this distance increases with time at an increasing speed. This relation between the recession velocity, $v$, and the relative distance of two galaxies, $d$, became known as the Hubble law given by

$$v = H_0 \times d \tag{1.1}$$

where $H_0$ is the value of the Hubble constant today, currently thought to be $H_0 = 71 \pm 4$ km s$^{-1}$ Mpc. The Hubble diagram shown in Figure 1.1 displays the measured distance-redshift relations using Type Ia supernovæ as standard candles. Such a Hubble diagram is the most direct evidence that the universe is expanding.

### 1.1.2 Big Bang Nucleosynthesis

The second strong piece of evidence towards the hot Big Bang model is the prediction of light elements' abundances from Big Bang Nucleosynthesis (BBN). The epoch of BBN refers to a period of time of about one second after the Big Bang. Before that, the universe was too hot for bound nuclei to form. As the universe expanded and cooled below a temperature of $\sim 10^{10}$ Kelvin (1 MeV), the first lightest nuclei such as Hydrogen, Deuterium, Helium-4, and Lithium-7 started to form. As the universe further cooled below T$\sim$30 keV, nuclear reactions ceased and the primordial abundance became fixed. These abundances remained unchanged until the first stars began converting the primordial elements into heavier elements. The initial fraction of these primordial nuclei is entirely calculable and depends only on the baryon-to-photon ratio, $\eta = n_b/n_\gamma$. Therefore, measurements of primordial element abundances directly test our prediction about the contents and dynamics of the very early universe. Figure 1.2 shows the predicted primordial abundance of the light elements as a function of the baryon-to-photon ratio. The calculated predictions agree well with the best observations of $\eta \simeq 5 \times 10^{-10} \rightarrow \Omega_b \simeq 0.04$. Note that because this value differs from the matter density $\Omega_m = 0.3$, it suggests evidence for non-baryonic dark matter.

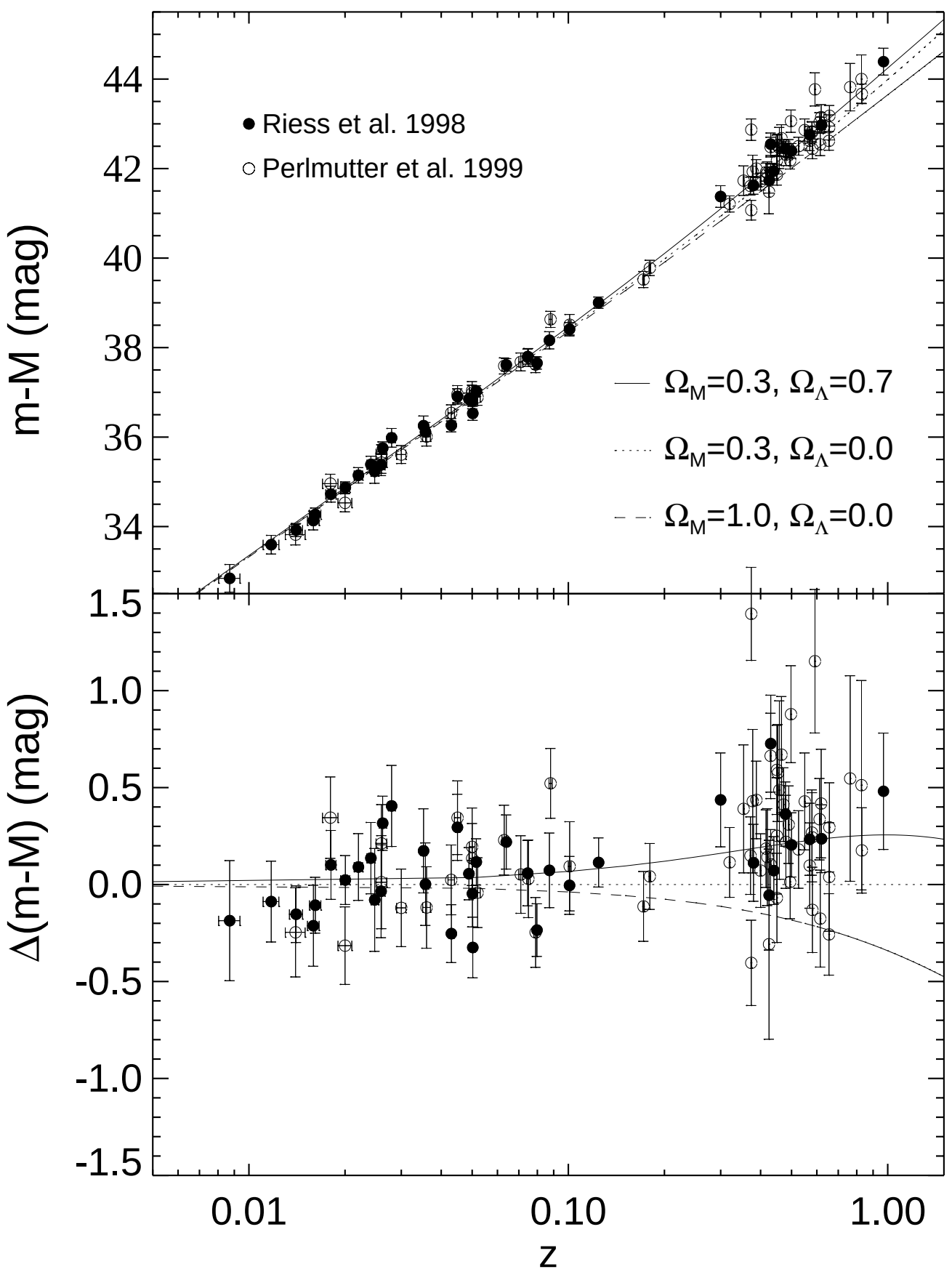

Figure 1.1: From [32]. Hubble diagram using distant type-Ia supernovæ as standard candles. Top panel shows apparent magnitude (an indicator of the distance) vs. redshift. Lines show the predictions for different energy contents of the universe. Bottom panel plots the residuals, making it clear that the high redshift supernovæ favor a Λ-dominated universe over a matter-dominated one.

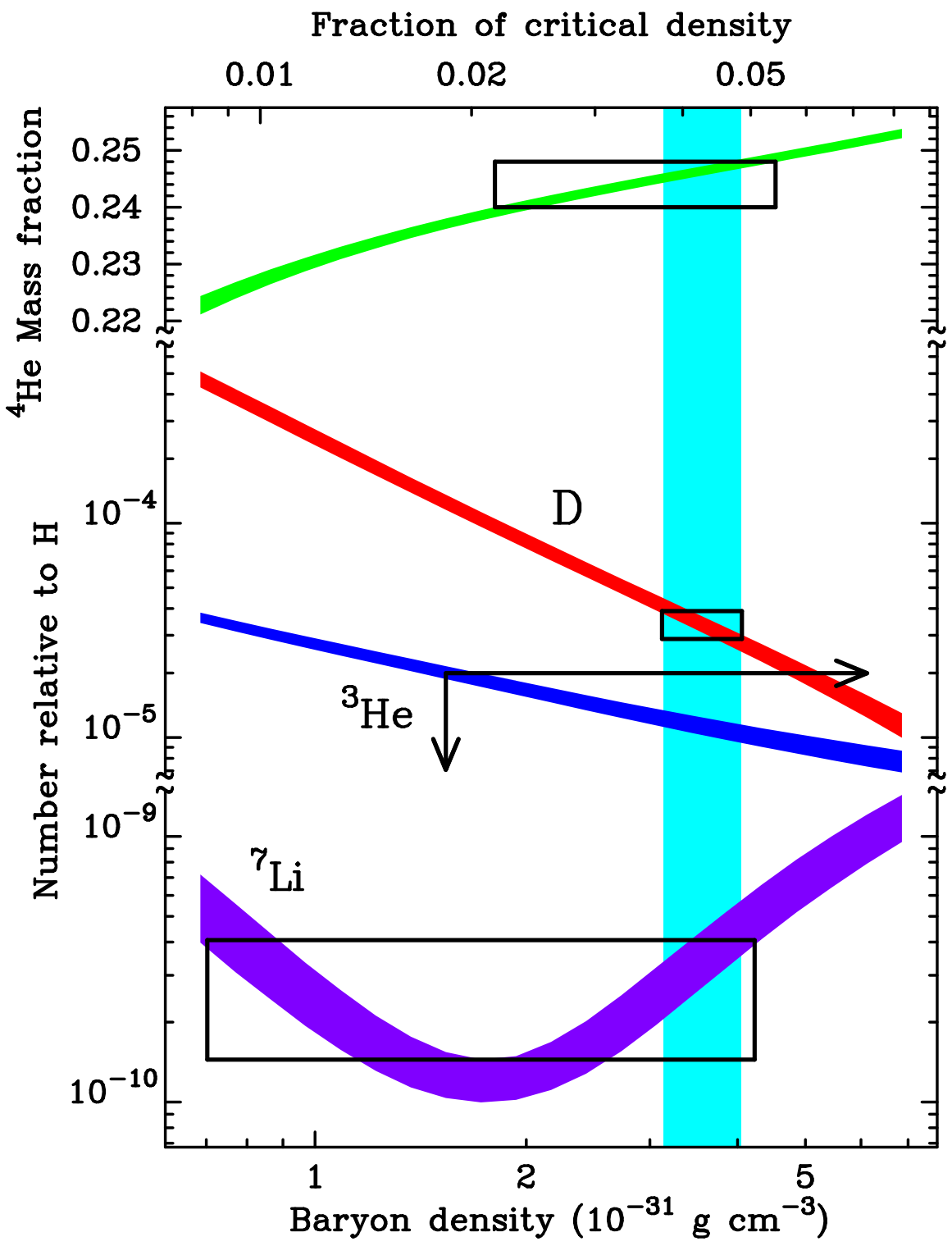

Figure 1.2: Predictions of light element abundances from Big Bang nucleosynthesis and current measurements from [20]. The four curved lines are the calculated predictions of the initial fraction of $^4$He, Deuterium, $^3$He, and $^7$Li relative to Hydrogen, as a function of $\eta$. The boxes are the current best measurements. Since the primordial abundances are a function of the initial baryon-to-photon ratio, an observation of any of the abundances provides a constraint on the baryon density. The best measurement to date of light element abundance is that of deuterium from the Ly-$\alpha$ absorption line in the spectra of QSO. This constrains the baryon density to $\rho_B = 3.6 \pm 0.4$ * ($10^{-31}$ g/cm$^3$),($\Omega_B$ h$^2$ = $0.019 \pm 0.0024$) [19].

## 1.2 The Cosmic Microwave Background Radiation

The third and most convincing evidence for the hot Big Bang model is the Cosmic Microwave Background (CMB) radiation. The CMB is the relic radiation from the epoch when the universe was hot enough to keep matter and radiation in thermal equilibrium. The CMB now fills space with a homogeneous distribution of $\sim$400 photons/cm$^3$. The early universe was a bath of photons, electrons, protons, and ionized light nuclei kept in thermal equilibrium by the constant Thompson scattering of photons and electrons. At redshift $z_{dec} = 1089$ (at a time of 379 000 years), the universe had cooled to $\sim$2000 K (0.3 eV), allowing the combination[1] of electrons and protons to form the first Hydrogen atoms[2].

[1] curiously this is called recombination

[2] Recombination does not happen at the binding energy of the Hydrogen atom, 13.6 eV, because there are $10^{10}$ more photons than baryons and the photons have a broad energy distribution. The recombination thus happens at an energy when the tail of the distribution starts having a small enough interaction rate.

This decoupling corresponds to the last interaction between the CMB and matter (apart from foregrounds, see Section 1.4). Because this period is relatively short ($\Delta z_{dec} = 195$), it is usually referred to as the Last Scattering Surface (LSS). After the LSS, the universe becomes optically transparent. The CMB photons then propagate freely, while conserving all the properties acquired during the last scattering event (their blackbody spectrum distribution, anisotropy, and polarization). The only modification to the CMB photons is a cooling of their temperature due to the expansion of the universe.

The CMB was unexpectedly discovered in 1965 by two Bell Telephone Labs radio engineers, Arno Penzias and Robert Wilson [118]. While trying to understand the precise noise properties of their horn antenna, they found a ~3 K isotropic source of white noise that could not be attributed to any instrumental or terrestrial cause. It was presented as a measurement of excess antenna noise [119]. With the help of Dicke and Wilkinson in Princeton University [30], the excess noise was correctly interpreted as the leftover radiation from the hot and dense state of the universe.

Since its discovery, the CMB has been the object of exhaustive measurements of the three characteristics of this electromagnetic radiation: its spectrum, spatial distribution, and polarization.

### 1.2.1 Spectrum

Although the CMB photons are red shifted by the expansion of the universe, their spectrum is preserved. Therefore, the fact that the CMB spectrum matches that of the best known black body curve is a strong confirmation of its thermal origin. The intensity of the CMB photons is described by the Planck radiation law given by

$$I_\nu = \frac{2h\nu^3}{c^2} \frac{1}{e^{h\nu/kT} - 1} \tag{1.2}$$

where $I_\nu$ is in units of Watts m$^{-2}$ Hz$^{-1}$ rad$^{-2}$. The Planck blackbody law is fully characterized by its temperature. The CMB is currently at a temperature of $2.725 \pm 0.002$K [40, 102], as measured by the FIRAS instrument aboard the COBE spacecraft in 1989. This temperature corresponds to a peak wavelength of 1.8 mm and frequency of 160 GHz. Figure 1.3 shows the excellent agreement between the measured spectrum and the theoretical blackbody curve at the correct temperature.

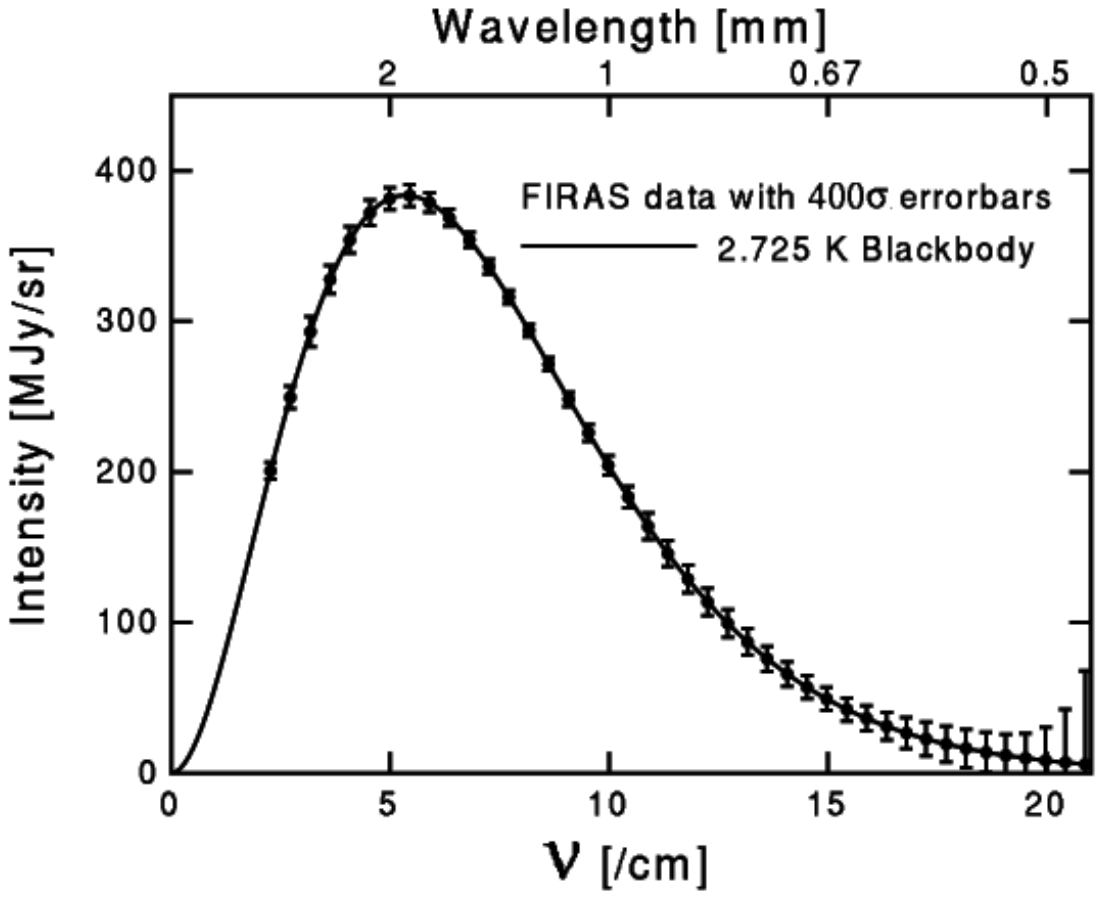

Figure 1.3: Blackbody curve and CMB spectrum measured by FIRAS [40, 102]. The exquisite agreement is an inevitable confirmation of the thermal origin of the CMB. Any alternative theory must confront this direct measurement.

### 1.2.2 Spatial distribution

The CMB radiation was observed to be isotropic until the DMR instrument on the COBE satellite [129, 13, 146, 85] detected intensity fluctuations on the sky. The largest anisotropy signal is the dipole effect with an amplitude of $2.353 \pm 0.024$ mK [14], which is simply the doppler shift of the CMB photons relative to the motion of the earth in the galaxy. The primary temperature anisotropies in the CMB reported by COBE are only $\sim$30 $\mu$K. After COBE's detection of anisotropies, a multitude of ground and balloon experiments continuously improved the angular coverage and precision of the temperature anisotropy measurement. The current best temperature anisotropy measurements will come from the upcoming satellite missions, WMAP and Planck. An up-to-date review of the pre- and post-WMAP experimental data is provided in [18].

The spatial distribution in the observed temperature of the CMB photons reflects the conditions on the surface of last scattering and thus provides a direct probe of the parameters describing the contents ($\Omega_m$, $\Omega_{tot}$, $\Omega_\Lambda$, $\eta$) and the dynamics ($H_0$, $\sigma_8$, $n_s$, $r$) of the universe. This connection is valid for two reasons: First, the CMB photons have free streamed since last scattering with relatively little interaction with other free electrons; second, the formation of the CMB anisotropies happened in a linear regime, allowing cosmologists to make strong predictions about the properties of the observed anisotropies based on relatively simple physics. These two facts make the CMB anisotropy one of the most powerful probes of the early universe. The reader is referred to [66] for an excellent review of the physical effects that relate the condition at LSS to the temperature we measure today.

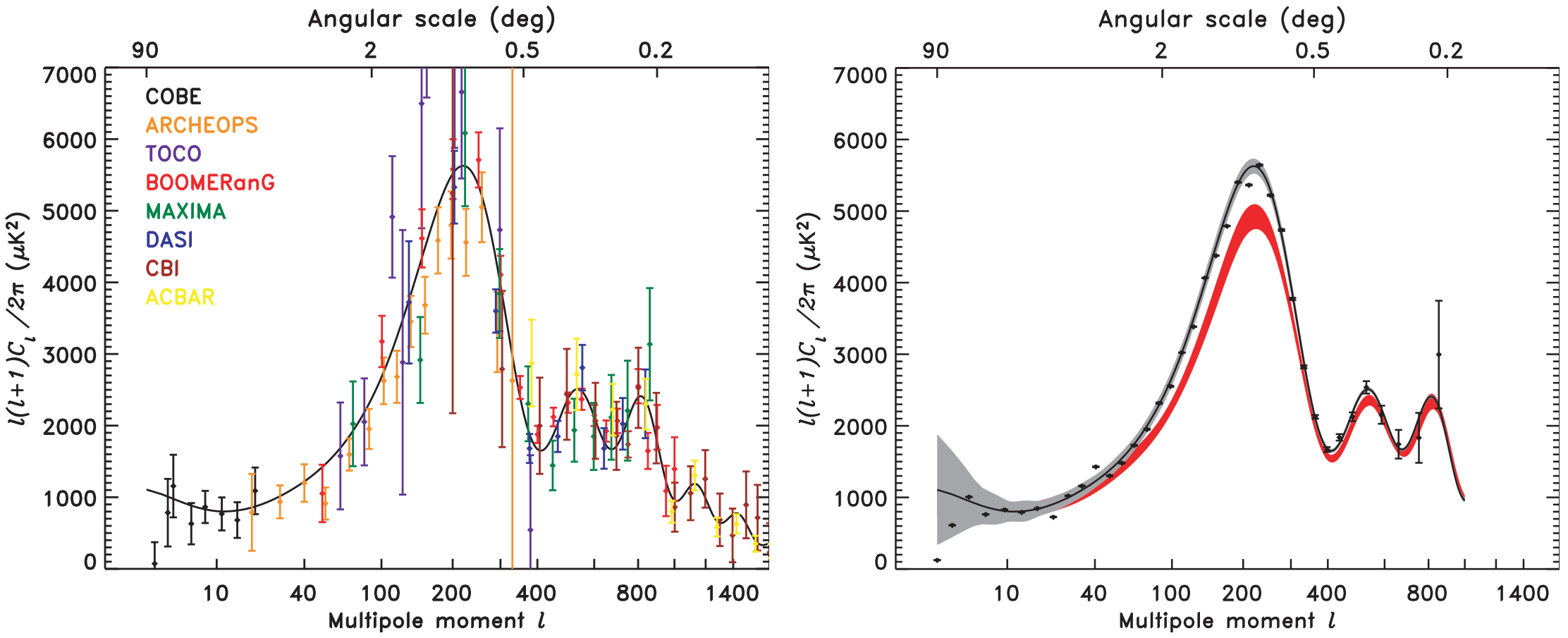

Figure 1.4: From [62]. *left*: Compilation of recent CMB power spectrum measurements (excluding WMAP) compared to the best fit ΛCDM model from the first-year WMAP data. Data points include noise and cosmic variance uncertainty but not calibration uncertainty. *right*: WMAP power spectra (black points) compared to the average spectra of the pre-WMAP experiments (red band). The best fit ΛCDM model is plotted with a $1\sigma$ cosmic variance error band (gray).

The spatial distribution of the CMB temperature is conventionally described in terms of the angular power spectrum $C_\ell$ versus the angular multipole moment $\ell$. The angular power spectrum quantifies the variance in the fluctuations of a certain angular size averaged over the whole sky. The temperature anisotropy is written as

$$\frac{\Delta T(\theta,\phi)}{T_{avg}} = \sum_{\ell=1}^{\infty}\sum_{m=-\ell}^{\ell} a_{\ell m} Y_{\ell m}(\theta,\phi) \ , \tag{1.3}$$

and the angular power spectrum $C_\ell$ is defined as

$$\langle a_{\ell m} a^*_{\ell' m'} \rangle = C_\ell \ \delta_{\ell\ell'} \ \delta_{mm'} \ . \tag{1.4}$$

If the $a_{\ell m}$ are gaussian, then the angular power spectrum encodes all the information contained in the map of the temperatures. Equation 1.4 assumes that $C_\ell$ is obtained by averaging the variance of $a_{\ell m}$ over many realizations of the sky. Because only a single universe is available to measure, the average is instead taken over the $2\ell+1$ independent $m$ modes available for each angular scale $\ell$, yielding

$$\widetilde{C}_\ell = \langle |a_{\ell m}|^2 \rangle = \frac{1}{2\ell+1}\sum_{m=-\ell}^{\ell} |a_{\ell m}|^2 \ . \tag{1.5}$$

This produces an inherent uncertainty in the measurement, called cosmic variance equal to

$$\mathrm{Var}(\widetilde{C}_\ell) = \frac{2}{2\ell+1}\widetilde{C}_\ell^2 \ . \tag{1.6}$$

Figure 1.4 shows an example of such a temperature power spectrum along with many measurements from ground- and balloon-based experiments, and the WMAP satellite. A complete derivation of the formula for $C_\ell$ starting from $\Delta T$ is found in [33]. There are many reviews of CMB anisotropies. A few of my recent favorites are [65, 66, 136, 98], while an up-to-date guide of CMB anisotropy resources is a useful pedagogical starting point[144].

## 1.3 Polarization

The chronological order in the CMB discoveries follows the technological improvements of millimeter-wave receiver sensitivity. The blackbody spectrum was identified first because it only requires measuring a temperature of $\sim$3 K. The detection of temperature anisotropy came 25 years later because the peak of the temperature signal is at a much smaller level ($\sim$80 $\mu$K). At a level of $\sim$5 $\mu$K, the anisotropy in the CMB polarization was just detected in 2002 by the DASI team [89, 96], after an incrementally decreasing list of best upper limits. Figure 1.5 summarizes polarization searches and results. This section provides a heuristic description of the evolution of the density and velocity perturbations that give rise to the temperature and polarization anisotropies.

### 1.3.1 Evolution of Perturbations

In order to appreciate the connection between the observed spatial fluctuations and the properties of the primordial universe, it is useful to give a qualitative (and simplified) overview of the origin and evolution of the perturbations that gave rise to the CMB fluctuations.

The hypothesis of inflation [133, 48, 97, 1, 51] was proposed in the early 1980's to resolve many problems with the standard Big Bang picture at the time. Three of the most prominent problems to which inflation suggests an answer are:

**The horizon problem:** How can the temperature of two causally disconnected regions at the time of recombination be observed to be precisely the same, to better than a part in $10^5$ ?

**The flatness problem:** Why does the measured total density today suggest the seemingly unnatural tuning to unity ($\Omega_{tot} - 1 \leq 10^{-59}$) at Planck time?

**The perturbation problem:** How do you generate larger-than-horizon density perturbations to give rise to the large scale structure visible today?

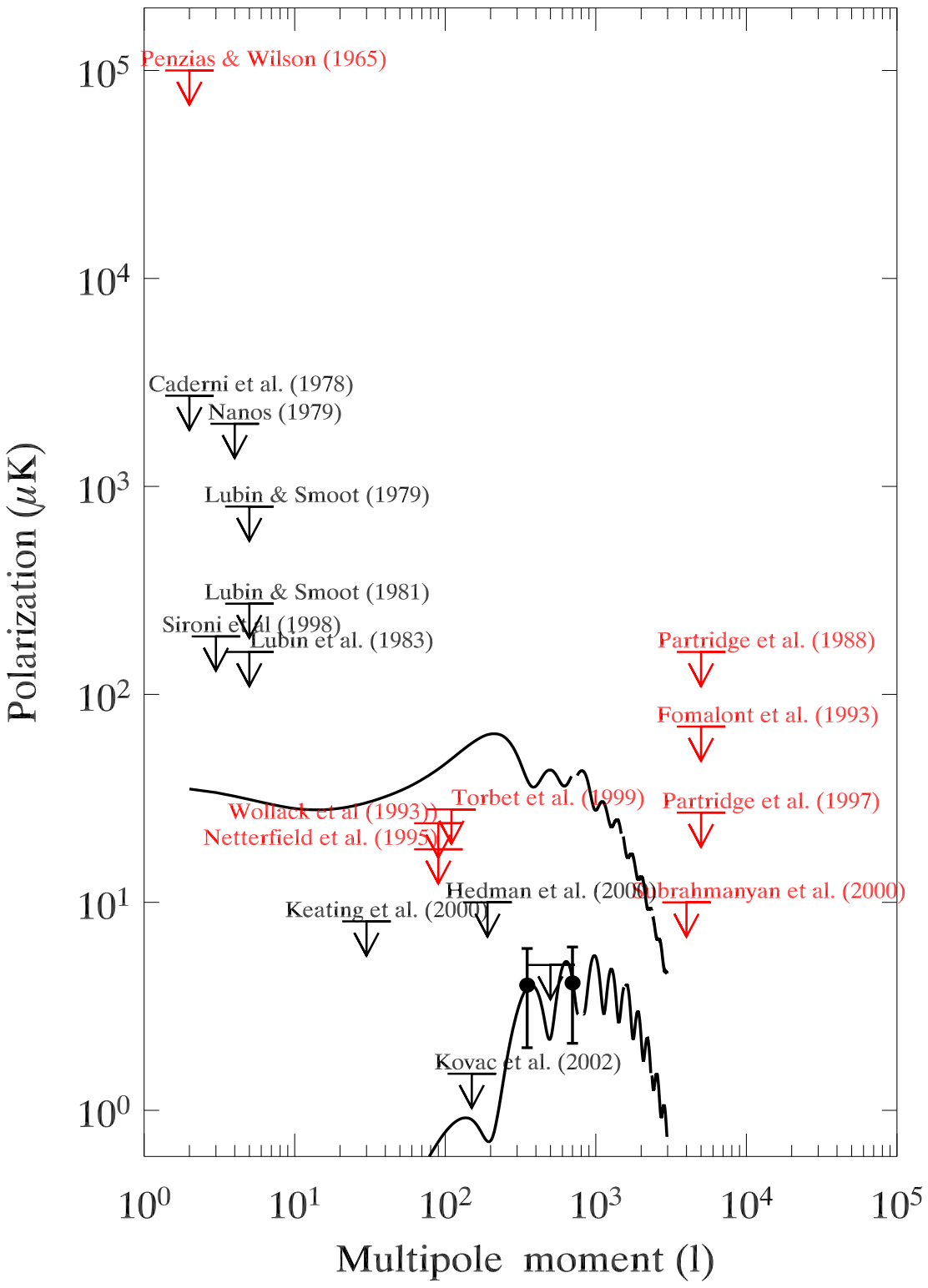

(a) Status of CMB polarization measurements as of March 2004. The curves are the predicted temperature and E-mode power spectra for a Λ-Cold-Dark Matter model as a function of multipole moment $\ell$. The points indicate the level of the upper limit placed with an instrument dedicated to polarization (black) or temperature anisotropy (red). The four lowest points are the DASI [89] detections.

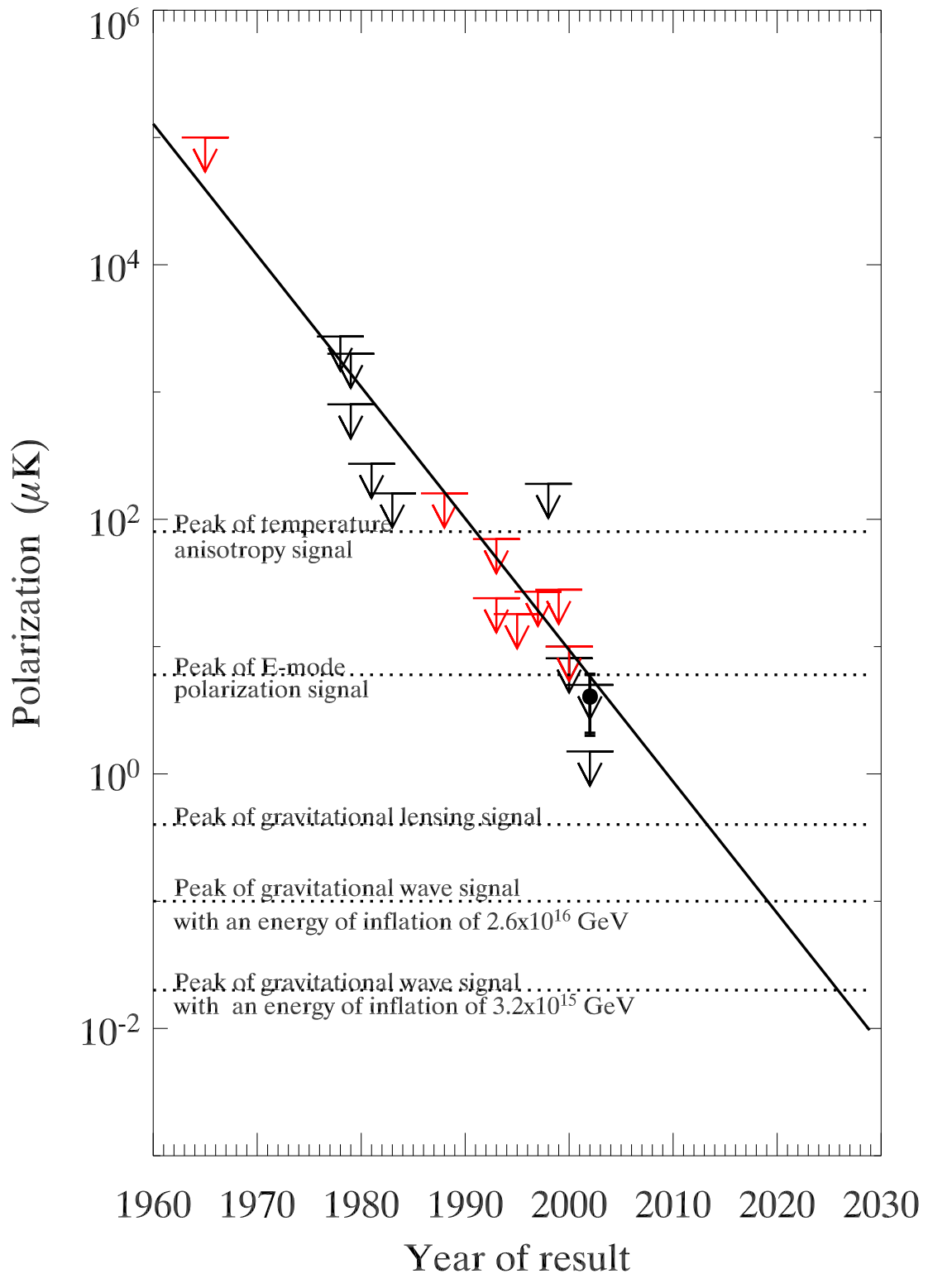

(b) Moore's law of CMB polarization result. A power law where the lowest detectable signal halves every 3 years fits surprisingly well. It predicted a detection of the E-mode polarization in 2002. If this trend continues, the optimistic gravitational wave B-mode signal will not be detected until 2020. This plot neglects all foregrounds which will complicate these measurements.

Figure 1.5: Experimental status of polarization measurements as of March 2004.

The theory of inflation provides a scenario that resolves these riddles. An excellent recent review of the inflation scenario is presented in [99]. According to a simplified version of this scenario, the early universe (t $< 10^{-30}$) underwent a period of exponential, superluminal expansion during which the quantum fluctuations of the scalar field which drove this expansion grew in size to become larger than the horizon. These inflated quantum fluctuations became the seeds for density perturbations to grow with respect to the homogenous universe.

To understand what happens to these initial density perturbations after inflation but before recombination, it is useful to think of the fluctuation at different scales in terms of different Fourier modes each evolving independently. This assumption is warranted because the density perturbations are small and evolve linearly. The evolution of each mode then depends on the wavelength of that mode $\lambda(t)$ compared to the size of horizon at that time. The horizon is the distance a photon could have travelled since the Big Bang and simply grows as $t$. The wavelength of each mode grows as the scale factor $a(t)$, that is as $t^{1/2}$ or $t^{2/3}$ for radiation- or matter-dominated universe. As long as $\lambda(t)$ is much greater than the horizon scale, the mode remains "outside" of the horizon and does not evolve because the region is causally disconnected. When the universe has grown sufficiently that the horizon scale exceeds the wavelength of a mode, that mode "re-enters the horizon". That region of space comes in to causal contact. What happens to that region then depends on the forces that start acting on it.

Prior to recombination, the universe is filled with a hot "soup" of electrons, protons, neutrinos, light nuclei, cold dark matter, and mostly photons, referred to as the baryon-photon fluid. The baryon-photon fluid is tightly coupled by frequent Thompson scattering of photons on free electrons. There are effectively two forces acting on this fluid. The baryons gravitationally attract each other, while the radiation pressure of the compressed photons resists the gravitational infall. The result is to set up an acoustic oscillation in the baryon-photon fluid of a region that has come into causal contact. At recombination, the modes that are captured at maximum compression and rarefication display a characteristic peak in the observed CMB temperature power spectrum (Figure 1.4). Because the largest modes enter the horizon last, the first peak in the CMB angular power spectrum correspond to the mode that had just enough time to reach maximum compression before recombination. Figure 1.6 provides an illustration of the acoustic oscillation process. A good overview of the physical effects which translate the conditions of the baryon-photons fluid at decoupling into the temperature measured by experiments is provided in [134].

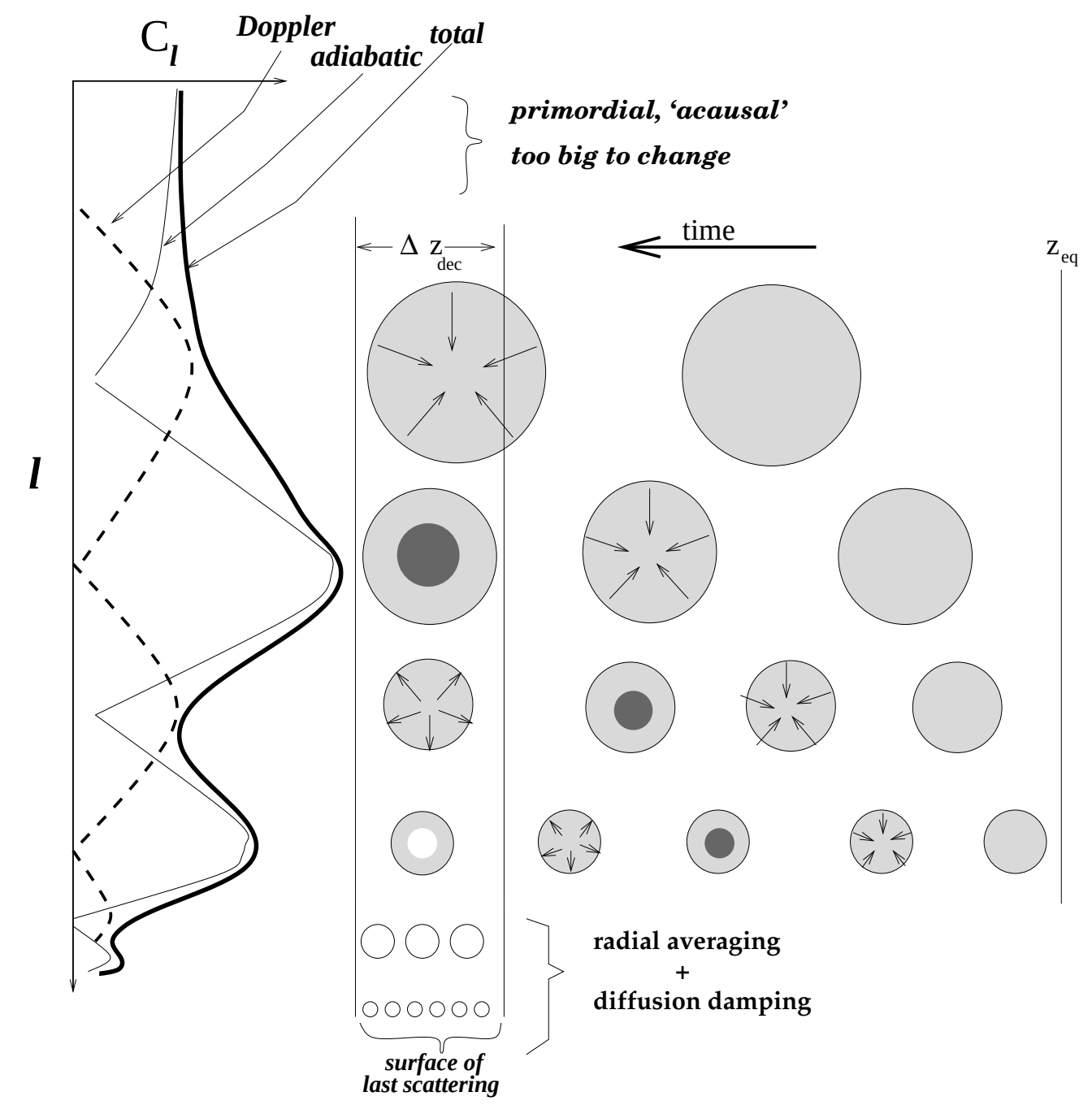

Figure 1.6: Figure by [99]. Acoustic oscillations of different modes in the baryon-photon fluid generate characteristic peaks in the angular power spectrum. The observer is on the left. The gray spots are initial over-densities seeded by inflation. The smallest scales have re-entered the horizon earliest so they have a longer time before recombination and go through more oscillations. The first peak in $C_\ell$ corresponds to the mode which has just reached its maximum compression at recombination. The second corresponds to the mode which has just reached maximum rarefication. The peaks in the polarization spectrum correspond to a phase when the fluid reaches maximum velocity, while the peaks in the temperature spectrum correspond to a phase of extremum density. As a result, the temperature and polarization angular power spectra are 90°out-of-phase.

### 1.3.2 Origin of Polarization

Polarization of the CMB is naturally created by the Thompson scattering of photons off electrons at the moment of decoupling. The cross-section of radiation scattered by a charged particle of charge $e$ and mass $m$ is given by [69]

$$\frac{d\sigma}{d\Omega} = \left(\frac{e^2}{mc^2}\right)^2 |\hat{\epsilon} \cdot \hat{\epsilon}'|^2 \ , \tag{1.7}$$

where $\hat{\epsilon}$ and $\hat{\epsilon}'$ are the unit vectors indicating the linear polarization of the incident and scattered photon[3]. Equation 1.7 predicts that only a quadrupole distribution of initially unpolarized radiation will generate a linearly scattered polarization. Figure 1.7 illustrates the scattering mechanism for a quadrupolar radiation incident on a single electron.

[3]Equation 1.7 is usually written in terms of the total Thompson cross-section $\sigma_T = \frac{8\pi}{3}\left(\frac{e^2}{mc^2}\right)^2$ equal to $0.665 \times 10^{-24}$ cm$^2$ for electrons.

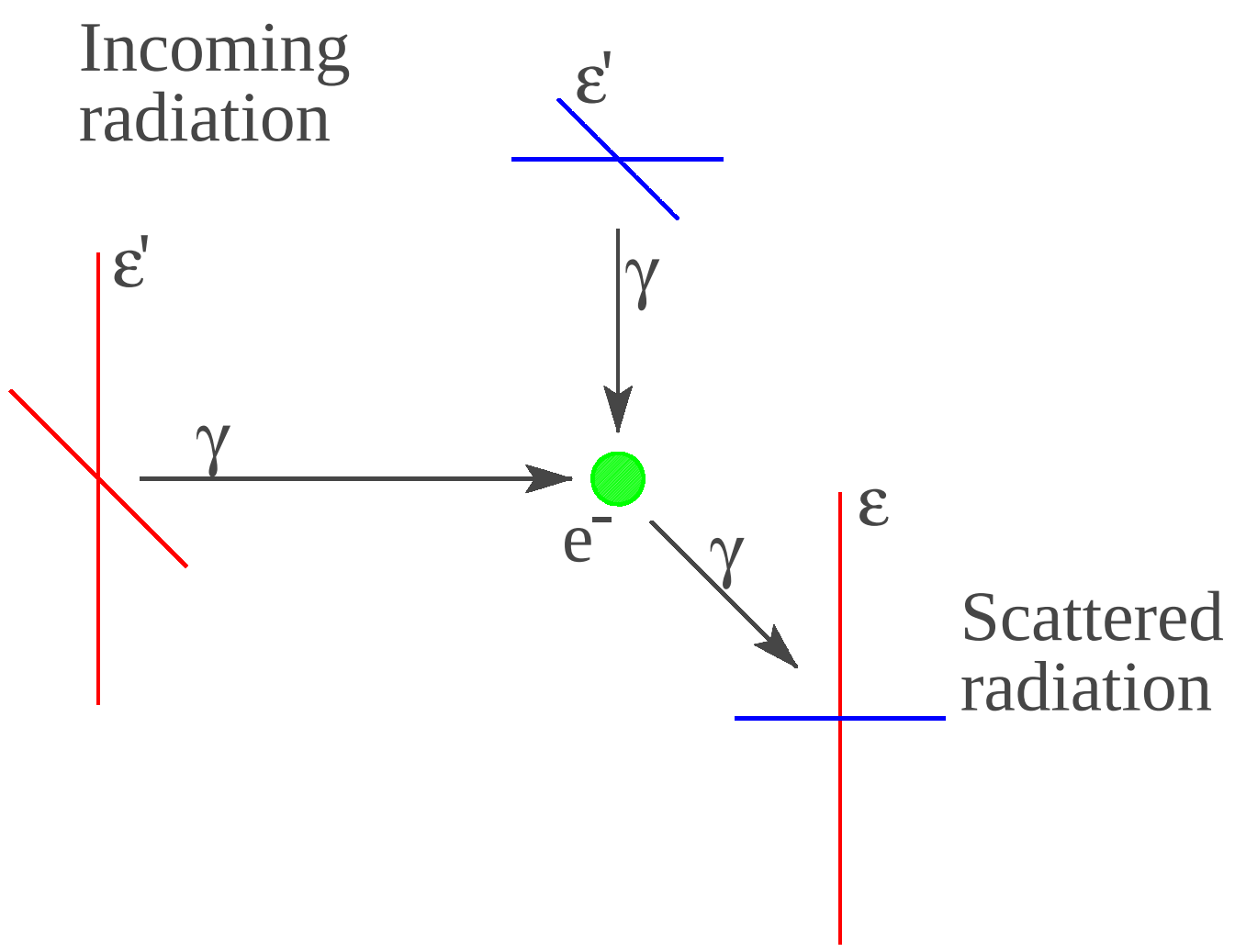

Figure 1.7: A quadrupolar intensity distribution incident on an electron of the primordial plasma generates a linearly polarized scattered photon. The incoming radiation is not polarized. The size of the lines and their color are redundant and indicate their intensity. This effect is a result of the transverse nature of light. Only the component of the incident radiation polarized in a direction perpendicular to the scattered direction is picked up by the scattered photon. A monopole or dipole distribution of the incoming radiation does not generate a polarized photon.

During the epoch of recombination, the electrons can produce a net linear polarization because of local quadrupole moments generated by doppler-shifted radiation. The doppler-shift is the result of local velocity fields in the baryon-photon plasma. Let's consider a single spatial mode of the density field to illustrate this effect. The density fluctuations of that mode are shown in Figure 1.8(a). Assuming the mode is growing so the fluid is moving towards the density crest, we look at the behavior of the fluid near a crest (zoom in Figure 1.8(a)). An electron on the crest will see radiation doppler-boosted perpendicularly from the crest but not along the crest. Using Figure 1.7 as reference, the scattered radiation will be polarized parallel to the crest and perpendicular to the wave vector. Similarly, a photon in a trough sees a radiation with a reduced intensity perpendicular to the trough and uniform along it, leading to a polarized pattern perpendicular to the trough. If the amplitude of the density perturbation is decreasing, the direction will be rotated by 90°. Finally, if the density perturbation is at an extremum, the fluid velocity will be zero leading to a minimum in the polarization pattern. In general, when the scattered polarization direction is either parallel or perpendicular to the wave vector of the perturbation, the polarization pattern is referred to as a curl-free, or an E-mode pattern [152, 153, 67]. When the polarization direction is at ±45° to the wave vector, as in Figure 1.8(b), the polarization pattern is called a B-mode. The decomposition of a polarization field into a sum of E- and B-modes has become the standard of analyzing polarization maps. Note that a B-mode pattern cannot be created by density perturbations and its presence would be a distinctive signature of primordial gravitational waves from inflation [127, 126, 77].

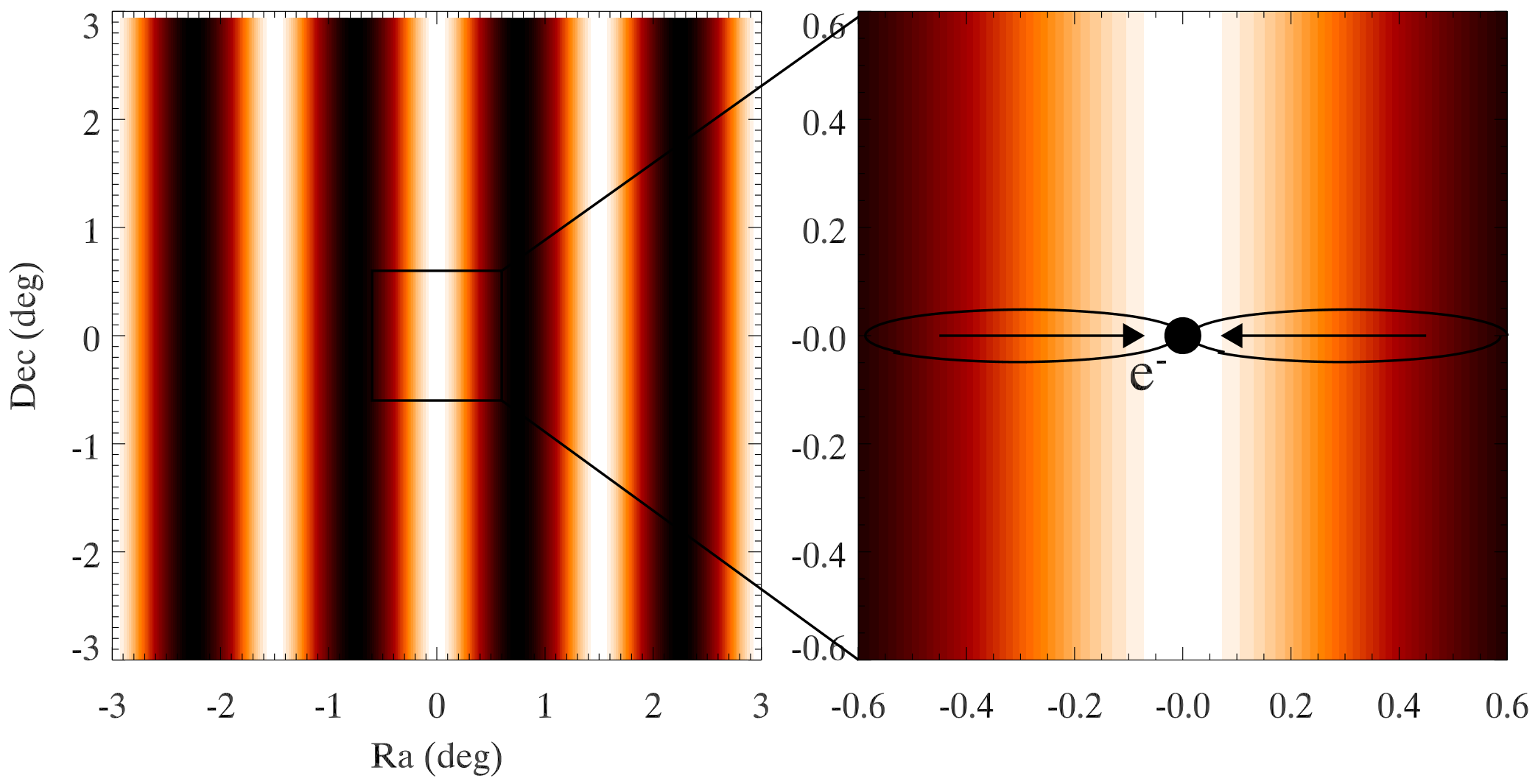

(a) A single Fourier mode of a density perturbation on the sky. The mode is shown growing with white area representing a crest of high density. The zoom of the central region shows that an electron located on a crest will see radiation doppler-boosted perpendicularly to the crest. The scattered radiation will therefore be linearly polarized parallel to the crest. Conversely, an electron located in a trough will emit linearly polarized radiation perpendicular to the trough. The velocity distribution in the fluid will generate the E-mode polarization pattern shown in the figure below.

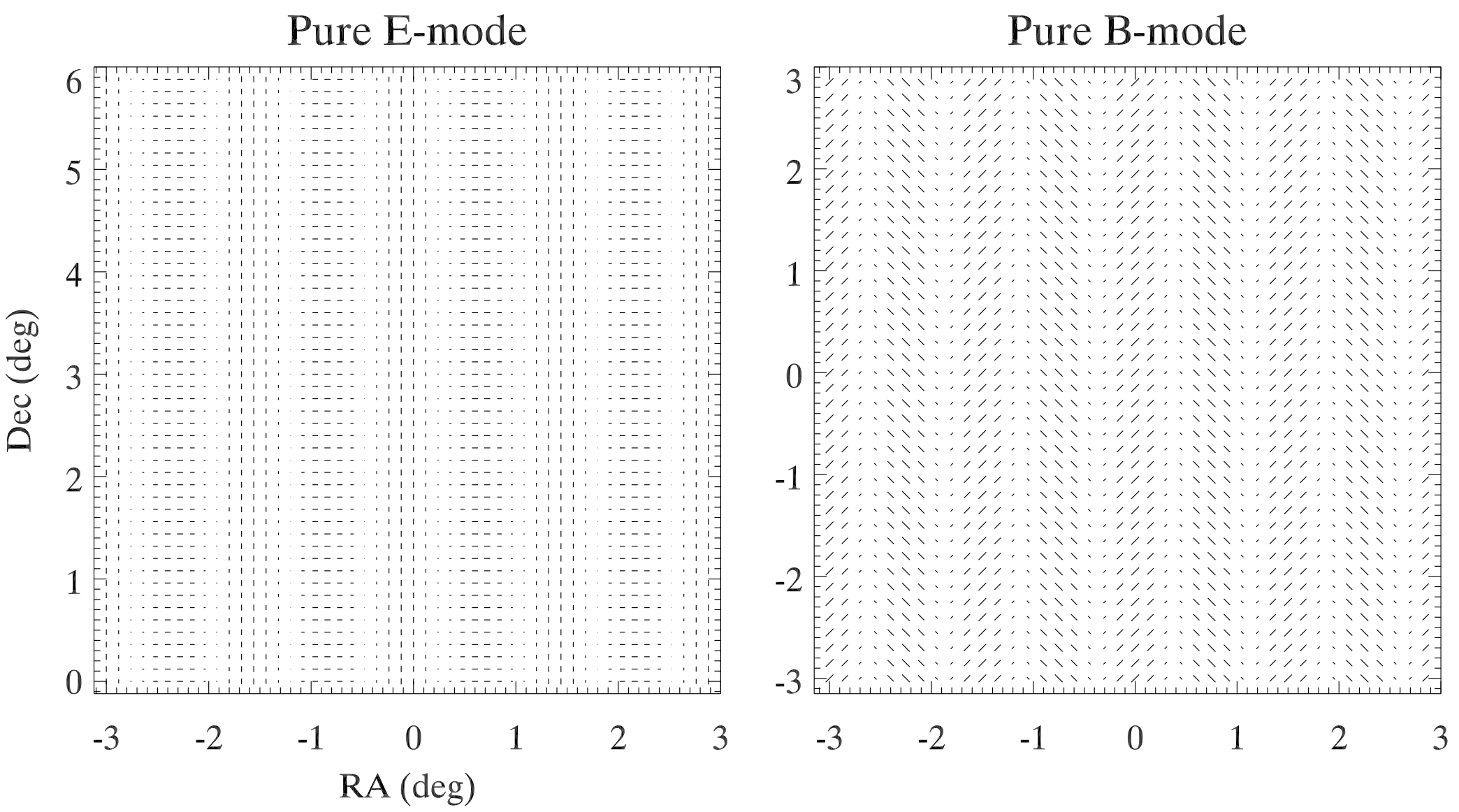

(b) *left*: E-mode polarization pattern resulting from the Thompson scattering of the density perturbation mode above.*right*: B-mode polarization pattern. This figure is for illustrative purpose only as a B-mode polarization pattern cannot be created by a scalar perturbation and is only generated by tensor perturbations.

Figure 1.8: E and B mode polarization pattern from a single Fourier mode density perturbation.

## 1.4 Foregrounds

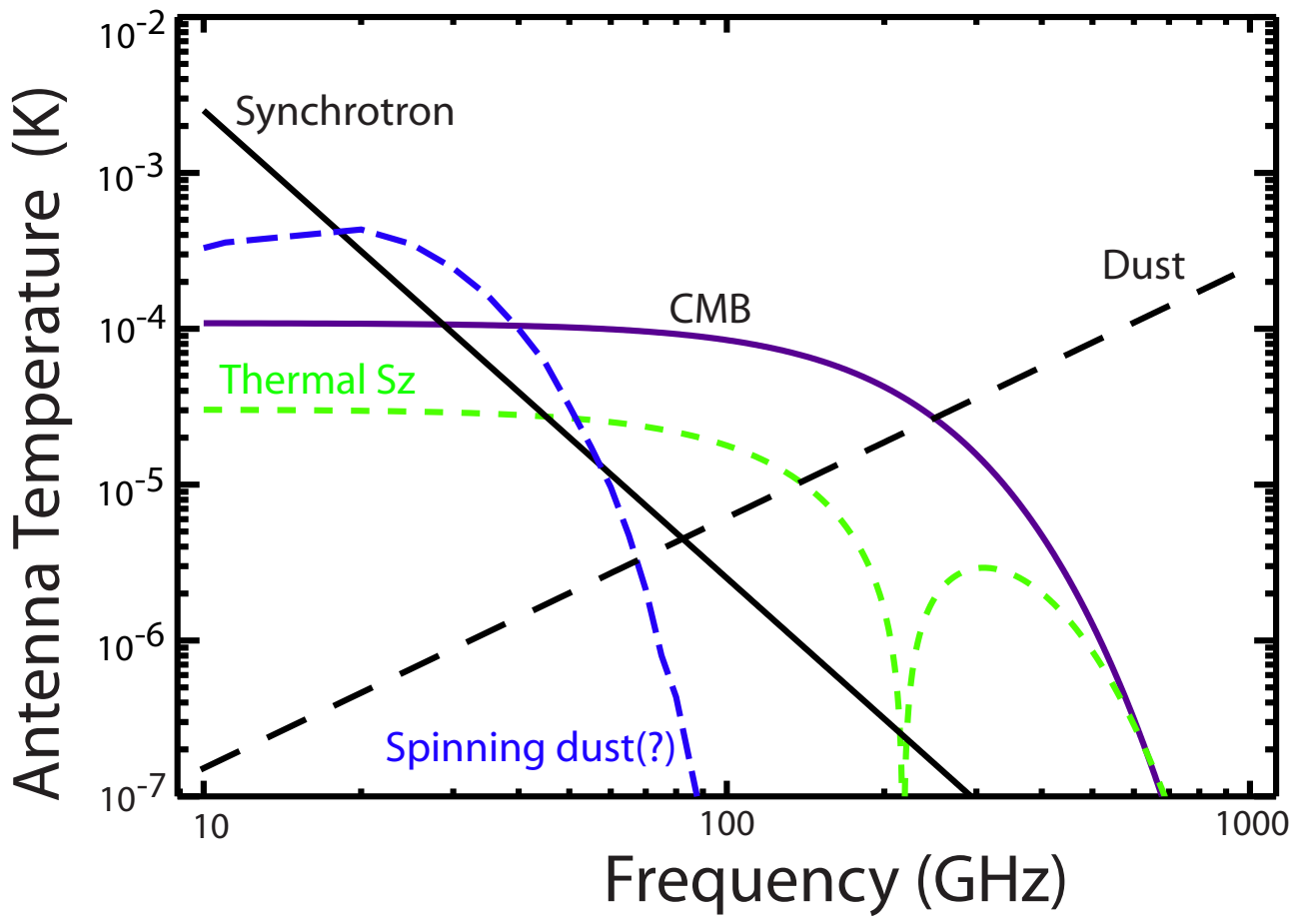

Figure 1.9: Figure by N. Ponthieu [121]. Spectrum of the various foreground emissions for angular scales of 8′ at the galactic anti-center. Although this plot is only for unpolarized emission, it shows that there is a window between 50 and 100 GHz where the CMB dominates the foreground emissions. The emission is also very position dependant. Synchrotron emission is thought to have the largest polarized contribution.

Foreground radiation is the potential contamination of the cosmological signal by microwave-emitting junk other than CMB. The principal foreground sources are dust, free-free, and synchrotron emission from galactic and extragalactic sources. The spectral dependance of these foreground sources is illustrated in Figure 1.9. Note that the foreground emission is brighter than the CMB at certain frequencies. *Synchrotron* is the radiation emitted by electrons when they spiral in the magnetic field of a galaxy. *Free-free* (or bremsstrahlung) represent the radiation emitted by a charged particle when it decelerates from the interaction with another particle. Finally, *dust* in the interstellar medium can emit a thermal radiation because it is at a non-zero temperature. An important aspect of these foregrounds is that they are polarized and can interfere with CMB polarization measurements. Although the intensity distribution of these various emission mechanism is well studied, practically nothing is known about their polarization properties. We provide, as a qualitative guide of the strength of the foregrounds signal near NCP, the WMAP reconstructed maps for the synchrotron, free-free, and dust emission at 90 GHz (Figure 1.10).

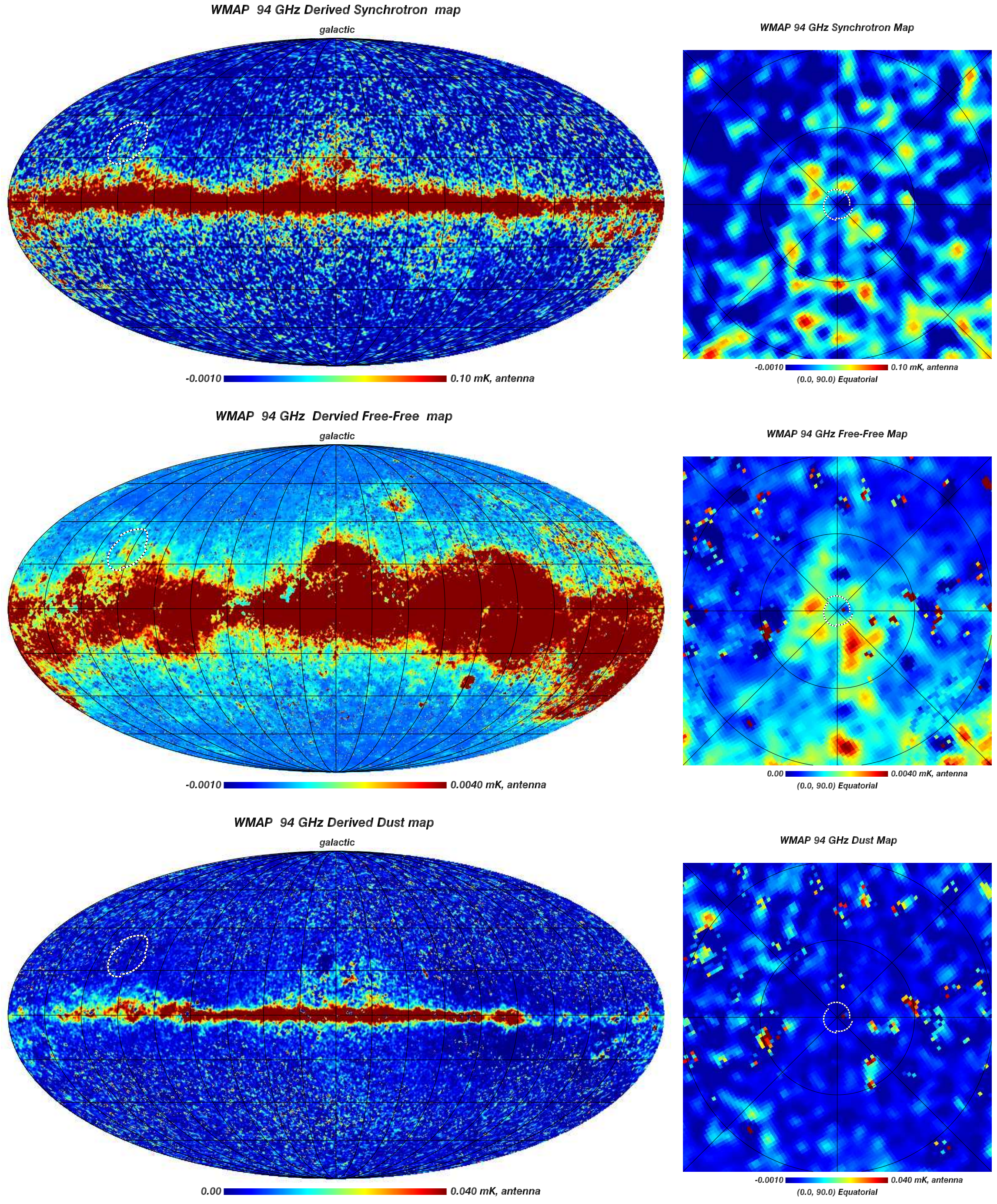

Figure 1.10: *left*: WMAP Maximum Entropy Method derived foreground full sky galactic maps [15] for synchrotron (top), free-free (middle), and dust (bottom) emission. The white circle in the Mollweide full sky maps is a $10^\circ$ radius centered on the NCP. The NCP is located at galactic coordinates of $(l, b) = (123^\circ, 27^\circ)$ with the galactic center located at $(0^\circ,\ 0^\circ)$. The prior foreground observations used for these models are available at `http://lambda.gsfc.nasa.gov/product/map`. *right*: The gnomic projections show a region $20^\circ$ on a side, where the central white circle represents the CAPMAP scanned region. The unpolarized temperature contribution from each of these foregrounds averaged in the CAPMAP region is $3.8 \pm 1$ $\mu$K for synchrotron, $2.1 \pm 0.1$ $\mu$K for free-free, and $30 \pm 11$ $\mu$K for dust. The polarized fraction is not measured yet but is expected to be 10 to 300 times smaller.

## 1.5 Personal Contributions

The complexity of CAPMAP is such that it can only be the fruit of a collaboration of many people. However, the instrument is still of a relatively small scale such that many sub-systems can be completely designed, developed, and tested by a single person. Throughout this dissertation, most efforts are attributed to the ubiquitous "we" in reference to the CAPMAP collaboration, whose members are listed below.

My personal effort on the experiment was concentrated in three areas: I was the principal graduate student responsible for integrating and testing the receivers in the first CAPMAP cryostat. I designed and assembled the first cryostat and associated sub-systems (windows, internal structure, wiring). I tested many of the receiver components individually.

I coordinated the refurbishment effort and optical testing of the 7-meter Crawford Hill antenna, which included repairing many of the broken electrical circuits and key parts of the antenna and taking care of the overall maintenance of the antenna. I designed and implemented the tertiary mirror with which we first test the antenna's beam properties. I also started the use of GRASP8 to simulate the antenna beam pattern.

I was a core member of the CAPMAP03 and CAPMAP04 field campaigns (although less during the CAPMAP04 due to the writing of this dissertation) which each involved four months of non-stop action. I was also responsible for a sub-set of the CAPMAP03 data analysis (see Chapter 6 and 7).

**CAPMAP collaboration**

- Denis Barkats (Graduate student), Princeton University, NJ, USA
- Colin Bishoff (Graduate student), University of Chicago, IL, USA
- Phil Farese (Post-doctoral fellow), Princeton University
- Josh Gundersen (Faculty), University of Miami, FL, USA
- Matt Hedman (Post-doctoral fellow), University of Chicago
- Lewis Hyatt (Graduate student), Princeton University
- Jeff McMahon (Graduate student), Princeton University
- Dorothea Samtleben (Post-doctoral fellow), University of Chicago
- Suzanne Staggs (Faculty) , Princeton University
- Eugenia Stefanescu (Graduate student), University of Miami
- Keith Vanderlinde (Graduate student), University of Chicago
- Bruce Winstein (Faculty), University of Chicago

## Chapter 2

# Instrument Overview

CAPMAP's scientific mission is to make a strong detection of the CMB polarization power spectrum at angular scales where it is expected to peak ($\theta \sim 10'$, $\ell \sim 1000$). Although temperature anisotropy measurements are well-established, making a reliable measurement of the CMB polarization anisotropy is still an experimental challenge.

The instrument must be built specifically to (1) have the statistical sensitivity to detect the faint signal, (2) have a small and clean enough beam to probe arcminute angular scales, (3) control systematic effects at least down to the level of statistical errors, and (4) be minimally contaminated by signals other than the polarized CMB. This chapter provides an overview of how these general guidelines translate into specific experimental choices regarding the receiver, optics, focal plane layout, and observing strategy. The details of the polarimeter and its characterization are presented in Chapter 3. Although some attention is given to the optical design in this chapter, a detailed treatment of the horn and lens design will be presented in a future thesis [107]. Chapter 4 is devoted to the description of the telescope optical, mechanical, and electrical systems. The features of the first observing season (CAPMAP03) are presented in Chapter 5, along with specific characteristics of the observing site. The last two chapters deal with the data obtained during the 2002-2003 winter season. Chapter 6 presents the calibration data while Chapter 7 glances at how the main CMB data is processed into final maps useable for power spectrum estimation.

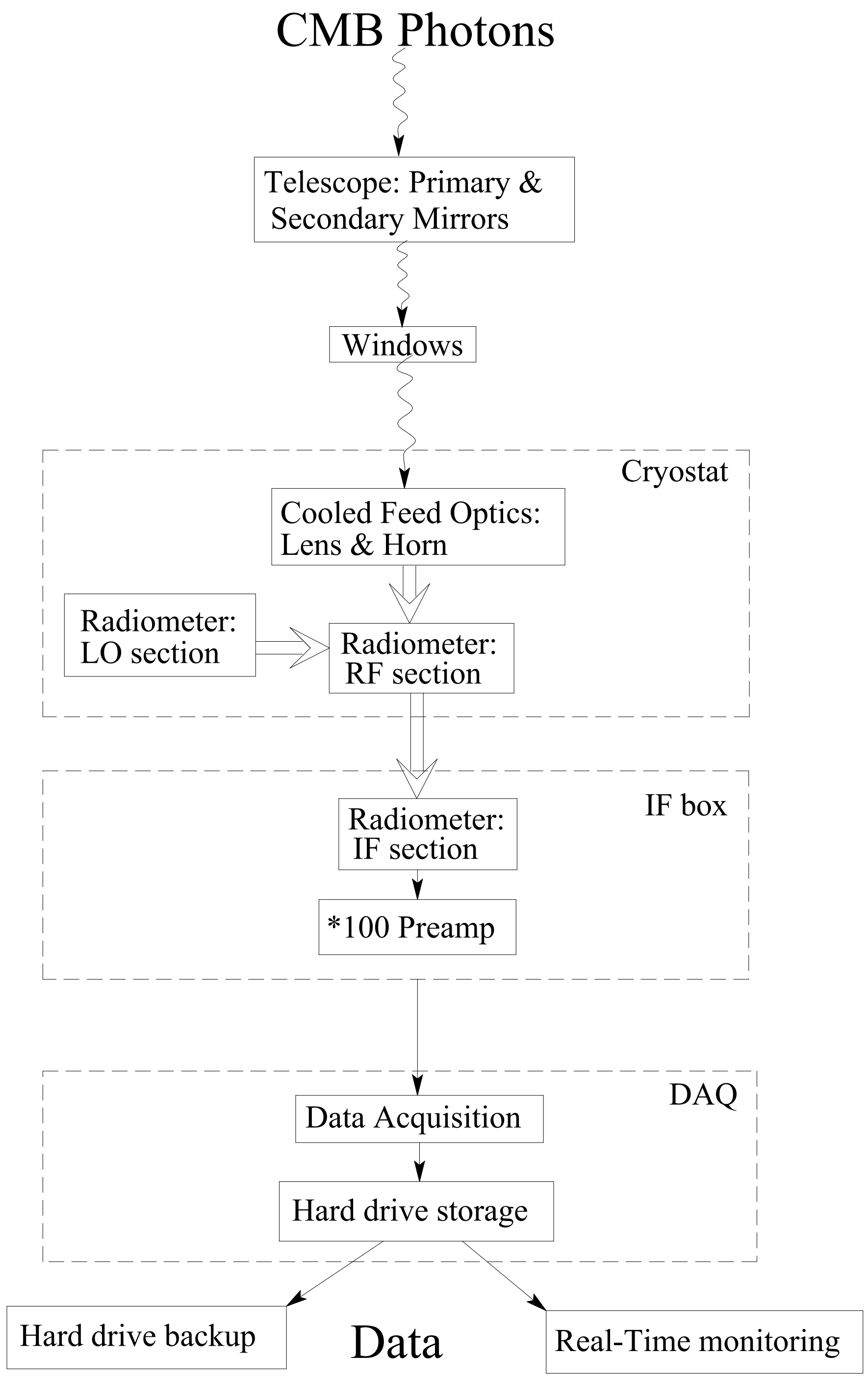

Figure 2.1: Block diagram of photon path. The signal is transmitted via three different methods: free space propagation (squiggly line), waveguide or coaxial transmission line (double line), and analog or digital signal (straight line).

## 2.1 Design Parameters

The full CAPMAP instrument is a 16-element array of phase-switched, heterodyne, correlation polarimeters using cooled MMIC HEMT as front-end amplifiers. To allow for discrimination of possible foregrounds, twelve of the receivers are at W-band (84-100 GHz) and four at Q-band (35-45 GHz). Each receiver observes a single Stokes parameter. The receivers are coupled to the sky via the 7-meter Crawford Hill antenna, in Holmdel, NJ. The narrow beam needed to illuminate the antenna is produced by a profiled corrugated horn feeding into a high-density polyethylene (HDPE) lens which further collimates the beam. The feed optics are cooled. The combination of feed and telescope optics produces a $4'$ and $6'$ beam for the W- and Q-band receivers respectively. The observing strategy is to scan the telescope in azimuth across the North Celestial Pole (NCP) at a fixed elevation. The Earth's rotation makes the scan cover a small cap ($< 2°$) centered on the NCP.

The CAPMAP instrument is designed to be fielded in a staged deployment. In the winter 2002-2003, four W-band receivers in a single dewar made observations. The purpose of the first observing season is to investigate and improve any unforseen systematic effects for the second year, but also to demonstrate the viability of the new experiment. The full 16 receivers, distributed into four dewars will be deployed for the the winters 2003-2004 and 2004-2005. The rest of this thesis will focus on the details of the first version of the instrument, hereafter CAPMAP03. However, many choices were made so the first year instrument could be used with minimal changes for the next observing season.

Table 2.1 lists the performance parameters for W- and Q-band CAPMAP receivers as well as a PIQUE [60, 59] receiver for comparison. The senfac parameter, $\sigma_E$, estimates the experimental weight for a receiver (or an array of receivers) after a season of observation from the system noise, bandwidth, and expected integration time using Equation 2.12. If all the data were collected from a single pixel, the error for that pixel would be, $\sigma_E$. If instead the integration time is spread evenly over $N_{pix}$ pixels, the error for each pixel is $\sigma_E \times \sqrt{N_{pix}}$. For CAPMAP03, we had four receivers with a sensitivity of $\mathcal{S} = 1000\ \mu\text{K}\sqrt{s}$ each observing for 250 hours ($9\times10^5$ sec), then $\sigma_E = 1000/\sqrt{9 \times 10^5 \times 4} \approx 0.5\ \mu$K. It is useful to keep in mind that many factors tend to decrease the signal or increase the expected noise (tighter cuts, atmospheric absorption of signal, removal of scan synchronous signal). In PIQUE, for example, the final measured $\sigma_{pix}$ exceeded the predicted one by a factor of 1.5. We use this factor in Table 2.1 to estimate a realistic $\sigma_{pix}$ for CAPMAP.

| Characteristic | PIQUE[a] (W-band) | CAPMAP[b] (W-band) | CAPMAP[b] (Q-band) | CAPMAP03 first year | CAPMAP04[g] full system |
|---|---|---|---|---|---|
| $T_{sys}$ | 140 K | 120 K | 40 K | - | - |
| $\Delta\nu$ | 11 GHz | 14 GHz | 8 GHz | - | - |
| $t_{int}$ | 250 hrs | 250 hrs | 250 hrs | 250 hrs | 500 hrs |
| N modes[c] | 1/2 | 17/20 | 17/20 | - | - |
| Beam FWHM | $12'$ | $4'$ | $6'$ | - | - |
| ${N_{pix}}^d$ | 24 | 300 | 170 | 600 | 1300 |
| Ideal senfac, ${\sigma_E}^e$ | 1.9 $\mu$K | 1.15 $\mu$K | 0.51 $\mu$K | 0.57 $\mu$K | 0.2 $\mu$K |
| Ideal ${\sigma_{pix}}^e$ | 9.3 $\mu$K | 19.9 $\mu$K | 6.64 $\mu$K | 13.9 $\mu$K | 7.2 $\mu$K |
| Realistic ${\sigma_{pix}}^f$ | 14.6 $\mu$K | 39.2 $\mu$K | 9.9 | 19.1 $\mu$K | 10.8 $\mu$K |
| $C_\ell$ peak | 2 $\mu$K | 6 $\mu$K | 6 $\mu$K | 6 $\mu$K | 6 $\mu$K |

Table 2.1: CAPMAP single receiver sensitivity. All numbers are in antenna temperature. To convert the antenna to thermodynamic temperature, multiply by $(e^x - 1)^2/(e^x x^2) = 1.251$ at W band and 1.044 at Q band.
[a]The PIQUE performance numbers are as measured during observation [56].
[b]The CAPMAP receiver performance numbers are conservative estimates (see Chapter 3 for details).
[c]PIQUE measured the difference between two points across the NCP and thus lost one out of two measured numbers. CAPMAP measures 20 points across an arc and loses 3 modes by subtracting the mean, slope, quadratic from each scan.
[d]Number of pixels observed, $N_{pix} = (\pi r^2)/\theta^2_{FWHM}$. Note that each receiver of CAPMAP03 observes a slightly different number of pixels (A =332, B=580, C=324, D=476)
[e]Ideal senfac and $\sigma_{pix}$ are calculated from the receiver performance using $\sigma_E = T_{sys}/\sqrt{\Delta\nu\ t_{int}\ N_{modes}}$ and $\sigma_{pix} = \sigma_E \times \sqrt{N_{pix}}$. See text for details.
[f]The realistic senfac is derived by calculating the senfac from the actual PIQUE data set, which accounts for all non-idealnesses. The CAPMAP realistic senfac is then scaled from the PIQUE one with the appropriate improvement ratio from the ideal senfac.
[g]The column gives the senfac for two conservative seasons of observing with all 16 receivers (12W and 4Q).

Figure 2.2 gives a graphical representation of the angular resolution and frequency coverage of most of the polarization experiments of the last several years. As expected, experiments have been designed to cover the the region where the CMB polarization signal is the strongest and the contamination from foregrounds is the smallest. Note that CAPMAP is the only instrument probing the high angular scales around 90 GHz with coherent receivers. Table 2.2 is compilation of the most recent polarization experiments.

CAPMAP is a second generation experiment. It takes advantage of microwave-polarimetry techniques and know-how learned from its precursor experiment, PIQUE. Like PIQUE, CAPMAP uses a phase-switched heterodyne correlation polarimeter. CAPMAP operates at the same frequencies, 90 and 40 GHz and is tested and calibrated in similar ways. Although the receiver design is similar to PIQUE, many parts of the instrument are completely new (cryostat, microwave amplifiers, optics, data acquisition chain, telescope) and therefore require people and time to integrate, test, and troubleshoot. To improve sensitivity, CAPMAP uses newly available low noise microwave amplifiers (MMIC HEMT[1] as opposed to discrete HEMT) cooled to 20 K. To overcome the noise limitation of a single receiver, CAPMAP combines the data from many receivers observing at once. The noise of the array of $N$ detectors scales as $\Delta T_{min} \propto 1/\sqrt{N}$. New calibration and integration procedures were devised to be able to field this large array in so little time (see Section 3.3).

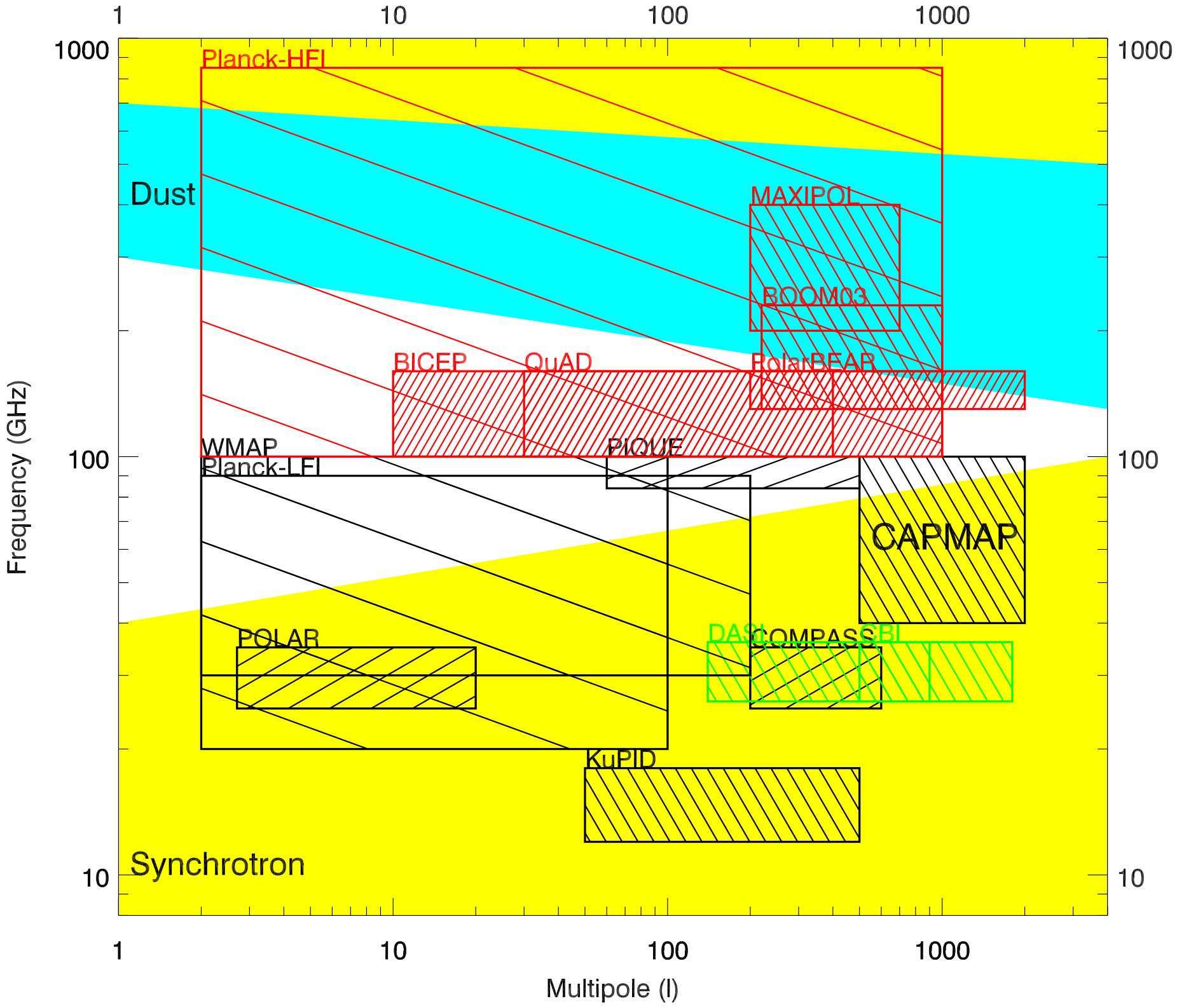

Figure 2.2: Angular multipole vs frequency coverage of past, current, and planned polarization experiments. Descriptive parameters of these experiments are listed in Table 2.2. Experiments are distinguished by detection techniques: black (correlation polarimeter), red (bolometer), green (interferometers). The experimental coverage is overlaid on top of a "middle-of-the-road" foreground prediction from [135].

CAPMAP is expected to have enough sensitivity to detect E-modes. The E-mode $C_\ell$ error is estimated using the approximation [70, 84, 78]:

$$\sigma(C_\ell^E, l) = \sqrt{\frac{2}{(2l+1)f_{sky}}}\left(C_\ell^E + f_{sky}\frac{4\pi\mathcal{S}^2}{t_{int}}e^{l^2\sigma_b^2}\right) , \tag{2.1}$$

where $f_{sky}$ is the fraction of sky observed, $\mathcal{S}$ is the instrument sensitivity in $\mu\text{K}\sqrt{s}$, $t_{int}$ is the total integration time in seconds, related to the integration time per pixel $t_{pix} = t_{int}/N_{pix}$, $\sigma_b = \theta_{fwhm}/\sqrt{8\ln 2}$ is the beam sigma in radians, and $\theta_{fwhm}$ is the beam FWHM

[1]provided by Todd Gaier at JPL.

listed in Table 2.1. Using conservative estimates for the performance of CAPMAP03, $\mathcal{S} = 1000\ \mu\text{K}\sqrt{s}$, $t_{int} = 250$ hrs, $\theta_{fwhm} = 0.06°$, and $\text{f}_{sky} = 4.5\times10^{-5}$, we obtain the $C_\ell$ band power predictions shown in Figure 2.3.

| Experiment | Frequency (GHz) (#) | Beam FWHM | Site | Detection technique | Date | Ref. |
|---|---|---|---|---|---|---|
| | | | PAST | | | |
| POLAR | 30(1) | $7°$ | WI (43°N) | CP[a], axial spin | -2001 | [80] |
| COMPASS | 30(1) | $20'$ | WI (43°N) | CP, NCP az. scan | -2002 | [36] |
| PIQUE | 40(1), 90(1) | $30'$ $15'$ | NJ (40°N) | CP, NCP az. chop | -2001 | [56] |
| Polatron | 90(1) | $2.5'$ | CA (37°N) | B[b] + 1/2 wave plate | - | [120] |
| VLA | 8.4(1) | $6'\ '$ | NM (34°N) | I[c] | - | [117] |
| | | | ON-GOING | | | |
| CAPMAP | 40(4), 90 12) | $6'$ , $4'$ | NJ (40°N) | CP array, NCP az. scan | 2003- | |
| KuPID | 15(1) | $12'$ | NJ (40°N) | CP, NCP az. scan | 2004- | [47] |
| CBI | 30(13) | $4.5'$ | Chile (23°S) | I | 2001- | [21] |
| DASI | 30(13) | $20'$ | South Pole | I | 2001- | [96] |
| BOOM03 | 145(8), 245(4), 345(4) | $9'$ , $7'$ | South Pole | PSB[d], LDB[f] | 2003 | [112] |
| Maxipol | 140(12), 440(4) | $10'$ | US balloon | B + 1/2 wave plate | 2002- | [73] |
| WMAP | 22, 30, 40(2), 60(2), 90(4) | $13'$ | Space, L2 | pseudo CP, full sky | 2001- | [86] |
| | | | PLANNED | | | |
| QUIET | 90 | $4'$ | Chile | Array of CP | 2006 | [43] |
| QUaD | 100(12), 150(19) | $6.3'$ , $4.2'$ | South Pole | PSB + 1/2 wave plate | 2005 | [23] |
| BICEP | 100, 150 (48) | $1°$, $0.7°$ | South Pole | PSB + Faraday rotators | 2005 | [79] |
| PolarBear | 150(300) | $3'$ | CA (37°N) | TES B + 1/2 wave plate | - | [139] |
| BaR-SPOrt | 32(1), 90(1) | $30'$ ,$20'$ | South Pole | CP, LDB | ? | [154] |
| SPOrt | 22(1), 32(1), 90(1) | $7°$ | Space, ISS | CP, 80%, full sky | ? | [27] |
| Planck LFI | 30(4), 44(6), 70(12) | $33'$ , 24, $14'$ | Space, L2 | pseudo CP, full sky | 2007 | [110] |
| Planck HFI | 100(4),143(12), 217(12) 353(6), 545(8), 857(6) | $9.2'$ , $7,'$ , $5'$ | Space, L2 | PSB, full sky | 2007 | [91] |

Table 2.2: Snapshot as of winter 2004 of past, current and planned CMB polarization experiments in chronological order. Note the definite trend towards incoherent detectors. All upper limits and detection to date come from coherent detectors. Experiments' web sites listed at `http://cosmology.princeton.edu/∼dbarkats`.

[a]CP: Correlation polarimeter.

[b]B: Spiderweb bolometer.

[c]I: Interferometer.

[d]PSB: Polarization sensitive bolometers.

[f]LDB: Long Duration Balloon flight.

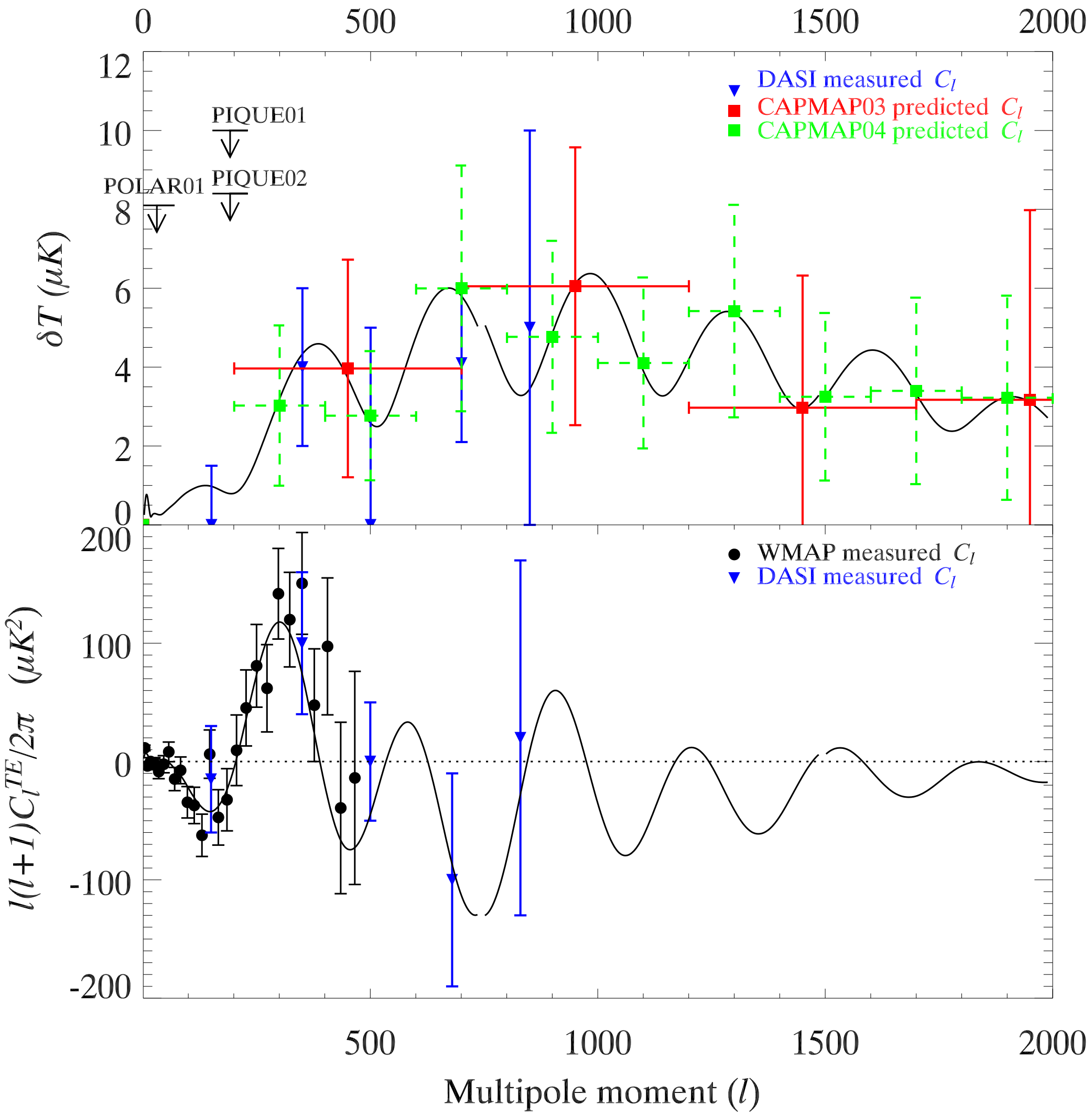

Figure 2.3: CMB E-mode (top) polarization and TE correlation (bottom) detections and upper limits. The CAPMAP03 (CAPMAP04) band power predictions are conservative. They assume an instrument with $1000\mu\mathrm{K}\sqrt{s}$ ($200\mu K\sqrt{s}$), $t_{int} = 250$ hrs (500 hrs), $\theta_{fwhm} = 4'$ , and $\mathrm{N}_{pix} = 600$ (1300) and are plotted in red (green) squares. The lines are the WMAP [14] E-mode and TE theoretical $\Lambda$CDM fit to the data (black circles). The recent DASI [89] detections are plotted in blue triangles, while the WMAP data are black circles [86]

## 2.2 Polarimeter

**Stokes Parameter** The 4 Stokes parameters, I, Q, U, and V fully describe the polarization state of an electromagnetic wave. The reader is referred to [90] for an exhaustive review of Stokes parameters.

Qualitatively, I is a measure of the total intensity of the radiation, while Q and U measure the power carried in linear polarized radiation. V describes the circular polarization. For unpolarized radiation, Q, U, and V are zero. Q and U are defined with respect to a local reference system. If we define two orthogonal systems, a-b and x-y rotated 45° with respect to each other, as in Figure 2.4, then Q is defined as the difference between the power carried by orthogonal components in the x-y system and U as the difference between the power carried in the orthogonal components in the a-b basis. Q and U are given by

$$Q = E_x^2 - E_y^2 = 2\ E_a E_b \qquad \text{and} \qquad U = 2\ E_x E_y = E_a^2 - E_b^2\ . \tag{2.2}$$

The rotation transformations of the Q and U Stokes parameters follow that of a spin-2 vector. If $E_x$ and $E_y$ are two orthogonal components of the electric field in the x-y basis, then if the a-b basis is rotated by an angle $\alpha$ with respect to x-y, the components of the electric field in the a-b basis are:

$$\begin{pmatrix} E_a \\ E_b \end{pmatrix} = \begin{pmatrix} \cos\alpha & \sin\alpha \\ -\sin\alpha & \cos\alpha \end{pmatrix} \begin{pmatrix} E_x \\ E_y \end{pmatrix} \tag{2.3}$$

Therefore, the Stokes parameters in the two bases are related by:

$$\begin{aligned} Q_{ab} &= \langle |E_a| \rangle^2 - \langle |E_b| \rangle^2 && (2.4) \\ &= |\cos\alpha E_x + \sin\alpha E_y|^2 - |-\sin\alpha E_x + \cos\alpha E_y|^2 && (2.5) \\ &= (E_x^2 - E_y^2)\cos 2\alpha + 2 E_x E_y \sin 2\alpha && (2.6) \\ &= Q_{xy}\cos 2\alpha + U_{xy}\sin 2\alpha && (2.7) \end{aligned}$$

The result is the same for $U_{ab}$ with $\alpha \to \alpha + \pi/4$:

$$U_{ab} = Q_{xy}\sin 2\alpha + U_{xy}\cos 2\alpha \tag{2.8}$$

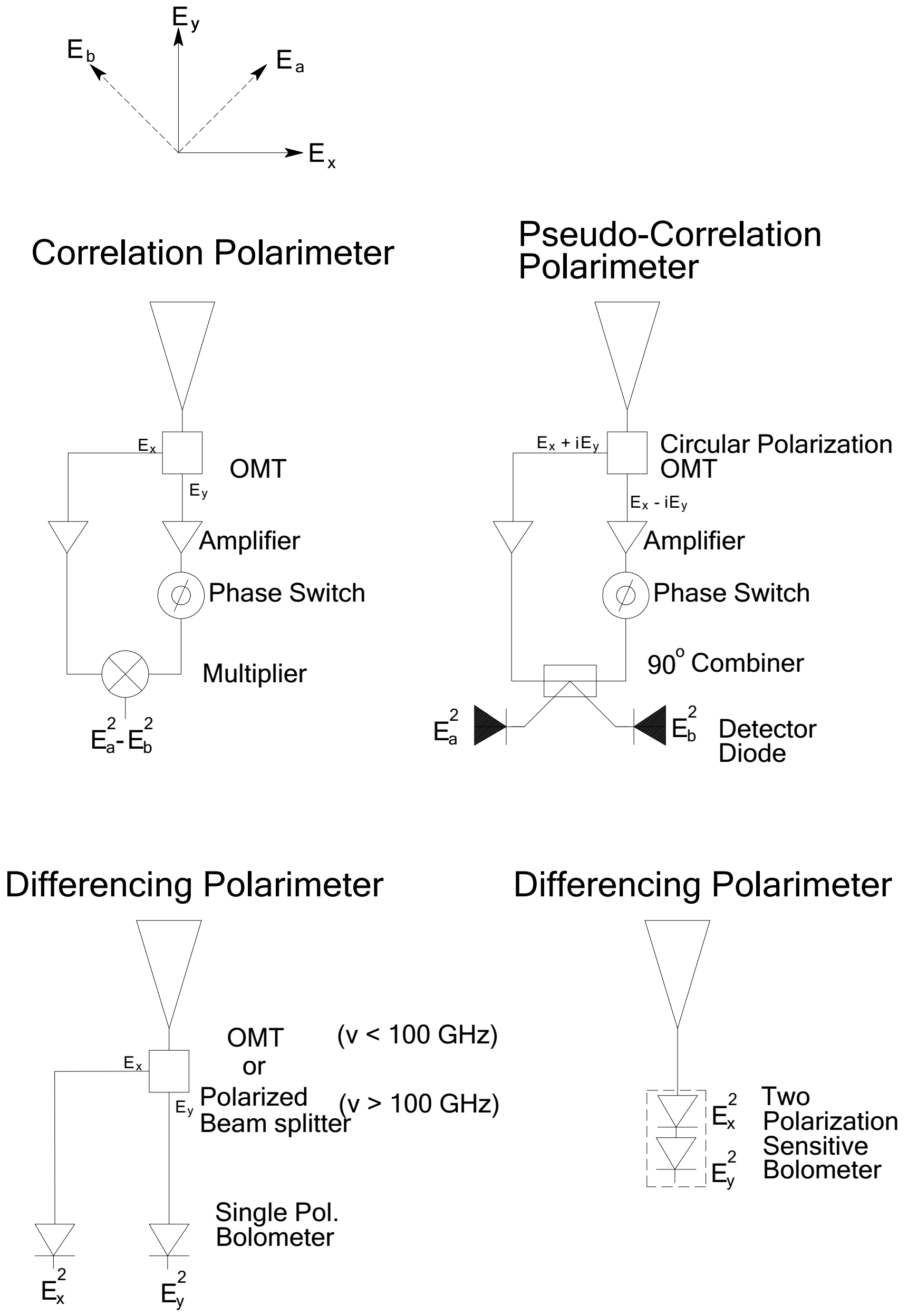

Figure 2.4: Schematic diagram of CAPMAP correlation polarimeter polarization detection technique (top left). The reader is referred to [113] for an excellent review of the Jones matrix formalism to interpret the measurement of each type of polarimeter. CAPMAP measures a single Stokes parameter but has the capacity, just like a pseudo-correlation polarimeter, to measure both Stokes parameter from the same feed horn simultaneously. One advantage of the WMAP-style pseudo-correlation polarimeter is that it is homodyne and thus avoids all the complication associated with an LO. The downside however is the tougher phase-matching requirements and the need for a broadband 180°phase switch. A standard bolometer differencing assembly as well as the newer Polarization Sensitive Bolometer polarimeter are shown in the bottom. Although these types seem simpler than correlation receivers, they have to deal with the large common-mode unpolarized signal and cannot rely on phase switching to modulate the output to obtain a flat noise spectrum. Other modulation techniques employed are 1/2-wave plate rotation [50], Faraday rotators [79], and mechanical rotation of the whole receiver assembly [80].

When observing polarization on the sky, we use the NCP as reference point for a coordinate system with respect to which Q and U are defined. The IAU [68] convention states that a radially polarized signal has positive Q. Positive U is rotated 45°counter clockwise from positive Q. Figure 2.5 illustrates this convention.

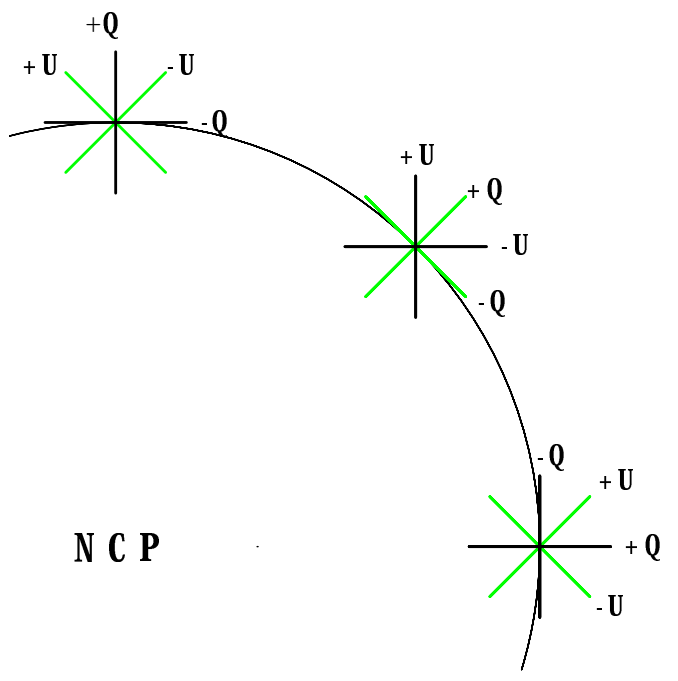

Figure 2.5: IAU convention of Q and U Stokes parameter near NCP. Figure by M. Hedman

**Correlation polarimeter** The role of the polarimeter is to extract the minuscule polarized signal from the large unpolarized microwave radiation of the atmosphere, the receiver, and to a smaller extent the CMB itself.

A direct way to measure polarization is to detect the power carried in each of two orthogonal components of the radiation and difference them. Although simple, this scheme is far from ideal. Even if the input radiation were completely unpolarized, the difference between the two large detected powers is likely to drift due to gain variations in each arm and thus generate a spurious polarized signal. Correlation polarimetry provides a particularly clean technique for measuring small signals on top of large unpolarized backgrounds by taking advantage of the information carried in the correlated phase of polarized waves.

A correlation polarimeter measures polarization not by differencing two signals but by multiplying them. Figure 2.4 illustrates the method and compares it to the differencing scheme used by bolometers. In a correlation polarimeter, the radiation is split into two orthogonal components by the orthomode transducer (OMT). The OMT is oriented with one arm parallel to the local horizontal so that the detection basis is at 45° from the local horizontal and vertical. $E_x$ and $E_y$ are then amplified, filtered, and down-converted separately in each arm of the receiver. The phase of the time varying electric fields is conserved. Upon reaching the multiplier, the two signals are combined to produce a voltage proportional to U:

$$V_{out} \propto \langle E_x E_y \cos \Delta\phi \rangle \tag{2.9}$$

Non-idealities in the multiplier result in additional terms (see Section 3.1.5) such that

$$V_{out} = \langle g(E_a^2 - E_b^2)\cos\Delta\phi + \epsilon_1 E_x^2 + \epsilon_2 E_y^2 \rangle \tag{2.10}$$

where $\Delta\phi$ is the average phase difference between the two arms, $g$ is the polarized responsivity (where $g \gg \epsilon_1$, $\epsilon_2$, and the $\langle\ \ \rangle$ denotes time average over the sampling period of the detector.

What are the benefits of a correlation polarimeter? The output DC voltage of the multiplier is proportional to one of the Stokes parameters and is ideally null when the input is unpolarized. This is an important point because we know that the small polarized signal rides on top of an unpolarized background $10^8$ times larger. In addition, because the phase is conserved, the polarized output can be modulated simply by using a device which switches the phase in one of the arms from $0°$ to $180°$. The modulation has two purposes. The first one is to discard the systems's small non-ideal sensitivity to the intensity in each arm, $\epsilon_1 E_x^2 + \epsilon_2 E_y^2$. Upon phase switching, only the desired polarized output will switch sign (not the total intensity because it is squared) and the real polarized signal can be extracted by digitally locking on to this modulated signal. The second advantage of phase-switching is that it provides a way to circumvent the $1/f$ fluctuations of the noise from the front-end amplifiers and thus to integrate down the random gaussian noise from the receiver. This technique is called Dicke-switching after its inventor [29]. By phase switching at a certain frequency $f_{switch}$, the amplitude of the polarized output will only be affected by noise fluctuations at frequencies greater than $f_{switch}$ or at time scales less than the modulation period. The smallest polarized signal measurable by an unmodulated correlation polarimeter is expressed as

$$\Delta Q_{min} \propto T_{sys}\sqrt{\frac{1}{\Delta\nu\ t_{int}} + \left(\frac{\Delta G}{G}\right)^2}\ , \tag{2.11}$$

where $T_{sys}$ is the total system temperature (which includes contributions form the sky, receiver, and CMB), $\Delta\nu$ the receiver RF bandwidth in Hz, $t_{int}$ the integration time in seconds and $\left(\frac{\Delta G}{G}\right)^2$ the fractional gain variation of the receiver on time scales of the integration time. The latter would vanish for an ideal radiometer system [72]. As can be seen from Equation 2.11, for a radiometer with $\Delta\nu = 10$ GHz and $t_{int} = 1$ second, fractional gain variations on the order of $\frac{\Delta G}{G} \approx 10^{-5}$ will degrade the radiometer performance.

In CAPMAP, the $1/f$ knee of the amplifiers is at ∼1 kHz and the phase switch frequency is 4 kHz. As long as the modulation is at a frequency higher than the $1/f$ knee of the amplifiers, the smallest measurable signal will be dominated by the first term in Equation 2.11 and can be made as small as desired given enough integration time. The smallest measurable signal[2] is then given by

$$\Delta Q_{min} \propto T_{sys}\sqrt{\frac{1}{\Delta\nu\ t_{int}}} = \frac{\mathcal{S}}{\sqrt{t_{int}}}\ . \tag{2.12}$$

Another advantage of a HEMT-based correlation polarimeter is its ease of use. Because the noise of the front-end amplifiers reach their lowest level at ∼30 K, there is no need for liquid cryogens. Although nowadays, low-temperature cryostats reaching ∼100 mK are becoming available, the necessary 20 K can easily be reached with a mechanical refrigerator. Those require minimal maintenance and can run uninterrupted for many months in a row.

Finally an important advantage of the correlation polarimeter compared to other incoherent detection techniques is that it can measure both Stokes parameters simultaneously within the same feed at no extra cost in sensitivity [156, 155]. This is a consequence of the fact that once the front-end amplifiers have provided the large amplification in the signal at the expense of their added noise, the signal can be split many times to detect Q and U without the added quantum noise per mode.

In the end the goal is to get a flat noise spectrum as in Figure 2.6.

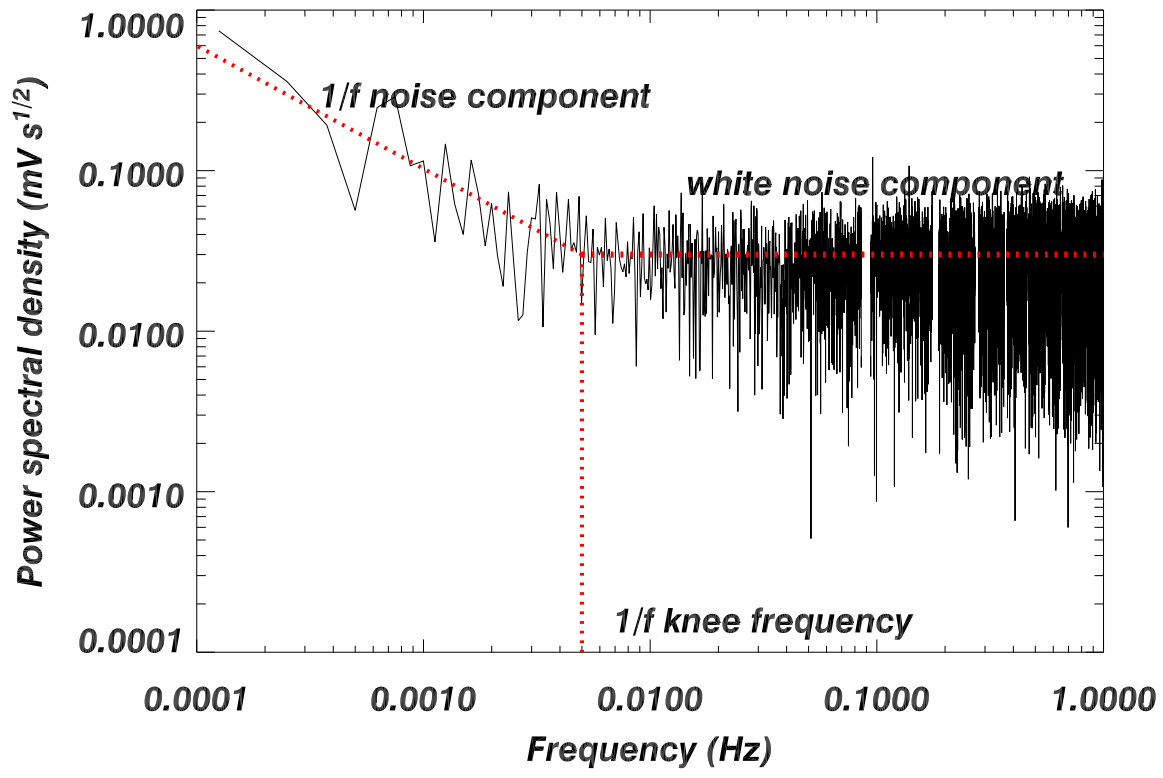

Figure 2.6: Typical noise power spectrum of a phase-switched polarization channel showing the white noise plateau and a $1/f$ component at low frequencies. For CAPMAP receivers, the $1/f$ knee is at a few mHz. The residual $1/f$ noise is removed by the scan strategy position chop.

A correlation polarimeter is insensitive to the relative gain in the two arms because these gains enter as a common factor to the final polarized output in Equation 2.9. However, the relative phase shift between the two arms needs to be carefully controlled. The response of the polarimeter is reduced by a factor of $\cos\Delta\phi$ for phase shift $\Delta\phi$ in the two arms. This is a known peculiarity of correlation polarimeters and phase matching the two arms accounts for a major part of the receiver's integration (see Section 3.3.1).

[2]See Section 3.3.3 for a discussion of the exact pre-factor in Equation 2.11 and 2.12.

Another disadvantage of a correlation polarimeter is that the detection is performed at IF frequency. This down-conversion is necessary because the mixers which are used as multipliers in the IF chain have never been tested at the RF frequency band. The LO signal needed by the mixers to perform the down-conversion adds a large amount of complexity, size, and unexpected problems to the receivers. A correlation polarimeter which could detect directly in the RF band would be a large improvement (see for example the pseudo-correlation polarimeter in Figure 2.4).

## 2.3 Optics

The optics of the CAPMAP experiment can be separated into two parts: the telescope and the feed system. The telescope optics is detailed in Section 4.2. Because the telescope already exists and cannot be modified (without much more time and money), the task at hand is to design a feed system which will couple the receivers to the telescope while fulfilling a set of requirements:

- The final beam size must be $4'$ or smaller.
- The sidelobe and spillover pickup must remain less than the final noise per pixel.
- The feed system must be small enough to be closely packed around the focal point.

These requirements have immediate consequences on the characteristics of the feed system.

### 2.3.1 Beam Size

The choice to observe spherical harmonics up to $\ell = 1500$ sets the beam size to be $4'$ or smaller. Finite angular resolution causes a smoothing of the CMB signal at high angular scales, equivalent to an exponential decrease of the signal as $\ell$ increases. This effect is quantified with a beam window function $B_\ell^2 = e^{\ell^2\sigma_b^2}$ used in Equation 2.1. The degradation of the E-mode polarization signal due to four possible beams sizes is shown in Figure 2.7. A choice of $4'$ is ideal as the signal is only reduced by a factor of 0.88 at $\ell = 1000$, where the E-mode peaks.

The smallest beam in Figure 2.7 is $1.5'$, which is the minimum beam size attainable with the Crawford Hill 7-meter antenna at 90 GHz. This is the diffraction limited resolution of an optical element of diameter $D$ at a wavelength $\lambda$: $\theta_{diffraction} \simeq \frac{\lambda}{D}$. With a $1.5'$ beam size, the E-mode polarization signal is unaffected until much higher angular multipole. The telescope therefore has the potential to probe the very high angular scales. To first order, the scientific goals of the experiment are not limited by the size of the telescope, which is not always the case. Other properties, however, have more direct influence on the CAPMAP choice of $4'$.

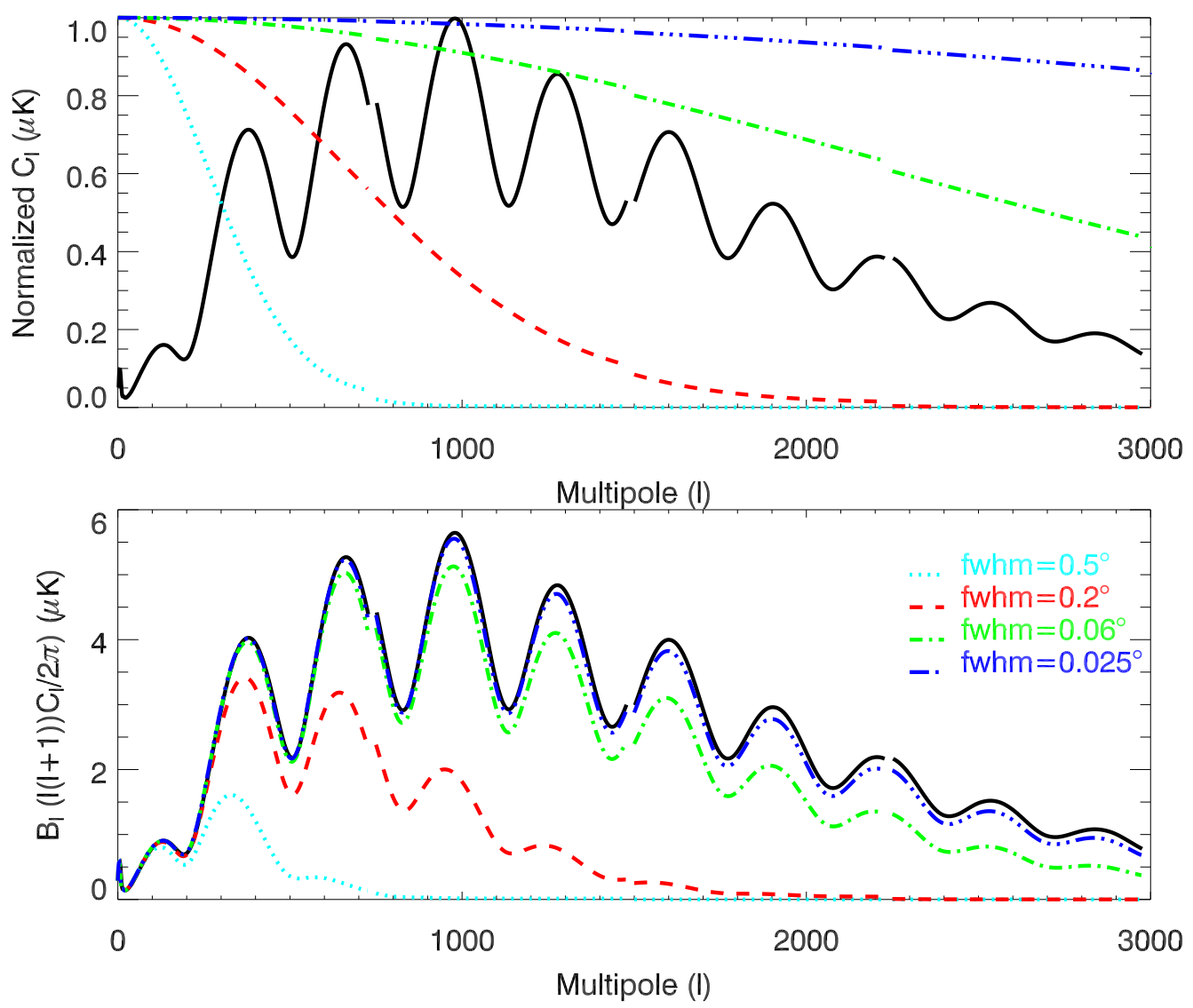

Figure 2.7: *top*: Beam window function for four beams with different full widths at half maximum (fwhm) normalized to a $C_\ell^E$ spectrum. $B_l^2$ not $B_l$ is plotted. The CAPMAP beam is chosen to be $4'$ ($0.06°$) so the signal is only reduced by a factor of 0.8 at $\ell = 1500$. *bottom*: The reduction in signal strength due to the four beams. Note how a beam size larger than $12'$ (ie. PIQUE beam) greatly reduces the signal from the higher peaks.

### 2.3.2 Angular Resolution and Control of Spillover

To think about radio telescope beam patterns like a radio engineer, one usually thinks of the feed horn as a source instead of a receiver, with the signal propagating towards the sky. Reciprocity guarantees that the receiving and transmitting antenna pattern are the same. Once the desired beam size is established, we determine the illumination of the telescope from the feed horn.

The feed optics beam pattern is approximated by a perfect gaussian. With this assumption, the main parameter to optimize is the width of the feed system's beam at the reflectors. Increasing this width increases the illumination at the edge of the reflectors (edge taper) which in turns decreases the beam size on the sky (better angular resolution). Decreasing the edge taper will produce a larger beam because it is equivalent to using a smaller mirror. Changing the edge taper has a second consequence. Decreasing the edge taper decreases both the amount of energy that falls outside of the reflector (spillover) and the amount which can diffract from the edge of the mirror. Because the radiation in the spillover and the sidelobes collect energy from sources $10^8$ times warmer than the CMB polarization signal[3], it is extremely important to minimize the total pickup from these sources. Indeed a total pickup of several Kelvin would most likely induce a large scan-synchronous polarized pickup.

CMB polarization experiments typically minimize pickup from sidelobes and spillover in two ways. At the expense of angular resolution, reflectors are under-illuminated with an edge taper of approximately -30 dB. This means that any radiation originating from the ground and diffracting from the edge of the mirror into the receiver will be a factor of $10^3$ weaker than that from the main beam in the sky. This fraction is usually still too high so CMB experiments are also outfitted with large metallic ground screens which redirect stray radiation from the sidelobes and spillover back onto the cold sky (see for example the PIQUE experiment [56] ground screen design).

| Secondary edge illumination[a] | Gaussian feed FWHM[b] | Gaussian feed aperture[c] | Primary top edge illumination[d] | Primary bottom edge illumination[d] | Telescope beam FWHM |
|---|---|---|---|---|---|
| -0.5 dB | 25° | 1.0 cm | - | - | - |
| -10 dB | 5.5° | 4.5 cm | - | - | 0.030° |
| -20 dB | 3.92° | 6.3 cm | -18 dB | -37 dB | 0.034° |
| -30 dB | 3.36° | 7.4 cm | -31 dB | -55 dB | 0.039° |
| -40 dB | 2.77° | 9.2 cm | -40 dB | -74 dB | 0.043° |
| -50 dB | 2.48° | 10.0 cm | -51 dB | -92 dB | 0.048° |
| -60 dB | 2.26° | 11.2 cm | -61 dB | -110 dB | 0.053° |

Table 2.3: Secondary mirror illumination versus telescope beam size. Values are calculated using the physical optics simulation (GRASP8) of a purely gaussian feed at the focal plane of the telescope.
[a]Taper at a 5.06° half angle from the central ray.
[b]The gaussian beam size of an ideal horn is calculated from gaussian optics propagation using the secondary illumination.
[c]The hypothetical aperture of horn scaled from the PIQUE horn using the ratio of their FWHM. The horn is scaled by keeping the flare half-angle constant and scaling the aperture and the length. The PIQUE horn has a 3.5° half flare angle, a 25° FWHM, and is 5.6 cm long.
[d] The edge tapers are calculated for purely gaussian illumination. The real feed horn sidelobe pattern (shown in Figure 2.8(b)) would increase these edge tapers.

CAPMAP is not in the typical regime described above because the desired beam size is already larger than the minimum beam of the antenna. In addition, the top of the antenna sits ∼15 meters above the ground making such a large ground screen impractical. The solution to minimize sidelobe and spillover pickup is to aggressively under-illuminate the reflectors. The rest of the mirror effectively serves as ground screen. Table 2.3 tabulates the calculated beam size which varies linearly as a function of secondary illumination in dB. The reflector illuminations and beam size are calculated using a physical optics simulation package (GRASP8 [4]). The simulation's only input is an ideal gaussian beam pattern defined by an edge taper on the secondary. An example of the far field beam pattern of the telescope with a -30 dB secondary edge taper is shown in Figure 2.8(a). Using Table 2.3, a $0.05°(3')$ final beam size is chosen, satisfying both the low illumination requirement and the scientific constraint. However, such a choice for the beam size means that the secondary has to be illuminated with a 2.3° FWHM beam. Section 2.3.3 addresses the technical details of creating such a narrow beam.

[3]The ground, trees, and control building radiate most at microwave frequencies.

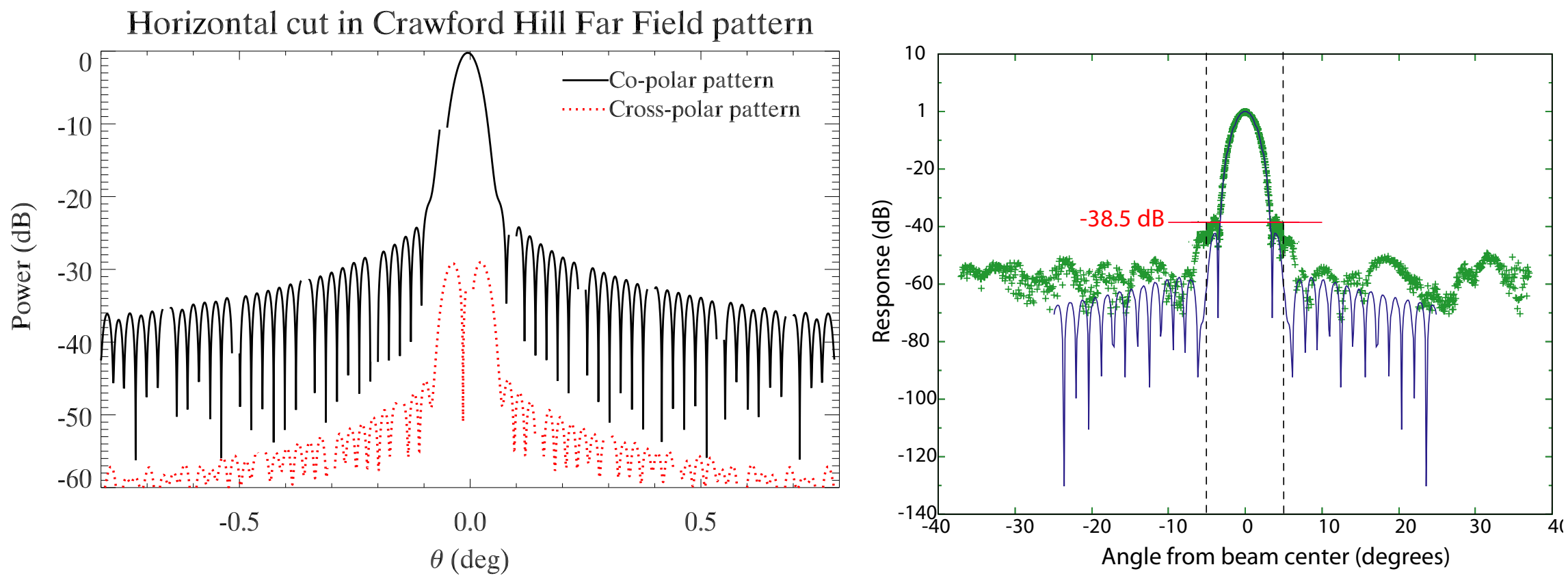

(a) (b)

Figure 2.8: (a) Far-field beam co- and cross-polar beam patterns of the 7-meter antenna at 90 GHz. The input to the simulation is a gaussian beam with a -30 dB edge taper (-30 dB at 5.06° at the edge of secondary). The co-polar beam pattern is normalized to 0 dB. It otherwise peaks at 74 dBi. The cross-polar pattern in two-dimensions is dipolar with a null line along the vertical. The simulation does not extended more than 1° from the peak because the calculation time scales as $(\frac{D}{\lambda}\sin\theta)^2$ where D is the reflector diameter, $\lambda$ the wavelength and $\theta$ the angle from the beam center up to which the simulation should converge. The beam pattern presented here took ∼1 hour to calculate on a 700 MHz computer.
(b) Beam map of CAPMAP03 horn and lens on top of predicted beam pattern. The measurement was made from the testing range on the physics dept roof, where beams can be mapped down to ∼50 dB noise level. The vertical dashed lines show the position of the top and bottom edge of the secondary mirror. The lens used was not anti-reflection grooved so its sidelobe response is probably a factor of 3 dB better than a grooved version. Figure by J. McMahon

[4] We use the free but limited student edition of GRASP8 available from http://www.ticra.com.

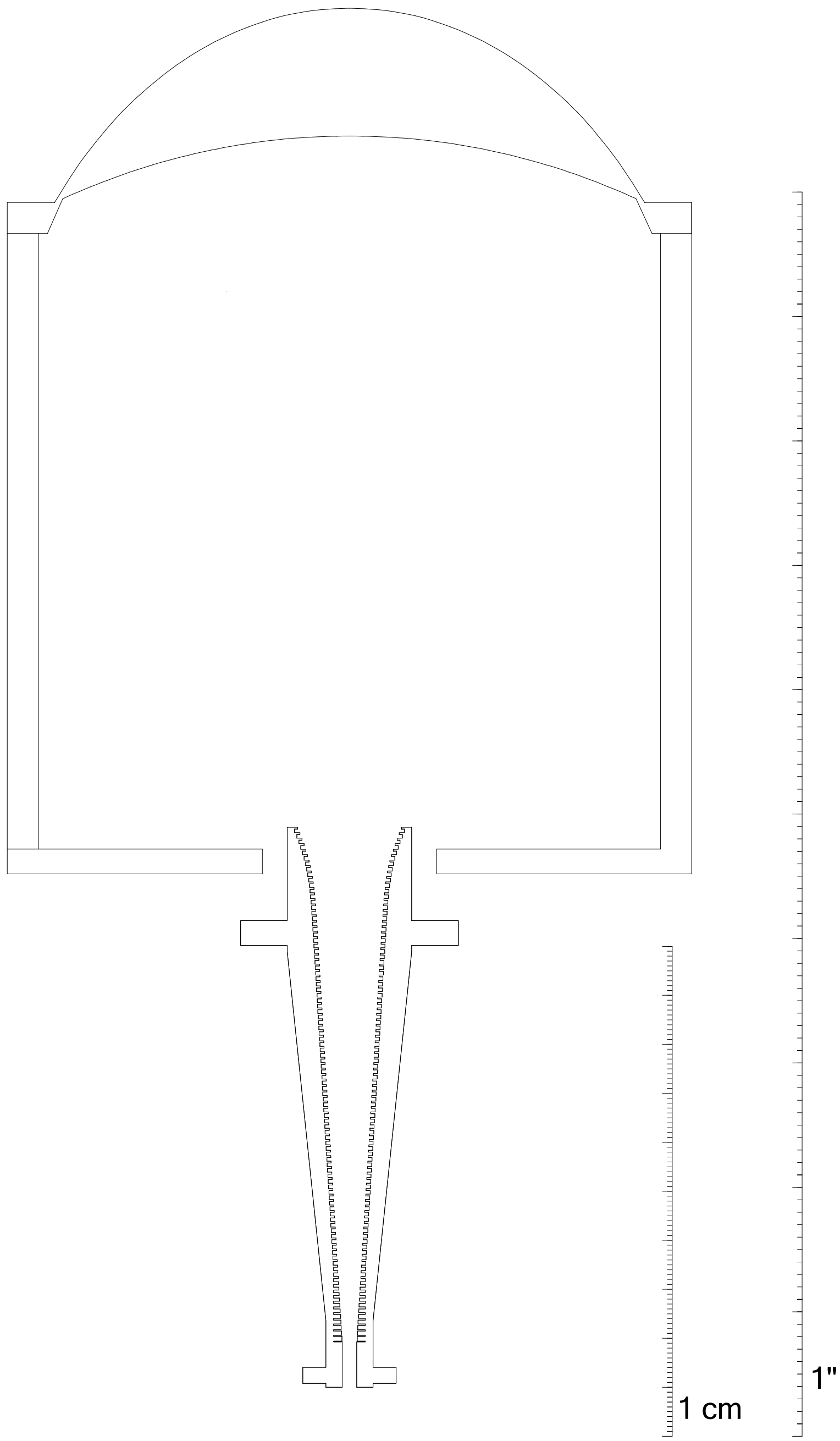

Figure 2.9: Cutaway of CAPMAP horn, lens shroud and lens. Drawing is to scale. The lens is 4.83" diameter, with a spherical back surface of radius $R_1 = 5.5$" and front surface radius described by $R_2 = \frac{F(n-1)}{n-\cos\theta}$, with respect to the phase center at the horn aperture. The focal length is $F = 6.68$" and $n = 1.518$ the index of refraction of the HDPE. Lens grooving is not shown. The shroud is made of 0.3" thick, 5.5" diameter Al tube. The inside surface of the shroud is lined with a 1/8" layer of honeycomb eccosorb to capture stray radiation. The lens shroud is mechanically and thermally stood off from the horn mounting plate by four 3/4" diameter G-10 pillars (not shown). The horn has a 1" aperture and is 4.5" long. It is electroformed in copper, then gold plated. It is mounted using the eight clearance holes on the top ring. Refer to Section 2.3.3 for discussion.

### 2.3.3 Feed Horn and Lens

The design of the feed optics plays a critical role. The gaussian beam pattern assumed in Table 2.3 is an idealization. The deviation from the gaussian case and thus the unwanted sidelobes and beam distortions depend on the details of the feed system. In addition to the 2.3° beam size requirement, the real feed must be compact and not too massive. The feed optics must be cryogenically cooled, and the feed optics' footprint must allow at least four receivers (16 for the whole experiment) to fit within the focal area and more specifically inside a dewar. Moreover, the sidelobes from the feed must be very low so as not to dominate the telescope sidelobe performance.

As hinted in Table 2.3, scaling the PIQUE feed horn (25° FWHM) to 2.3° results in a 60-cm long horn with a 12-cm aperture, which would be lossy, heavy, and difficult to manufacture. During the CAPMAP design phase, a system comprising a horn plus a large tertiary mirror was investigated as an early feed system attempt (see Section 4.2.1). It performed well with a single receiver but required too large a tertiary mirror for multiple receivers.

To satisfy all the design criteria, feed optics consisting of a compact profiled horn coupled to a lens were chosen. A scale drawing of the horn and lens assembly is shown in Figure 2.9. These feed optics were designed by J. McMahon and more details will be discussed in [107, 106]. Some parameters are detailed in the caption of Figure 2.9. The horn is electroformed copper which has been gold plated to minimize ohmic losses. The lens is a meniscus lens for which all the refraction happens on the front surface (telescope side). The back surface is spherical, matching the radius of curvature of the beam from the feed horn. The lens is made of HDPE and positioned a distance 5.5" above the horn aperture (rather than directly on it [96]). The lens is connected to the horn by a hollow aluminium tube (the shroud). Note that after optimization, the resulting final aperture (lens diameter = 12.2 cm) is the same as estimated for a single horn in Table 2.3.

The feed system is optimized to have a gaussian beam and a very low first sidelobe, predicted and measured 38 dB below the main peak. Predicted and measured beam patterns are shown in Figure 2.8(b). The gaussian beam expansion from the feed to the primary is plotted in Figure 2.10. To reduce reflections, the lens is cut with concentric circular grooves, 1/4 wave deep and parallel to the lens optical axis on both surfaces (actually receiver A and D lenses are grooved on back surface only). In retrospect, we find that these anti-reflection groovings increased the first sidelobe and caused a birefringent reflection [142] that was responsible for most of the quadrupolar cross-polar pattern described in Section 6.3.1. The horn is cooled to $\simeq 20$ K and the lens to $\simeq 150$ K by conduction through the shroud, adding a total of 6 K to the system temperature.

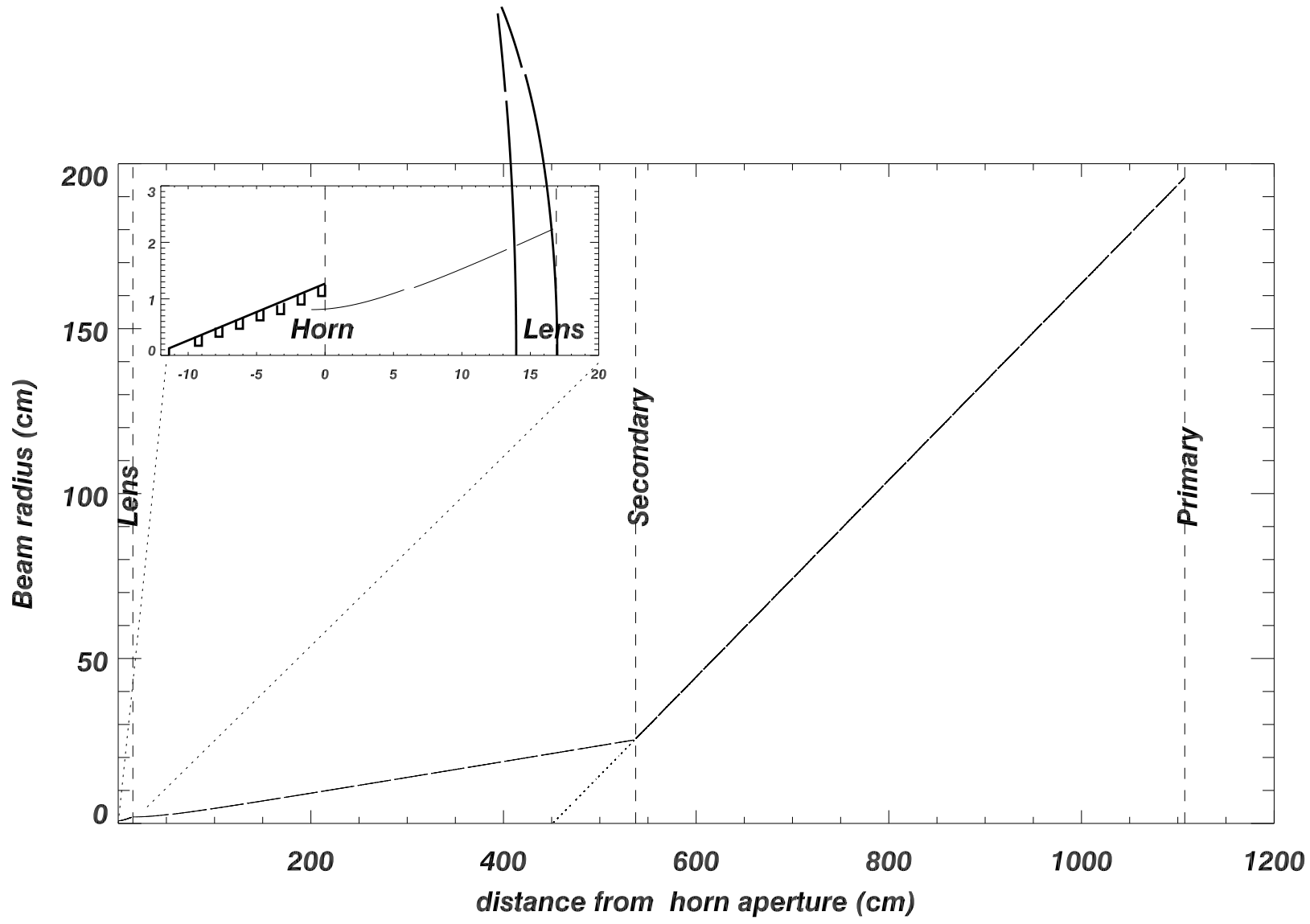

Figure 2.10: Beam expansion as a function of the distance from horn aperture using gaussian beam optics propagation [44]. All distances are to scale. The curvature of the field is only visible in the closeup of the horn (inset) because it is the near-field region. The origin of the beam is at the phase center, $\sim$1 cm behind the horn aperture. The expansion to the secondary and primary mirrors follow simple ray tracing optics with curvature only very close to the beam waists. The line from the secondary to the primary originates at the primary virtual focus.

## 2.4 Focal Plane and Scan Strategy

The arrangement of the receivers in the focal plane is a complicated task because it brings together many interleaved requirements, detailed below.

**Structural constraints:** The total area available in the focal plane[5] is limited to a 2 meter diameter circle around the focal point. The volume taken by the receivers is approximately that of a tube of diameter 7" and height 22". The geometry of the dewar and the fridge sets a lower limit on how closely the receiver can be packed together.

**Optical constraints:** To minimize beam distortions and aberrations, the receivers should be packed as close to the focal point (FP) as possible. However, this is the weakest constraint. Extensive studies were done using Grasp8 (physical optics software) and Code V (ray tracing software) to simulate the effect of moving the receivers away the FP. From Grasp8, we found that the beam remained essentially symmetric (vertical vs. horizontal cut in the far field beam pattern and horizontal vs vertical polarization of the feed pattern) for horns located as far as 40 cm away from the FP, as long as the feed horns were pointed at the center of the secondary. Incorrect tilt of the horns caused large illumination asymmetry on the mirrors which in turn was visible in the sidelobe pattern. This result was further confirmed by Code V. The strehl ratio[6] for points within 40 cm of the FP were better than 0.99. A spot diagram for the final CAPMAP03 receiver position is shown Figure 2.11 to illustrate that the aberration remain small compared to the diffraction beam size.

[5]The focal "plane" is of course somewhat curved, so each receiver is tilted 1.7° towards the secondary. The aperture of each lens is still in the same plane.

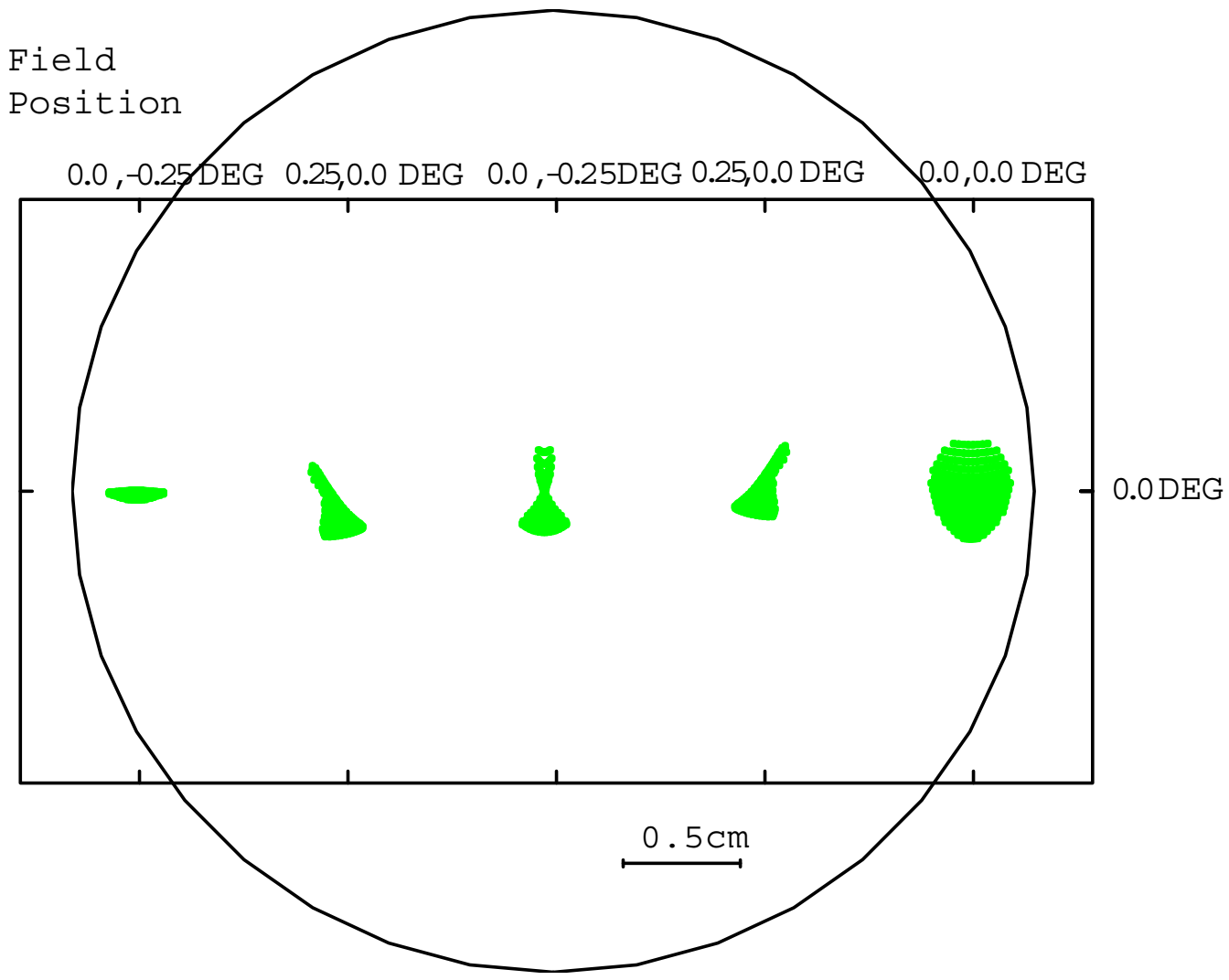

Figure 2.11: Spot diagrams of the Crawford Hill antenna. The spot diagram are calculated using the CODE V software[25]. The five spots displayed are the dispersion of $\sim 300$ rays distributed over the whole primary and focused to (from right to left) the focal point and the four CAPMAP03 horn positions. The large circle is the diffraction beam size of the telescope for the central spot diagram. This agrees with the fact that the beam distortions are very small away from the focal point. The strehl ratios for the four CAPMAP03 beam positions are better than 0.99. Simulations done with the help of M. Niemack.

An additional concern to consider when dealing with a polarization (as opposed to a simple total-power) instrument is the small polarized signal, caused by the reflection and thermal emission of radiation from a metallic mirror surface with a finite conductivity [31, 123, 52]. The amplitude of this polarized signal is proportional to

[6]the ratio of the intensity at the image point (the origin of the reference sphere is the point of maximum intensity in the observation plane) in the presence of aberration, divided by the intensity that would be obtained if no aberration were present

the difference between the temperature of the mirror and the temperature of the sky. Because of the symmetries in the off-axis telescope, a different polarized signal will be injected in the different Stokes parameters measured by the detectors. Figure 2.12 displays the expected telescope-induced polarization as a function of position in the focal plane in the three cases: Q (difference between the power carried by the vertical and horizontal components of the electric field) or U (difference between the power carried by the components of the electric field oriented at 45° from horizontal) is measured with horns tilted towards the center of the secondary, and U measured with non-tilted horns. Given the expected position of the horns in the focal plane for CAPMAP03, the polarized offset should be smaller than 100 mK.

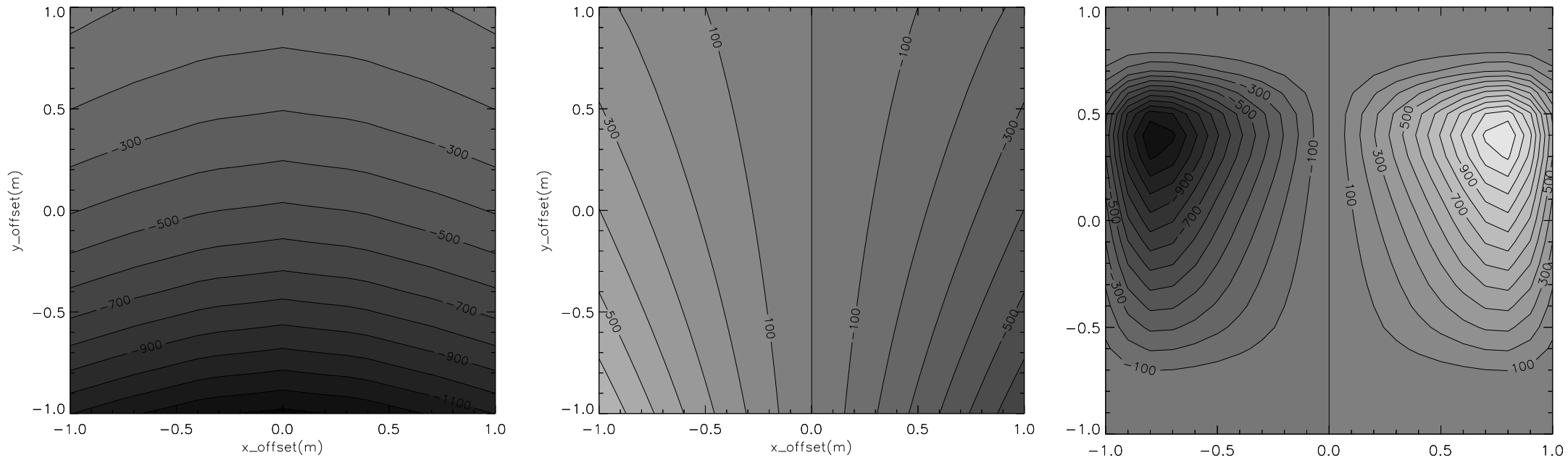

Figure 2.12: Induced polarization from the telescope mirrors (in mK) for a 1×1 meter area in the focal plane. Contours are spaced by 100 mK. The simulation assumes a sky temperature of 40 K, an isotropic mirror temperature of 300 K, an aluminium resistivity of $\rho = 4$ $\mu\Omega$ cm at 90 GHz. The dominant polarized signal originates from the secondary. Left and center plots are for a horn measuring Q and U respectively and tilted towards the secondary. Right plot is for a horn measuring U in the case of the horns not tilted (oriented parallel to the central ray). The induced polarization effect is symmetric with respect to the telescope's plane of symmetry. Figures and simulations by M. Hedman

**Scientific constraints:** As a rule of thumb, an experiment's sensitivity to CMB fluctuations is statistically optimized when a signal-to-noise ratio of unity in each pixel is achieved. This means that given a beam size, the area of sky covered should be large enough to minimize sample variance, yet small enough to remain in the $S/N = 1$ regime. In addition to this coverage requirement, the scan strategy should: take advantage of the redundancy available from the multiple receivers, observe Q and U equally, and cross-link pixels (pass through the same point through different directions).

The chosen focal plane layout for CAPMAP03 consists of four horns located 6" on either side of the focal point arranged in a diamond pattern as in Figure 2.13. Given the 61 cm/° plate scale, calculated from studies in Section 2.3, the four horns are spaced 0.25° equidistant from the central ray on the sky. The geometric arrangement of the horn satisfies all of the above requirements and also allows for either of two planned scan strategies: the azimuth scan [145] or the ring scan. The ring scan consists in scanning the telescope in a circle such that each receiver sees the NCP sequentially. The ring scan was not used for the primary CAPMAP03 observations.

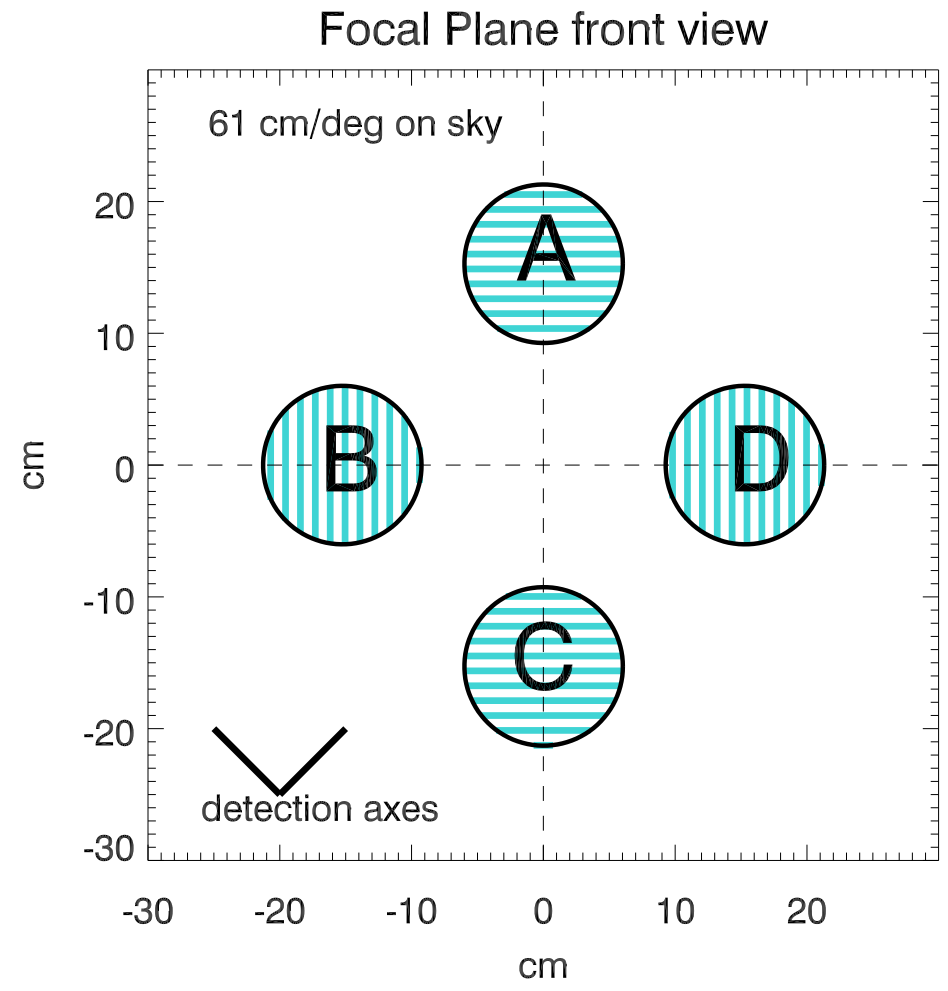

Figure 2.13: View of the focal plane layout in the CAPMAP03 dewar as seen from the secondary. The cross-hatch indicates the direction of the E-plane polarization of the main arm of the OMT. The resulting detection axes are at 45°from the OMT axes. Each circle represents a lens. The receivers are tilted 1.7° inwards in the focal plane to point towards the center of the secondary. From ray tracing, it is evident that the positions of the beams on the sky with respect to the center of the array is reversed in both the up-down and left-right orientation. As seen from the focal point, the arrangement of the beams on the sky has A on the bottom, B on the left, C on the top and D on the right.

The azimuth scan sweeps the telescope in azimuth across the NCP at constant elevation (see Figure 2.16). The constant elevation prevents the $\sec(\theta)$ total power modulation from the atmosphere. Figure 2.14(a) shows the path traced out by the four horns, in a single scan across the NCP. The optimal amplitude of the scan is determined later. The redundancy for this scan is excellent since the top-bottom and left-right horns will observe the same points twelve hours apart. In addition the left-right will cover the same pixels in the middle within the same scan. These redundant measurements are very desirable because they allow the creation of null data sets by differencing the data with appropriate time lags. The null data sets are used to verify the quality of the data.

Figure 2.14(b) shows the integration time of the azimuth scan strategy as a function of ring radius. The discrete steps show that the inner potion is observed by the two left-right horns, the central portion by all four horns and the outer annulus only by a single horn.

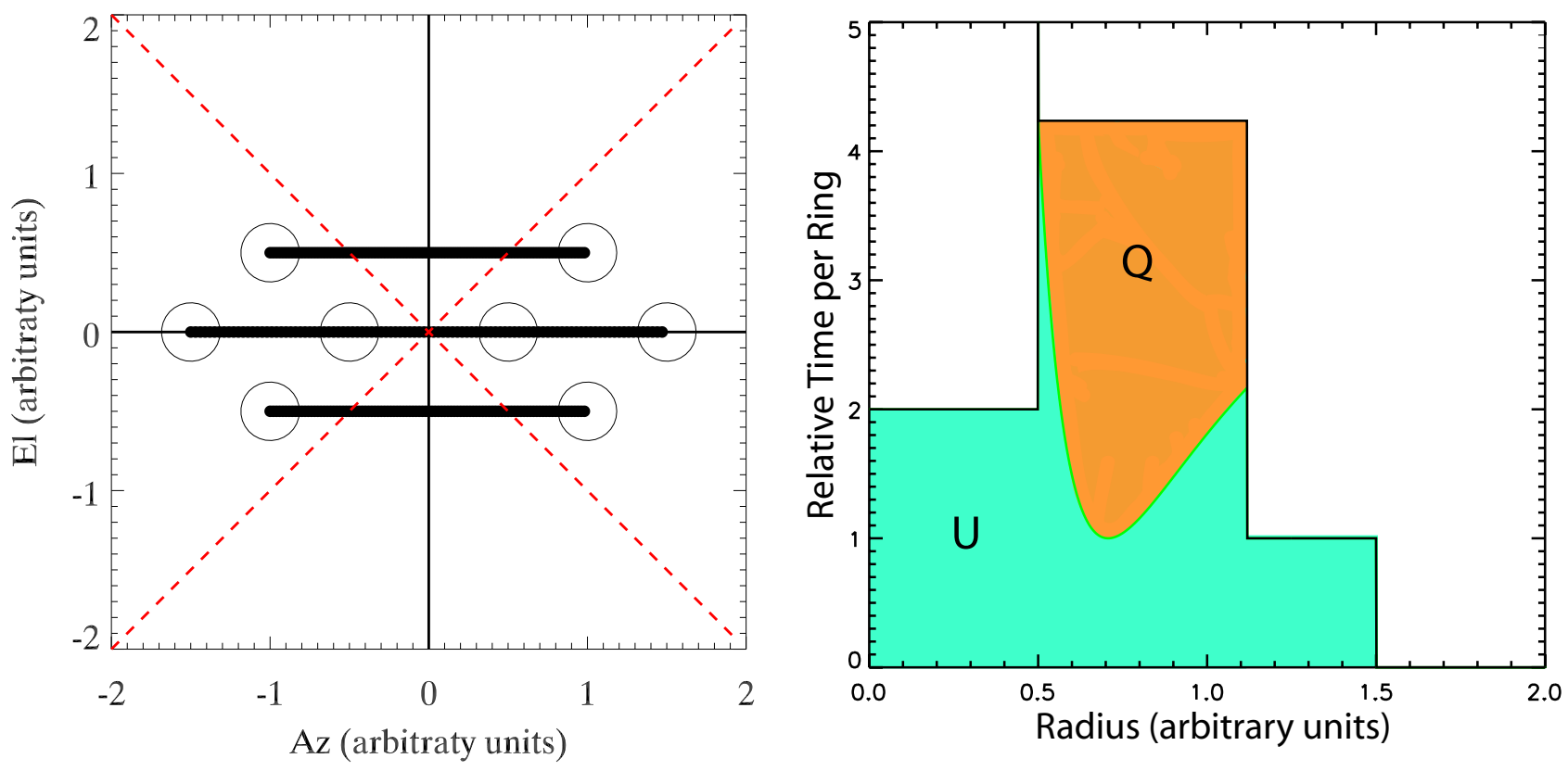

(a) (b)

Figure 2.14: (a) CAPMAP03 azimuth scan pattern on the sky. The axes are in arbitrary degrees and the size of the throw is determined in Figure 2.15. Each horn is designated by a circle. The NCP is at the origin. A horn on the solid lines measures U alone, on the dashed red lines measures Q alone, and otherwise measured a combination of Q and U.
(b) Coverage from the CAPMAP03 scan strategy as a function of cap radius. The black line shows the time spent in each declination ring around the NCP, normalized to the time spent by a single beam sweeping a unit radius. The green curved line shows how much time is spent observing Q (above line) and U (below line). The final radius used for CAPMAP03 is 0.75°. Study by Matt Hedman [54]

The curved line delimits the observed Q and U fraction. The Q-U partition of this scan is approximately 60%-40% with only Q coverage in the central annulus. The scan strategy assumes a constant velocity scan which is feasible given the telescope velocity acceleration limits.

The redundancy and Q-U coverage of this scan strategy assumes a perfect positioning of the beams with respect to NCP. As described in Chapter 6, the respective position of the beams matched predictions but the whole array was pointed low with respect to the NCP. Although this caused all redundancy of the scan to disappear, this greatly improved the cross linking of the scan and made the Q-U coverage more even over a larger region.

In the previous section, the radius of the final scan region remained undetermined. To estimate the optimal size, we calculate the area covered by the scan, divided by the beam solid angle to get the number of pixels observed, and verify that the signal-to-noise ratio per pixel approaches unity. From the numerical factors for CAPMAP03 given in Table 2.1, the signal-to-noise per pixel is $S/N = S/(\sigma_{pix}) = S/(\sigma_E \sqrt{\pi * r_{cap}^2/\Omega_{fwhm}}) \simeq 0.3/r_{cap}$. Choosing $r_{cap} = 0.75$ gives a S/N per pixel of 0.4. This agrees reasonably well with simulations [38] of an array with varying sensitivity levels scanning a cap of varying radii. Figure 2.15 illustrates some of the results. The error on the recovered polarization power spectrum is plotted vs cap radius for different $\sigma_E$. In the sensitivity regime of CAPMAP03 (top line), the optimal cap size has a radius of approximately 0.5°. An azimuth scan strategy was chosen for CAPMAP03 with a scan radius of 0.5°, a 0.25° separation between each horn and the focal point, and a ∼10 second period scan. The 10 second scan period was chosen to be much faster than the $1/f$ knee of the receiver, but slow enough to be well below the maximum speed of the telescope, to prevent straining the drive system.

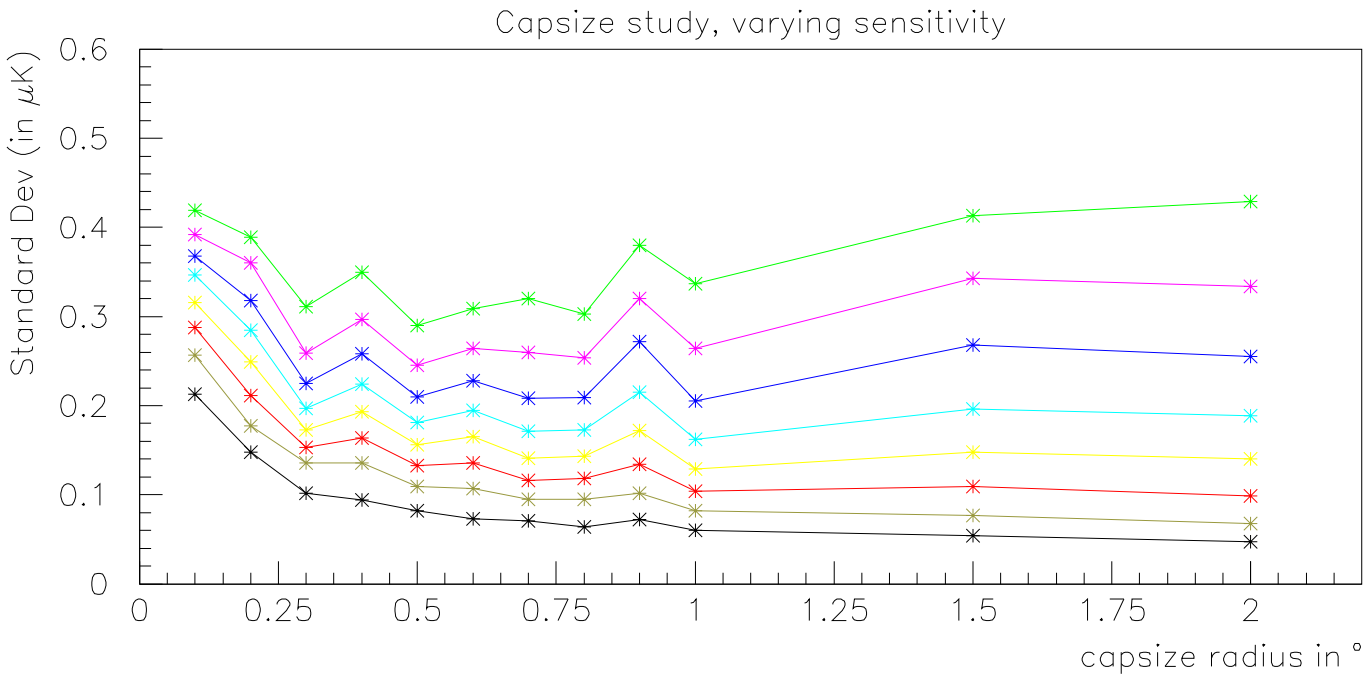

Figure 2.15: Error on the recovered power spectrum as a function of cap size and varying sensitivity. The simulations perform 100 realizations of the sky assuming a Wang theory power spectrum [141] and observed assuming a 0.05° FWHM beam and sensitivities varying from $\sigma_E = 0.1$ (bottom line) to $\sigma_E = 0.8$ (top line). For the expected sensitivity for CAPMAP03 ($\sigma_E = 0.8$), the errors have a broad minimum for cap radius between 0.5° and 0.8°. The simulations are done with no scan synchronous signal removed which may affect the results. Simulations and figure by Liam Fitzpatrick [38].

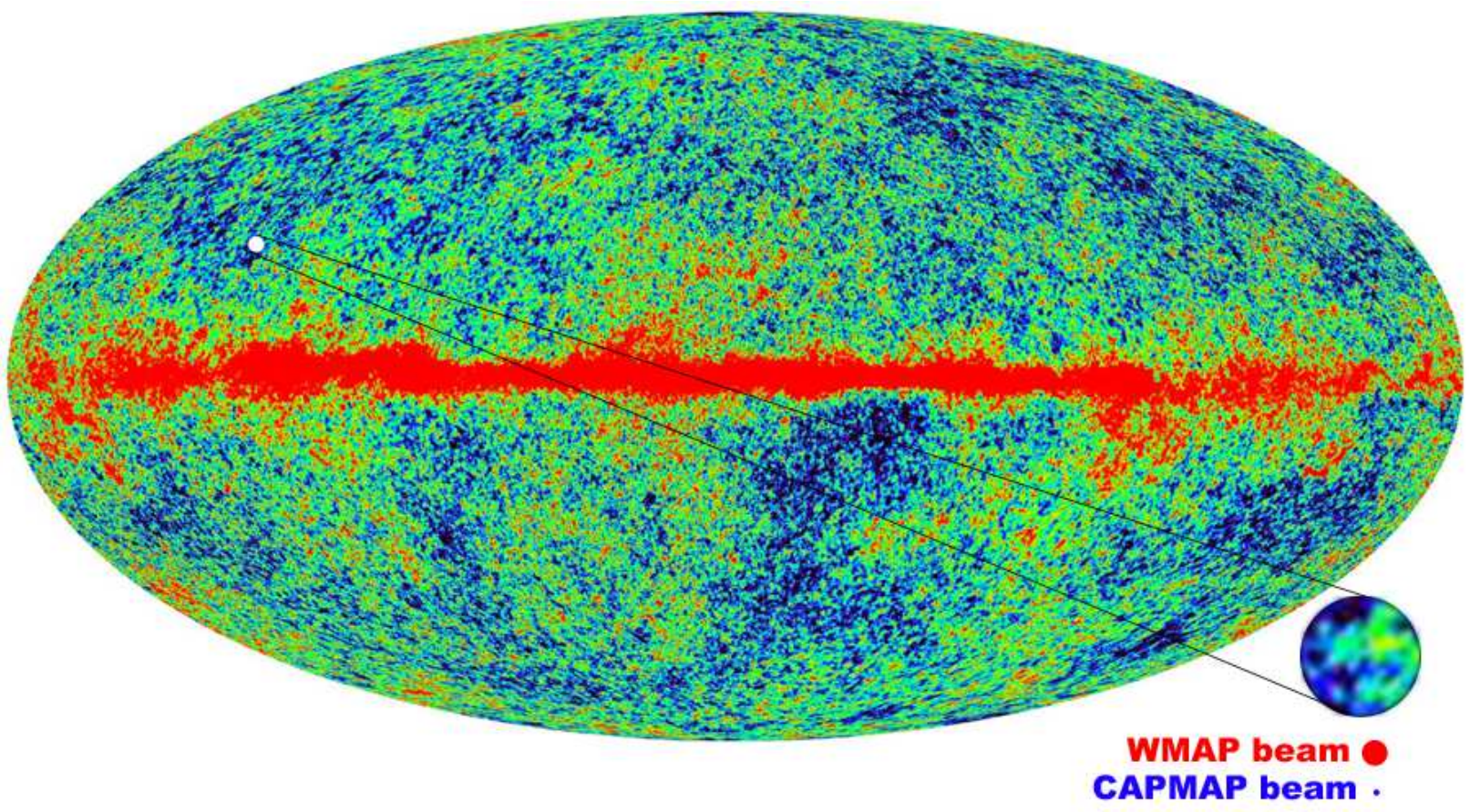

Figure 2.16: CAPMAP observes the white region of sky. Overlaid on top is the WMAP 94 GHz full sky map [93] in galactic coordinates. CAPMAP maps a 1° diameter cap centered on the NCP. Inset shows the zoomed WMAP region along with the WMAP 94 GHz beam and CAPMAP W-band beam scaled to the size of that region.

# Chapter 3

# The Polarimeter

This chapter describes in detail the W-band radiometers used by CAPMAP to extract the small polarized CMB signal. The components of the polarimeters and their functionality are summarized in Section 3.1. Section 3.2 considers the thermal and mechanical environment of the cryostat designed to keep the receivers cold. The tests done on the receivers to verify their basic functionality before fielding are presented in Section 3.3.

## 3.1 Description

The CAPMAP receivers are phase-switched heterodyne correlation polarimeter (Section 2.2), shown in Figure 3.1 and Figure 3.2. This section details how the radio frequency (RF) radiation from the sky in W-band (84-100 GHz) is amplified, filtered, then mixed down to an intermediate frequency (IF, 2-18 GHz) before being detected by the multipliers. Each receiver naturally separates into three sections: the RF, LO, and IF sections. A component list is provided in Appendix D. The special attention needed to integrate four receivers together into a single cryostat is described in Section 3.1.3. Table 3.1 provides a summary of the gains or losses through each component of the receiver.

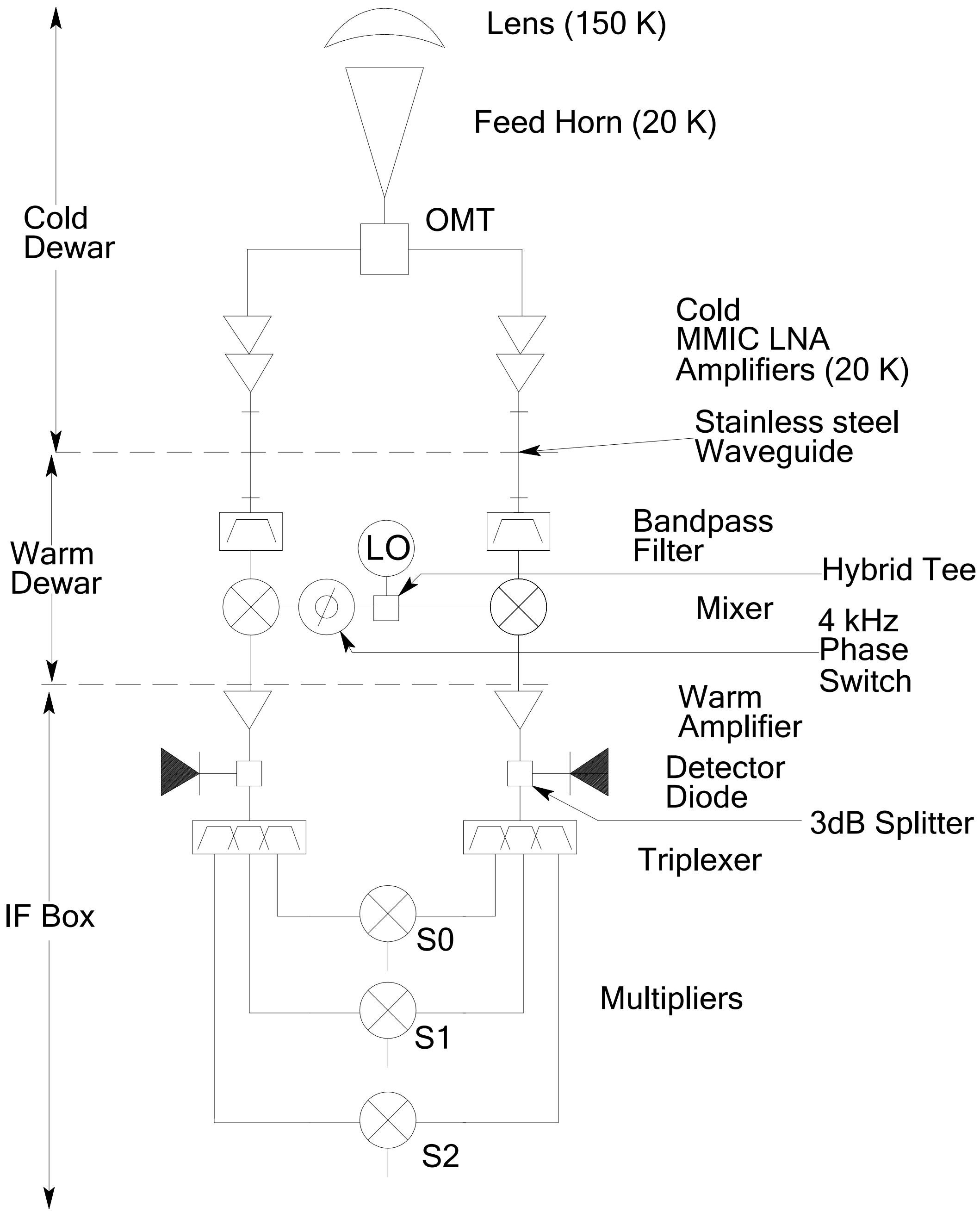

Figure 3.1: Schematic diagram of the CAPMAP correlation polarimeter.

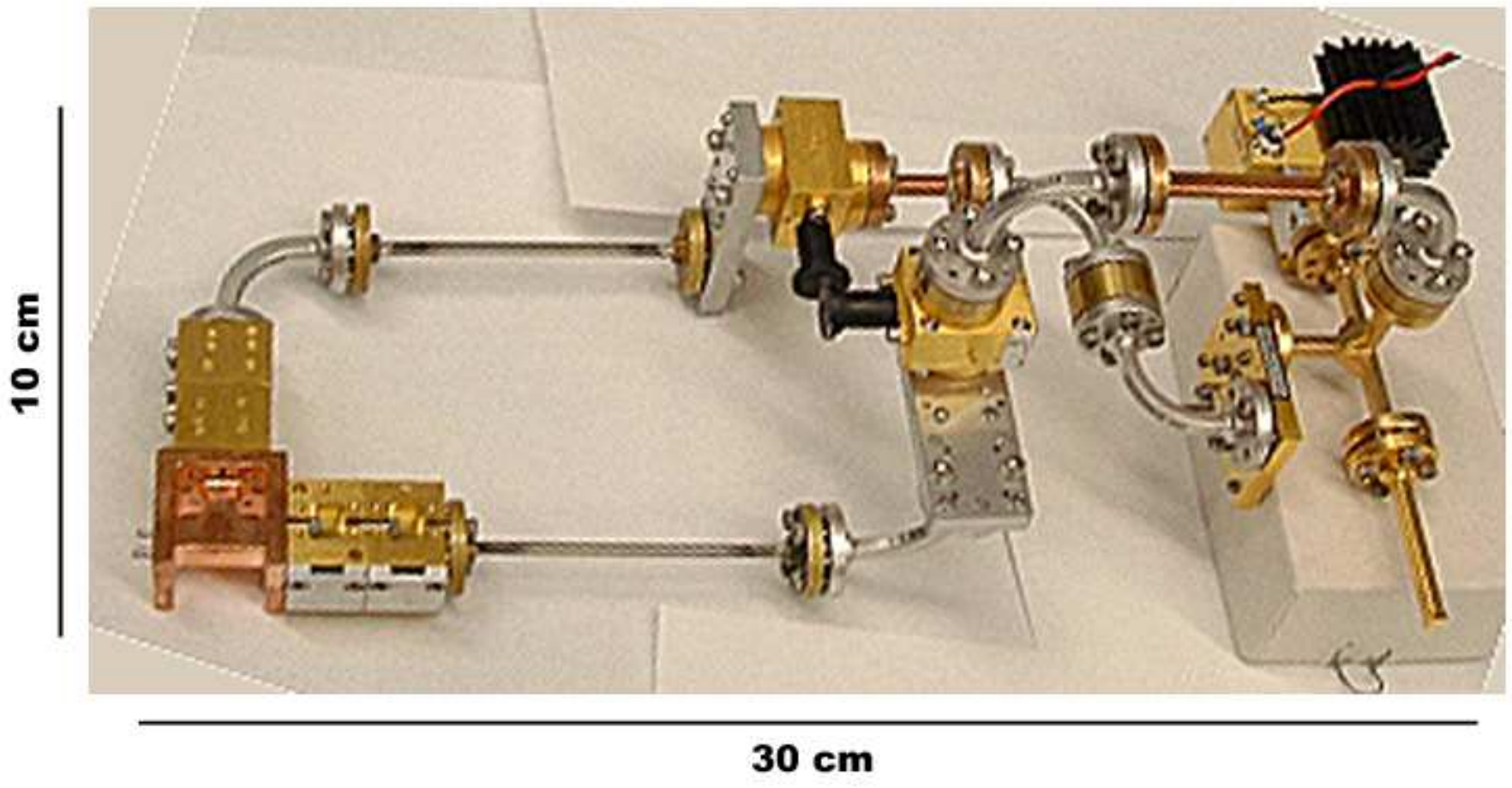

Figure 3.2: Photograph of the RF and LO sections of an actual W-band polarimeter. Detailed views with component names are shown in Figures 3.3 and 3.7, and Figure 3.10 for the IF section. A drawing of the dewar housing the four receivers is displayed in Figure 3.12. An entire W-band radiometer (without horn or lens) is 12" long and about 4 wide. The wires on the top right power the LO. In the CAPMAP03 season, a single LO operated the four radiometers. The leftmost part of the receiver is cooled to 20 K and the thermal break happens at the center of the 3" stainless steel (SS) waveguide. The receiver past the end of the SS waveguide is mechanically and thermally held at room temperature. The IF coaxial lines (not shown) attach to the two mixers which are capped off in the figure.

### 3.1.1 RF Section

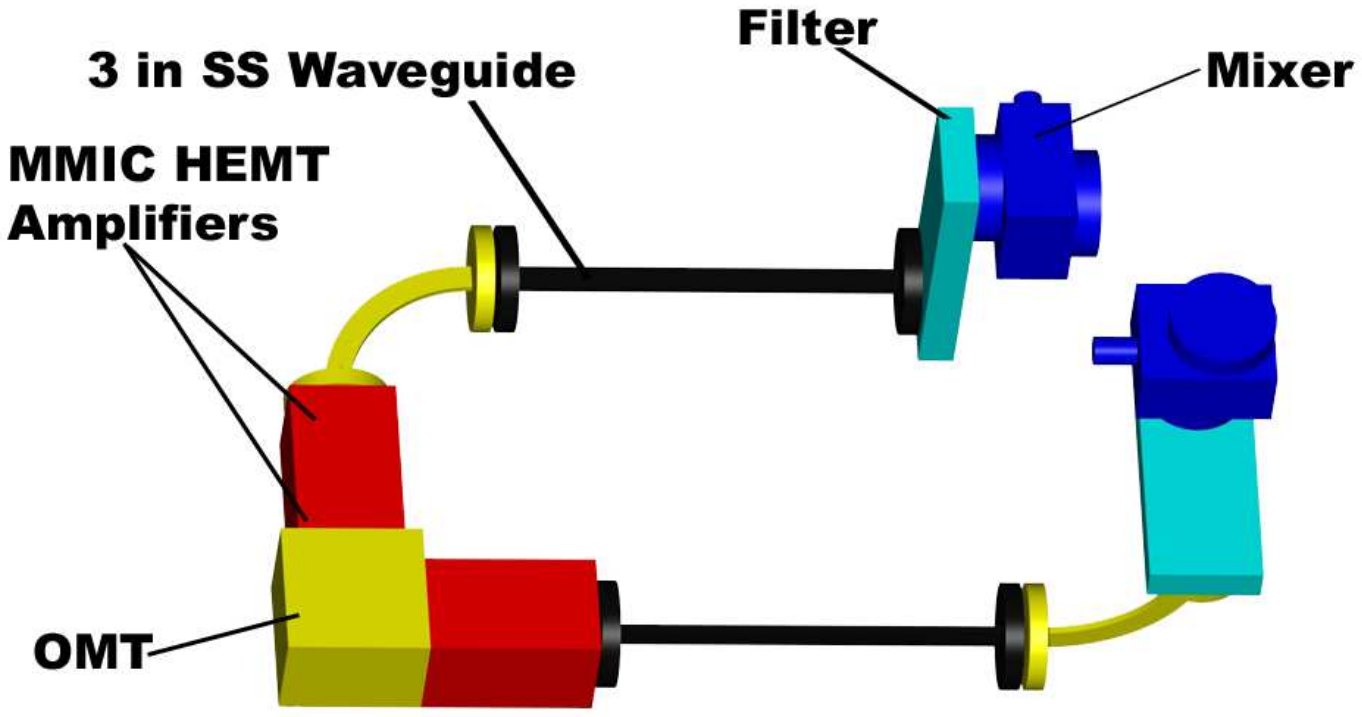

Figure 3.3: Scale 3D drawing of an RF section of a CAPMAP polarimeter. The RF section is shown in the same orientation as the receiver picture in Figure 3.2. The OMT attaches directly to the horn.

The RF section comprises the part of the receiver from the orthomode transducer (OMT) to the mixers. The radiation is collected by the corrugated feed horn (see Section 2.3) which couples a single transverse electric mode ($TE_{10}$) into WR-10 circular waveguide at the throat of the horn. The OMT then splits the radiation into two orthogonal polarization. The OMT is oriented on the sky such that it couples the vertical and horizontal polarized component

| Component | Physical Temperature[a] | Gain[b] | Contribution to the Noise Temperature[c] | Polarized Signal Level[d] | Total Power Signal Level[d] |
|---|---|---|---|---|---|
| Vertex Cab window | 300 K | - 0.005 dB | 0.4 K | $1.1\times10^{-12}$ $\mu$W | $1.1\times10^{-5}$ $\mu$W |
| Dewar window | 270 K | - 0.02 dB | 1.2 K | " | " |
| Lens | 160 K | - 0.08 dB | 3 K | " | " |
| Horn | 30 K | - 0.05 dB | 0.3 K | " | " |
| OMT | 30 K | - 0.05 dB | 0.3 K | $1.0\times10^{-12}$ $\mu$W | $1.4\times10^{-5}$ $\mu$W |
| MMIC HEMT 1 | 30 K | +22 dB | 50 K | $1.5\times10^{-10}$ $\mu$W | $3.9\times10^{-3}$ $\mu$W |
| MMIC HEMT 2 | 30 K | +22 dB | 0.3 K | $2.5\times10^{-8}$ $\mu$W | $6.1\times10^{-1}$ $\mu$W |
| H-bend | 30 K | - 0.3 dB | 0.01 mK | $2.3\times10^{-8}$ $\mu$W | $5.6\times10^{-1}$ $\mu$W |
| Stainless Steel wg. | 200 K | - 2 dB | 2 mK | $1.4\times10^{-8}$ $\mu$W | $3.5\times10^{-1}$ $\mu$W |
| Filter | 300 K | - 3 dB | 14 mK | $7.0\times10^{-9}$ $\mu$W | $1.7\times10^{-1}$ $\mu$W |
| Mixer | 300 K | - 12 dB | 30 mK | $4.4\times10^{-10}$ $\mu$W | $1.0\times10^{-2}$ $\mu$W |
| Attenuators | 300 K | - 10 dB | - | $4.4\times10^{-11}$ $\mu$W | $1.0\times10^{-3}$ $\mu$W |
| IF amplifier | 300 K | +45 dB | 0.45 K | $1.4\times10^{-6}$ $\mu$W | 30 $\mu$W |
| Splitter | 300 K | - 3 dB | - | $7.0\times10^{-7}$ $\mu$W | 15 $\mu$W |
| Attenuators[e] & Detector diode | 300 K | 0.1 mV/$\mu$W | - | n/a | 1.5 mV |
| Triplexer | 300 K | - 5 dB | - | $2.2\times10^{-7}$ $\mu$W | n/a |
| Attenuators & Multiplier | 300 K | 1.2 mV/$\mu$ W | - | 0.26 nV | n/a |
| ×100 Pre-amp | 300 K | +20 dB | - | 26 nV | 150 mV |
| DAQ | 300 K | $2^{30}$ bits/V | - | 27 bits | $1.6\times10^{8}$ bits |

Table 3.1: Summary of a CAPMAP03 polarimeter microwave properties.
[a] The physical temperature is the typical temperature of the components during the CAPMAP03 season.
[b]The gain and loss are conservative estimates, averaged over the band. These typical values come from actual measurements and components specifications (except for the horn). Differences between the two arms or among individual receivers are not noted.
[c]The contribution of each component to the total noise budget of the receiver is the effective temperature of that component, referenced to the primary mirror. An active element (amplifier) with a noise figure of $F_{dB}$ has a noise temperature in Kelvin of

$$T = (F_{linear} - 1) * 290$$

where the noise figure has been converted to linear using

$$F_{dB} = 10 * \log\ F_{linear}, \qquad \text{or equivalently} \qquad F_{linear} = 10^{F_{dB}/10}.$$

The noise temperature of a passive component (attenuator) at a temperature $T_{physical}$ in Kelvin with attenuation $L_{linear}$ is

$$T = (1 - L_{linear}) * T_{physical}$$

where attenuation is defined as the ratio of the outgoing to the incoming power. A -6 dB attenuator at 300 K has a noise temperature of 224 K. Finally, the effective temperature of a component which has an intrinsic noise temperature of $T_{noise}$, located after various elements whose total gain and loss add up to $G_{eff}$ is

$$T_{effective} = T_{noise}/G_{eff}$$

[d] The power level is calculated using $P = G_{eff} k_B T_{eff} \Delta\nu$, assuming a detector bandwidth of 16 GHz, a sky noise of 50 K, and a polarized signal level of 5 $\mu K$. The gain and noise temperature are the effective values up to that component.
[e] Each receiver's specific gains are different and the attenuator settings were tuned accordingly.

which means the detector is sensitive to the power difference between two diagonal polarizations, rotated 45° from vertical, as described in Section 2.2. The OMT has a typical return loss in each arm of -19 dB and an arm-to-arm power isolation of -35 dB averaged over the 84-100 GHz band. The arm-to arm isolation directly converts into an instrumental polarization[1] because it couples a signal reflected from one arm into the other which then appears as a correlated signal to the multiplier. Therefore a high isolation is equivalent to a good rejection of unpolarized signals appearing as correlated signal.

From the OMT down to the multiplier in the IF section, the signals in the two arms each proceed through a similar sequence of components to maintain the phase matching requirement. The only asymmetric components are the OMT and the phase-switch, accounted for in Section 3.1.3. The rest of the components are well enough phase matched a-priori that we can pick any two of them for a receiver. For all four receivers, the largest waveguide path length which had to be added to correctly balance the phase in both arms was 20 mils (0.5 mm). Precision phase tuning is carried out with in-line phase trimmers in the IF section (see Section 3.1.4).

Directly after the OMT are two Monolithic Microwave Integrate Circuit (MMIC) High Electron Mobility Transistor (HEMT) low noise amplifiers (LNA) in series which provide $\sim$22 dB of gain each at the price of a $\sim$50 K increase of the system temperature. More details on the amplifiers are provided in Section 3.1.2. The LNA are connected directly to the OMT even though the side arm of the OMT increases the size of the receiver. This geometry is preferred over an increase in system noise due to lossy waveguide bends. As shown in Table 3.1, the gain of the amplifiers is enough to make the noise due to loss through the following components insignificant. The two LNA are attached directly together. Although an early version of the LNA had a positive return loss out of band that caused standing waves to form between two such devices, the problem was fixed before the CAPMAP receivers were built. Each LNA is heat sunk to the fridge 20K stage via a 0.1”$\times$1” thick, 5” long flexible copper strap. The LNA have another thermal link to the 20K stage though the OMT via the horn mounted on the top 20 K plate (see Figure 3.12). The largest temperature difference observed between any LNA and the 20 K stage is 3 K.

After the LNA, the radiation is sent through a 3” stainless steel (SS) waveguide to the warm sections of the receiver. The center of the SS waveguide in each arm is heat sunk to the 70 K stage and at the bottom to a 300 K plate. Note the asymmetry in the H-bends: cold in one arm and warm in the other. The differential thermal contraction causes a phase

[1]Instrumental polarization is a polarized signal detected even though the input to the receiver is unpolarized.

shift, which is repeatable and is accounted for during the cold phase tuning of the receivers.

The signal comes out of the SS waveguide into a band defining filter. The −1 and −15 dB band-edge frequencies are (84,104) GHz and (82,107) GHz respectively. The typical insertion loss of the filters is -3 dB. Even though the band is defined by the filter, the ultimate limitation on the bandwidth comes from the OMT. The filter band edges are defined to limit the bandpass to where the OMT has acceptable return loss.

The last RF component is a single-balanced mixer which uses the strong 82 GHz continuous wave from the Local Oscillator (LO) to convert the weak RF signal down to 2-18 GHz. The mixers multiplies the RF and LO signal using two diodes [122], generating the sum and difference of the LO and RF frequency. A low-pass filter on the output means only the difference emerges on the coaxial IF port. The IF signals are carried out of the dewar through long phase-matched flexible coaxial aluminum cables into the IF box.

### 3.1.2 MMIC HEMT Amplifiers

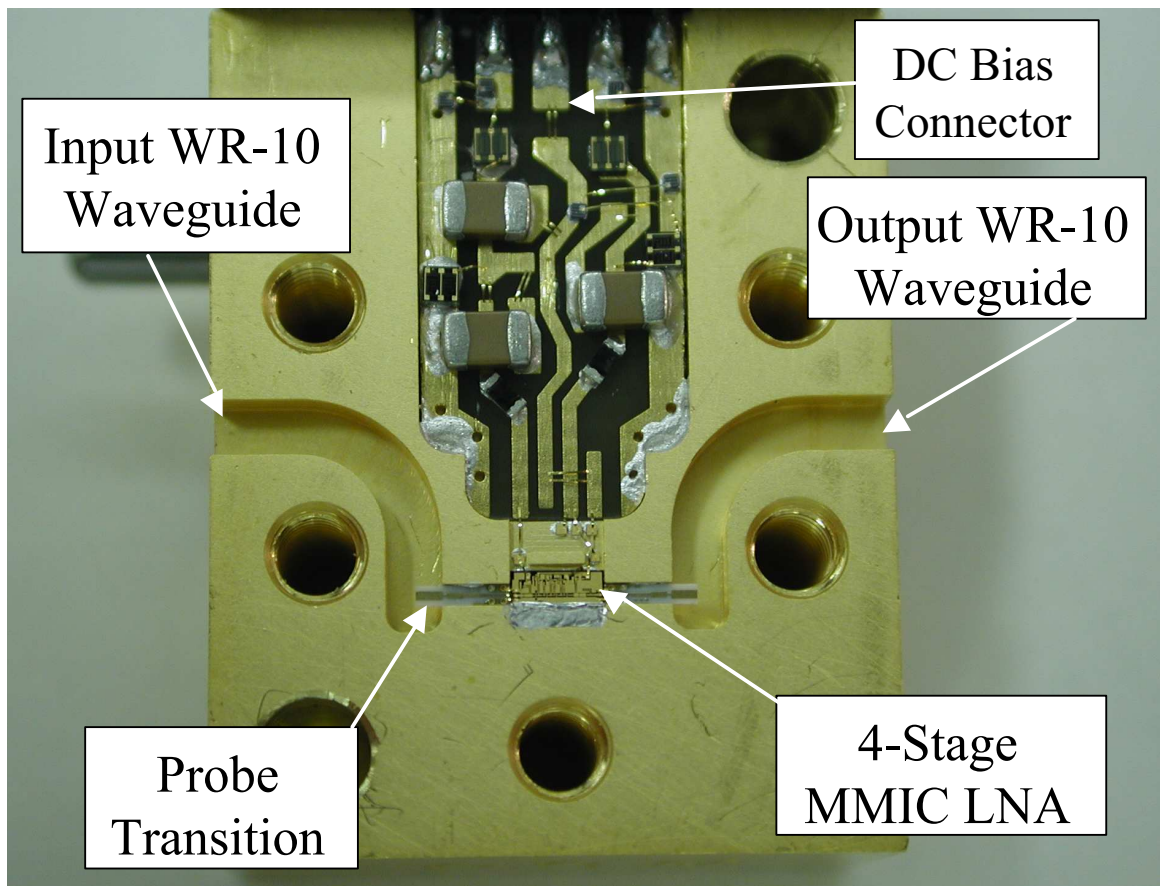

Figure 3.4: Photo of one half MMIC LNA housing. The input and output waveguides are visible. The small circuit in the lower center is the actual amplifier. The housing require a precision finish (Appendix B) to ensure a good seal of the two faces when the device is closed.

The first stage amplification is provided by cooled Monolithic Microwave Integrated Circuit Low Noise Amplifiers (MMIC LNA)[2]. These devices are based on cryogenic InP HEMT (High Electron Mobility Transistor) technology. They are used as front-end amplifiers for many scientific instruments [94, 43] at operating frequencies up to 240 GHz. Their low noise temperature, high gain, low power dissipation, small size, and controllable phase make them excellent amplifiers for a CMB polarization instrument.

[2]MMIC LNAs provided by Todd Gaier, JPL.

Figure 3.4 shows a W-band MMIC LNA housed inside its metal body. The microwave amplifier itself is the 2×0.8 mm circuit in the center. The specific amplifiers used for CAPMAP have four stages of transistors in series connected together such that a single drain and two gates are sufficient to bias it. The amplifiers are designed to be cooled because their noise temperature  decreases with their physical temperature. Figure 3.6 shows that the noise temperature is a steep function of the physical temperature. Below 35 K physical temperature, further cooling does not improve the noise performance.

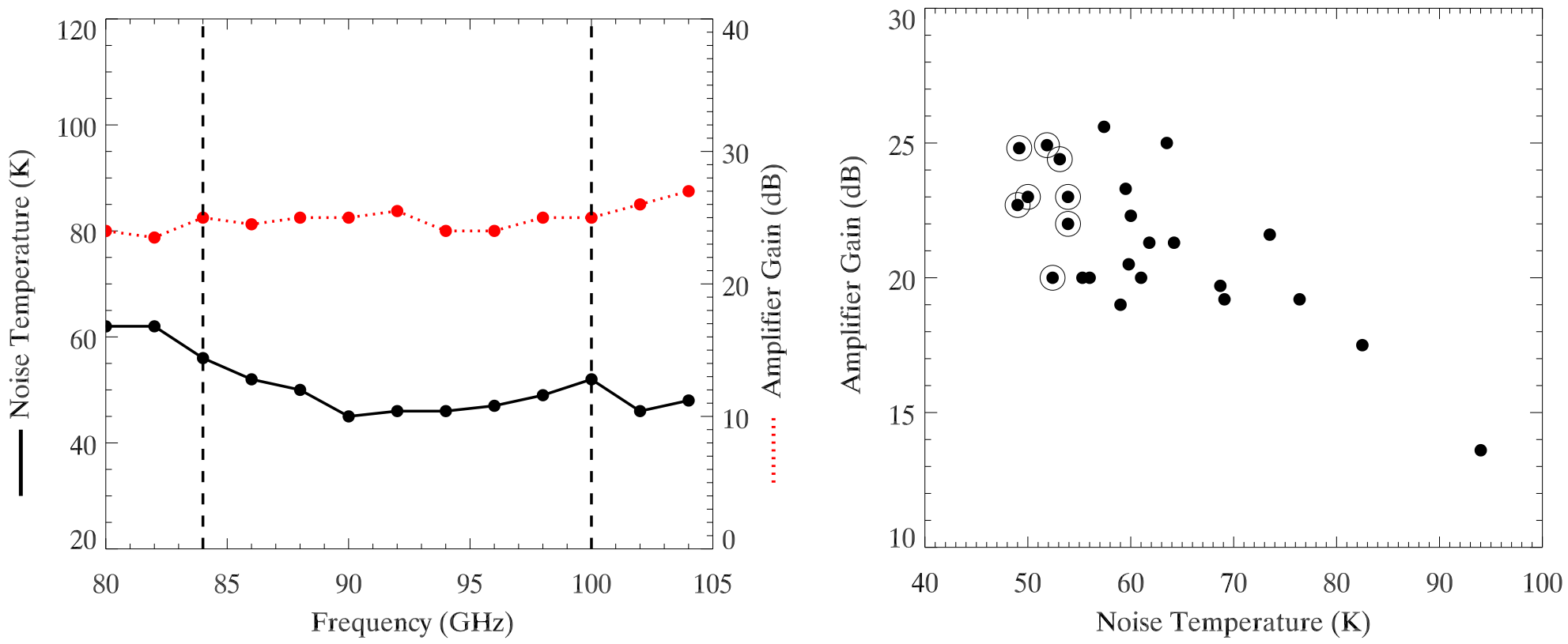

Figure 3.5: (a) Typical gain and noise performance of the W-band MMIC amplifiers across the 84-100 GHz band. The typical band-averaged gain is 22 dB and the noise temperature 55 K.
(b) Histogram of performance for the 26 devices delivered for CAPMAP03. The eight quietest device (circled) were chosen as the first amplifiers in the eight receiver arms.

The performance of these amplifiers is excellent. Figure 3.5(a) shows a typical gain and noise temperature as a function of frequency in the CAPMAP W-band. Gains of 25 dB with noise temperatures of 55 K are typical at cryogenic temperatures. Figure 3.5(b) shows a histogram of the first 26 amplifiers delivered to operate CAPMAP03. The eight quietest amplifiers (circled) were chosen as the first amplifiers in the receiver chain. Noise performance is less critical for the second amplifier in the chain.

The amplifiers can be tuned in-situ. They are voltage biased. The drain and two gate biases are adjusted, which varies the drain current, causing a change in the gain or the noise over the band of interest. An example of the strong dependency of the noise temperature on the drain current is shown in Figure 3.6. The bias is adjusted to provide at least 20 dB of gain and less than 60 K for noise for the front-end amplifiers. Typical bias settings yield a power dissipation on the order of a few mW, which is a negligible load for the cryogenic system.

The amplifier bodies (see Figure 3.4) were built in the Princeton machine shop and finished in Princeton. Appendix B summarizes the steps to produce an amplifier body ready to be integrated. The housings were subsequently sent to JPL where the MMIC and biasing circuits were installed and the amplifiers tested cryogenically to determine the optimal bias settings. The phases of any of the 26 first devices were similar enough that any pairs could be chosen when balancing receivers.

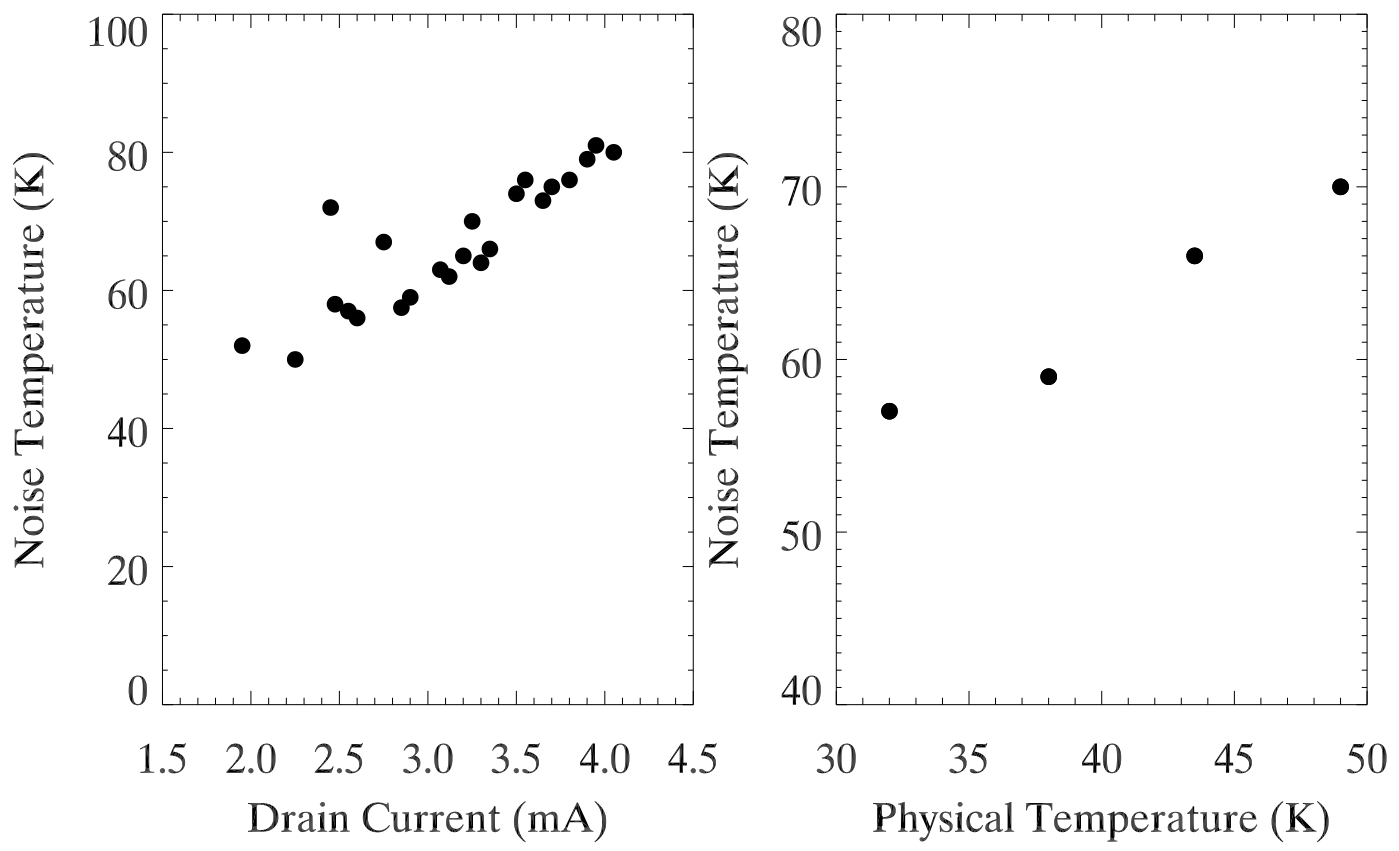

Figure 3.6: Noise temperature dependence on the drain current (left) and the physical temperature (right) for CAPMAP MMIC LNAs.

### 3.1.3 LO Section

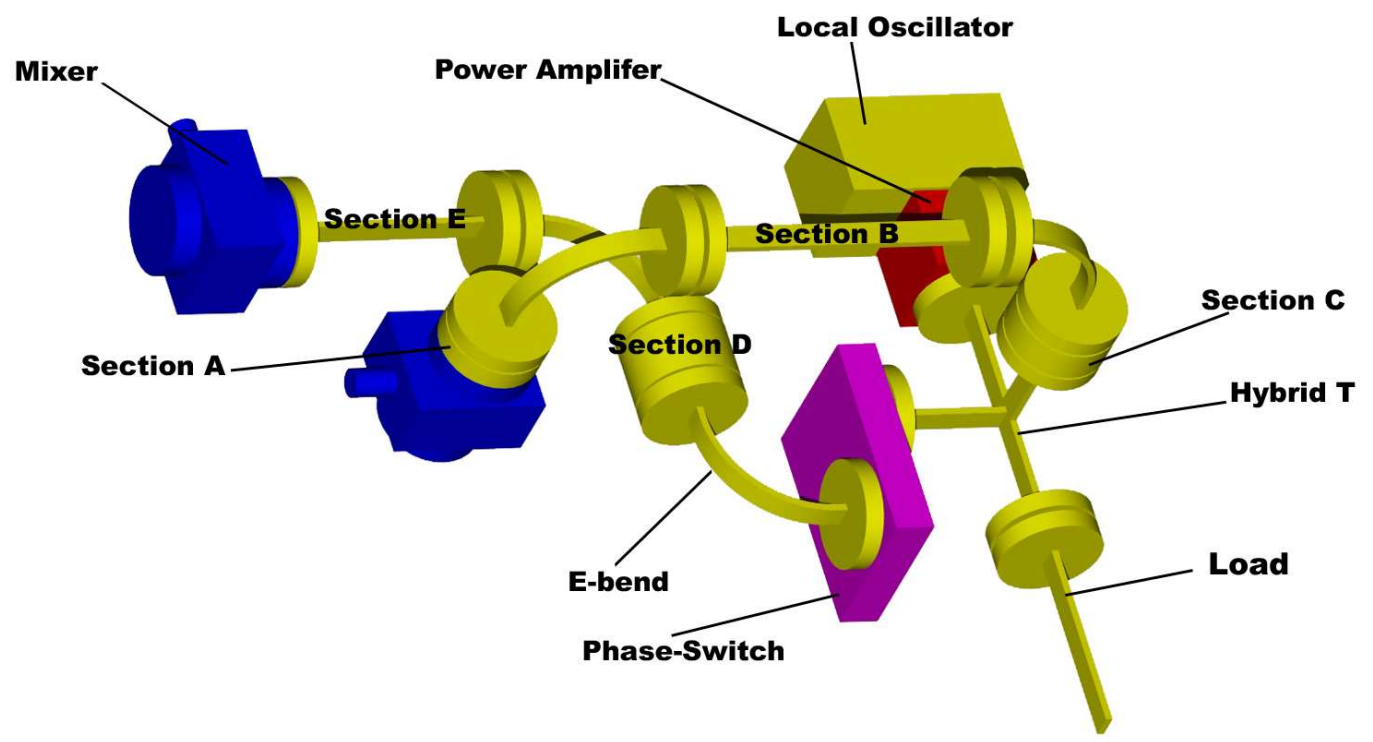

Figure 3.7: 3D drawing of the LO section of receiver. The Local Oscillator is only present in one of the receivers (Section 3.1.3)

The LO is a narrow-band high-power source at 82 GHz. Figure 3.7 shows a drawing of the LO section. The LO signal comes in through the power amplifier and is split between the two lines by the hybrid T. The lower line in the figure contains the phase switch which modulates the LO signal at 4 kHz by alternatively inserting 180° and 0° of phase. The rest of the path to the mixer is waveguide, composed entirely of standard E-bends[3] and custom-built straight sections[4]. The geometry of the two arms is as symmetric as possible to deliver similar power to the mixers. At the same time, the arrangement of the bends and custom sections allows enough flexibility to account for differences in the path length of the two arms in any of the three axes. The length of the custom section has to be adjusted prior to the final phase tuning (see **A-Priori Phase Matching**).

The choice to operate the mixers and filters warm largely simplifies the LO section. The LO section is housed entirely in the dewar, at room temperature. This reduces the length of waveguide necessary to bring the LO signal to the mixers, and obviously the need for waveguide vacuum feed-through. The thermal break happens in the stainless steel lines in the RF section so the LO waveguide is made of copper, less lossy than stainless steel. The LO dissipates ∼2 Watts and the power amplifier ∼0.5 Watts. It is important to isolate these from the cold section of the dewar. All the warm parts of the receiver remain under the bottom of the 70 K shield (the pies). Having the LO in the dewar also leaves the LO distribution to the receivers completely flexible (see **LO Distribution System**).

**LO Power Level** The mixers require at least 10 dBm[5] of power to operate in the regime where they are insensitive to variations in the LO power. The lossy elements between the power amplifier and the mixer are listed in Table 3.2. Therefore, if we want about 10 dBm at the mixers, we need roughly 19 dBm going into the hybrid tee. The LO alone can provide approximately 17 dBm of power. Thus, a power amplifier is needed. The power amplifiers are custom built[6] and provide about 10 dB of gain with a 1 dB compression output point of 90 mW (19.5 dBm), and a maximum output of 21.5 dBm (see Figure 3.8). Therefore, the minimum power necessary at the input of the power amplifier is approximately 9 dBm. Note that the phase switch attenuation is slightly different for the two states of the phase switch. The current delivered to the phase switch can be tuned around the nominal ±15 mA values to balance the power delivered to the mixer. Balancing the 4 kHz offset helps improve the rejection of polarization channels to total power.

**LO Distribution System** The LO sits inside the dewar for the various reasons listed above (less waveguide plumbing to connect to the mixers, more flexibility to position to four receivers, no need for waveguide feed-through). There are however, downsides to having a

[3]by Aerowave.
[4]Built by the Princeton machine shop.
[5]0 dBm = 1 mW
[6]by Todd Gaier, JPL

| Component | Loss |
|---|---|
| Hybrid T | 5 dB |
| Phase switch | 2.3 dB |
| 2 E-bends | 0.6 dB |
| 1" Cu waveguide | 0.2 dB |
| Total | 8.1 dB |

Table 3.2: Lossy elements in the LO section. The numbers are conservative estimates, taken from measurements (hybrid T) and data sheets.

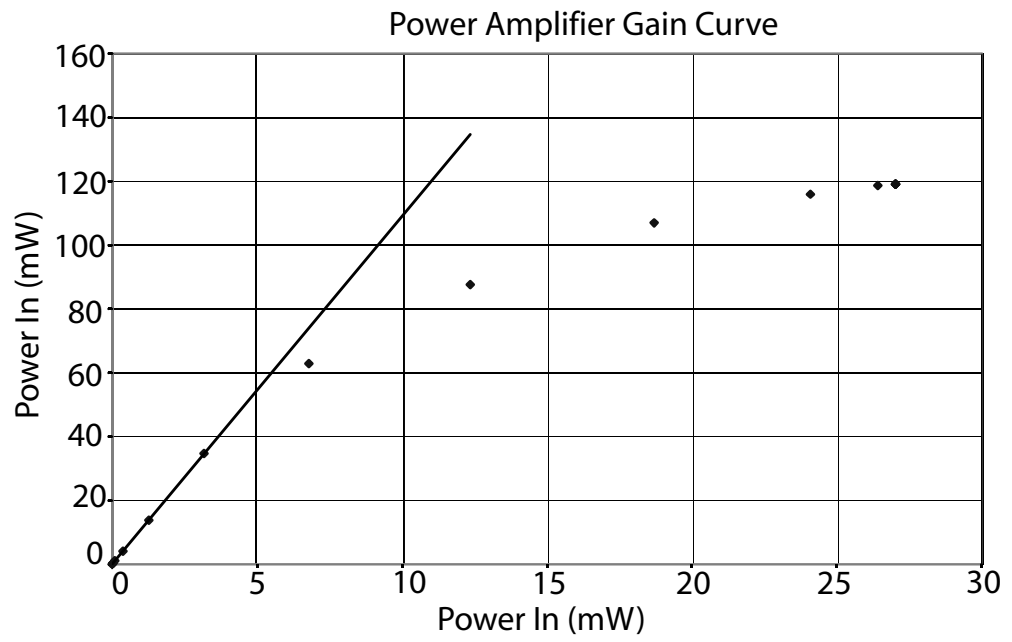

Figure 3.8: Power amplifier gain curve. The output is linear until approximately 50 mW output power. The maximum output is 120 mW. Figure and tests by P. Hamlington.

100 mW, 82 GHz source alongside receivers sensitive to $\mu$W power levels at 84-100 GHz.

Initially, each receiver in the dewar had its own LO. Each LO was tuned to operate at 82 GHz but the four differed from one another by a few 100 MHz because each LO's frequency is both temperature dependent (~5 MHz/°C) and bias-voltage dependent (200 MHz/V). We discovered that the LO radiation was leaking out of the LO-section waveguide joints, propagating throughout the reflective environment of the dewar and coupling back optically through all four horns into the RF sections. The band-defining filter attenuation is ~-15 dB at 82 GHz, so the LO radiation was mixed down and amplified by the IF amplifiers. The IF amplifiers have gain all the way down to DC. Because of the strength of these narrow-band, mixed-down LO signals, they rebiased the IF amplifier.

We became aware of this problem by probing the IF signal coming out of one receiver's mixer with a spectrum analyzer. In the 100 to 400 MHz region, we found 3 sharp peaks, ~40 dB above the IF noise level, whose frequency shifted with the temperature and voltage of the three other LOs. These peaks were then mixed down LO radiation from the three other LOs. The LO radiation leaked *out* through the waveguide joints and was coupled *in* through the feed horn because covering the horn of a receiver made the lines in its IF spectrum disappear, but covering any of the other horns did not. In addition, putting pieces of eccosorb in and out of the LO section region strongly affected the levels of the three peaks. We concluded that the LO waveguide joints, even correctly tightened, leaked at a -30 to -40 dB level. This effect was already known to WMAP [71].

We opted for two radical solutions to solve this problem. The first one was to minimize leakage of radiation from the LO waveguide joints. Since any kind of mechanical modulation of the dewar (ie. 1.2 Hz vibration from the fridge) can affect this freely propagating LO signal and thus produce a direct modulation of the IF output, it is crucial to keep the amount of LO power leaking into the dewar to a minimum (this statement remains true even for a single LO in a dewar). The eight waveguide joints in the LO section were each enclosed in a 1/8" thick strip of eccosorb (adhesive backing VF-30 eccosorb strips) and then encased with aluminum tape. To monitor the residual strength of the leaking LO signal, we placed an 82 GHz total power receiver with a horn pointing in the general direction of the LO sections. The taping was carefully repeated on each receiver until the peaks were not visible on the spectrum analyzer (reduction of the leaked LO signal by 40 dB).

The second solution was to eliminate all but one of the four LOs. This decision required a major redesign of the LO distribution system. Instead of attaching one LO per receiver, a single LO is fed to each of the four receivers via a distribution system which connects to each receiver in series. The E and H port of the first hybrid tee naturally connect to the two arms of that receiver as shown in Figure 3.7. We measured the transmission coefficients of a few hybrid tees and found that the co-linear port only attenuates the signal by 8 dB[7]. This meant that the leftover LO signal from one receiver's hybrid tee could be sent to the next as long as the power amplifier in front of each hybrid tee provides enough gain to satisfy the 19 dBm requirement stated above. The loss through the hybrid tee and through the 6" of waveguide necessary to connect each receiver adds up to 9 dB.

The final configuration of the LO distribution system is shown in Figure 3.9. The input to C's power amplifier is therefore $21.5 - 9 = 12.5$ dBm, enough to provide the correct amount of power to that receiver's mixer and to the following receivers. The receivers are connected together in this manner and the gain of each power amplifier is tuned to give the same amount of power to each receiver.

**A-Priori Phase Matching** All active components but the OMT and phase switch are the same in both arms. Components were purchased in phase-matched pairs and their relative phase over the RF band was specified to be within 10°. The absolute phase is irrelevant. The length of each arm of the receivers is therefore balanced a priori. The remaining relative phase between the arms are expected to be small and can be corrected with waveguide shims. The final phase tuning is presented in Section 3.3.1. Additional details about choice of twists and turns of the LO section are provided in Appendix C.

[7] A hybrid tee is a poor version of a magic tee. A magic tee provides a very good isolation between its co-linear ports (> 30 dB isolation) and only 3 dB loss from the co-linear to the E and H ports.

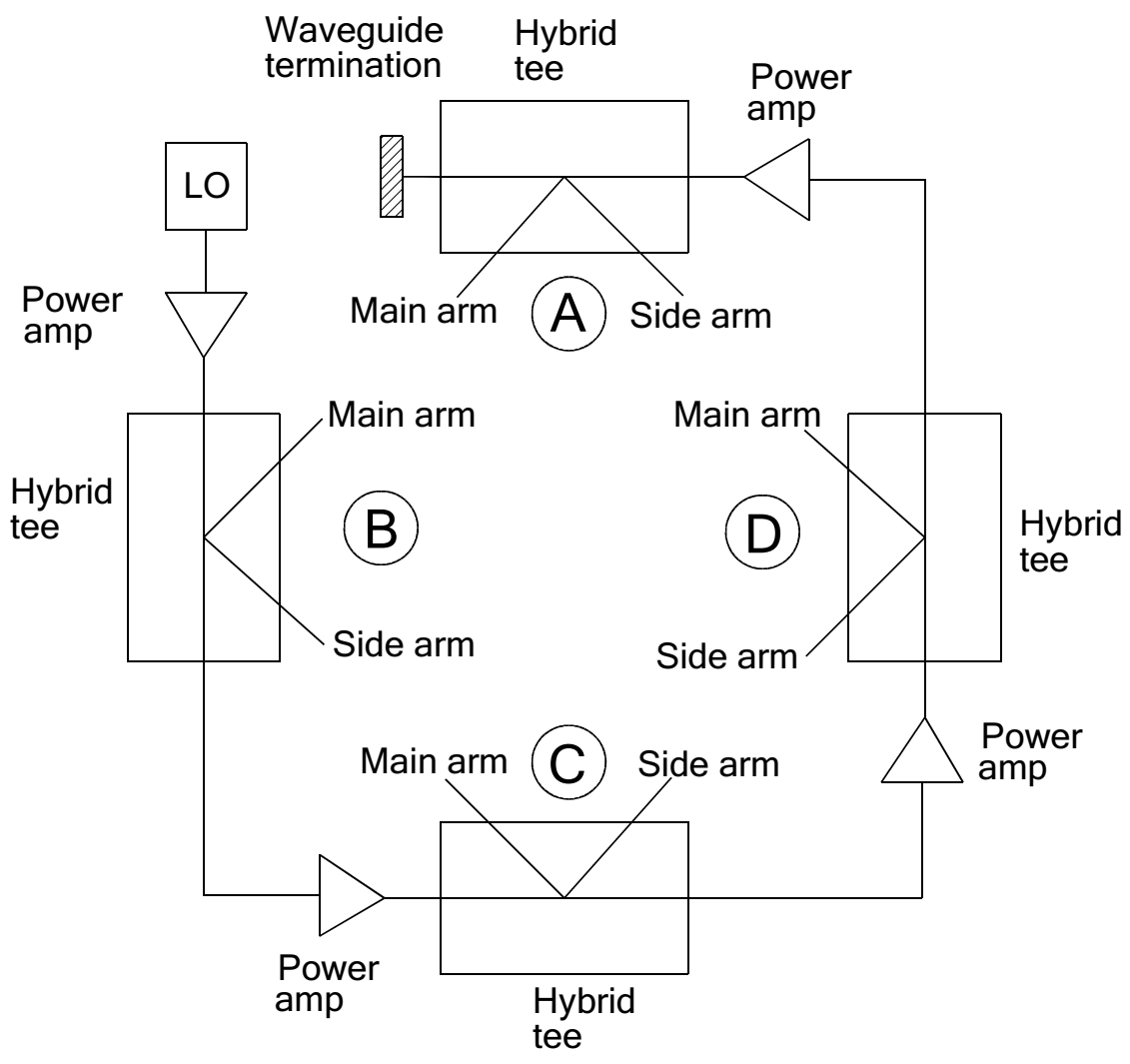

Figure 3.9: Final LO distribution configuration for CAPMAP03, seen from above. The IF box is near receiver A. Each receiver is connected to the next power amplifier via a custom-made ~6" copper waveguide.

### 3.1.4 IF Section

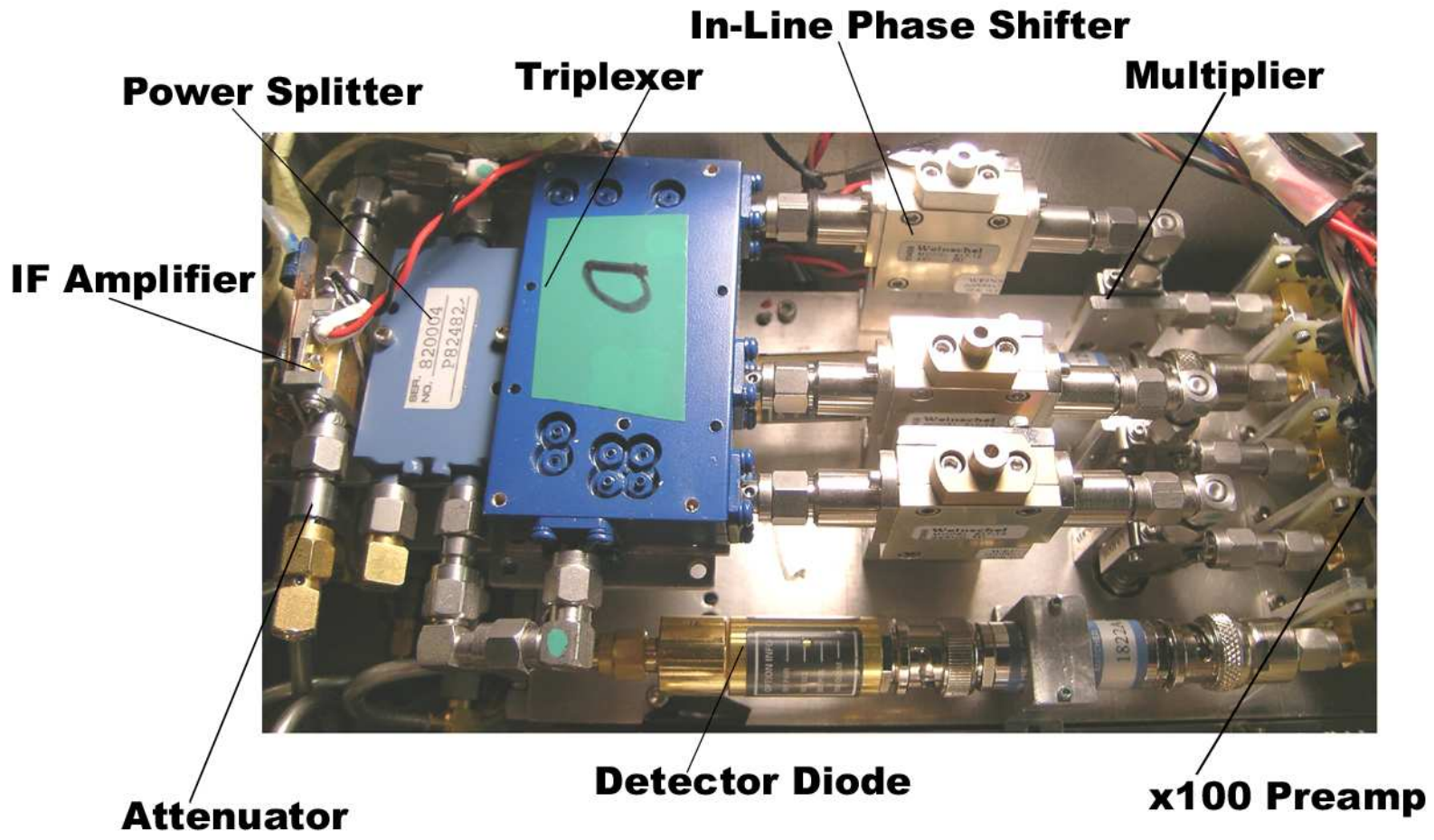

Figure 3.10: One arm of a W-band receiver IF module. Individual components are described in the text. All connections to the preamp are SMA. Only a small part of the preamp card is visible here.

The IF section of the radiometer provides the final amplification and filtering of the modulated, down-converted (2-18 GHz) radiation before the detection by the multipliers.

A single IF module is shown in Figure 3.10. For CAPMAP03, four such modules were housed in an RF-tight 9”×23” box (IF box). At these frequencies the signals are carried by coaxial lines rather than waveguide. The eight IF signals (the two arms of the four receivers) exit the dewar and enter the IF box via hermetic bulkhead feed-throughs. Each IF arm contains the same components so apart from imperfections between pairs of devices, the IF section is phase-matched prior to any tuning. A single arm of the IF chain consists of the following components:

**IF amplifier** It provides an additional ∼45 dB of gain over the 2-18 GHz band (the amplifier actually has gain all the way to DC). The amplifier's effective noise temperature is insignificant compared to that of the cold amplifiers. The IF amp's maximum noise figure is 2.5 dB which translates into a noise temperature of 225 K. The maximum loss through the components between the MMIC amplifier and the IF amplifier is 17 dB (see Table 3.1). Given 44 dB gain for the two MMIC amplifiers, the net gain before the IF amplifier is 27 dB, which means that the contribution of the IF amplifier to the total system temperature is 0.4 K. The IF amplifiers dissipate ∼2 W each and have a small surface area which make them particularly temperature sensitive to their environment. With the IF box closed, they run as warm as 60°C but remain below 30°C with the IF box open. The IF amplifiers are temperature-controlled by servoing thermoelectric coolers mounted on the outside of the entire IF-box. This is critical because the IF amplifiers display a strong temperature dependant gain (dGain/dT $\simeq$ 0.5 %; see section 6.5.1). The CAPMAP03 IF amplifiers were designed for cryogenic use and lacked protection circuitry. Early during integration, two were damaged and returned to the manufacturer to have protection circuitry installed. The protected IF amplifiers operate from 12 V bias supplies rather than the 8 V needed by the original IF ampplifiers.

**Power Splitter** In each arm, the signal is split into two parts. Half of the radiation continues to the rest of the IF chain and the other half is detected via the detector diodes. The total power signals detected in each arm (D0 and D1) provide a way to characterize the properties of the receiver during observations (see Chapter 6). The detector diodes' typical gain is 0.1 mV/$\mu$W and they have a square law response for output voltages up to 10 mV. Attenuators are placed before the detector diodes as needed to keep their responses in the linear regime.

**Filter Bank** The remaining radiation from the power splitter is fed into a filter bank

(triplexer) which splits the 2-18 GHz band into three equal sub-bands. These sub-bands are referred to as S0 (2 to 7 GHz), S1 (7 to 12.7 GHz), and S2 (12.7 to 18 GHz). The purpose of this splitting is both to provide frequency discrimination of the detected polarized signal and also because the large bandpass is more easily phase-matched over small segments of the whole band (see Section 3.1.5).

**In-Line Phase Shifter** These introduce a frequency-dependent phase in the coaxial lines by "stretching" the coaxial line length. For a line length change of $\delta L$, the phase shift introduced at a wavelength $\lambda$ is

$$\delta\phi = k\delta L = \frac{2\pi\ \delta L}{\lambda} \tag{3.1}$$

The phase shifter is only in one of the arms. The phase shifter can only change the line length from 78.5 mm (fully screwed in) to 67.9 mm (fully screwed out), which allows a phase change of 60° for S0, 120° for S1 and 180° for S2. The procedure for phase tuning with these is described in Section 3.3.1.

**Multiplier** Each sub band is converted into a DC voltage (0 to 100 MHz) by the multiplier. The output of the multiplier is proportional to the product of the two inputs as long as the two inputs remain small. The specific properties of the multiplier, relevant to its behavior as a correlating device are discussed further in Section 3.1.5

**Preamp** Each 4 kHz-modulated polarization channel and each total power channel is sent to a differential op-amp amplifier (×100 gain) followed by a buffer.

In addition to SMA connectors to connect the IF components, coaxial attenuators are added before the IF amp, between the IF amp and the detector diodes, and between the IF amp and the multiplier to keep these active components within their specified linear regimes. Typical attenuation values are provided in Table 3.1. The signals exits the IF box through filtered connectors and are conveyed to the data acquisition system, described in Section 5.2.

### 3.1.5 Multiplier

For a correlation polarimeter, the multiplier acts as the detector which converts the radiation into a voltage. The multiplier is a double-balanced mixer which yields the broad-band product of its two inputs using four diodes in a bridge configuration [53, 6]. The use of a mixer as an analog continuum correlator has many advantages [114] over standard correlators : it is wide-band, has a good rejection ratio of uncorrelated noise, and is extremely simple to operate. Some of its limitations, however are described below.

If $E_1$ and $E_2$ are the instantaneous time varying electric fields (at 2-18 GHz) at the input to the two arms of the multipliers, then the output voltage is given by

$$V_{out} = \pm g \, \langle E_1 \; E_2 \; \cos\Delta\phi\rangle + \epsilon_1 \, \langle E_1^2\rangle + \epsilon_2 \, \langle E_2^2\rangle \; , \tag{3.2}$$

where the $\langle \quad \rangle$ denotes a time average on time scales long compared to the typical period of the IF electric fields, $\Delta\phi$ is the relative phase shift of the two arms, and $g$, $\epsilon_1$, and $\epsilon_2$ are response coefficients. The $\pm$ shows that the polarized output is modulated at 4 kHz. For completely uncorrelated inputs the $\langle\cos\Delta\phi\rangle$ term averages to zero because the electric fields have random phases, so only the total power terms remain. This is the property we are interested in because it allows the extraction of a small correlated input (the CMB polarization signal) from a large uncorrelated noise (the detector and sky noise).

It is already known [53] that Equation 3.2 is only valid for correlated inputs below 10 $\mu$W. For larger correlated inputs, the response stops behaving linearly. Typical values of the response coefficients [6], averaged over the 2-18 GHz band are: $\overline{g} = 1.2 \pm 0.2$ mV/$\mu$W, $\overline{\epsilon_1 + \epsilon_2} = 0.01 \pm 0.02$ mV/$\mu$W.

Another limitation of the multiplier is that the *correlated* response coefficient $g$ can also be compressed if the uncorrelated inputs are too high. We find that for *uncorrelated* noise at the input of the multiplier larger than 100 $\mu$W, $g$ starts saturating. This turns into a direct limit to the maximum amplification before the multiplier.

The mixers used for CAPMAP as multipliers are not specifically designed to be symmetric in their two arms. As a result, the relative phase shift between the arms has a component that cannot be removed by a variable line length (ie. the phase shift does not follow Equation 3.1). The relative phase shift due to a typical multiplier device alone is shown in Figure 3.11 as a function of frequency. The figure shows that although the intrinsic phase shift due to the multiplier cannot be removed over the whole band, it varies slowly enough in frequency that it can be removed by adding an appropriate line length in each of three frequency sub bands.

Finally, the most obvious limitation is the fact that such a multiplier only operates at IF frequency. A multiplier at RF frequency would circumvent the need for an LO and highly simplify the receiver design. Unfortunately, no such mixer has been sufficiently characterized at W-band to operate as a wide-band analog correlator. This seems like a natural avenue for future coherent polarimeter research.

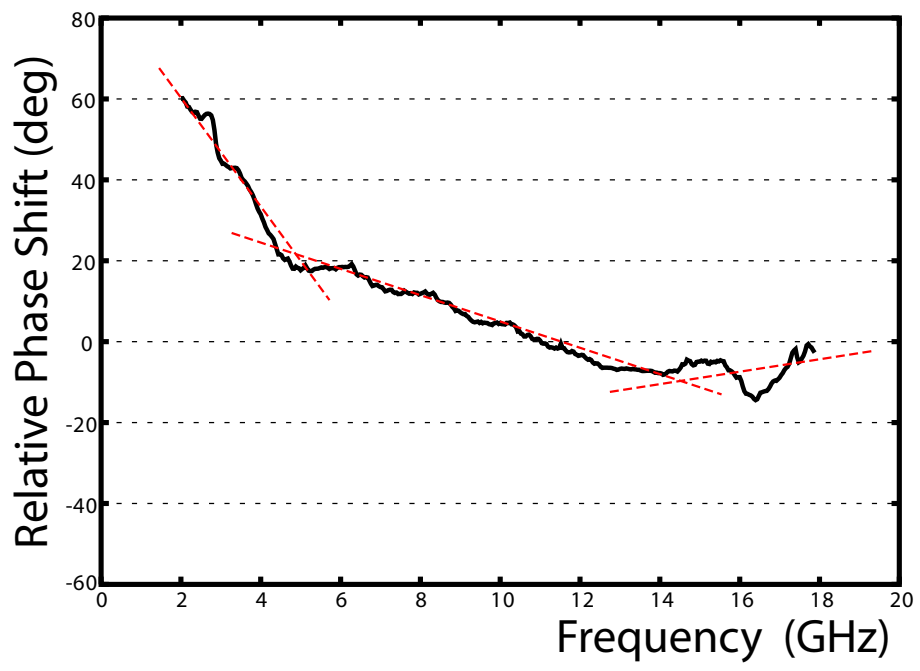

Figure 3.11: Intrinsic phase difference between the two arms of the multiplier itself. The phase shift can only be removed in small regions but not for the totality of the band. The red dashed lines show for example that the phase can be removed if the full IF bandpass is divided in three sub-bands

## 3.2 Thermal and Mechanical Considerations

To minimize the receiver noise temperatures, the physical temperature of the amplifiers is lowered to below 30 K. To reach such low temperatures, the receivers is enclosed in a large cryostat and evacuated to ~0.1 mTorr[8] to minimize convective heat loads. This section summarizes the design of such a cryostat and associated mechanical structure.

### 3.2.1 Cryostat

The receivers are cooled with a closed-cycle mechanical refrigerator. The fridge operation is based on the compression and expansion of Helium gas. Compared to liquid cryogen system, it has the advantage that it can operate indefinitely because the cooling fluid is continuously recycled. The CAPMAP cryogenic cooler is composed of a two-stage mechanical fridge (1020 CP) and a compressor (1020 R) from CTI[9] cryogenics [28].

The design of the dewar is based around the geometry of the mechanical fridge (cold head). A picture of the CAPMAP03 dewar during installation of the receivers as well as a CAD drawing are shown in Figure 3.12. The structural dimensioning of the dewar follows closely the NRAO cryogenic design guidelines [12].

The dewar is 22" diameter by 23" high cylinder without the cold head or the IF box. The vacuum shell is made of two sections of 3/32" rolled stainless steel sheet welded hermetically at the seam[10]. Stainless steel was chosen over aluminium to make the welding easier. The bottom vacuum shell has two 4.75" diameter hermetic ports, one next to receiver D to attach vacuum gauges, and the other next to receiver A to connect to the IF box. The top and bottom lids are 0.6” thick Aluminium disks with 4 holes for electrical vacuum feed-through in the bottom and for the microwave windows on the top. The full dewar with radiometers weighs ∼300 lbs.

[8] 1 Torr = 1 mm of Hg = $\frac{1}{760}$ atmosphere
[9] `http://www.helixtechnology.com`
[10] An error in the material order caused the shell to actually be twice as thick.

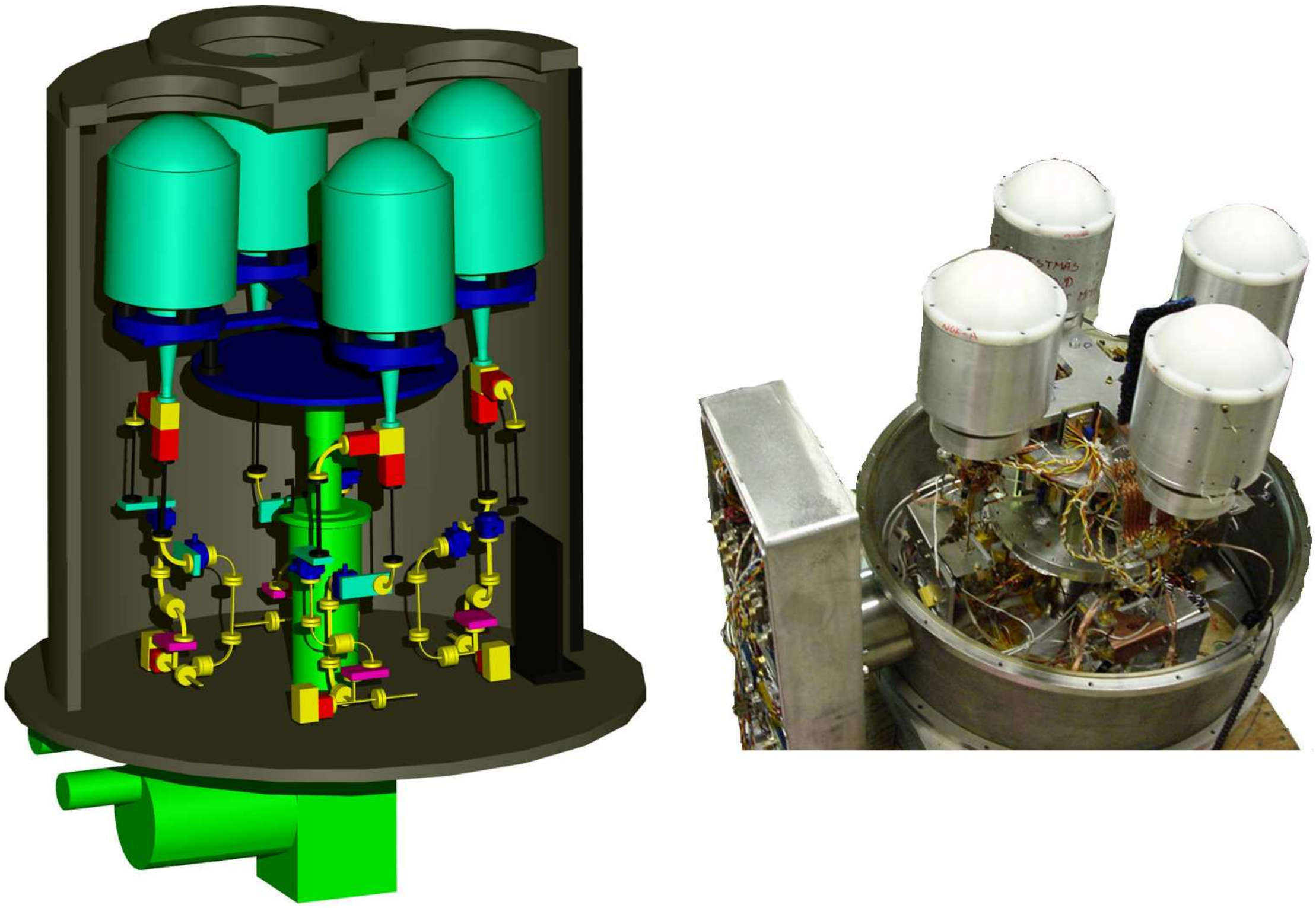

Figure 3.12: *left*: Cutaway view of the cryostat with the four W-band receivers. The only inaccuracy is the LO distribution system which does not represent the final configuration (described in Section 3.1.3). The 70 K radiation shield is not shown nor are the thermal or electrical connections. The green central cylinder in the cold head. The black “T” on the bottom vacuum lid connects the three thin copper straps which maintain the receivers in place. *right*: Photograph of the open CAPMAP03 cryostat with 4 W-band receivers. The receivers are labelled A, B, C D, counter-clockwise with A closest to the IF box.

The internal dewar structure naturally divides into three sections at 20 K, 70 K, and room temperature (300 K). The cold head 20 K stage is attached to two 20 K plates (colored blue in Figure 3.12). The top plate is a symmetric cross which provides hard mounting point for the four horns. The cross is supported by four copper pillars on the bottom 20 K plate. The 16 LNAs are heat sunk to the bottom plate, nominally the coldest point in the dewar. A 70 K radiation shield (omitted for clarity in Figure 3.12) protects all the 20 K components up to the top of the lens shrouds from the 300 K radiative loading of the vacuum shell. The outside of 70 K shield is covered with 20 layers of superinsulation to bring radiative loading to a value acceptable for the fridge (see Section 3.2.3). The radiation shield is connected to the cold head 70 K stage by four pie-shaped plates which form a solid disk around the 70 K stage when attached together. The stainless steel waveguide passes though the radiation shield via slots cut through the pies. The section below the pies is warm. Apart from the attachment at the horn, each receiver is mechanically mounted by three thin flexible copper straps connected from the bottom lid of the dewar to the two filters and the power amplifier. This provides enough structural strength to prevent the receivers from shaking but is flexible enough to allow for thermal contraction of the cold section. A single receiver, including the optics, weighs $\sim$4 kg. The cold head has a zero torque requirement. To relieve the weight on the cold head, four G10 pillars are installed from the pies to the lower 20 K plate, and four "bumpers" from the top of the 70 K lid to the vacuum shell. These prevent torque from being applied to the cold head when the dewar is tilted.

### 3.2.2 Window Design

The microwave window is a necessary part of any radio astronomy cryostat and has been the subject of many previous studies [88, 24, 87, 92, 81]. The window is designed to allow the radiation to enter the dewar but must be built to have the following properties:

1. It must be impermeable to air to keep a good vacuum in the dewar.
2. It must withstand an atmosphere of pressure difference.
3. It must be as transparent as possible at the frequencies of interest.
4. (optional but attractive) It should filter the unwanted radiative load.

Properties 1 and 2 may be obvious but property 3 is particularly important because all the elements upstream of the cold amplifier cause a direct increase to the system noise temperature, as well as attenuate the signal. These four requirements only leave a select few materials as candidates for microwave windows.

The CAPMAP03 windows are 6" diameter circular openings and are mounted to the dewar as shown in Figure 3.13. The CAPMAP03 windows are made of two materials: a 1/8" thick Gore-tex[11] layer used to support the pressure differential between atmospheric pressure and the vacuum. A second layer on top, made of 0.00065" thick UBS-2 (polypropylene), provides an air-tight barrier. Gore-tex is an open cell material, made of expanded PTFE (polytetrafuoroethylene). The Gore-tex material used (RA 7957, also used as a radome material) was chosen because it is physically thin, yet strong (it has an very high tensile modulus so it will stretch but not break), and has a low microwave attenuation. A

[11]fabricated by W.L Gore (`http://www.goreelectronics.com/products/specialty/Radome.html`), distributed by Atlantech Distribution Inc, 68 Brunswick Avenue, Edison, NJ.

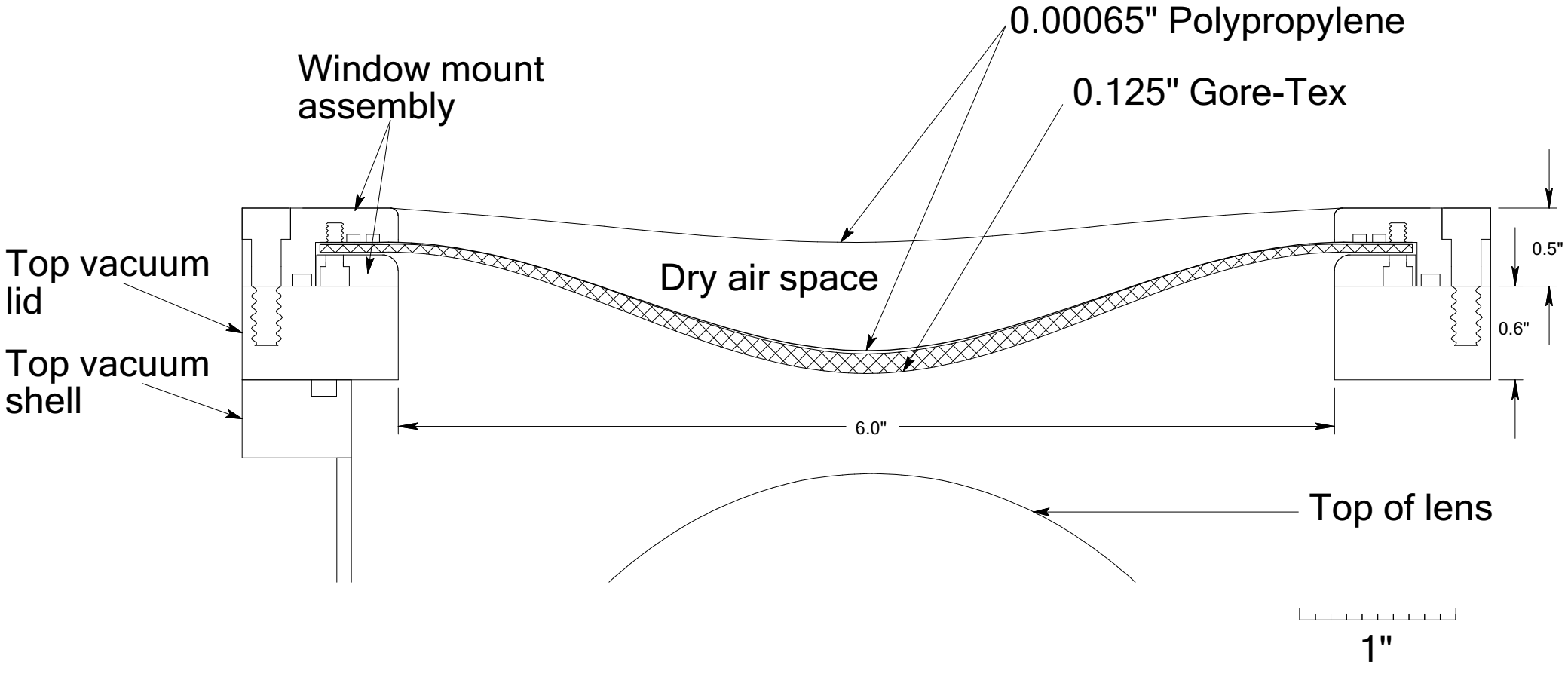

Figure 3.13: The CAPMAP vacuum window consists of a 0.65 mil polypropylene film on top of a 1/8" Gore-tex layer. The Gore-tex supports the atmospheric pressure while the polypropylene provides the vacuum barrier. Two O-rings on top of the polypropylene create a hermetic seal with the window mount. The materials are carefully chosen for their mechanical and microwave properties. Upon testing, this type of window imploded twice because the Gore-tex slipped inside the mount. The mount was modified to apply more pressure to the small rim clamping the Gore-tex down. As a side note, another common and possibly better material for microwave transmission windows is the closed cell foam called Zotefoam PPA-30 (Zotefoam company,Walton KY, 4109).

thickness of 1/8" provides the best compromise between a reasonable window deflection and a small microwave attenuation. A 0.75" $\pm$0.1" deflection was measured at the center of the window. Gore-tex has a dielectric constant $\epsilon_r \simeq 1.2$ [87], a loss tangent $\tan\delta = 0.0005$ which yields [92] a skin depth[12] of $\alpha = \lambda/(2\pi\sqrt{\epsilon_r}\tan\delta) = 77$ cm at $\lambda = 3$ mm. This is equivalent to a transmission loss at 100 GHz of -0.15 dB/in. The 1/8" thick Gore-tex film should therefore only add $T_{noise} = (1 - L)T_{physical} \simeq$1 K to the noise budget.

However, Gore-tex is permeable to air, which is why a layer of polypropylene is used on top. The material used is UBS-2[13], a replacement for the no longer available HR500/2S previously evaluated as an attractive vacuum window material [81]. UBS-2 is a laminate of biaxially oriented polypropylene with 0.0001" layers of polyvinylidene chloride on both side. The sealing properties of this kind of UBS-2 are 3 gm/m$^2$/24 hours at 38° C and 90% relative humidity, and 4.56 cc/m$^2$/24 hr at 23° C, 0% Rh. UBS-2 is also used for the dry space window to provide a volume of dry air to prevent formation of dew on the cold vacuum windows. The air in the dry space is circulated in a closed cycle through a tube of drierite[14] to remain dry.

[12] The distance where the incident power has fallen by e$^{-1}$

[13] Sample and specification provided by AET (`http://aetnets.com`)

[14] `http://www.drierite.com/`

The last window between the receiver and the telescope is the vertex cab window. It is a 3×2 meter opening to the outside to allow radiation to enter the room. It is closed with a white 1" thick Extruded Polystyrene (EPS) layer to allow the cab to be thermally controlled and prevent humidity from entering. The window can be covered with a protective metal sheet during periods of bad weather. The white EPS is similar to the material used for the COMPASS secondary support [36]. The material's transmission properties was tested by chopping six 1" layers in front of a receiver. The material causes a -0.005 dB/inch attenuation, only adding 0.4 K to the system temperature.

### 3.2.3 Cryostat Thermal Loading

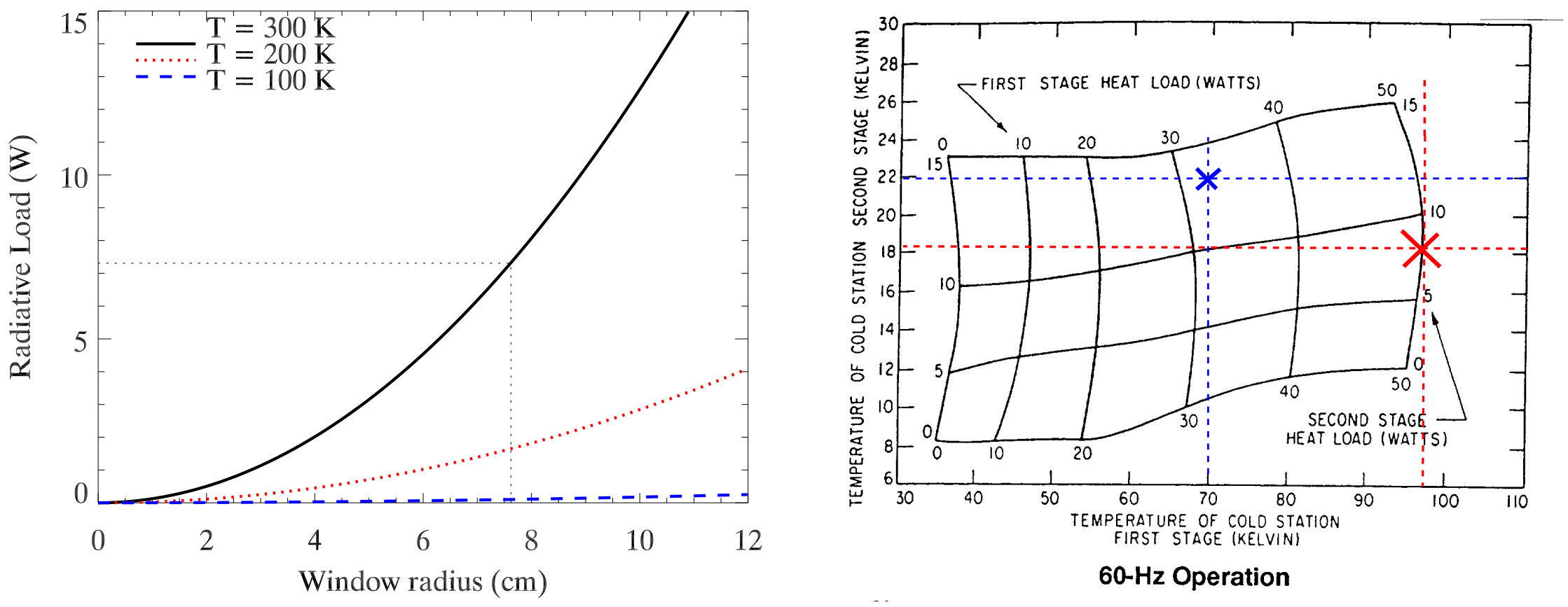

(a) (b)

Figure 3.14: (a) Radiative Loading from cryostat window. Assuming the window is at 300 K, and emits in the whole spectrum as a blackbody, the thermal loading on the 70 K stage due to a single CAPMAP 6.26 cm radius window is 7.3 Watts. This provides an upper limit on the thermal loading from the window because the real window is certainly colder (frost often forms on it in the lab when the dry space is not present) and Goretex is a good IR block at wavelengths greater than 50 $\mu m$ [39].
(b) 1020 cold head with 1020 compressor loading curve from [28]. The red and blue crosses show respectively the predicted and measured heat loads in the CAPMAP03 dewar. The diagram can be read in two ways. Either choose the temperature of the two stages on the axes and read off the required heat loads to produce such temperature at the intersection of the curves. Or from a given heat load, determine the operating temperature. Note that the discrepancy between predicted and measured heat loads indicate a thermal short between the 70 and 20 K stage. This thermal short was actually discovered post-season, in the G10 pillars to the lens shroud.

The total thermal loading on the inner dewar parts can be estimated as the sum of the convective, conductive, and radiative loads. The thermal load from convection is usually assumed to be negligible when the pressure is below 0.01 mTorr. From empirical evidence[15]

[15]The dewar temperature started to exponentially increase when the pressure reached 1 mTorr.

| Thermal Conductivities (W $m^{-1}$ $K^{-1}$) | | |
|---|---|---|
| | $\overline{\kappa}$ <br> $T_1 = 300\ K$ <br> $T_2 = 77\ K$ | $\overline{\kappa}$ <br> $T_1 = 77\ K$ <br> $T_2 = 4\ K$ |
| 18-8 stainless steel | 12.3 | 4.5 |
| G10 fiberglass | 0.21 | 0.17 |
| Manganin | 17 | 8 |
| Aluminum (6063-T5) | 200 | 220 |
| Copper (OFHC) | 420 | 800 |
| Brass (70 Cu, 30 Zn) | 81 | 26 |

Table 3.3: Mean thermal conductivities of common materials used in the CAPMAP03 dewar. The definition of $\overline{\kappa}$ is given in the text. Values are taken from a compilation in [148], originally from [143].

with the CAPMAP03 cryostat, we (guess)-estimate a convective load of 1 W at ~1 mTorr.

Assuming the emissivity of the radiating surfaces are both much smaller than unity, the standard formula [3] for radiative heat transmitted between two surfaces A1 and A2 (with area A) at temperature $T_1$ and $T_2$ ($> T_1$), with $n$ layers of superinsulation in between is

$$\dot{Q}_{rad} = \frac{1}{n+1}\ A\ \sigma\ (T_2^4 - T_1^4) \tag{3.3}$$

where A is in $m^2$, $\sigma = 5.67 * 10^{-8}$ W $m^2$ $K^4$, T is in Kelvin, and $\dot{Q}_{rad}$ in Watts. We recover the standard Stephan-Boltzmann relation in the case of no superinsulation.

Since the thermal conductivity of materials varies with temperature, the heat flow through a bar of material of constant cross section $A$, length $l$, and a temperature of $T_1$ and $T_2$ at each end, is obtained by the following integral:

$$\dot{Q} = \frac{A}{l} \int_{T_1}^{T_2} \kappa(T)\ dT\ , \tag{3.4}$$

where $\kappa(T)$ is the thermal conductivity. Because there are certain favorite temperature in cryogenic work (room temperature: 300 K, liquid nitrogen temperature: 77 K, liquid helium temperature: 4 K), it is simpler to define the heat flow between two temperature $T_1$ and $T_2$ within these ranges as

$$\dot{Q} = \frac{A}{l} \bar{\kappa}_{21} \Delta T_{21}\ , \tag{3.5}$$

where

$$\bar{\kappa}_{21} = \frac{1}{T_2 - T_1} \int_{T_1}^{T_2} \kappa(T) dT \tag{3.6}$$

The mean thermal conductivities of common materials used in the CAPMAP dewar are listed in Table 3.3.

Estimates of the conductive and radiative thermal loads in the CAPMAP03 cryostat are presented in Table 3.4. These are rough estimates and are used to verify that the thermal loads are within the operating capacity of the mechanical fridge. Comparison with the actual loads during the experiment is shown in Figure 3.14(b). The loading curve for the 1020 mechanical fridge translates a measured temperature of the 20 K and 70 K stage into a specific heat load on the fridge. CAPMAP03 operated close to the maximum loading for the CTI 1020 cryocooler system.

| Source | Absorber | A ($m^2$) | l (m) | A/l (m) | $T_1$ (K) | $T_2$ (K) | #[a] | Load on 20K stage[a] | Load on 70K stage[a] |
|---|---|---|---|---|---|---|---|---|---|
| Window | Lens | $1.8e^{-2}$ | - | - | 290 | 150 | 4 | - | 29 W |
| Window[c] | Horn | $1.2e^{-2}$ | - | - | 290 | 20 | 4 | 1 W | - |
| Vacuum shell | 70K shield | $8.5e^{-1}$ | - | - | 300 | 100 | 1 | - | 18.5 W[b] |
| 70K shell | Horn | $7.5e^{-3}$ | - | - | 100 | 20 | 4 | 0.17 W | - |
| 70K shell | OMT | $2.5e^{-3}$ | - | - | 100 | 20 | 4 | 0.04 W | - |
| 70K shell | 20K plates | $5.0e^{-1}$ | - | - | 100 | 20 | 1 | 2.6 W | - |
| SS waveguide | - | $2.2e^{-6}$ | 0.038 | $5.7e^{-5}$ | 300 | 70 | 8 | - | 1.4 W |
| SS waveguide | - | $2.2e^{-6}$ | 0.038 | $5.7e^{-5}$ | 70 | 20 | 8 | 0.14 W | - |
| G10 bumpers | 70K shield | $1.2e^{-4}$ | 2.54 | $5.0e^{-3}$ | 300 | 100 | 4 | - | 1.2 W |
| G10 pillars | 20K plate | $5.0e^{-4}$ | 0.13 | $3.9e^{-3}$ | 70 | 20 | 4 | .13 W | - |
| G10 pillars | Lens shroud | $2.8e^{-4}$ | 0.019 | $1.5e^{-2}$ | 100 | 20 | 16 | 3.2 W | - |
| Wiring | 70K stage | $3.0e^{-8}$ | 0.46 | $6.8e^{-8}$ | 300 | 70 | 112 | | 0.01 W |
| Wiring | 20K stage | $3.0e^{-8}$ | 0.36 | $8.0e^{-8}$ | 70 | 20 | 112 | 0.003 W | |
| Cu wires[d] | 70K stage | $7.0e^{-7}$ | 0.46 | $1.7e^{-8}$ | 300 | 70 | 112 | - | 8.0 W |
| Total | | | | | | | | 7 W | 50 W |

Table 3.4: Thermal load calculation for the CAPMAP03 dewar. The thermal conductivities used for these calculations are in Table 3.3. In most cases, the thermal loads are conservative calculations so as to provide upper limits on the heat load the fridge will have to absorb.
[a]The total load in the last two columns is the individual load multiplied by the number of similar elements.
[b] With 20 superinsulation layers around the 70K shield, the heat load is $\frac{390}{21} = 18.5$ W
[c] Assuming the IR emission of the window through the lens only contributes 5% to the horn's IR loading.
[d] This is the thermal load from hypothetical 24 gauge Cu wire connected directly from the warm stage to the 20 K stage. Instead, the cryostat has manganin wires (see Appendix D)

## 3.3 Instrument Characterization

Calibration of the newly assembled instrument is a necessary task. The purpose is to convert the output voltage of the receiver into the observed quantity (one of the Stokes parameter in thermodynamic temperature units). The unfamiliar reader is referred to [124, 116, 90] for a detailed overview of the use of temperature units in radio-astronomy.

The in-lab characterization is needed to measure the following properties of the receivers:

- $\Delta\nu$, $\nu_c$ — the bandpass and central frequency of the radiometer.
- $\mathcal{R}_p$ — polarization channel gain.
- G, $T_{rec}$ — total power channel gain and noise temperature.
- $\mathcal{S}$ — polarization channel sensitivity (noise characteristics)

The philosophy of the in-lab characterization has evolved since the PIQUE [56] receiver was built. For PIQUE, the receivers had to be fully characterized before being fielded because their overall performance and behavior were still unknown. Therefore the phase, responsivity, and sensitivity were carefully measured before deployment. For the CAPMAP receivers, this requirement has been relaxed. In an effort to compromise between the slow but exhaustive testing of each receiver prior to deployment and the characterization performed entirely during the observing season, many new in-lab tests were devised. Figure 3.15 describes these new methods and compares their relative merits to the standard tests performed on the PIQUE receiver. These tests are designed to be fast, non-intrusive ways to test all the receivers in the dewar at once. Although the new methods are more prone to systematic errors, they serve as an indication that the receivers are operational. The final characterization is gathered from observations.

The four CAPMAP03 receivers have an additional peculiarity. They were individually assembled and fully characterized in Chicago [55]. But many components were broken during the shipping to Princeton and during integration into the dewar (two IF amps, two phase switches, two mixers, and four IF lines). As a result, the characteristics gathered during the tests in Chicago could not be used. The receivers were therefore re-phase-matched and the new bandpass measurements are presented in Section 3.3.1. The gains and sensitivities of the polarization channels, described in Section 3.3.2 and 3.3.3, were derived solely from actual data taken during the observing season.

| Test | Property Measured | Technique Description | Feasability | Systematics |
|---|---|---|---|---|
| Narrow-band test | phase-match<br>central frequency $\nu_c$ (GHz)<br>bandwidth $\Delta\nu$ (GHz) | 1/ Bolt an OMT rotated 45º from the polarimeter's OMT and inject a swept narrow band signal from a sweeper outside the dewar through a hermetic waveguide feed-through. Measure the out-of-phase response by adding 90º of phase with a variable phase-shifter in the LO line. Finely tune the phase with the phase-shifters in the IF section. (see Hedman, 2002) | - Most invasive.<br>- Requires an accessible LO phase shifter.<br>- Requires opening dewar for each receiver tested. | - Cleanest method. |
| | | 2/ Inject the swept signal into the receiver via a horn mounted on top of the receiver's window. The injected signal is linearly polarized at 45º from the polarimeter's OMT. Measure the in and out-of-phase signals by switching legs a 90º coupler in the IF line. Fine tuning of phase as in method 1. | - Semi invasive.<br>- Requires bolting and unbolting SMA coax connector. | - Standing wave between horns.<br>- Injected signal rotation and movement between in and out-of-phase measurement. |
| | | 3/ Signal injected as in method 2. The out-of-phase signal is measured by switching to a circularly polarized injected signal, equivalent to adding a 90º phase in one of the arms. Fine tuning of phase as in method 1. | - Least invasive.<br>- Requires a linear-to-circular polarizer. | - Same as method 2. |
| Polarized gain measurement | $\mathcal{R}_p$ (V/K) | 1/ Two-load test: Two cryogenic temperature controlled waveguide loads are attached to an OMT rotated 45º from the polarimeter's OMT. The injected polarized signal, $Q_{in} = (T_1 - T_2)/2$ is compared to the measured polarized signal to derive the responsivity. (see Hedman, 2002) | - Most invasive.<br>- Requires a pair of cryogenic loads for each receiver to be tested simultaneously. | - Loss through injector system.<br>- Thermometer calibration. |
| | | 2/ Miniplate test: A small plate is placed on top of the receiver's window at 45º such that the receiver's beam falls on a Liquid Nitrogen load. The plate is then rotated continuously through 360º to produce a sinusoidally varying polarized signal. The amplitude of the sin wave, proportional to the load temperature, the plate temperature and the varying angle of the mirror can be calculated from first principles. The measured amplitude is compared to the predicted one to derive the responsivity. | Least invasive. Signal is coupled through the window without affecting the receiver. | - Resistivity of miniplate material.<br>- Load temperature.<br>- Optical efficiency. |
| | | 3/ Chopper plate test: Test is similar to miniplate except that the receiver's beam is coupled to the sky via a large plate installed on the telescope. The plate nutatates back and forth to produce a variable signal. Responsivity is derived as in miniplate test. | Semi-invasive. CMB observations must be paused during test. | - Same as miniplate method. |
| | | 4/ Sky point source: A point source of known polarized signal is observed during the observing season. The comparison of the measured output to the expected signal yields the responsivity. | Semi-invasive. CMB observations must be paused during test | - Theoretically cleanest method.<br>- Requires accurate knowledge of source polarization. |
| Total-power gain measurement | G (V/K)<br>$T_{rec}$ (K) | 1/ Two-load test: with the same two cryogenic loads, the total power injected into the receiver, $T_{in} = (T_1 + T_2)/2$, is compared to the measured total-power signal to derive the total power gain. | Most Invasive. | - Same as polarized two load test. |
| | | 2/ Y-factor test: Two loads at two temperatures (typically 300K and 77K) are placed on top of the receiver's window in a fast succession. The total power gain and noise temperature are derived from the measured signal difference. | Least Invasive. | - Temperature of Loads.<br>- Optical efficiency.<br>- Possible compression |
| Sensitivity | $\mathcal{S}$ (K$\sqrt{s}$) | 1/ Circular Load test: A temperature-controlled cryogenic circular waveguide load is bolted to the receiver's OMT. With the load temperature-stable, the RMS of the polarization channel is measured to yield a sensitivity using the polarized gains. This test also gives a measured of the polarization to total power rejection ratio. | Most Invasive. | - Loss through injector system.<br>- Thermometer calibration. |
| | | 2/ Warm load: Same as in the circular load test except that a room temperature load is placed on top of receiver's window. This test is not ideal because it is hard to control the temperature of a room temperature load to a few mK. | Least Invasive. | - Stability of the warm load temperature. |
| | | 3/ Sky observations | Ultimate test | -Accurate knowledge of polarized gains |

Figure 3.15: Summary of different calibration techniques for the CAPMAP instrument.

### 3.3.1 Phase Matching

A correlation polarimeter performs well as long as the phase between its two arms is well balanced. The receivers were built with this constraint in mind (see section 3.1.3) so they are close to phase-matched prior to any tuning. This section describes how the relative phase between the arm is deduced from the *in-phase* and *out-of-phase* response to correlated input. The actual phase tuning is done with the easily accessible in-line phase shifters in the IF box.

A narrow-band signal is injected in each radiometer via a horn outside the dewar. The horn is aligned such that the linearly polarized signal is injected equally in each arm of the polarimeter (ie. the input polarized direction is rotated 45° from the axes of the OMT). The narrow band signal is swept from 82 GHz to 102 GHz in 30 seconds. The input frequency and response of the receiver are continuously recorded at 100 Hz with same DAQ system used during the season. During these tests, the IF box is stood off from the dewar to insert a 90° hybrid splitter in each arm before the IF amps. The 90° splitter can be set to add either a 0° phase (to measure the in-phase response, $in(\nu)$) or 90° phase (to measure the out-of-phase response, $out(\nu)$).

If the ideal response of the system is $A(\nu)$ and the relative phase shift between the two arms is $\phi(\nu)$, then the in-phase response of the polarimeter is $in(\nu) = A(\nu)\ \cos(\phi(\nu))$ and the out-of-phase $out(\nu) = A(\nu)\ \sin(\phi(\nu))$. From the in- and out-of-phase measurements, it is simple to extract $A(\nu)$ and $\phi(\nu)$ and to calculate the effective bandwidth, $\Delta\nu$ and the central frequency $\nu_{eff}$ using the formulae:

$$A(\nu) = \sqrt{in(\nu)^2 + out(\nu)^2}, \qquad \phi(\nu) = \arctan\left(\frac{out(\nu)}{in(\nu)}\right) \tag{3.7}$$

$$\Delta\nu = \frac{\left(\int A(\nu)\ d\nu\right)^2}{\int A(\nu)^2\ d\nu}, \qquad \nu_c = \frac{\int A(\nu)\ \nu\ d\nu}{\int A(\nu)\ d\nu}. \tag{3.8}$$

The frequency-dependent response and phase of receiver A after phase tuning is presented in Figure 3.16. The resulting parameters are listed in Table 3.5. Note that the response is not corrected for the varying input power level of the injector system, which partly explains the ~10 dB level shift between S0 and S2. The average value of the relative phase shift $\langle\cos\phi\rangle$ is also computed in each sub band to evaluate how the performance of the receivers is reduced by the non-zero phase shifts. In the end, the average receiver bandwidth is ~13 GHz instead of the ideal 16 GHz. This reduced bandwidth is caused in part by the non-zero phase shifts but also by the non-flat shape of the bandpass. The level shifts visible in the derived response in Figure 3.16 are a natural consequence of the integrated spectral slopes in the gain of the various amplifiers.

| Statistic | S0 | | S1 | | S2 | | Total |
|---|---|---|---|---|---|---|---|
| | Pre[a] | Post[b] | Pre[a] | Post[b] | Pre[a] | Post[b] | |
| **Radiometer A** | | | | | | | |
| Ideal bandwidth $\Delta\nu_A$ (GHz) | 3.4 | | 4.2 | | 3.5 | | 11.1 |
| In-Phase bandwidth $\Delta\nu_a$(GHz) | 3.0 | 3.37 | 4.2 | 4.5 | 3.3 | 3.6 | 10.5 |
| $\langle\cos\phi\rangle$ | 0.86 | | 0.93 | | 0.95 | | |
| Band center $\nu_c$ (GHz) | 87.2 | 86.9 | 91.3 | 91.5 | 96.3 | 96.4 | |
| **Radiometer B** | | | | | | | |
| Ideal bandwidth $\Delta\nu_A$ (GHz) | 4.9 | | 5.0 | | 4.7 | | 14.6 |
| In Phase bandwidth $\Delta\nu_a$(GHz) | 4.7 | 4.8 | 4.7 | 4.4 | 4.5 | 4.8 | 13.9 |
| $\langle\cos\phi\rangle$ | 0.90 | | 0.92 | | 0.91 | | |
| Band center $\nu_c$ (GHz) | 86.5 | 86.4 | 91.3 | 91.2 | 96.9 | 97.3 | |
| **Radiometer C** | | | | | | | |
| Ideal bandwidth $\Delta\nu_A$ (GHz) | 4.9 | | 5.3 | | 3.9 | | 14.1 |
| In Phase bandwidth $\Delta\nu_a$(GHz) | 5.0 | 4.7 | 5.3 | 5.0 | 3.6 | 2.8 | 13.9 |
| $\langle\cos\phi\rangle$ | 0.91 | | 0.93 | | 0.84 | | |
| Band center $\nu_c$ (GHz) | 86.7 | 86.8 | 91.6 | 91.6 | 96.7 | 96.3 | |
| **Radiometer D** | | | | | | | |
| Ideal bandwidth $\Delta\nu_A$ (GHz) | 5.1 | | 5.3 | | 4.8 | | 15.2 |
| In Phase bandwidth $\Delta\nu_a$(GHz) | 4.3 | 4.7 | 5.0 | 4.5 | 4.5 | 4.0 | 13.8 |
| $\langle\cos\phi\rangle$ | 0.87 | | 0.84 | | 0.91 | | |
| Band center $\nu_c$ (GHz) | 86.9 | 86.6 | 91.8 | 91.2 | 96.7 | 96.8 | |

Table 3.5: Pre- and Post-season measurements of the frequency characteristics of the four receiver, for the three sub-bands S0, S1, and S2.

[a]Pre refers to the measurement performed in lab before the observing season after the final phase tuning was settled. The measurement is made with the IF box stood-off from the dewar and the 90°hybrid in the IF lines to produce the out-of-phase measurement.

[b] Post refers to the in-phase measurement on the dewar at the telescope, after the observing season.

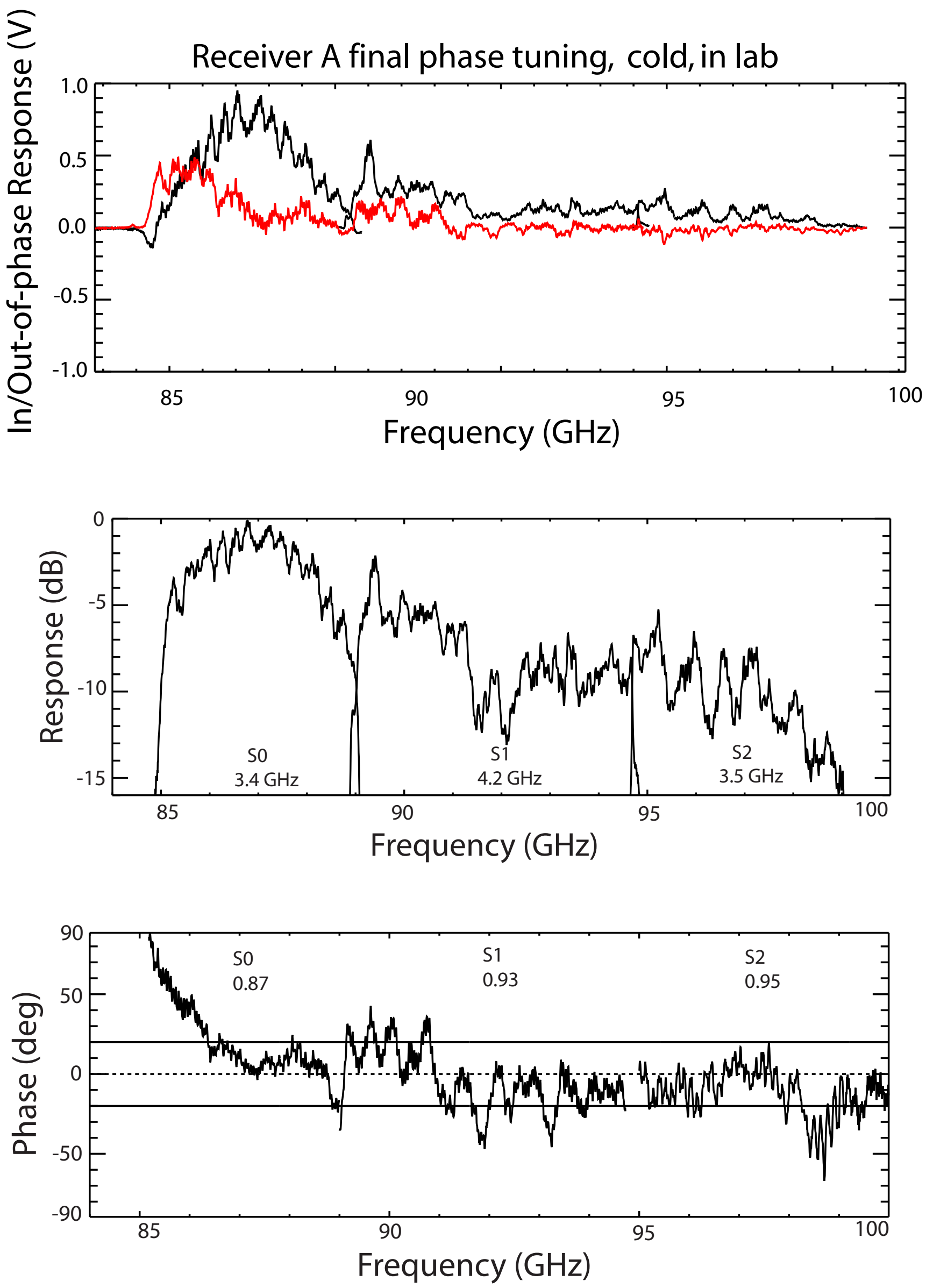

Figure 3.16: Receiver A phase measurements as function of frequency. The phase and response are typical of other receivers as well, although A has the smallest total bandwidth.
*Top*: Raw data for determining the phase properties of the receiver, showing the in- (black) and out-of-phase (red) response of receiver A after the best phase tuning is reached. To zero the relative phase, the out-of-phase response is tuned to be as close to zero as possible with the IF phase shifters. The phase of the S0 sub band for this particular receiver could have been improved but the IF phase shifter was already at the extremum of its play.
*Middle* Ideal response $A(\nu)$ for receiver A derived from the in- and out-of phase measurements above. The calculated bandwidth is printed for the three sub-bands.
*Bottom* Relative phase $\phi(\nu)$ for receiver A after final phase tuning. The value of $\langle \cos\phi \rangle$ is printed for each sub-band.

### 3.3.2 Polarized Gains

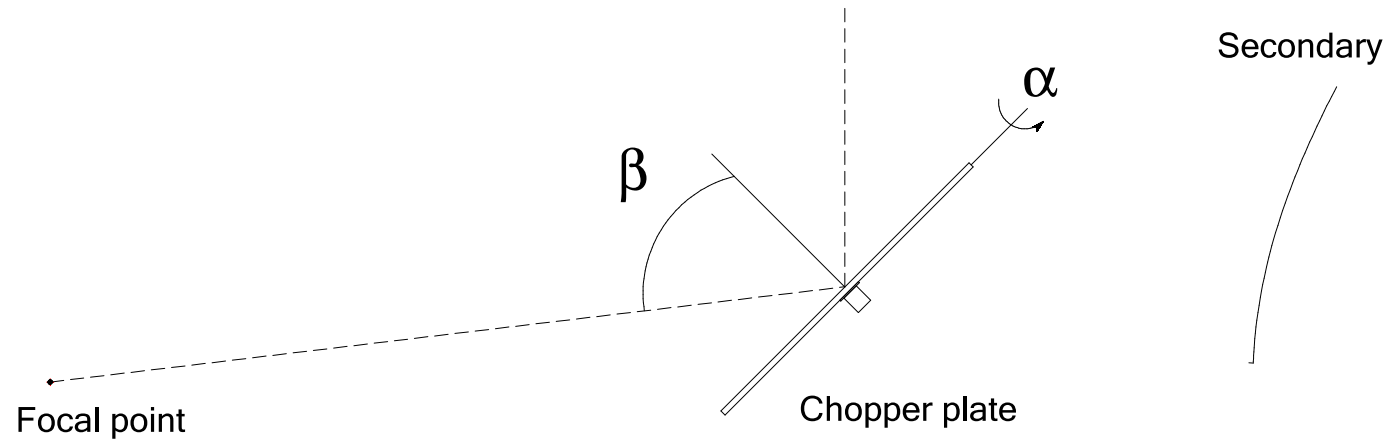

Figure 3.17: Chopper plate geometry on the telescope. The plate is rotated so as to redirect the beam towards zenith. For clarity, the telescope structure and chopper plate mounting are not shown.

Although many calibration techniques were used during the integration and observations with CAPMAP03 (see Figure 3.15), the primary polarized calibration was provided by the chopper plate test. The chopper plate is a nutating aluminium plate which injects a controlled polarized signal in the radiometer. The theory is that the thermal emission from the plate and the reflected radiation from the sky are weakly polarized due to the finite conductivity of aluminium. This effect is described in more detail in [31, 123, 26]. The plate is installed on the telescope as shown in Figure 3.17 and is nutated about the chopper's axis of symmetry by an angle $\alpha$ to produce a modulated polarized signal. The size of the polarized signal is given by the formula [61]:

$$Q_T(\alpha) = \sqrt{16\pi\rho\epsilon_0 f}\,\tan(\beta)\;\alpha\;(T_{plate} - T_{sky})\;, \tag{3.9}$$

where $\rho$ is the DC resistivity of the aluminium plate (4 $\mu\Omega\cdot$cm), $\epsilon_0$ is the permittivity of free space ($8.85\times10^{-12}$ in SI units), $f$ is the effective central frequency of the three sub bands (87, 91.5, 97 GHz for S0, S1, S2 respectively), $T_{plate}$ and $T_{sky}$ are the physical temperature of the plate and the noise temperature of the sky, and $\alpha$ and $\beta$ are the angles shown in Figure 3.17.

It is important to note the following convention. If two different single-polarization receivers measure the two orthogonal components of the electric field (eg. $\mathrm{E}_x$ and $\mathrm{E}_y$) in a single pixel, it is standard to assign two different estimates of the effective temperature on that spot, $\mathrm{T}_x$ and $\mathrm{T}_y$. Then a correct estimate of the unpolarized intensity would be $T = (T_x + T_y)/2$. Similarly, the correct definition for the Q stokes parameter is $Q_T = (T_x - T_y)/2$, not just the difference between $\mathrm{T}_x$ and $\mathrm{T}_y$. To be more concrete, we say that if a source is polarized vertically such that it emits $E_x^2 = 100$ mK in one of the detector axis, then we define the detected signal as $Q = 50$ mK. Note that this causes the smallest measurable polarization signal to be factor of $\sqrt{2}$ smaller than the definition for the smallest measurable temperature signal given in [90] (see Section 3.3.3 for details).

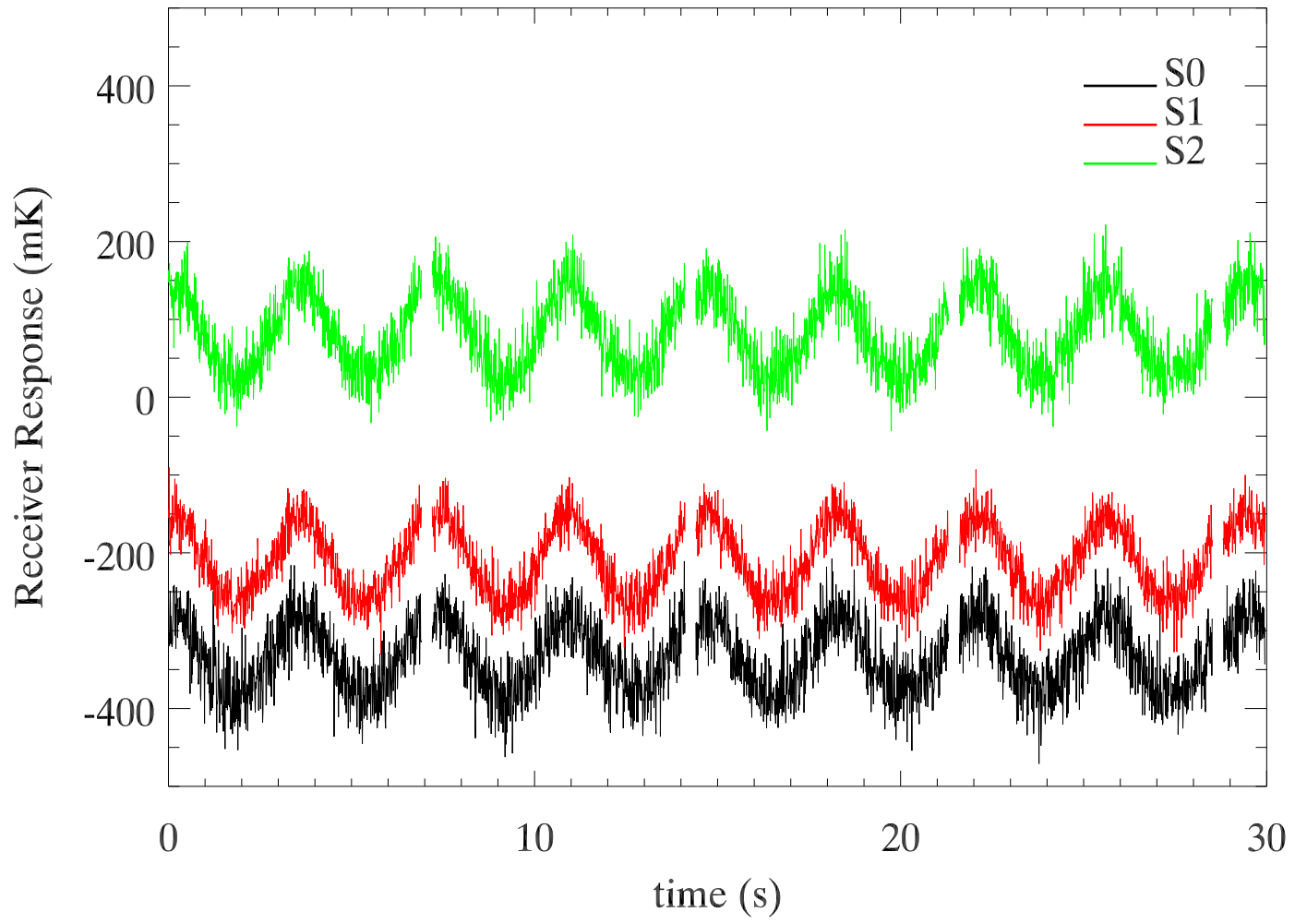

Figure 3.18: Receiver B polarization channel response during chopper plate calibration. The sinusoidal oscillation is a result of the chopper plate nutation.

During the observing season, four separate chopper-plate data sets (labelled C331, C519a, C519b, C519c) were obtained operating the chopper plate for 30 minutes each. The $\alpha$, $\beta$, $T_{plate}$, and $T_{sky}$ parameters are listed in Table 3.6 for each of the runs. $T_{plate}$ is recorded from outside thermometers. $T_{sky}$ is derived from the total power channels and the estimated receiver temperature from sky dips (see Section 6.5.1). For each run, the data is segmented into 40-second periods and a sinusoidal function of the form $V(t) = A_0 + A_1 \cos(\omega t + \phi)$ is fit to each segment of data. For each run and each channel, the fitted amplitudes $A_1$ are averaged together to produce a single amplitude. The ratio of the measured amplitude to the predicted amplitude, given in Equation 3.9, yields an estimate of the polarized gain, tabulated in Table 3.7. The gains for S0, S1, S2 have been reduced by a factor of 0.83, 0.81, 0.79 respectively to reflect a thermodynamic temperature unit. The gains also account for the clock correction of each channel (see Section 5.2).

| Parameter | Data run | | | |
|---|---|---|---|---|
| | C331 | C519a | C519b | C519c |
| $\alpha_{max}$ (rad) | 0.092 | 0.092 | 0.153 | 0.046 |
| $\beta_A$ | 50.1° | 45.5° | | |
| $\beta_{B,D}$ | 51.8° | 47.2° | | |
| $\beta_C$ | 53.6° | 48.9° | | |
| $\tan(\beta)_A$ | 1.196 | 1.018 | | |
| $\tan(\beta)_{B,D}$ | 1.272 | 1.080 | | |
| $\tan(\beta)_C$ | 1.354 | 1.146 | | |
| $T_{plate}$ | 281 K | 299 K | | |
| $T_{sky}$ | 30 K | 43 K | | |
| $T_{IF}$ | 40.3 K | 39.2 K | | |
| $T_{LO}$ | 17.5 K | 18.2 K | | |

Table 3.6: Parameters for individual Chopper-Plate runs

| Run | As0 | As1 | As2 | Bs0 | Bs1 | Bs2 | Cs0 | Cs1 | Cs2 | Ds0 | Ds1 | Ds2 |
|---|---|---|---|---|---|---|---|---|---|---|---|---|
| C331 | 21.0 | 15.8 | 14.2 | 12.1 | 5.1 | 9.7 | 16.1 | 8.1 | 9.9 | -23.6 | -21.2 | -18.4 |
| C519a | 22.4 | 16.7 | 15.0 | 14.6 | 5.8 | 9.8 | 18.8 | 9.4 | 10.7 | -25.7 | -24.2 | -19.1 |
| C519b | 21.8 | 16.5 | 14.5 | 15.4 | 6.2 | 10.5 | 19.5 | 9.9 | 11.3 | -26.4 | -25.0 | -19.8 |
| C519c | 19.0 | 14.5 | 12.7 | 14.5 | 5.5 | 9.5 | 18.4 | 8.9 | 10.4 | -25.4 | -23.3 | -19.2 |
| Combined | 21.0 $\pm$1.5 | 15.9 $\pm$1.0 | 14.1 $\pm$1.0 | 14.2 $\pm$ 1.4 | 5.6 $\pm$0.4 | 9.9 $\pm$0.5 | 18.2 $\pm$1.5 | 9.1 $\pm$0.8 | 10.6 $\pm$0.6 | -25.3 $\pm$1.2 | -23.4 $\pm$1.6 | -19.1 $\pm$0.6 |

Table 3.7: Derived polarized gains (in mV/K) from the chopper-plate tests. The values in the table are in thermodynamic units, converted from R-J by applying the appropriate correction factor for S0, S1, and S2 (0.83, 0.81, 0.79). The values also include the clock phase correction. No systematic error for the uncertainty in the predicted signal is included. Note that these gains are with the IF box at 40° C, and the LO at 18° C. Analysis by M Hedman.

The errors for the combined estimate in Table 3.7 are derived from the standard deviation of the four runs and do not include any systematic uncertainty. The leading source of systematic uncertainty [57] in the chopper plate gains measurements is the estimation of the resistivity of the plate. Assuming the plate is made of Aluminium 6061 [128], we use a resistivity $\rho = 4\ \mu\Omega$ cm. The resistivity of pure aluminium sets a lower bound of $\rho = 2.6\ \mu\Omega$ cm but can be much higher with more impurities in the metal. If we conservatively assume a 50% uncertainty on the estimate of the resistivity of Aluminium, this would translate into a 20% systematic error in the expected signal. Systematic errors from other sources are minor. A 1 GHz shift in the effective frequency causes a 1% change in the estimated polarized signal. The uncertainty in the sky temperature determination is 5 K [58], which corresponds to a 3% uncertainty in the predicted polarized signal. Finally, the error on the angles $\alpha$ and $\beta$ can be estimated from the consistency of the different runs.

### 3.3.3 Sensitivity

The gain of the polarization channel is necessary to convert the measured voltage into a unit of temperature. But this is not the whole story. A receiver may well have a high gain but its very high noise would make it incapable of measuring any signal. The noise property of the receiver is typically described by its sensitivity, $\mathcal{S}$, equal to:

$$\mathcal{S} = \frac{T_{sys}}{\sqrt{\Delta\nu}} \quad , \qquad \Delta Q_{min} = \mathcal{S}/\sqrt{t_{int}} \tag{3.10}$$

where $T_{sys}$, $\Delta\nu$, and $t_{int}$ are the receiver's system noise in Kelvin, RF bandwidth in GHz, and integration time in seconds. The sensitivity is a measure of the smallest detectable signal for a given integration time. A 1 mK$\sqrt{s}$ sensitivity means that a signal of 1 mK can be detected as a $1\sigma$ deviation from the noise with a 1 second integration, a 100 $\mu$K signal detected in 100 seconds, and so on. The sensitivity is intimately related to the noise power spectrum of the receiver. If the noise power spectrum is white, the intrinsic random noise of the receiver can be integrated as long as necessary to reach a certain signal level. This is not true if the receiver noise spectrum contains $1/f$ noise. When characterizing a receiver, one wants to ascertain that the sensitivity is in good agreement with $T_{sys}$ and $\Delta\nu$ both determined separately, and that the noise power spectrum is white at least for frequencies greater than the scan frequency.

**Factors of $\sqrt{2}$** The standard equation for the smallest measurable temperature receiver is defined by [90] as

$$\Delta T_{min} = K \frac{T_{sys}}{\sqrt{\Delta\nu t_{int}}} , \tag{3.11}$$

where K is a constant which depends on the type of receiver. For a simple total-power receiver, $K = 1$, and $K = \sqrt{2}$ for a correlation receiver. This is because a correlation polarimeter fundamentally measures the difference between two signals. The expression for the smallest measurable *polarization* signal then also differs from that of the smallest measurable *temperature* by $\Delta Q_{min} = \Delta T_{min}/\sqrt{2}$ because the polarization is defined as $Q_T = (T_x - T_y)/2$ (see Section 3.3.2). These two factors then yield Equation 3.10, the true measure of the smallest detectable polarization signal for a given integration time[16].

**Receiver Noise Temperature** The total power channel temperature is a good tracer of the polarization channel receiver temperature and has the advantage that it is a directly accessible quantity using the Y-factor method. The receiver's total power response, $V_h$ ($V_c$) is measured while looking at a hot (cold) load at temperature $T_h$ ($T_c$). The two loads are provided by a room temperature eccosorb foam and a piece of eccosorb foam in $LN_2$. With this setup, the noise temperature and gain of each arm of the receiver is given by

$$T_{rec} = \frac{T_h - YT_c}{(1-Y)}, \text{ with } \quad Y = \frac{V_h}{V_c} \tag{3.12}$$

$$G = \frac{T_{cold}}{(T_{rec} + T_{load})}. \tag{3.13}$$

The total power channel's receiver temperature and gain, measured in lab prior to deployment, are presented in Table 3.8. Although they are rather high, they agree reasonably well with the receiver temperature measured with sky-dips and Jupiter observation (see Section 6.5.1).

[16] An additional factor should be accounted for when estimating the sensitivity if the scan strategy differencing strategy discards some of the information. For instance, the PIQUE [60] experiment differences the data taken from two points on the sky. The sensitivity is therefore multiplied by $\sqrt{2}$ for a given integration time because half of the information is discarded. For CAPMAP, this factor is only $\sqrt{3/20}$ because only the mean, slope, and quadratic are removed for every 20 points in each azimuth scan (see Chapter 7).

| TP Channel | AD0 | AD1 | BD0 | BD1 | CD0 | CD1 | DD0 | DD1 |
|---|---|---|---|---|---|---|---|---|
| $T_{rec}$ (K) | 102 | 166 | 76 | 68 | 161 | 96 | 99 | 76 |
| Gain (mV/K) | -1.4 | -2.1 | -2.3 | -1.6 | -2.4 | -1.3 | -4.5 | -3.6 |

Table 3.8: Total-power gain and receiver noise temperature from in-lab Y-factor two-load test. The two loads are eccosorb foam at 290 and 77 K. The offset from the DAQ is removed from both measurements. The results are comparable to other methods used during the season (see Section 6.5.1). The one caveat with the Y-factor method is that the 290 K warm load may have saturated the detector diode, even though it could remain linear while looking at the 40 K sky.

The sensitivity is measured by analyzing a five hour stretch of data during which the sky temperature is stable. The sensitivity at integration times equal to the sampling period (10 ms) is calculated from the average of the variance of all adjacent 10-second segments. The sensitivity at 1 Hz is just one tenth that at 100 Hz, according to Equation 3.10. The 10-second averaging of the variance is large enough to obtain a precise measure of the variance (1000 numbers), yet small enough that the power spectra are still white-noise dominated. The measured sensitivities are listed in Table 3.9.

| Sensitivity | S0 (mK$\sqrt{s}$) | S1 (mK$\sqrt{s}$) | S2 (mK$\sqrt{s}$) | Total[a] (mK$\sqrt{s}$) | $T^b_{rec}$ (K) |
|---|---|---|---|---|---|
| A | 3.26 | 2.13 | 2.68 | 1.48 | 101 |
| B | 3.35 | 2.05 | 2.00 | 1.35 | 109 |
| C | 2.94 | 2.06 | 3.49 | 1.51 | 128 |
| D | 3.10 | 2.22 | 2.49 | 1.46 | 121 |

Table 3.9: Sensitivity derived from sky observations. The sensitivities are calculated from the variance of 10 s of data. These numbers are then converted into mK using the gains from the chopper plate tests (see Section 3.3.2). The sensitivities are therefore in thermodynamic units. During the 5 hour data segment chosen for this measurement, the sky temperature, measured by the 8 TP channels, is 50 K.
[a]Total sensitivity is the sum of the S0, S1, and S2 sensitivity, added in quadrature.
[b]System temperature derived from $T_{rec} = \mathcal{S} \times \sqrt{\Delta\nu} - T_{load}$ where $\mathcal{S}$ is the total sensitivity, $\Delta\nu$ the bandwidth from Table 3.5 and $T_{load} = T_{sky} = 50$ K.

The power spectral density of the same stretch of data is shown in Figure 3.19. The raw polarization channels show two components: on short time scales (high frequency), the power spectrum is white, meaning that the gaussian random noise of the receiver is the dominant source of noise. On time scales longer than 300 seconds (~3 mHz), the polarization channel has $1/f$ structure, due to the finite response of the polarization channel to the total power fluctuations. The relevant feature is that the polarization channel power spectra remains white on time scales longer than the 8 s scan period. This ensures that the offset, slope, and quadratic removal performed at the telescope scan frequency will indeed eliminate the residual polarization drifts.

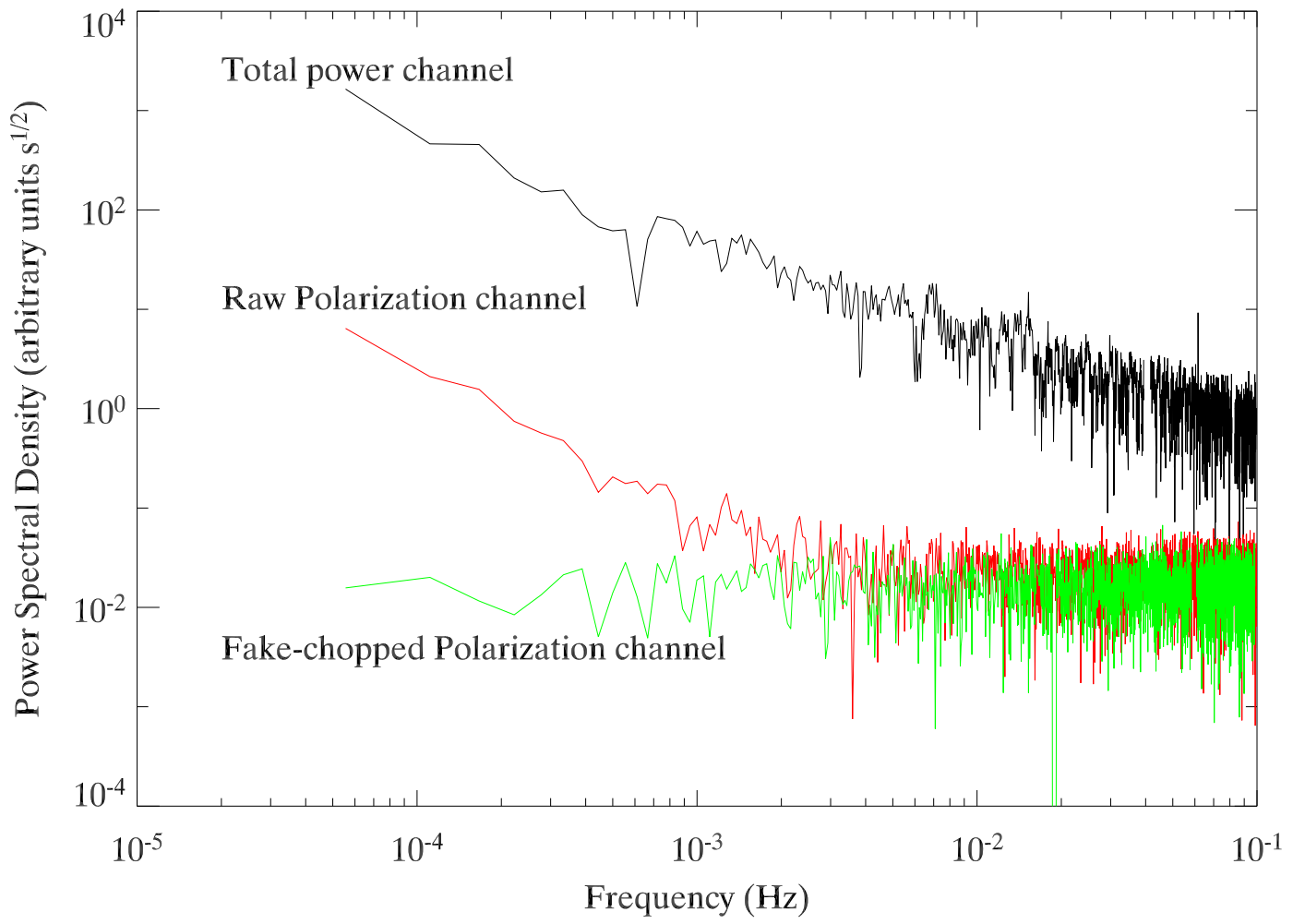

Figure 3.19: Power spectral density of the same 5-hour stretch of data used to calculate the sensitivities in Table 3.9. The spectral density of the total-power and raw polarization channel are calculated by simply taking the Fourier transform of the raw data. The fake-chopped polarization channel is generated from the raw polarization channel by differencing adjacent one-second stretches of data. Note how the fake-chopping removed the residual $1/f$ noise of the polarization channel on long times scales.

One way to simulate the scan synchronous offset, slope, and quadratic removal is to fake chop the polarization channel data by differencing adjacent one-second segments of data. This is shown as the "fake-chopped polarization channel" in Figure 3.19. The fake-chopped signal remains white to the longest measurable time scales (2.5 hours). An equivalent way to verify that the smallest measurable signal in that data scales as $1/\sqrt{t_{int}}$ is to measure the RMS (the square root of the variance) of the data averaged over increasingly long time periods. The results shown in Figure 3.20 demonstrate that the noise does "integrate down".

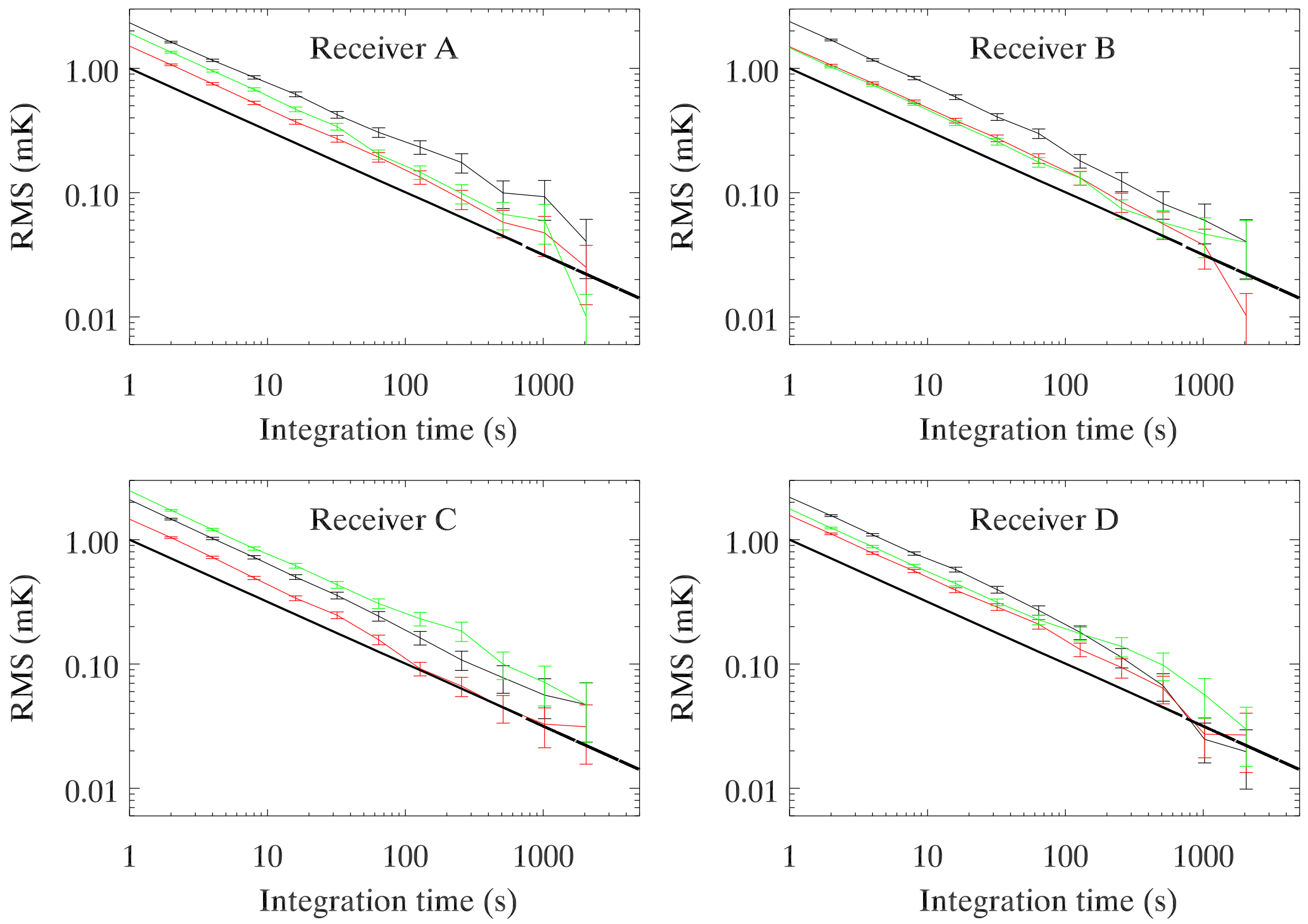

Figure 3.20: Integration down plots for all twelve polarization channels from data taken on the sky. A one-second fake chop was applied to the data. The figure shows the RMS of the fake-chopped data averaged over different time periods. The RMS expectedly scales as the inverse of the square root of the integration time. The sensitivity from this data can be read off from an integration time of one second and are tabulated in Table 3.9. The solid black line is equal to $1/\sqrt{t_{int}}$. Numbers are in thermodynamic units.

# Chapter 4

# The Seven Meter Telescope

Of every 20 photons which finish their journey in the receivers, one of them comes from the early universe. Collecting these photons is the role of the telescope, in this case the 7-meter antenna at AT&T Bell Laboratories' Crawford Hill Laboratory (now Lucent Technologies). Although it has imperfections, the antenna's optical design is well suited for measuring the faint microwave background signal. Perhaps, the most important advantage of the antenna is that it is otherwise completely unused and easily accessible, granting us full use of the antenna during the winter observing season.

The first part of this chapter gives a mechanical description of the telescope and the associated control systems. The second part describes the telescope optics, focusing on the properties relevant to CMB polarization observations. and an example of a tertiary mirror designed for early tests of the performance of the telescope.

### Historical Overview

The Crawford Hill 7-meter antenna is located in Holmdel, NJ[1]. It was originally built by Ford Philco to test polarization multiplexing with geosynchronous telecommunication satellites (COMSTAR) in K-band, but was also designed with radio-astronomy in mind. Soon after its completion in 1976, it came into use as a general purpose radio astronomy telescope. Early scientific projects with the antenna were the mapping of galactic CO molecules [4, 130, 83, 95], measuring $^{18}$CO, $^{13}$CO, and $^{12}$CO isotope ratios, mapping the CS molecule to find protostellar cores in star forming regions [131, 35], and searching for molecular gas outflows [111, 140]. The last major project (in 1995) with the 7-meter was the search for $^{14}$CO in the carbon star IRC+10216 [82]. Until 1998, it was used sporadically

[1]latitude: $40^\circ\ 23'\ 31''$N, longitude: $74^\circ 11'\ 10''$W, elevation: 119 meters, as measured by the local GPS receiver.

to train post-docs who wintered at the South Pole on the AST/RO telescope [132] which has a similar interface.

## 4.1 Mechanical Description

A comprehensive description of the antenna is given in the Bell Labs Technical Journal [22]. A summary of the distinctive features of the telescope is presented here.

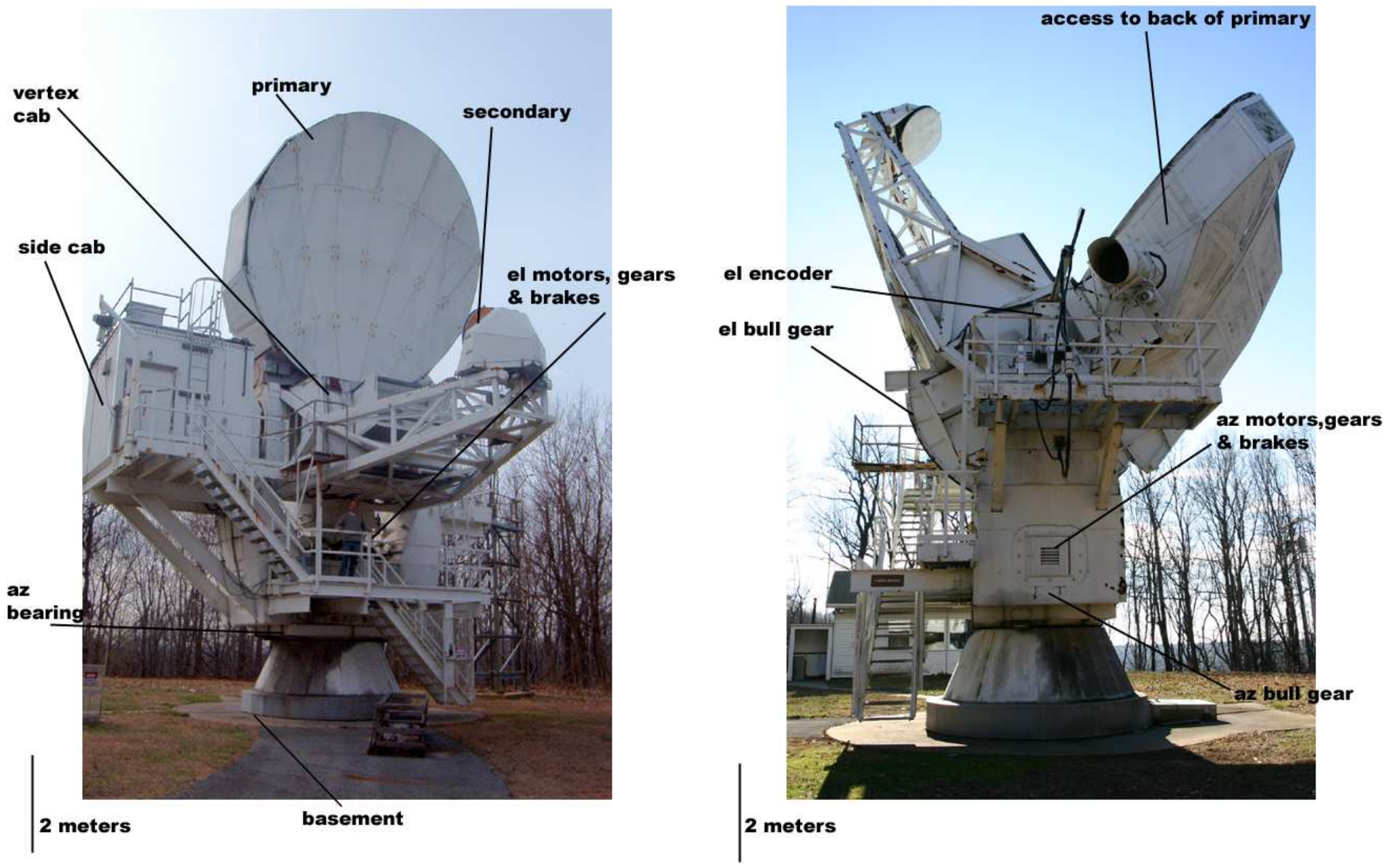

Figure 4.1: Picture of telescope showing the location of various elements. The azimuth encoder, azimuth bearing and cable wrap are located in the basement. The two trap doors in the right picture are to access the azimuth motors and the azimuth bull gear. The top trap door is only reachable when the telescope is pointed vertically, allowing access to the back structure of the primary.

- The telescope is mounted on a standard alt/az platform. It sits on a two meter diameter azimuth bearing and a yoke supports the vertex cab (see Figure 4.1). The vertex cab is built around a 3.2 m wide by 3.75 m deep by 2.3 m high steel box, which contains the Cassegrain focus. One wall of the vertex cab connects to the side cab via a 1.2 m diameter tube running through the elevation bearing. The side cab rotates in azimuth but not in elevation, and provides room for additional receivers at the Nasmyth focus. The side of the vertex cab facing the subreflector has a large rectangular window to allow entrance of the beam. The volume under the azimuth bearing contains the azimuth cable wrap which conveys the cables coming from the control building to the side and vertex cab. The cable wrap allows the telescope to rotate though 450° without straining cables. The elevation cable wrap is done through the elevation tube.

- The primary reflector is made of 27 surface panels arranged in 4 concentric rings. The panels are A354 aluminum castings. The surface accuracy of each panel was tested after completion of the antenna to be $\sim$ 50 $\mu$m. The surface error of the overall primary reflector was measured to be closer to 100 $\mu$m RMS. The secondary has a measured surface accuracy of 20 $\mu$m. The panels and the secondary are painted with 30 to 50 $\mu$m of white paint with a $TiO_2$ pigment to reflect heat. The back of the primary and secondary reflectors were sprayed with foam and enclosed inside a wood structure to use the large volume of air behind the reflectors as insulator. This helps control the differential thermal heating of the reflectors.

- For each axis, the drive system consists of a bull gear connected to two motors each with its separate speed reducer and pinions (see Figure 4.1). The motors are biased such that for low torque, they torque in opposite directions to eliminate backlash, and then one reverses to run in the same direction when more than 15 percent of the maximum torque is needed. Each of the four DC motors is rated to 3 horsepower. The maximum slew speed is $2°/s$ in azimuth and $1°/s$ in elevation with approximately $1.5°/s^2$ maximum acceleration in both axes. Each motor has a disk brake which is active when the telescope is in standby mode.

- The encoder system consists of two parts: the physical transducers located at the end of the azimuth and elevation shafts, and a digital position indicator. The transducers convert the mechanical angular displacement of the two axes into modulated sine and cosine signals. Those are sent to the digital position indicator, in the control room, which converts them into binary and binary coded decimal (BCD) outputs. The binary output is used by the OBS code to control the telescope motion, and the BCD output is sent to the CAPMAP data acquisition system (via the microcontroller (see Section 5.2.3)). The binary output is also used by the Nixie tubes which display the current azimuth and elevation on the front panel of the position indicator. The binary output has 20 bits plus sign for elevation and 21 bits plus sign for azimuth and the BCD 24 bits plus sign for elevation and 26 bits plus sign for azimuth (see Appendix E for a wiring diagram of the BCD outputs). The binary output's least significant bit represents 0.00017°. The conversion is done at a rate of 100 kHz.

## 4.2 Telescope Optics

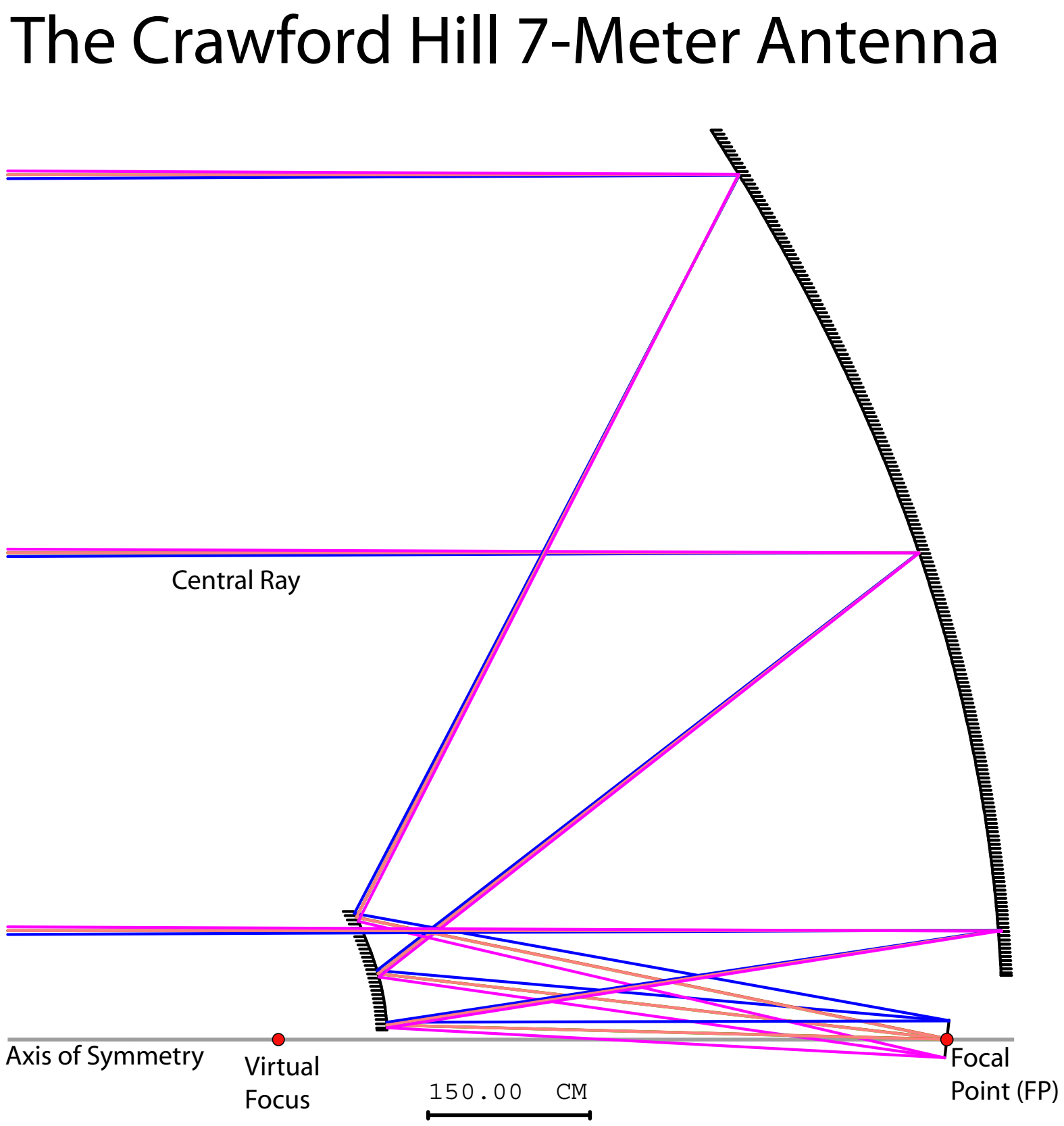

Figure 4.2: Side view scale drawing of the 7-meter telescope pointing at an elevation of 0°. All dimensions are in cm. A ray bundle is drawn from the Cassegrain focus to the secondary with an illumination half angle of 5.06°. The origin is at the vertex of the primary. The x-axis is the telescope's axis of symmetry.

The Crawford Hill antenna is the largest off-axis general use millimeter-wave antenna in existence. Its optical parameters are listed in Table 4.1. The primary is a 7-meter-diameter section of paraboloid, and the secondary a 1.2×1.8 meter oval section of hyperboloid. The primary and secondary are offset 120.74 cm and 8.258 cm respectively above their axes. Note that this simply means that each is the intersection of a conic section with a vertically offset cylinder. Both the primary's and secondary's axes of rotation are horizontal in Figure 4.2 and co-linear. The central ray[2] from the focal point (FP) to the subreflector makes an angle of 6.828° to horizontal. The subreflector subtends a illumination cone half angle of 5.06° in the vertical axis. From the virtual focus (see Figure 4.2), the primary subtends a cone of 26.75° half angle and the axis of this cone is tilted by 37.26° with respect to the horizontal. This geometry results in a 4.4 cm diameter area of the primary blocked by the secondary.

| Parameter | Value |
|---|---|
| Primary Diameter | 700 cm |
| Primary Focal Length ($f_1$) | 656.59 cm |
| Secondary: Diameter | 120 by 180 cm |
| : near focal length ($f_2$) | 86.87 cm |
| : far focal length ($f_3$) | 522.62 cm |
| Effective focal length | 3945.59 cm |
| Focal ratio (f/D) | 5.63 |
| Magnification | 6.01 |
| Measured plate scale | 61 cm/deg |

Table 4.1: Optical parameters of the 7-meter antenna. The plate scale is the measured value at the center of the focal plane. Far away from the focal point, the plate scale changes because the focal plane has a radius of curvature and is not really a plane.

Because the Crawford Hill 7-meter antenna was originally designed with millimeter radio astronomy considerations, it actually possesses many advantageous characteristics for CMB polarization observations:

**Sidelobe** The original design calls for low sidelobe levels. With a corrugated feed horn illuminating the primary at -20 dB edge taper, the side lobes at 1° off the main beam were observed to be less than -40 dB [22] at 90 GHz. This is a necessary characteristic because the pickup of radiation from the 300 K ground or other nearby warm objects can be a significant source of error when trying to detect temperature differences of a few $\mu$K. The low sidelobe level is due to the unblocked aperture design, the achieved surface accuracy of 100 $\mu$m RMS, and a low edge taper on the primary.

**Cross-Polarization** Cross polarization is an effect where the optical elements in front of the instrument rotate the polarization of the source. Systematic cross polarization can be caused by aperture blockage, mirror surface roughness, reflection at oblique incidences, and refraction from birefringent materials [44]. Because the Crawford Hill antenna was designed to test polarization multiplexing, it was specified to have low cross-polarization (better than -40 dB) in the main beam. This design parameter was partly confirmed by polarization channel beam map measurements made on Jupiter, an unpolarized source (see Section 6.3.1).

[2] An imaginary ray propagating from the Cassegrain focal point to the centers of both mirrors such that it leaves the antenna in a direction parallel to the optical axis.

**Focal Plane** A large focal plane is desired to accommodate many receivers simultaneously. Given the scope of the full CAPMAP experiment (16 receivers) and the size of a fully assembled receiver with its feed optics, an antenna with a large usable focal plane is necessary. The large focal plane is understood in the qualitative sense that receivers can be placed in a large area around the focal point without suffering much from beam distortions [52]. In its final configuration, CAPMAP has beams launched from a point 19" away from the focal point (vertically and horizontally) with negligible beam distortions (see Section 6.7). This is the result of the large focal ratio at the Cassegrain focus.

**Pointing** The 7-meter antenna is mounted on an alt-az platform and was designed to be able to point and track an object with a 10" maximum pointing error. Although the tracking capacity is not necessary for CMB observations, it proved very useful for scheduled calibration observations (see Chapter 6). Preliminary measurements of the pointing accuracy yield a maximum pointing error of 40" (see Section 6.4.1), which shows that the telescope still performs close to its designed expectations.

**Receiver cabs** The antenna has two temperature-controlled rooms for receivers installed at the two focal points (the Cassegrain and Nasmyth focus). The stability of the room temperature puts a less stringent requirement on the design of the cryostat temperature control. The indoor housing also facilitates the work on the receiver during periods of bad weather.

As mentioned in Chapter 1, the CMB polarization E mode power spectrum peaks at small angular scales ($\sim 10'$, $\ell \simeq 1000$) making narrow beams desirable. If the full aperture of the 7-meter were used, the smallest possible beam size at 90 GHz would be $\theta_{FWHM} = 1.5'$. This resolution is smaller than needed to probe the scales where the polarization peaks. In addition, in order to reduce the beam spillover and thus the increased ground pickup, the illumination on the primary and the secondary is aggressively tapered to -30 and -56 dB respectively. These two factors determine the final beam size of $\theta_{FWHM} = 4'$.

One inconvenience of the large effective focal ratio from the Cassegrain focal point is the small beam pattern needed to illuminate the antenna. In case of the -30 dB secondary edge taper, a $3°$ beam pattern is required. Given this narrow beam requirement and the mechanical constraint that the 16 CAPMAP receivers must fit inside the two meter diameter circular hole in the focal plane, the choice of feed optics was restricted to a tertiary mirror or a lens attached to a feed horn. The horn plus lens design was ultimately chosen (see Chapter 2). However, a warm total power receiver was initially constructed to provide first tests of the pointing and optical properties of the antenna. A special tertiary mirror was built to couple this receiver to the antenna. The design and fabrication of this mirror is presented below.

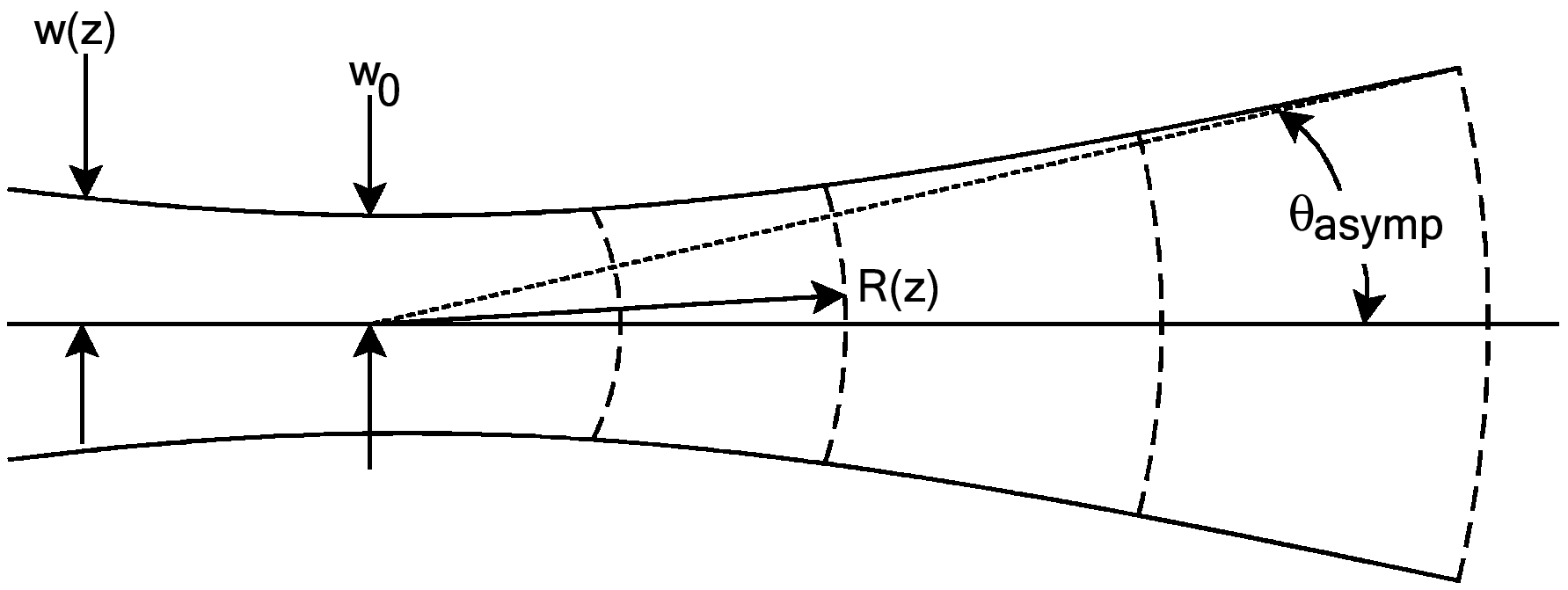

Figure 4.3: Schematic cut through a gaussian beam showing the equiphase surfaces (dashed lines), beam radius, beam waist, and radius of curvature.

### 4.2.1 Tertiary Feed Optics

#### Gaussian Beam Optics

Gaussian optics is a compromise between the simplicity of ray tracing and the computationally intensive physical optics calculation which include all the effects of diffraction. Gaussian optics includes the first order effects of diffraction by treating radiation as a light bundle with a radiall gaussian power distribution satisfying the Helmholtz wave equation. The unfamiliar reader is urged to refer to Goldsmith [44] for a comprehensive review. For reference, the main equations used to design the tertiary mirror are reproduced below. A pictorial description of the gaussian beam parameters is given in Figure 4.3.

The gaussian beam is assumed to be cylindrical with a gaussian cross section. Its power distribution goes as :

$$P(r) \propto Exp\left( - \frac{2r^2}{w^2} \right) . \tag{4.1}$$

The radius at which the field (not the power) reaches $e^{-1}$ is called the beam radius, $\mathbf{w}$, and the minimum of $\mathbf{w}$ is the beam waist, $\mathbf{w}_0$. A useful version of Equation 4.1 to convert from beam radius to edge illumination is

$$\frac{r_e}{w} = 0.3393\sqrt{T_e} \ , \tag{4.2}$$

where $r_e$ is the mirror radius and $T_e$ is the edge illumination in dB. The beam radius, as a function of the distance from the beam waist is:

$$w(z) = w_0\sqrt{1 + \left(\frac{\lambda z}{\pi w_0^2}\right)^2} \ . \tag{4.3}$$

Because surfaces of constant phase are spherical, they are described by a radius of curvature, $R$, which depends on the distance $z$ from the beam waist as

$$R(z) = z\left(1 + \left(\frac{\pi w_0^2}{\lambda z}\right)^2\right) \ . \tag{4.4}$$

One useful parameter is the confocal distance defined as

$$z_c = \frac{\pi w_0^2}{\lambda} \ , \tag{4.5}$$

which naturally divides the propagating beam into a near-field ($z \ll z_c$) and a far field region ($z \gg z_c$). In the near field, the beam remains collimated. In the far field, the beam radius asymptotes to a linear growth with $z$. It is useful to define the asymptotic growth angle:

$$\theta_{asymp} = \frac{\lambda}{\pi w_0} \ , \tag{4.6}$$

which is related to the usual far field FWHM of a beam by $\theta_{FWHM} = 1.18 \ \theta_{asymp}$. A thin lens or a conic mirror transforms a gaussian beam by focusing a beam waist $w_{01}$ at a distance $d_1$ from the lens into a new beam waist $w_{02}$ at a distance $d_2$. Because this is done by changing the phase of the wavefront but not the amplitude distribution, the beam radii must match at the lens leading to the following equality:

$$w_{01}\sqrt{1 + \left(\frac{\lambda d_1}{\pi w_{01}^2}\right)^2} = w_{02}\sqrt{1 + \left(\frac{\lambda d_2}{\pi w_{02}^2}\right)^2} \tag{4.7}$$

**Tertiary Design**

The feed optics has to meet two requirements: focusing the beam at the aperture of the feed horn and producing the desired beam size on the sky. We already know (Section 2.3) that a -56 dB secondary edge taper (5.06° half angle) produces a 4' beam on the sky. This requirement translates into a 2.41 cm beam radius at the Cassegrain focus. The tertiary mirror must therefore match the beam waist at the aperture of the horn to the beam waist at the Cassegrain focus. In this design, the telescope mirrors are approximated as thin lenses as in Figure 4.4.

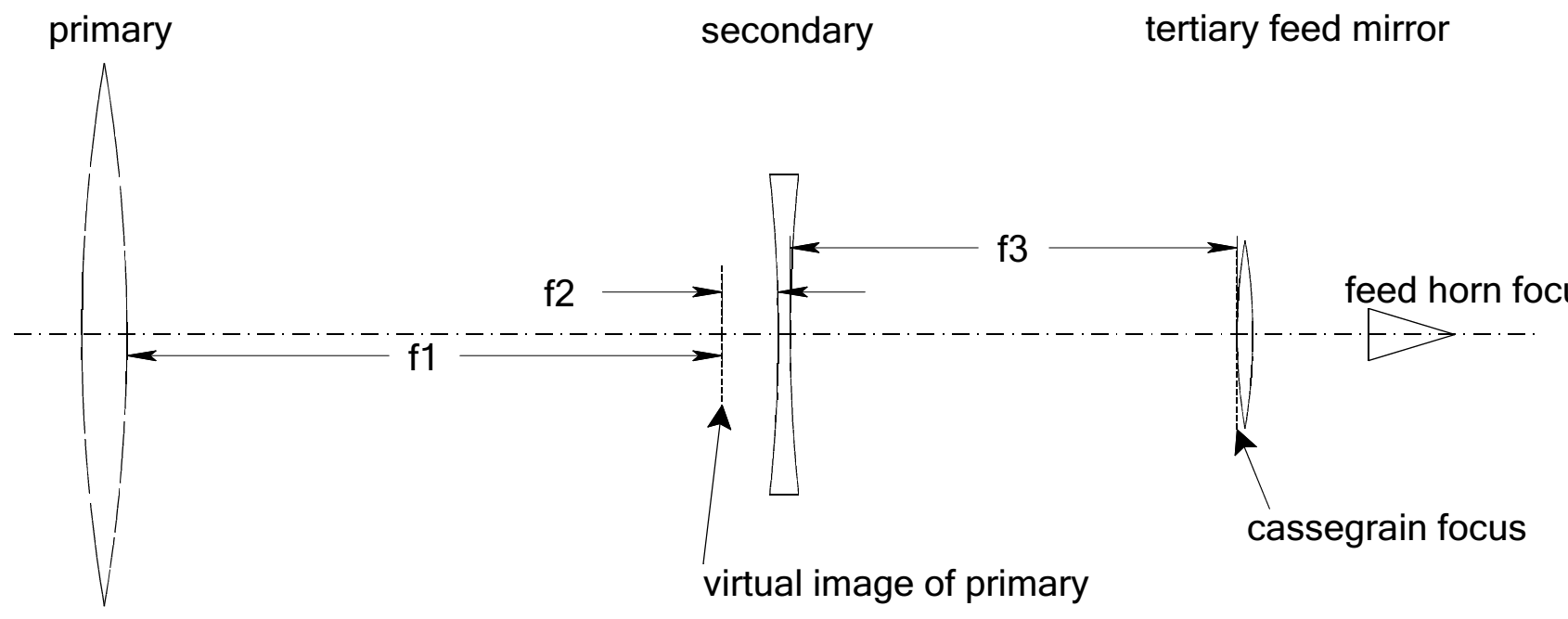

Figure 4.4: Thin lens approximation of the optics. The tertiary mirror refocusses the beam from the Cassegrain focus to the feed horn.

We start with a fiducial primary edge illumination and convert it using Equation 4.1 into a beam radius on the primary. We propagate the beam using Equation 4.3 to find the beam radius at the virtual focus which is also the near focus of hyperboloid. From the thin lens approximation (Equation 4.7) of the mirrors, we solve for the beam radius at the Cassegrain focus. This process can be repeated until the desired secondary illumination is obtained. The results for different edge illuminations are tabulated in Table 4.2. A faster but less intuitive way to derive the same result is to use the equation for the beam radius at the prime focus of a paraboloid of equivalent focal length to our system:

$$w_0^{cass} = 0.22\sqrt{\frac{T_e}{1\ \mathrm{dB}}}\left(\frac{f_e\lambda}{D}\right) \tag{4.8}$$

where $T_e$ is the edge taper, $f_e$ is the equivalent focal length, and D is the diameter of the primary. For the adopted primary edge illumination $T_e = 35$ dB, $D = 700$ cm, $\lambda = 0.3$ cm, and $f_e = 3945.59$ cm, the beam waist is $w_0^{cass} = 2.41$ cm. Note that this result matches the more intuitive calculation in the last row of Table 4.2. The characteristics of the spare WMAP [9] feed horn that this beam waist must be matched to are listed in Table 4.4.

Given the beam waist at Cassegrain focus, the feed horn beam waist, and the distance from Cassegrain focus to tertiary mirror which we chose for mechanical convenience to be $d_1 = 10$ cm, we can derive the distance from tertiary to feed horn using again Equation 4.7, $d_2 = 25.2$ cm. From Equation 4.4, we derive the incoming and outgoing radius of curvature of the tertiary mirror. The final parameter needed to fully specify an off-axis conic section is the half angle by which the beam is rotated. For mechanical convenience, we chose $\theta_i = 24°$. The parameters describing the tertiary mirror are summarized in Table 4.5. The definition of these parameters is listed in Appendix F. Figure 4.6 gives a side view of the mirror as configured in the receiver cabin.

| $T_P$ (dB) | $w_P$ (cm) | $w_0^{virtual}$ (cm) | $w_S$ (cm) | $T_S$ (dB) | $w_0^{cass}$ (cm) |
|---|---|---|---|---|---|
| -30 | 188 | 0.37 | 24.9 | -50 | 2.23 |
| -40 | 163 | 0.42 | 21.8 | -64 | 2.54 |
| -50 | 145 | 0.48 | 19.2 | -84 | 2.91 |
| -35 | 174.3 | 0.399 | 23.0 | -58 | 2.41 |

Table 4.2: Beam radii and mirror illumination for various initial primary illuminations. Subscripts P and S refer to the primary and secondary respectively. T is the edge illumination. $T_P$ is assumed and the rest of the parameters are calculated. The last row gives the final choice of beam parameters for the tertiary design.

| $w_0^{cass}$ (cm) | $d_1$ (cm) | $R_{incoming}$ (cm) | $d_2$ (cm) | $R_{outgoing}$ (cm) | $w_0^{horn}$ (cm) |
|---|---|---|---|---|---|
| 2.23 | 10 | 229.6 | 22.76 | 433.23 | 1.28 |
| 2.54 | 10 | 379.7 | 26.9 | 35.79 | 1.28 |
| 2.91 | 10 | 646.9 | 31.8 | 39.32 | 1.28 |
| 2.41 | 20 | 169.8 | 26.8 | 35.7 | 1.28 |
| 2.41 | 10 | 309.64 | 25.2 | 34.66 | 1.28 |

Table 4.3: Tertiary mirror incoming and outgoing radius of curvature for different initial primary illuminations. $w_0^{cass}$, $d_1$, and $w_0^{horn}$ are fixed and the rest of the parameters are calculated. The final design is tabulated in the last row.

| Parameter | Value |
|---|---|
| Aperture diameter | 4 cm |
| Length | 60 cm |
| Beam FWHM | 9° |
| Beam radius at aperture | 1.28 cm |
| Distance from aperture to phase center | 4.53 cm |

Table 4.4: Approximate characteristics describing the W-band WMAP feed horn used for the total power receiver. An exact description of the horn can be found in [9]. The beam radius at the aperture is calculated from the formula for a diffraction-limited feed horn in [44].

### Mirror construction

As suggested in [34, 148], it is preferable to make the mirror from jig plate as opposed to rolled plate because rolled plate has concentrated stress on the surface which is then likely to warp upon cutting. However, having had experience[3] cutting mirrors from rolled plate and because the cut in the plate is quite shallow (<1 cm), the mirror is cut from an 8”×9”×2” thick piece of 6061 Aluminum rolled plate. The size is certainly overkill, giving an edge illumination greater than -80 dB. The mirror has three 0.75” conical pockets machined in the back to accept 1" steel balls mounted at the end of a threaded rod. These allow the mounted mirror to be finely positioned.

[3]The Princeton machine shop built the MINT mirrors [41].

| Parameter | Value |
|---|---|
| $\mathrm{R}_1$ | 121.90 in |
| $R_2$ | 13.64 in |
| $\theta_i$ | 24° |
| $f_0$ | 11.16 in |
| e | 0.83 |
| a | 67.77 in |
| b | 37.26 in |
| $\theta_p$ | 174.86° |
| $\psi$ | 150.86° |
| $\theta_{tilt}$ | 29.13° |

Table 4.5: The parameters of the tertiary mirror. The mirror is an off-axis ellipsoid. The definition of these parameters is given in Appendix F.

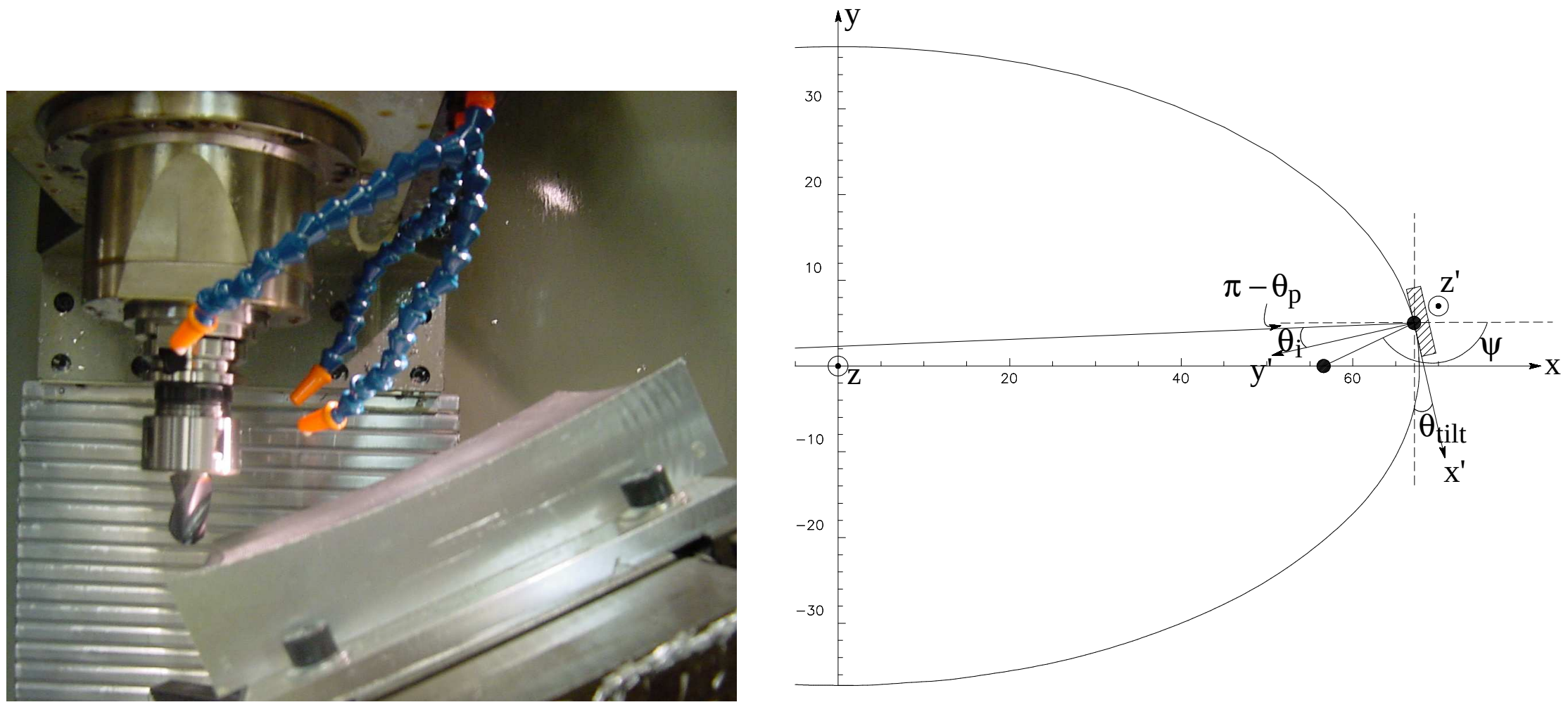

(a) (b)

Figure 4.5: (a) Picture of the mirror in the CNC mill during fabrication. Note the mirror is mounted at an angle to prevent being cut with the tip of the ball mill. (b) Diagram showing the section of ellipse the tertiary is cut from, along with useful angles. All numbers are in inches.

Machining an off-axis ellipsoid cannot be done on a lathe. Referring to Figure 4.5(b), the off-axis ellipsoid is symmetric with respect to the x'z' plane while the full on-axis conic section is symmetric around the x-axis. The mirror is therefore cut on a CNC Bridgeport mill in the Princeton Machine shop. Instead of specifying the cutter trajectory as a list of points, it is still possible to exploit the rotational symmetry of a conic section to generate an exact trajectory[4]. If the plate to be cut is placed as in Figure 4.5(b), with the x-axis pointing downwards, then the mill can cut horizontal circular arcs in the yz plane which will follow the exact mirror surface. Cutting circular arcs with the Bridgeport DX-32 controller is very simple. The CNC code used to machine the mirror is provided in Appendix F.

The mirror is cut with a 3-flute, carbide, 1" diameter ball end mill at a speed of approximately 2000 RPM. To prevent cutting with the bottom of the ball, the mirror is mounted on a tilted surface at an angle $\theta_{tilt} = 29.13°$, such that the xyz axes in Figure 4.5(b) coincide with the three axes of the CNC mill (see Figure 4.5(a)). The surface of the mirror in the code is specified such that the near vertex of the ellipse is the origin. Several passes were needed to obtain the final surface. Two roughing cuts brought the surface to within 200 and 20 mils of the final shape. During these roughing cuts, the horizontal feed rate was 20 and 10 inches per minute with a step size between each circular arc cut of 40 mils. Each roughing cut took approximately 3 hours. The final cut kept the same feed rate but the step size between arcs was reduced to 4 mils. This produced a smooth enough surface that the mirror did not require any more polishing for operation at $\lambda = 3$mm. The final cut took $\sim 6$ hours of continuous machining. After machining, the four corners facing the beam were flattened to place a 1/4" shoulder ball at each end. The shoulder balls serve as position reference points during the placement and alignment of the mirror.

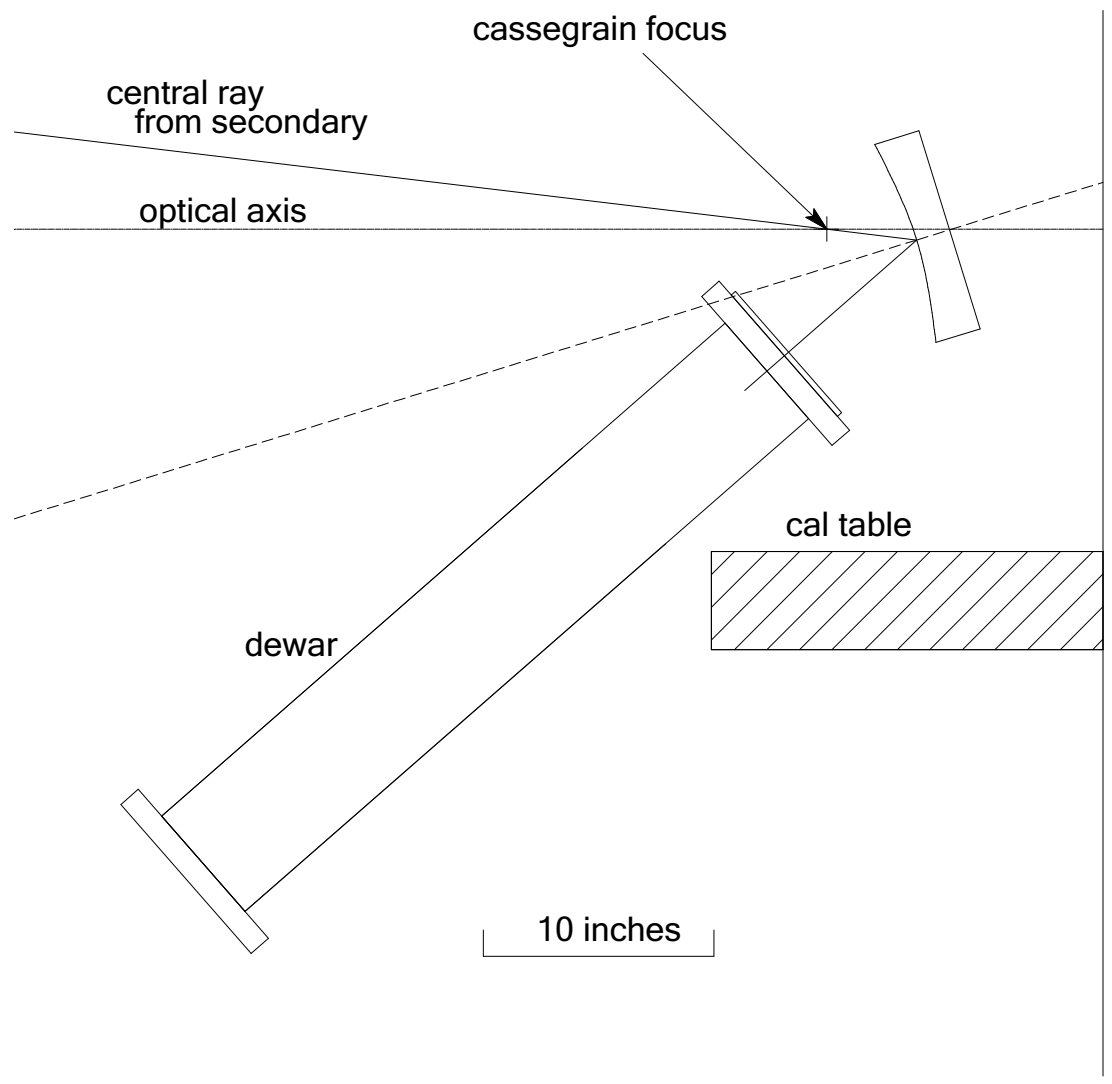

Figure 4.6: Side view of mirror feed system in receiver cab. The room temperature dewar mounting structure is not shown. The vertical line is the left surface of the vertex cab dividing wall. Drawing is to scale.

[4]The idea for the machining algorithm comes from Simon Dicker, post-doc at the University of Pennsylvania.

The alignment is first done using an outer micrometer and a digital protractor to roughly position the mirror with respect to the Cassegrain focus and the dewar. The final alignment is done with a 2-sided co-linear laser beam such that its two ends hit the Cassegrain focus and the center of the secondary mirror. The tertiary is then adjusted such that the laser beam reflects from its center to the center of the feed horn aperture.

The mirror feed was installed on the telescope on July 20, 2001. A good beam map was obtained on Jupiter on March 5, 2002. Considerable amount of work between these two dates was spent discovering and repairing new problems with the telescope drive system. With this single successful observation, the measured beam size was $FWHM_x = 0.073° \pm .02$, $FWHM_y = 0.082° \pm .04$.

## 4.3 Control System

The original antenna software control program,`obs`, was designed by Bob Wilson. In order to accommodate the new CAPMAP scan strategy and to allow remote telescope control, major changes were performed on `obs` over the period of two years prior to the CAPMAP03 season. The main modifications were made on the telescope control code and computer[5]. The OBS computer is now a 750 MHz computer running RedHat Linux 7.2. It was recently (Sept 2002) built to replace the former 100 MHz computer with 16 Mb of RAM running a Lynx real-time operating system. This upgrade to a newer and better known operating system was done by M. Yeh. A comprehensive description of how `obs` works and can be updated is now available [151].

### 4.3.1 Software Control

The Crawford Hill antenna can be controlled either manually in Slew mode (with the az/el slew knobs on the antenna control front panel) or in Remote mode via the OBS computer.

The OBS computer reads the position of each axis every 10 ms (see Appendix E) and applies the scaled difference as a velocity request to the hardware drive system. In order to keep the commanded velocity within the acceleration and deceleration limits of the telescope, the gain of the velocity feedback loop is compressed for velocities greater than $\frac{1}{8}$ ($\frac{1}{4}$) of the maximum azimuth (elevation) velocity. In addition, the acceleration is software limited to $1.5°/s^2$ in both axes. The azimuth and elevation velocity request loop is summarized in Figure 4.7.

[5]`obs` is the name of the control software and resides on the computer also called OBS

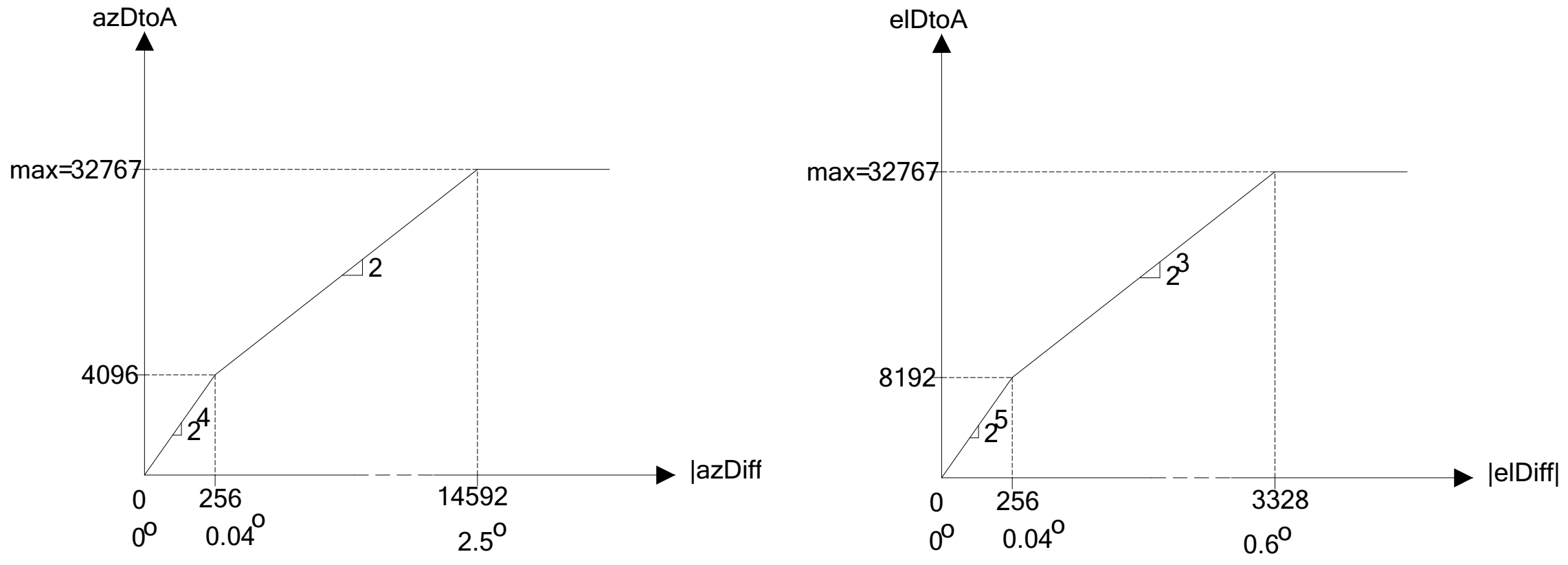

(a) (b)

Figure 4.7: Az and El velocity request servo loop parameters. azdiff and elDiff are the difference between the commanded and current position in encoder bits ($360° = 2^{21}$ bits). AzDtoA and ElDtoA are the requested velocity to the telescope is bits ($2^{15} = +10\text{V} = 2.16°/s$ in azimuth and $1.05°/s$ in elevation). In addition to the gain compression at large velocities, the software also constrains accelerations to be less than $1.5°/s^2$.

OBS communicates with the rest of the antenna hardware via 3 cards: a 24-bit digital I/O card for reading the encoders and sending a remote-enable bit to gain computer control of the telescope; a 16-bit analog output card to send the analog output velocity request voltages to the motors; and a GPS timing card. A more detailed description of their operation can be found in [151].

### 4.3.2 Hardware Control

The telescope is driven with 4 DC motors, two on each axis. The necessary hardware to control these motors is all located in the Antenna Control Rack. The control rack consists of a few separate subsystems: the AC power distribution panel, the power supplies, the SCR power amplifiers logic circuit board, the SCR power amplifiers themselves, the digital logic circuit card (the infamous “wire-wrap board”), and the antenna control front panel.

The SCR (Silicon Controlled Amplifiers) provide the DC currents to the motors. They are triggered to fire according to a complicated feedback loop. The feedback mechanism varies the current in the motors until the applied torque in the motors (measured by the tachometer) matches the requested torque. The current in the motors is changed by varying the phase at which the SCR are fired with respect to the 60 Hz cycle of the 240 VAC power. This feedback loop is controlled in the SCR logic circuit board. A more detailed description of the SCR power amplifiers and their logic circuit board is given in Appendix E.

The AC power distribution panel is a set of wiring breakouts, fuses, and circuit breakers located in the back of the antenna control rack. The control rack also houses 4 DC power supplies in the front. These are for various analog and digital interlock logic loops to control the antenna. All the digital logic circuits are located in the wire-wrap board. It controls the integrity of the various sub-systems coexisting on the antenna (brakes action, fuses faults, current limits, motion limits, power supplies, cable wrap, operation mode, and various interlock loops) and displays their status on the antenna control front panel. Circuit and wiring diagrams for these systems are located in the control room, with the rest of the telescope documentation. A maintenance guide to the telescope, summarizing the problems we encountered since we started using the antenna, their symptoms, and suggested solutions is available [7].

# Chapter 5

# The 2002-2003 Observations

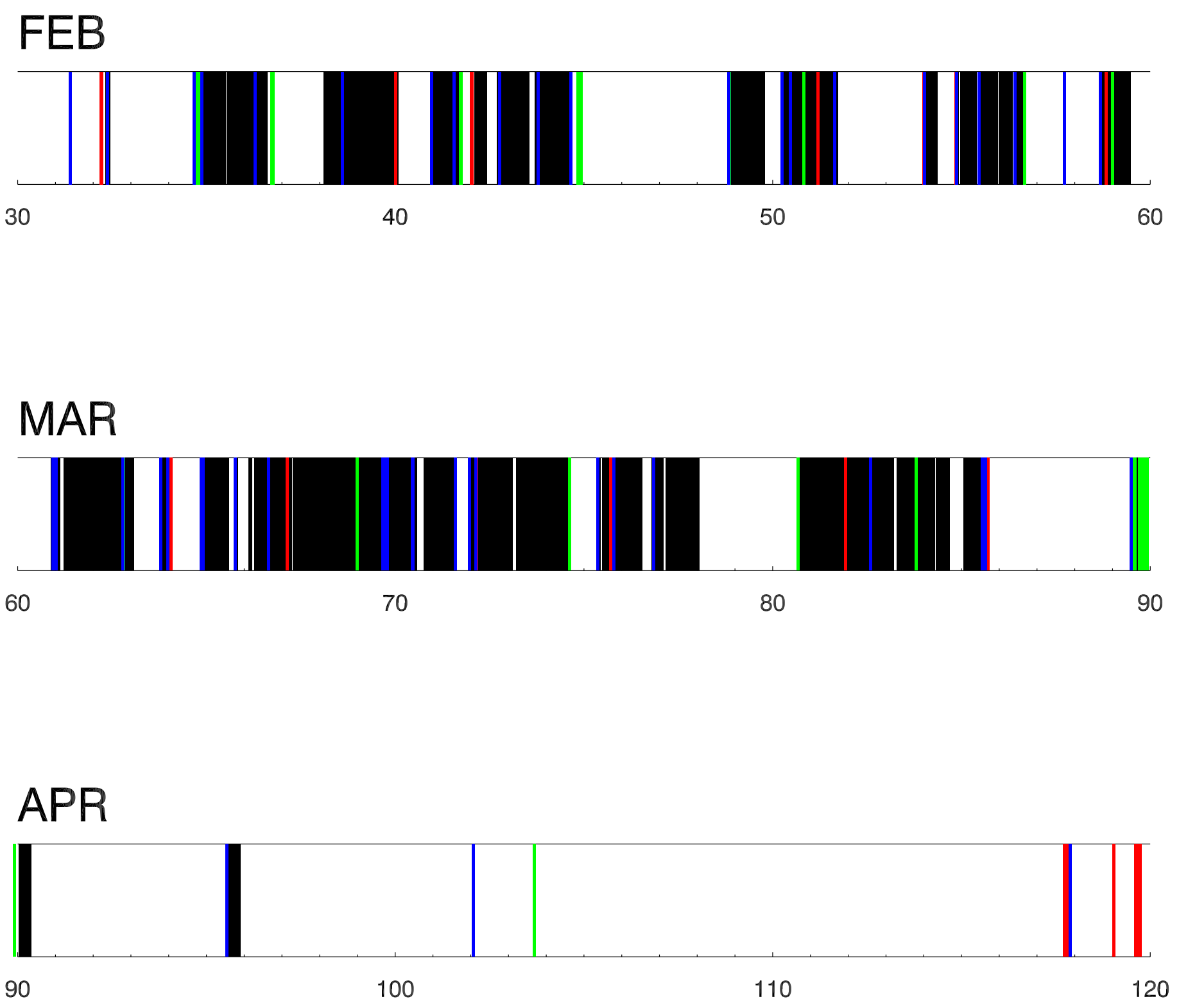

Figure 5.1: Time line of the 2003 observing season. The black blocks represent periods when the telescope was in CMB observing mode. The lines are scheduled observations of Jupiter (red), Taua (green), and sky dips (blue).

## 5.1 Summary of Observations

The CAPMAP03 instrument was brought to the telescope on Jan 16, 2003. The intensive CMB observing lasted from Feb 2 to April 6. The instrument remained on the telescope until June 18, 2003 for various post-season investigations. A breakdown of the time spent on various activities is presented in Table 5.1.

| Status | Amount of time |
|---|---:|
| Initial tests and tuning | 16 days |
| Post-season tests | 72 days |
| Active observing period | 63 days = 1500 hrs |
| Temperature regulation problems | 100 hrs |
| Telescope mechanical problems | 50 hrs |
| Scheduled calibration[a] | 30 hrs |
| Planet calibration scans[b] | 60 hrs |
| Bad weather | 720 hrs |
| Useable CMB observations | 540 hrs |

Table 5.1: Time summary of the CAPMAP 2003 winter season. Note that a total of 150 hours are lost at the beginning of the observing period due to electro-mechanical problems. Approximately 50% of the useable CMB observation go into final data maps.
[a]sky dips, chopper plate tests, Y-factor two-load tests and some ring mode (see Section 3.3).
[b]planet scans refer to observations of Jupiter, Taua, the Moon, and other celestial objects (see Chapter 6).

### 5.1.1 Instrument Deployment

Deployment is a small word for an extensive task. After integration and testing in the lab, the dewar, compressor, associated electronics boxes, and cables are brought to Crawford Hill.

**Dewar Mount** The dewar is the hardest item to install because it weighs ∼300 pounds. It is lifted with a crane to the vertex cab walkway and man handled through the vertex cab door onto the vertex cab cal table. The cal table is the old receiver mounting table (see Figure 5.2). The mount of the CAPMAP03 dewar comprises two legs on either side of the dewar (see Figure 5.2). The legs are made of 1” Al turnbuckle which thread into a 5” x-y translation block on the table side and a ball rod end on the other. The ball rod end slides into a 1” shaft attached to the dewar body 11.35” from the front lid. The two mounting points on either side act as a yoke allowing the dewar to translate in the x and z-axis, and pivot around the x-axis. The dewar center of gravity is behind the yoke so its weight rests on a third leg in the back of the dewar. After the initial alignment, we found that the two side legs deflected under gravity as the vertex cab rotated when the telescope elevation changed. Three cross bars were attached from the turnbuckle to the table to stiffen the mount. We visually inspected the mount upon tilting of the telescope vertically and could not see a displacement of the dewar. The telescope pointing solution (Section 6.4.1) puts an upper limit on the maximum dewar motion to be less that 0.72" for the whole range of telescope positions.

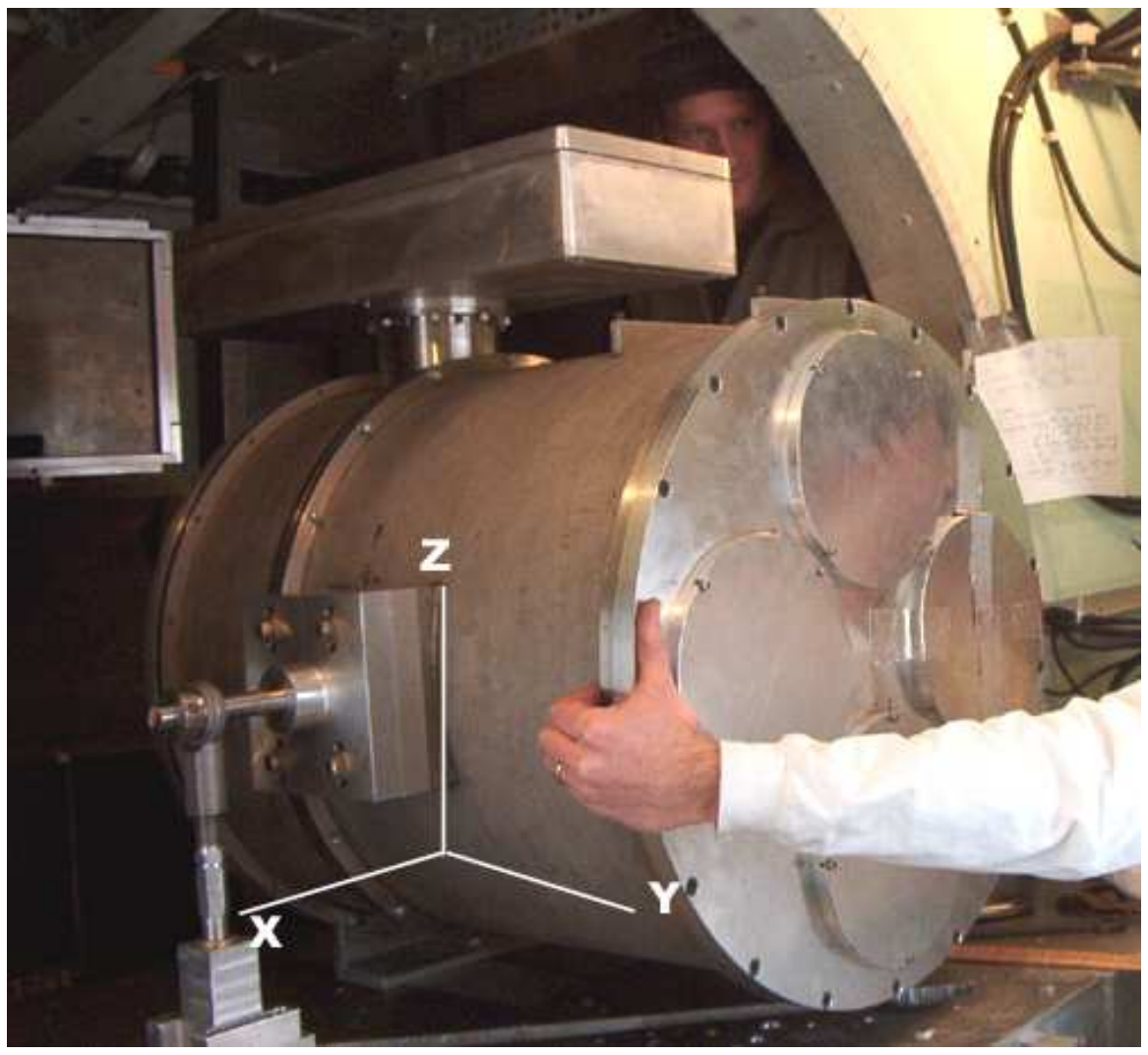

Figure 5.2: Geometry of the dewar on the cal table. The dewar is facing the secondary. The yoke style mount on either side of the dewar is not yet outfitted with the extra stiffening angled-brackets. See Figure 5.3 for a scale drawing of the focal plane area.

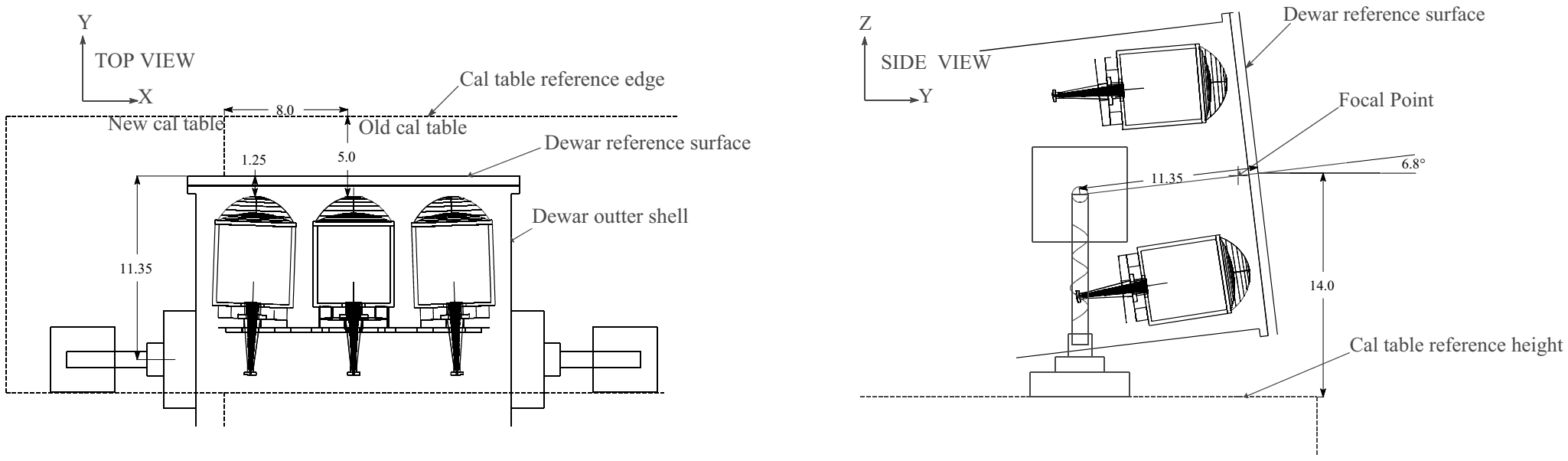

Figure 5.3: Reference surfaces and angles for dewar alignment.

**Dewar Alignment** The dewar is aligned using first a crude mechanical method to place it in the right position with respect to the focal point, then more finely with a laser to verify the tilt. The goal of the alignment is to place the focal point such that the plane perpendicular to the central ray (the focal plane) intersects the front of the lenses.

The reference surfaces for dewar alignment are shown in Figure 5.3. The dewar is placed within 1/8" of the correct position in the xy-plane and tilted at 6.83° with respect to horizontal. The final alinement of the dewar height (z-axis) is adjusted to within 1" and verified with a two-way laser which had been previously set up such that its two beams hit the center of the secondary and the focal point. Post season analysis of Jupiter beam maps (see section 6.4.2) indicate that the receiver array was positioned 1.2" too high.

The compressor is installed outside, 10 m away from the fridge. The Helium fridge lines are conveyed from the compressor to the cold head via the rotating tunnel which connects the vertex and the side cab. All the electronic support boxes (including the DAQ computer, ABOB, phase-switch box, and power supplies) sit in a 19"-wide rack in the back of the vertex cab ∼2 meters away from the dewar. The cables are routed, electrically checked and plugged into the dewar, IF box, ABOB and power supplies. After cursory verification that the receivers are still alive, the fridge is turned on. Although the receivers worked well right away, nominal CMB observing did not start until 16 days after deployment due to DAQ software issues, telescope mechanical failures, and encoder readout problems.

### 5.1.2 Highlights

The specific identity of the 4 receivers as they were finally fielded is presented in Table 5.2.

<table>
<tr><th>Receiver position</th><th>Arm position</th><th>Total power channel</th><th>Pol. orientation on the sky</th><th>Lens</th><th>IF amp</th></tr>
<tr><td rowspan="2">A</td><td>side</td><td>D0</td><td>vertical</td><td rowspan="2">not grooved</td><td rowspan="2">8 V</td></tr>
<tr><td>main</td><td>D1</td><td>horizontal</td></tr>
<tr><td rowspan="2">B</td><td>side</td><td>D0</td><td>horizontal</td><td rowspan="2">grooved</td><td rowspan="2">12 V</td></tr>
<tr><td>main</td><td>D1</td><td>vertical</td></tr>
<tr><td rowspan="2">C</td><td>main</td><td>D0</td><td>horizontal</td><td rowspan="2">grooved</td><td rowspan="2">8 V</td></tr>
<tr><td>side</td><td>D1</td><td>vertical</td></tr>
<tr><td rowspan="2">D</td><td>side</td><td>D0</td><td>horizontal</td><td rowspan="2">not grooved</td><td rowspan="2">12 V</td></tr>
<tr><td>main</td><td>D1</td><td>vertical</td></tr>
</table>

Table 5.2: Operational identity of the 4 receivers during the CAPMAP03 season. The lenses are all grooved on the back surface. Only the front surface is not grooved for receivers A and D.

**Scan Pattern** The instrument observed the sky in nominal CMB mode from Feb 2 (day 32) to April 6 2003 (day 95). In this mode (aka "scaz" mode), the center of the array scans back and forth in azimuth at constant speed while pointing at a constant elevation. In an effort to point the center of the array at NCP, the scan was centered at az = 359.97° and el = 40.3194°. We later realized that because of a sign confusion in the Jupiter pointing reconstruction (Section 6.4.2), the scan was centered ∼0.1° from NCP. Figure 5.4 shows the scan coverage for each receiver for the actual and ideal pointing of the scan center. Figure 5.5(a) shows the azimuth scan pattern as a function of time. In the initial scaz mode, the scanned region had an encoder radius of 0.75° (sky radius = 0.59°) with a period of 10 seconds exactly. However, we discovered an unexpected hiccup in the azimuth velocity visible in Figure 5.5(b). To stabilize the motion, the scan period was changed to 8 seconds and the scanned region to a sky radius of 0.53° on February 18 (day 48). This remained the nominal CMB scan mode for the rest of the season. Since then, the telescope has successfully completed a little over 500 000 scans. The only interruptions came from scheduled calibration, planet scans, and bad weather.

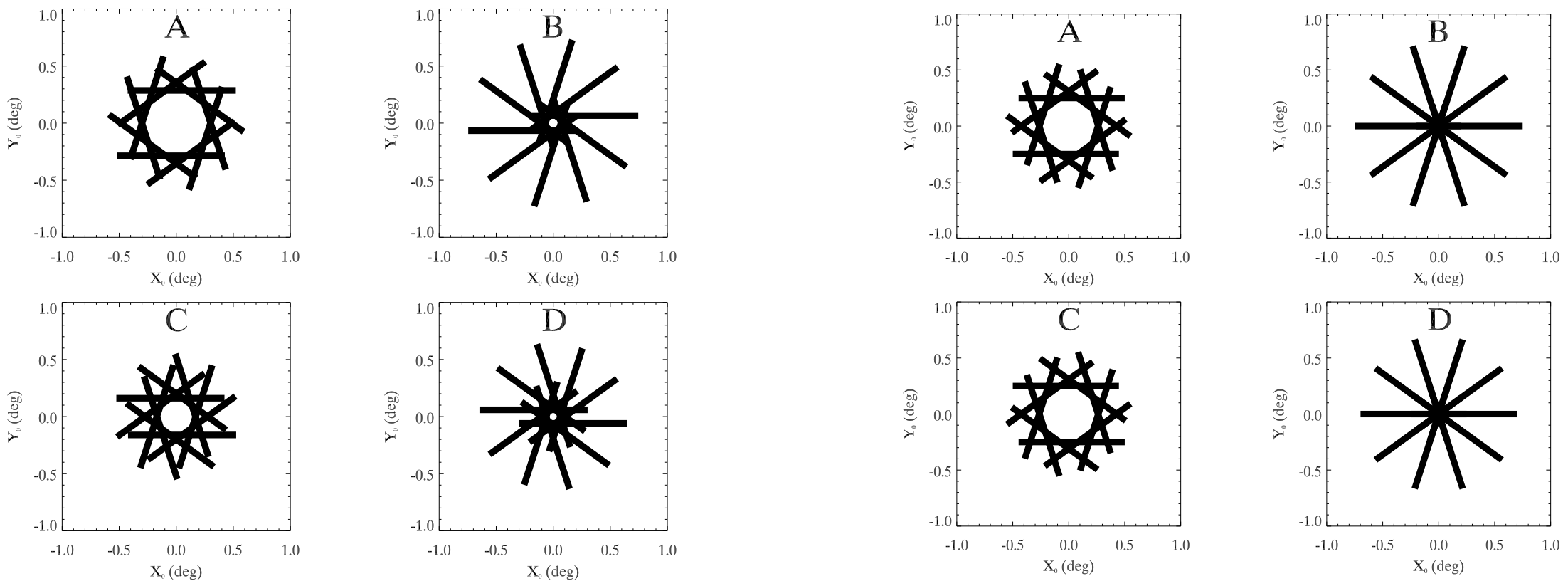

Figure 5.4: Scan pattern for the actual (left) and ideal (right) pointing. Each plot is centered on NCP. The plots show 10 individual azimuth scans separated by 2.4 hours. The width of each scan is the beam size (0.06°). By having the whole array mispointed with respect to NCP, the redundancy in the scan pattern is lost, but additional QU cross-linking of the receivers is gained.

**Cryostat Thermal Performance** After the initial in-situ cooldown of the cryostat on January 17, 2003, it stayed cold and under vacuum for six months without a single cryogenic problem[1]. Figure 5.6 shows the temperatures of the 20 K, 70 K, and warm stages in the dewar during the cooldown. The MMIC amplifiers, horns, and cold stage all reach approximately 22.3 K while the 70 K stage and the shroud reach 67 K. The dewar was pumped down to 1 mTorr before cooling and remained below 0.01 mTorr during the whole season, with no pumps attached to the dewar apart from the charcoal cryopump on the 20 K stage. Figure 5.7 displays the pressure and the temperature of the relevant dewar and external stages during the whole season. It is interesting to note that the dewar

[1]a singular event in the short history of experiments I've fielded.

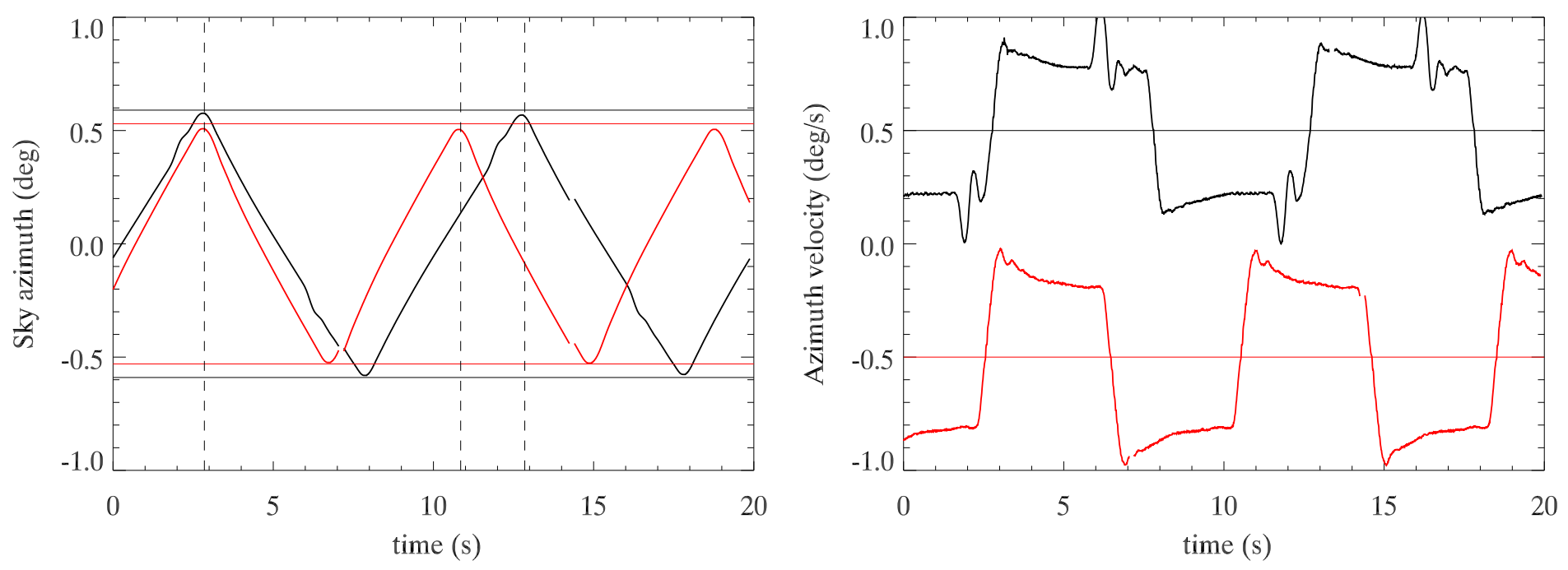

(a) (b)

Figure 5.5: (a) Initial (black) and nominal (red) azimuth scan pattern. The initial pattern scans a 0.59° sky radius region in 10 seconds but has an unexplained velocity glitch before the turn-around points. The scan period was changed to 8 seconds which also reduced the scan radius to 0.53°. (b) Azimuth velocity waveforms (in $^\circ/s$) for the initial (top) and nominal (bottom) scan. The plots are shifted by $0.5^\circ/s$ for clarity. The initial pattern clearly shows the velocity hiccup.

pressure follows an exponential decrease (with a halving period of 20 days). This behavior is consistent with constant rate of cyropumping on an outgassing process. The 20 K and 70 K stage are kept at $22.3 \pm .3$ K and $67 \pm 4$K, with no active regulation. The fluctuations on these stages are not from daily variations but from larger excursions in the heat loading, correlated with slow outside temperature changes such as the passage of a front. The vertex cab damps the daily outside variations because it is loosely regulated at 12° C and only responds to large excursions. For example, a sudden 5° C temperature increase on day 65.5 is visible in all cryogenic temperatures and pressure, and causes the LO and IF box to lose thermal regulation.

**Room Temperature Thermal Regulation** Although nominal CMB observation started on Feb 2, 2003, conservatively useable CMB data only starts on Feb 18. This is partly because of the motion hiccup problem described above, but also because of thermal regulation issues. Until Feb 18, neither the LO in the dewar nor the IF box were actively thermally regulated. Changes in temperature of the LO and the IF amplifiers cause the strongest gain variations (see Section 6.5.1), making these important to regulate. In addition, the IF box temperature was not correctly recorded until Feb 18. Thus the 150 hours of otherwise good CMB data until Feb 18 are discarded a priori. During the rest of the season, the IF box is regulated at 14.3° C and the LO at 32.5° C (see Figure 5.7). The LO is regulated using a switched 5W heating resistor. Both are regulated using an Omron[2] digital controller which toggles the heater (or cooler) current on and off with varying duty cycle to apply the appropriate power. The IF box is cooled with four Peltier coolers heat-sunk to radiators which can draw a maximum of 6 W each. Periods of high vertex cab temperature cause both devices to go out of regulation despite the vertex cab air conditioning. On April 1(day 90), in expectation of warmer spring temperature, the LO and IF regulation temperature were increased to 40° C and 17° C, although only two days of data were taken in these conditions.

[2]PN: E5GN, see also `http://oeiweb.omron.com/Products-Temp.shtm#32`

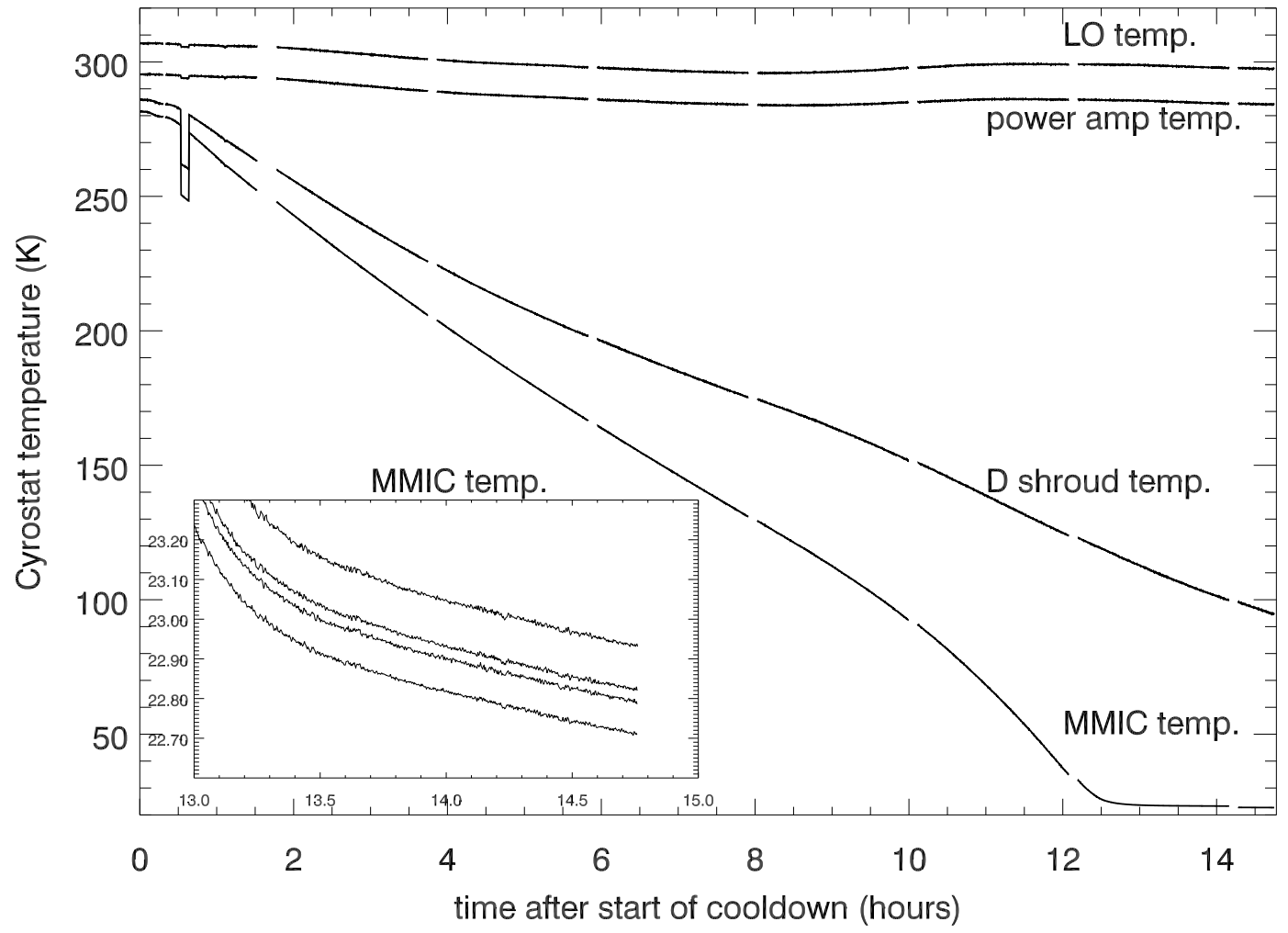

Figure 5.6: Cooldown temperatures of the 20 K (MMIC temp), 70 K (D shroud temp), and warm stage (LO and power amp temp) of the CAPMAP03 dewar using the 1020 compressor and 1020 cold head. The 70 K stage has not yet stabilized. The dewar is cooled down with metal covers on its windows to minimize thermal loading. The inset zooms in on the last two hours of the temperature for the four receiver's MMIC front stage amplifiers. They differ by ∼0.3K, the accuracy of the thermometers. This proves there is no azimuthal heat load distribution and the temperature in any receiver is a good indicator of the temperature in all receivers.

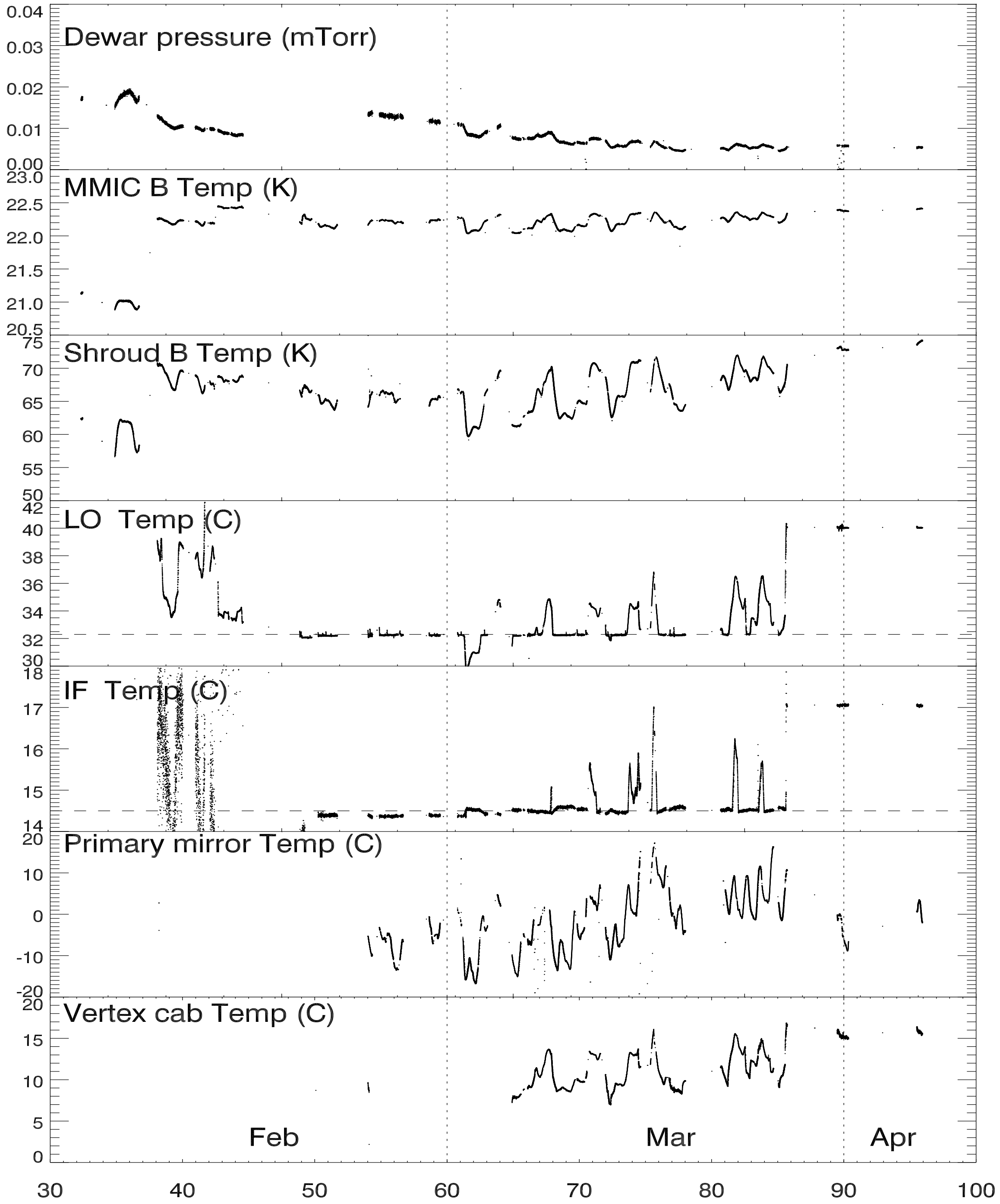

Figure 5.7: Dewar pressure, and cryogenic and external temperature during the 2003 winter season. The labels indicate the position of the thermometers. The IF thermometer is placed directly on one of the IF amplifiers. There are 8 mirror thermometers placed around the back of the primary mirror. The vertex cab thermometers is located on the electronics rack behind the dewar. The cryogenic sensors are Lakeshore DT-670 silicon diodes. The warm thermometers are all LM335 from National semiconductors. Problems in the ABOB during the early part of the season prevent several warm temperatures (including the IF and LO temperatures) from being recorded. One caveat was discovered after the season. All cryogenic temperature are actually 1 to 15 K warmer than recorded due to a miscalibrated current bias to the silicon diode temperature sensors (not 10 microamps).

## 5.2 Data Acquisition

The data acquisition is performed by a dedicated computer running a C program[3]. It is a PC running RedHat Linux 7.2[4]. The DAQ computer is physically located in the back of the vertex cab but is networked so it can be accessed from the control room. The interconnection between the data acquisition and the instrument is summarized in Figure 5.8 and 5.13. During standard observations, the data acquisition code (readout.c) records the telescope encoder position, GPS time, housekeeping and radiometer data channels, and one error flag. These data are all sampled at 100 Hz except the housekeeping channels which is sampled at 1 Hz, and written to a file every 30 minutes in binary format (see Appendix G). The acquisition of the radiometer data channels (the main data acquisition task) is described in Section 5.2.1. The encoder and housekeeping information are read out via separate interfaces, presented in Section 5.2.3.

### 5.2.1 Radiometer Data Acquisition

From the point where the signal is detected (in the multiplier for the polarization channels or in the detector diode for the total power channels), to where it is digitized in the DAQ, the signal must go through various conditioning steps to ensure that the digitized signals are a true representation of the analog signals. In the PIQUE experiment [56] for example, these conditioning tasks were each done in hardware, on separate electronic cards: an AC programmable gain, a demodulation card, and a Voltage-to-Frequency conversion card to act as an Analog-to-Digital converter. For CAPMAP, all these tasks are performed in a single commercial computer card[5]. This is a standard PCI card which contains 32 simultaneously sampled differential channels, each consisting of a programmable gain stage, an anti-aliasing filter, and a 24-bit Sigma-Delta A/D converter. One ICS card can accommodate up to 6 W-band receivers (5 differential channels per receiver). Before digitization, the signals are small (a few mV) analog voltages so they are carried from the IF box to the DAQ card on differential twisted shielded pairs of wires. The acquisition rates are defined as follows:

- **digitization rate: 100 kHz.** This is the initial rate at which the ICS card digitizes the modulated radiometer signals.

[3]the data acquisition code,“readout.c” was written by Keith Vanderlinde. Dorothea Samtleben was responsible for the design and implementation of the DAQ hardware (clock box, daq computer, and ICS card). Michelle Yeh and Ashish Gupta designed and wrote the Labview monitoring and DAQ-monitoring communication code.

[4]2.4 GHz processor with 1Gb of RAM.

[5]ICS-610 card: technical information available at `http://www.ics-ltd.com`

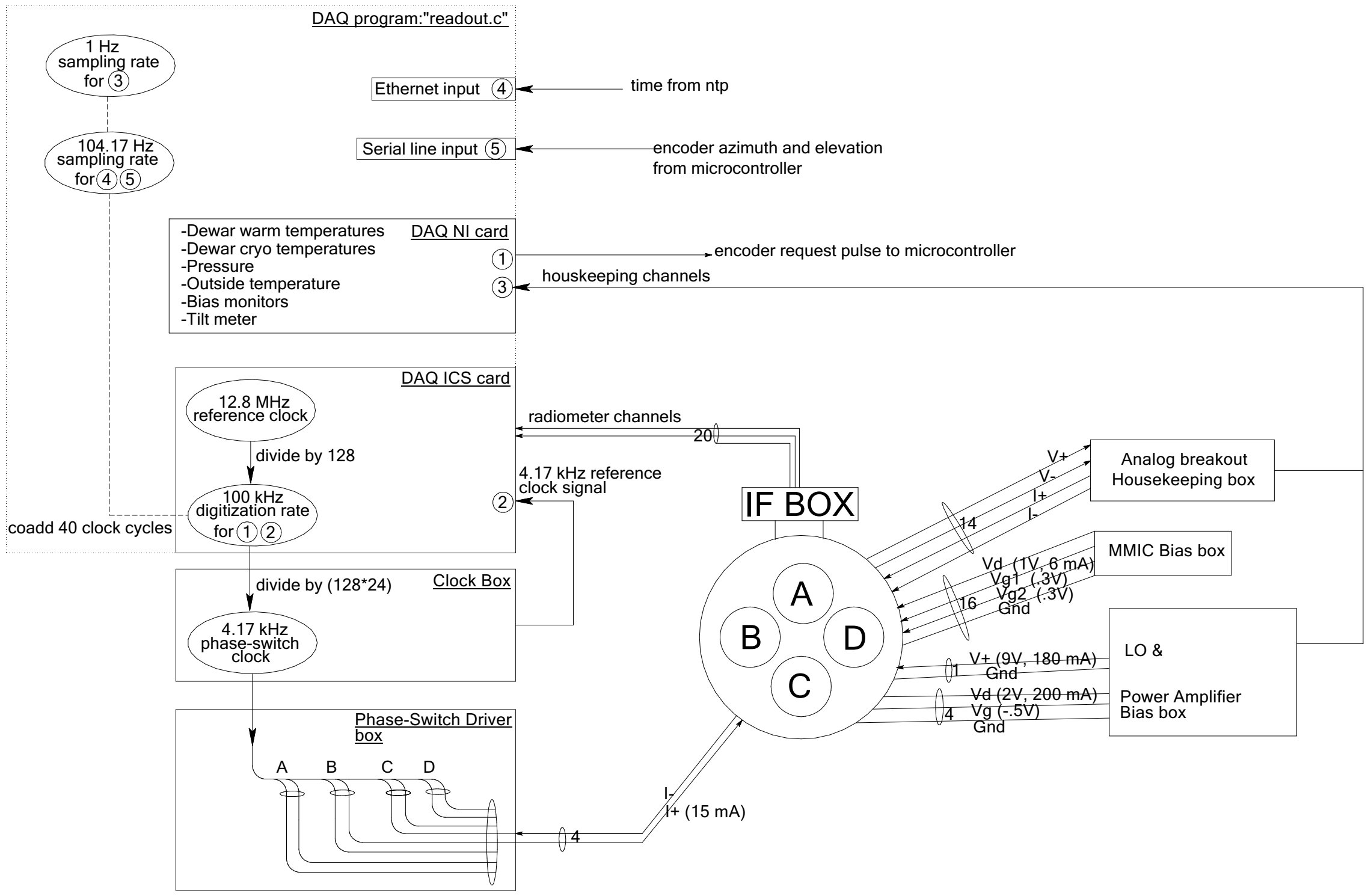

Figure 5.8: Instrument functional block diagram, showing the interconnection between the instrument electronics and receivers. Arrows indicate the direction of the signal. Where possible, the value of the typical bias voltages is indicated. All electronic boxes (ABOB, MMIC Bias box, LO bias box, Clock Box, DAQ computer, and power supplies) are mounted in the rack behind the dewar.

- **sampling rate: 100 Hz.** The final rate at which the radiometer signal data are recorded.
- **housekeeping rate: 1 Hz.** The housekeeping data are sampled once a second, at the start of a data frame (see Appendix G).
- **phase-switching rate: 4 kHz.** Frequency at which the polarization channels are phase switched in the LO line of the receivers. This is naturally also the frequency at which the polarization signals are demodulated in the DAQ code.

The acquisition scheme for a single data sample is illustrated in Figure 5.9. The ICS card fills up two buffers of 480 samples each acquired at 100 kHz. This represents exactly 40 phase switch cycles each with 24 samples. "readcout.c" then co-adds the 40 data values from each the 24 clock phases to produce 24 points. These 24 points are then demodulated (see below) into a single data sample for a polarization channel or averaged for a total power channel. These samples get recorded at 100 kHz/960 $\simeq$ 100 Hz ($\Delta t = 9.6s$).

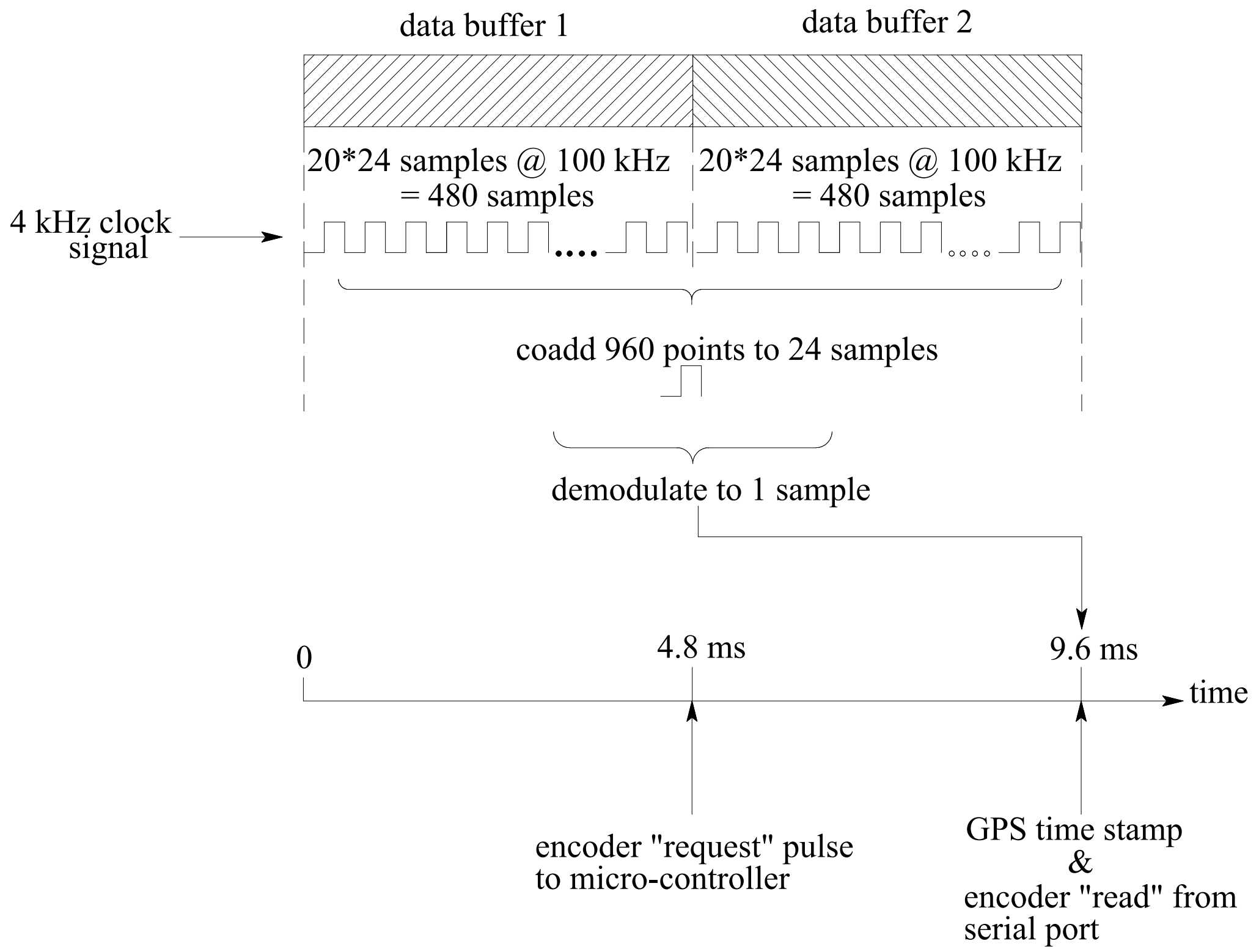

Figure 5.9: Timing diagram of the various data acquisition events for a single data sample. The final data rate is approximately 100 Hz. The recorded encoder position comes from the point where the DAQ requests the encoder sample, 4.8 ms earlier than the time stamp of the data sample.

The software demodulation performs the necessary lock-in on the 4 kHz modulation of the polarization channels. Refer to Figure 5.9 for a summary of timing signals. The master clock (a 12.8 MHz reference signal from the ICS card) which generates the 100 kHz digitization rate (by a factor of 128 down-converting) also generates the synchronous 4 kHz clock[6] which is used to drive the four phase-switch circuits. This clock is digitized and co-added just like a data channel. Readout.c then finds the extremum of the 24 points and derives the center of the square wave. The clock points are then separated into the 12 highest and lowest points. The mean of the 12 low data points is subtracted from the mean of the 12 high ones to generate a single demodulated data sample.

[6]the 4 kHz frequency is actually 100 kHz/24 = $4.16\overline{6}$.

### 5.2.2 DAQ Performance

Although the method to produce the digitization and phase switching rate effectively keeps them synchronized, it does not necessarily force them to have the same phase. Because of signal time delays, the rising edge of the down-sampled 4 kHz clock is not guaranteed to be in phase with the rising edge 100 kHz digitization rate as shown in Figure 5.10. Therefore, demodulating the radiometer signal by subtracting the mean value of the 12 highest points from the mean value of the 12 lowest points produces an inevitable error in the amplitude of the demodulated signal.

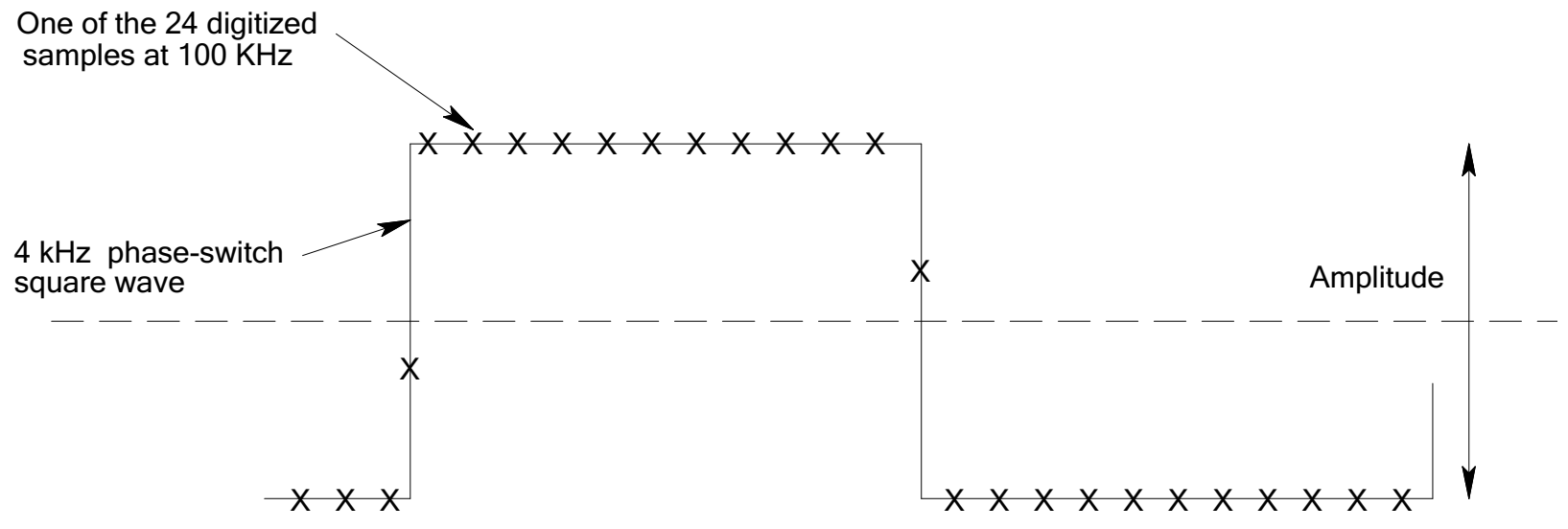

Figure 5.10: Example of a phase shift between the digitization rate and the phase-switch clock. The plot shows a single 4 kHz clock period during which 24 data samples are acquired (black crosses). Had the digitization rate and the 4 kHZ clock been in phase, the first digitized sample would have been at the top of the rising edge of the clock signal. Instead, two of the samples are acquired in the 4 kHz clock transition region. Given the demodulation algorithm described in the text, this effect gives rise to a underestimation of the demodulated amplitude of the data.

Let's take an example to clarify the situation. Assuming the modulated radiometer signal is digitized such that for the high state points, 11 are equal to 1 V and one to $\frac{1}{4}$V, and for the low state points, 11 are equal to $-1$ V and one to $-\frac{1}{4}$ V as in Figure 5.10. Performing the demodulation algorithm as described above would yield a demodulated amplitude $A = \frac{1}{12}\left(11*1 + 1*\frac{1}{4} - 11*-1 - 1*-\frac{1}{4}\right) = 1.875$. The demodulated amplitude with zero phase delay should have been $A = 2$. This clock phase delay has thus caused a 6.6% error in the demodulated amplitude.

Such phase shift between the digitization and the 4 kHz phase-switch clock occurred during the 2003 observing season [125]. This effect could be monitored both in real time with the 100 kHz clock data on Labview and off line because the 100 kHz data were also stored to disk. Until Feb 27 2003, the 4 kHz clock phase drifted continuously during observations at a rate of approximately one full cycle per day. This was eventually tracked down to the PECL→TTL translator card in the clock box [125] and repaired. However, a constant phase shift remained[7].

[7] The drifting was fixed but the phase shift skipped by a fixed amount at the start of "readout.c".

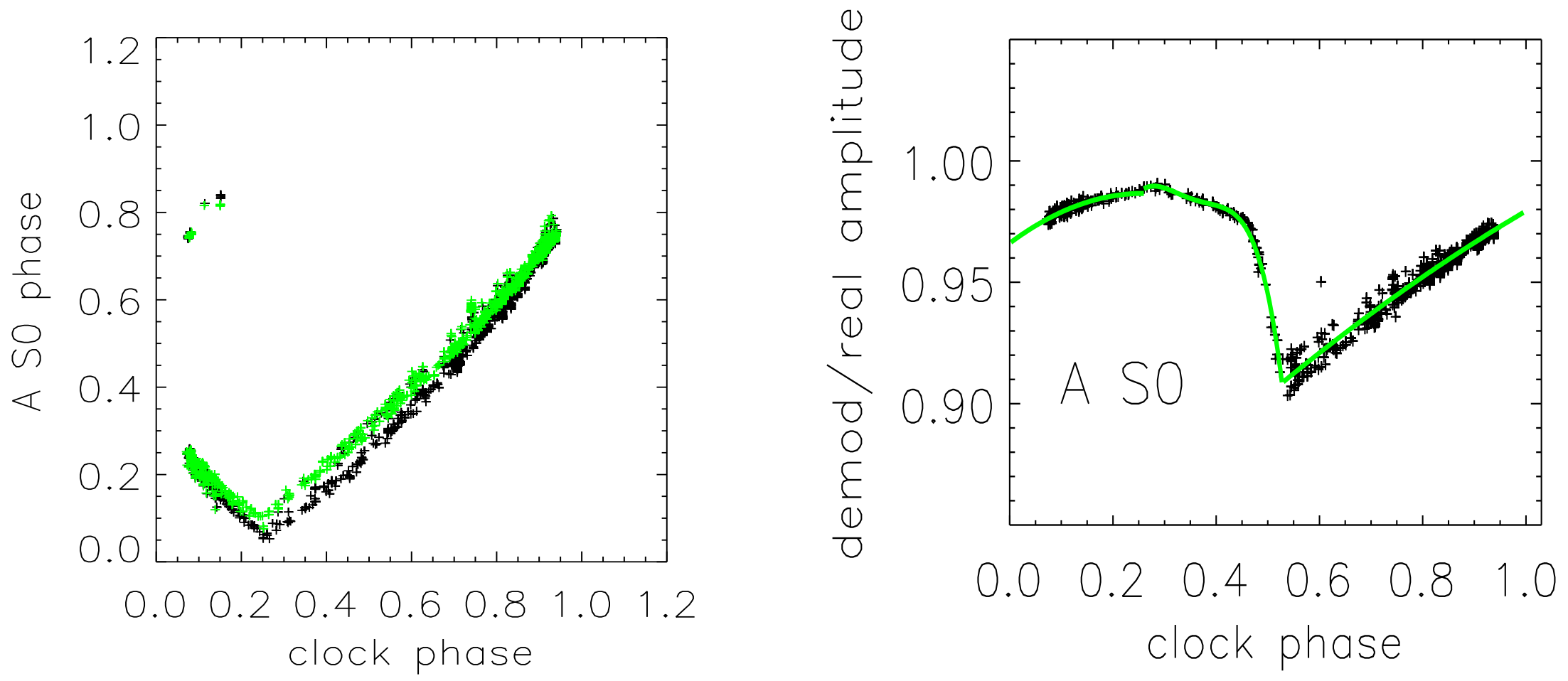

(a) (b)

Figure 5.11: (a) Correlation between the phase of the digitized clock signal and the phase of a modulated signal channel (AS0). Falling edge and rising edge (green) radiometer phase show the same behavior. All channels also behave similarly. The phase is defined as the distance of the in-between point relative to the amplitude. A point a quarter of the way up from the lower level state yields a phase of 0.25. The radiometer phase is very correlated to the clock phase. The correlation shows that the radiometer phase lags the clock phase by ~0.26 of a digitization period, which corresponds to a 2.6 $\mu s$ time delay, as expected.
(b) Derived relative gain of a polarization channel (AS0) as a function of the phase of the digitized clock. When the clock phase is 0.26 (so that radiometer channel phase is zero), the gain correction is closest to one. The gain correction can be as large as a 10% loss for some channels. Figures and analysis by D. Samtleben

In order to back out the correction factor for the amplitude of the demodulated signal, we need to know the phase of the clock during a specific run and the correlation between the digitized 4 kHz clock phase and the phase of each modulated polarization channel. During this 2003 season, both the raw modulated data and the demodulated data were recorded. The clock/signal phase correlation was measured for the entire season's data set [125] and is shown in Figure 5.11(a). The measured correlation between clock phase and the modulated data phase lets us derive a gain correction to recover the real amplitude of the demodulated signal (Figure 5.11(b)). The clock gain correction for the whole observing season is shown is Figure 5.12. The average gain correction for all channels for the whole seasons is 0.958 and can be as low as 0.90. The gain correction for different polarization channels only varies by as much as 3%. The fastest observed clock gain variation is 0.0025/hour (for day 55 for example). This translates into a 0.5 mK/hour offset variation on the channel with the largest offset (BS2 $\simeq$ 400 mK), or 0.55 $\mu$K change during a 4 second scan.

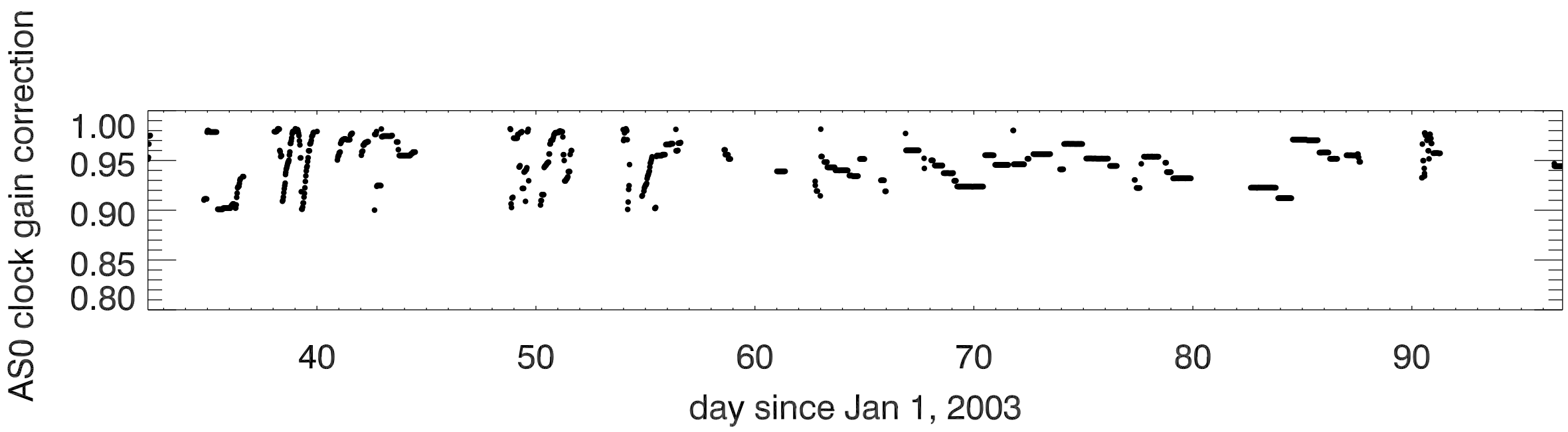

Figure 5.12: Relative gain correction for AS0 for all CMB observation during 2003 season. Note that after Feb 27 (day 55), the clock phase is constant during a run and only changes when the readout code is restarted. The largest clock gain variation observed is 0.0025/hour.

### 5.2.3 Encoder and Housekeeping Data Acquisition

The 47 housekeeping channels (Appendix G) are sampled at 1 Hz by a dedicated National Instruments A/D converter card[8]. The channels are acquired as singled-ended inputs with 16 bits resolution. The inputs are all bundled together in the ABOB and sent to the NI card in the DAQ computer via a 2-meter long 100 pin cable[9].

Getting the telescope position into the DAQ computer is a bit more tricky. As discussed in Section 4.1, the encoder's output is available at the position indicator in the control room, but the DAQ computer is located in the back of the vertex cab next to the receivers (see Figure 5.13). Various schemes were considered, such as sending the encoder bits from the telescope control computer via ethernet but the challenge remained to make sure that a given encoder sample was associated with the correct data sample. We settled on a handshake method where the DAQ computer requests the encoder information every at the 100 Hz sample rate (see Figure 5.9). On the other side, we use a micro-controller ($\mu$controller[10]) to process the encoder information and send it to the DAQ computer when requested. The $\mu$controller is a simplified version of a computer running at 22 Mips[11], with internal RAM and various useful other analog and digital peripherals. We particularly take advantage of the 8 bytes digital I/O ports to read in the BCD encoder inputs. The $\mu$controller comes with a development kit. This lets the user program the $\mu$controller in C language, burn the code in its flash memory, run it, and reprogram it until it functions as expected. This greatly reduces troubleshooting and testing time.

[8]NI 6033E, see `http://ni.com`
[9]P/N SH100100, shielded cable from National Instruments
[10]PN: C8051F020 from Cygnal systems, see `http://www.cygnal.com.`
[11]Mega instructions per seconds.

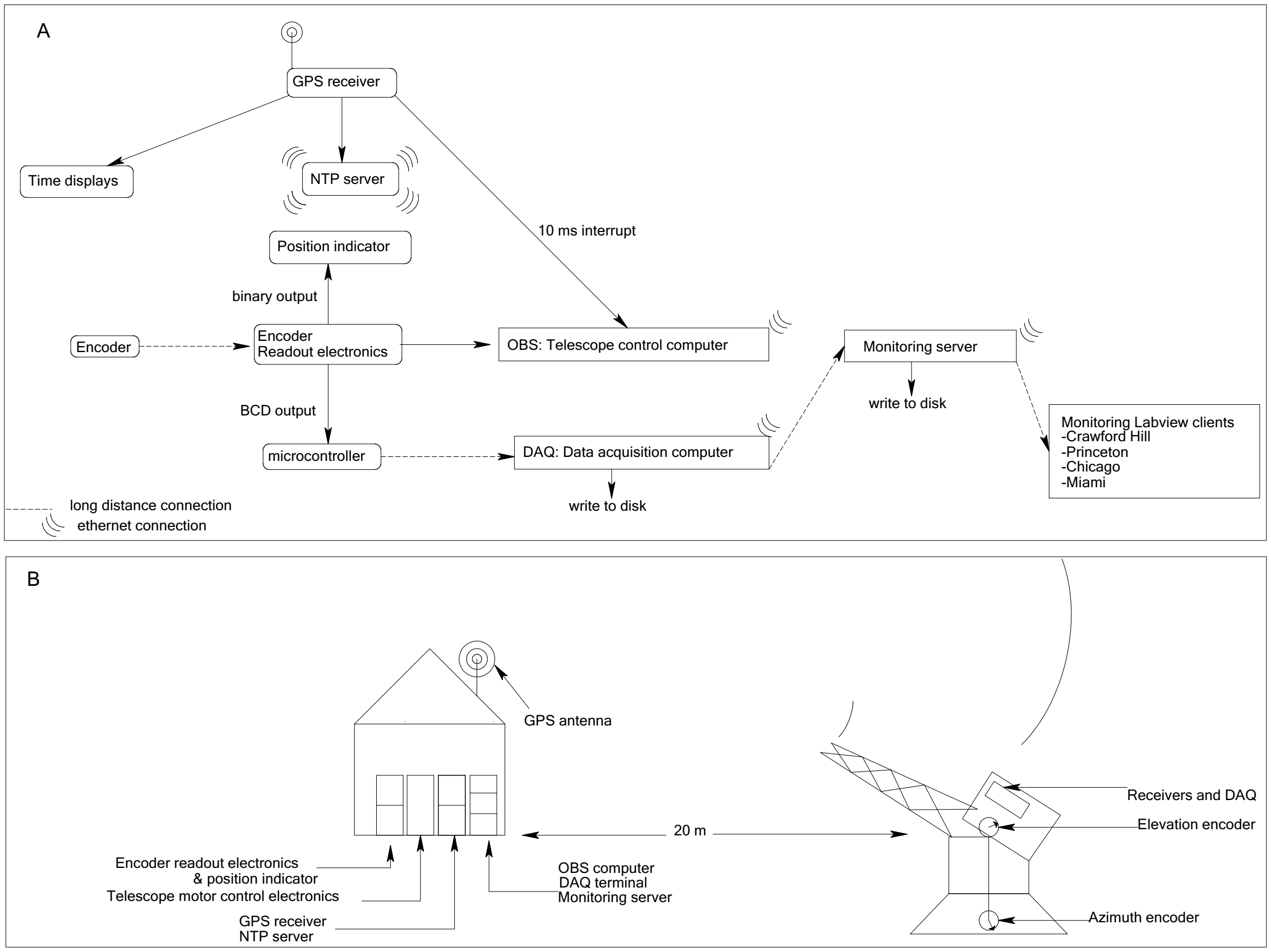

Figure 5.13: (a) Flow chart of the connections between the three computers which control the telescope (OBS), the data acquisition (DAQ) and the monitoring (MON). (b) Physical location of the different instrument parts at the telescope.

The $\mu$controller runs in an infinite loop presented in Figure 5.14 (see Appendix 5.2.3 for a listing of code and the wiring diagram). On the receiving end, the DAQ code, after having sent a request pulse, waits for a start byte to appear on the serial line and reads the following 8 bytes. If either the start or the end byte is not received, the value 555° is saved in the az and el position. The distance from the control room to the vertex cab via the telescope cable wrap is ∼50 meters, too far for the standard RS-232 serial communication protocol. A powered RS-232 to RS-422 converter[12] was installed at each serial interface to send the serial information via 2 differential signals, on two twisted wire pairs.

Although the $\mu$controller and receiving code were tested in Princeton before deployment, the acquisition of reliable encoder information was not achieved until Feb 7, 2003. We found that after working reasonably well at the telescope for some amount of time, the $\mu$controller would become flaky or stop working all together. After many unsuccessful trials at altering

[12]PN: 422pp9tb from BB electronics, see `http://www.bb-elec.com/`.

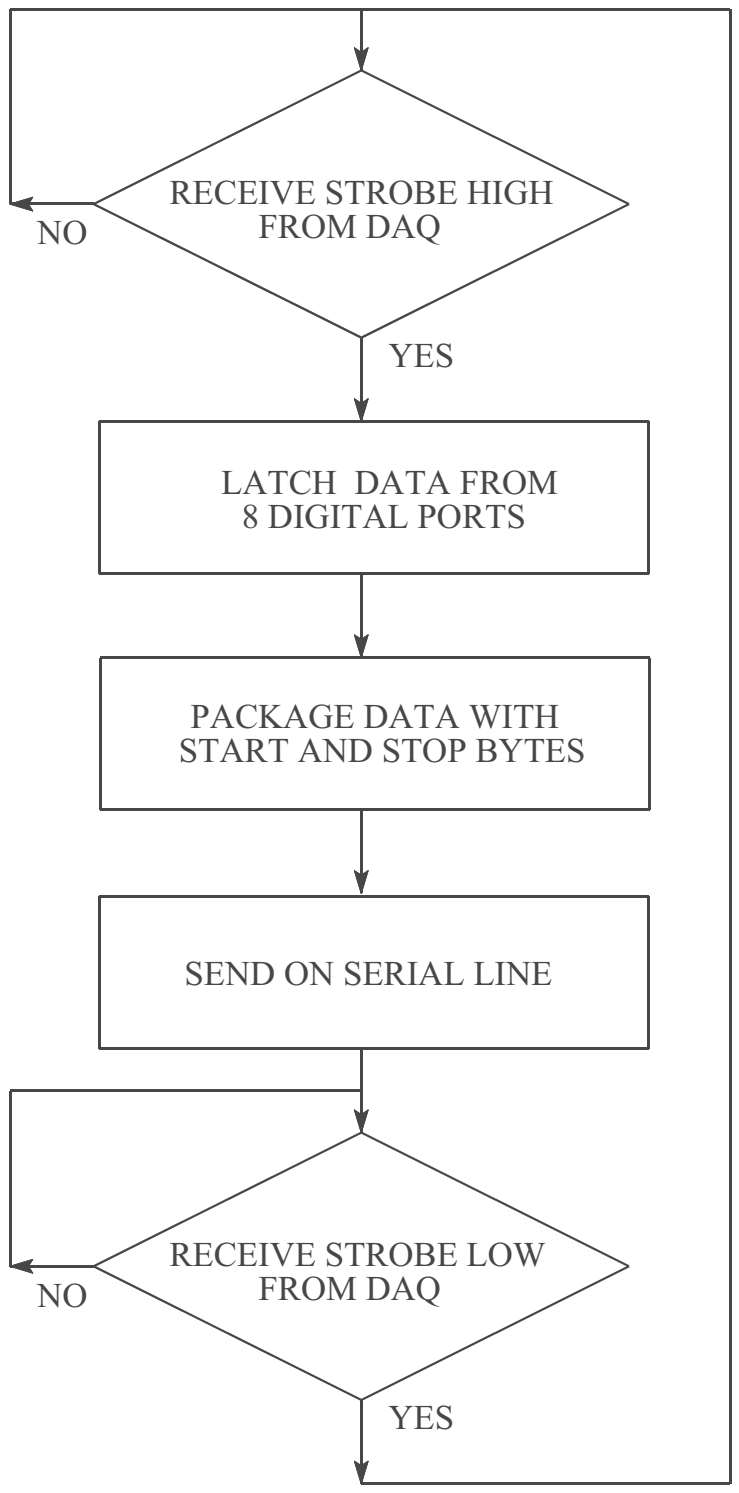

Figure 5.14: Micro-controller flow chart. The $\mu$controller waits for a strobe signal from the DAQ computer. The strobe signal is a 0.5 ms pulse sent from the DAQ at 100 Hz. When the strobe input goes high, the $\mu$controller latches the values on its 8 input digital ports (P0 through P7), where the 8 bytes of the BCD encoder are connected. Once latched, it sends them sequentially on its serial port, surrounded by a start (0xaa) and a stop (0xbb) byte, chosen specifically to be an impossible combination in BCD bytes. The serial port is configured as 8n1 protocol and runs at 115 kbps, thus sending the 10 bytes train in $\sim$0.7 ms. After the serial information is sent, the strobe pin waits for a low signal and the loop repeats.

the $\mu$controller code, we tracked down the problem to a known[13] hardware issue.

The 27 az and el BCD bits are fed into the $\mu$controller via a 2-meter long cable. The digital signals in those lines are updated by the position indicator electronics at a rate of 100 kHz, and thus switch between 0 and 5 Volts on even faster time scales. When probing the signals at the end of those lines, we found that on microsecond time scales, the signals could actually overshoot the $\mu$controller maximum digital voltage by up to 2V. It is expected that such large overshoots on the digital inputs can produce the strange symptoms we found, from overwriting part of the RAM where the code is stored to destroying the input. An individual diode protection circuit (see Figure 5.15(b)) was installed in series with each of the digital input. The $\mu$controller has performed flawlessly since, although the DAQ recorded it with an as-yet unexplained two-sample shift (see Section 6.3.2).

## 5.3 Crawford Hill Atmospheric Conditions

Needless to say, northern New Jersey is not a prime millimeter observation site. With respect to cloud cover, atmospheric opacity, atmospheric precipitable water vapor, and thus sky noise temperature at 90 GHz, Holmdel NJ, is certainly worse than many millimeter

[13] known to Norm Jarosik.

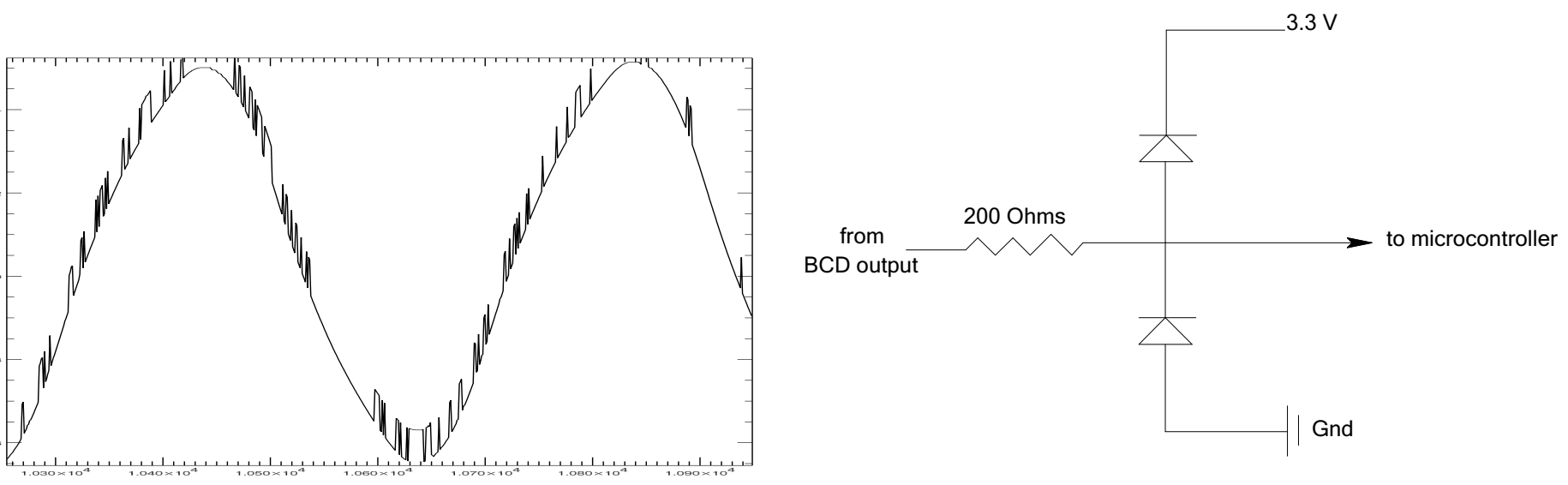

(a) (b)

Figure 5.15: (a) Az encoder readout vs time (ms). The sudden level shifts are artifacts of the $\mu$controller transient problems. (b) Diode protection circuit installed in series of each digital input. The diodes clamp the signal between $3.3 + 0.6$ V and $0 - 0.6$V. The 200 $\Omega$ resistor limits the current running through the diodes.

sites (Atacama Chile, Mauna Kea, South Pole). Nevertheless, given the low altitude and the fairly high water content of the atmosphere, the site has proved surprisingly good for polarization measurement at 90 GHz. The only reason CMB polarization experiments are even feasible in such a site is because at the level probed, the atmosphere is unpolarized [49]. The atmosphere acts as a variable source of extra noise adding linearly to the receiver noise temperature because the two are uncorrelated. The high system noise results in a decreased sensitivity, forcing the instrument to longer integration times to reach a desired level.

Three factors mainly affect the quality of observations: cloud coverage, atmospheric temperature, and atmospheric water content. Table 5.3 lists the observed characteristics for the 2003 or 2004 winter months.

At 90 GHz, apart from very high and thin cirrus clouds, any cloud coverage prevents all observations. Figure 5.16 shows that December, January, and February are the clearest months, with a perfectly clear sky nearly 40% of the time. The telescope is maintained such that it is observing in nominal CMB mode during every clear period. Clear sky is not sufficient to ensure the quality of the data taken during those periods.

Figure 5.17 shows models [46] of the total emission temperature of the sky at zenith. Absorbtion in the atmosphere has contributions from water vapor (two lines at 22 and 180 GHz), molecular oxygen (the saturated line at 60 and 120 GHz), and minor species such as $O_3$, CO, $CO_2$, and $NO_2$. That is why ground based observations only observe in windows located between these opaque regions. At 90 GHz, the emission from the atmosphere is at a local minimum. The bulk of the sky noise is due to the wings of the two $O_2$ lines, as expected given that at 100 m altitude, the telescope is looking through the full atmosphere. The inset in Figure 5.17 shows that a variation of the water vapor content in the atmosphere should affect the noise temperature in our three polarization sub bands. This agrees surprising well with the correlation between the measured PWV and zenith sky temperature listed in Table 5.3.

| Month | Fraction of the time the sky is clear[a] | Average temperature[b] High/Low (°C) | Average zenith sky temperature[c] (K) | 50% quartile PWV[d](mm) |
|---|---|---|---|---|
| Aug | 27% | 30/20 | - | - |
| Sep | 32% | 25/16 | - | - |
| Oct | - | 19/9 | - | 12.8 |
| Nov | 31% | 13/4 | - | 12.8 |
| Dec | 40% | 6/-2 | - | 7.0 |
| Jan | 38% | -3/-5 | 30 | 4.0 |
| Feb | 46% | 5/-4 | 38 | 4.8 |
| Mar | 22% | 10/1 | 48 | 9.6 |

Table 5.3: Summary of the winter months' observing properties at the Crawford Hill site.
[a]Archives from `http://cleardarksky.com` for the winter 2003-2004. See Figure 5.16 for the full cumulative histogram.
[b]Archives from `http://www.weatherunderground.com` recorded at Belmar-Farmingdale, New Jersey. Winter 2002-2003.
[c]Average of the 8 total power channel during the nominal CMB observations. The channel's receiver temperature has been removed [58], and the result divided by $\cos 40.3^\circ$ to refer to the zenith temperature. No data were taken with CAPMAP until Jan 17 2003.
[d]Averaged PWV from GOES-12 satellite hourly archive at `ftp://suomi.ssec.wisc.edu/pub/rtascii/tpwtext12` for the winter 2003-2004 (see Figure 5.18(b) for cumulative histogram). The data points are only sampled on a 100 by 100 km grid, but the flatness of NJ makes the values likely to be accurate.

The Precipitable Water Vapor (PWV) is the total height in mm of liquid water condensable from a column of atmosphere. At 90 GHz, it is good marker of the emissivity of the atmosphere so it is often used as a standard figure of merit to compare different millimeter wave observing sites. Figure 5.18 shows the time series of the PWV in Crawford Hill during the 2003-2004 winter along with its cumulative histogram. The cumulative histogram answers the question: What fraction of the time is the PWV better than a certain value? Choose the value on the y-axis and read the x-intercept of the given histogram. Again, the three months of Dec, Jan, Feb are consistently the best. PWV values as low as 1.6 mm were recorded and lasted for periods of 10 to 24 hours. This does not come close to any of the good observing sites which have PWV consistently lower that 1 mm (see Figure 5.19).

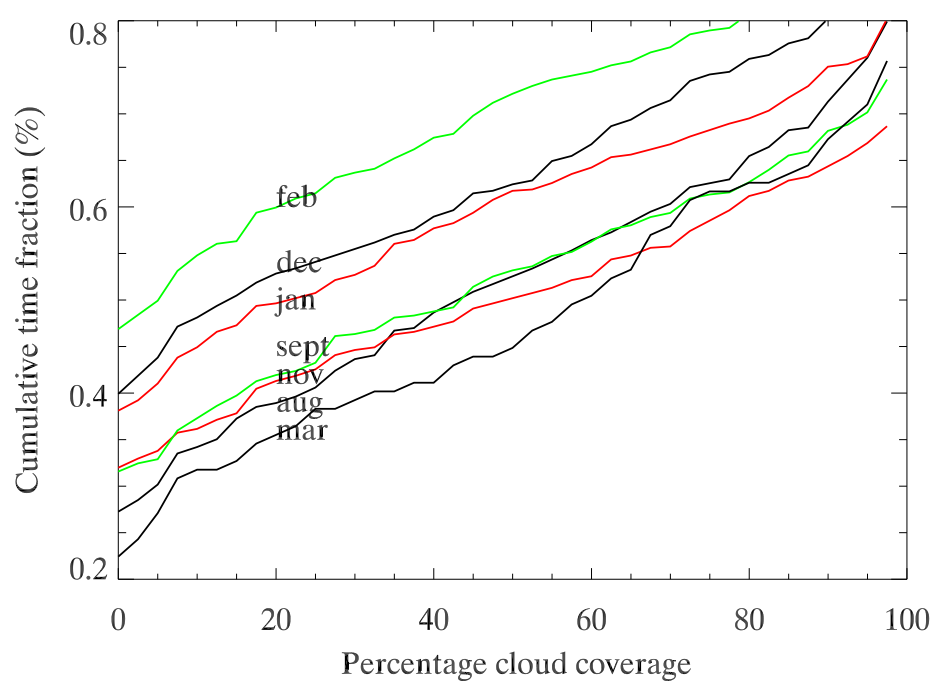

Figure 5.16: Cumulative histogram of sky cloud coverage for the 2004 winter season.

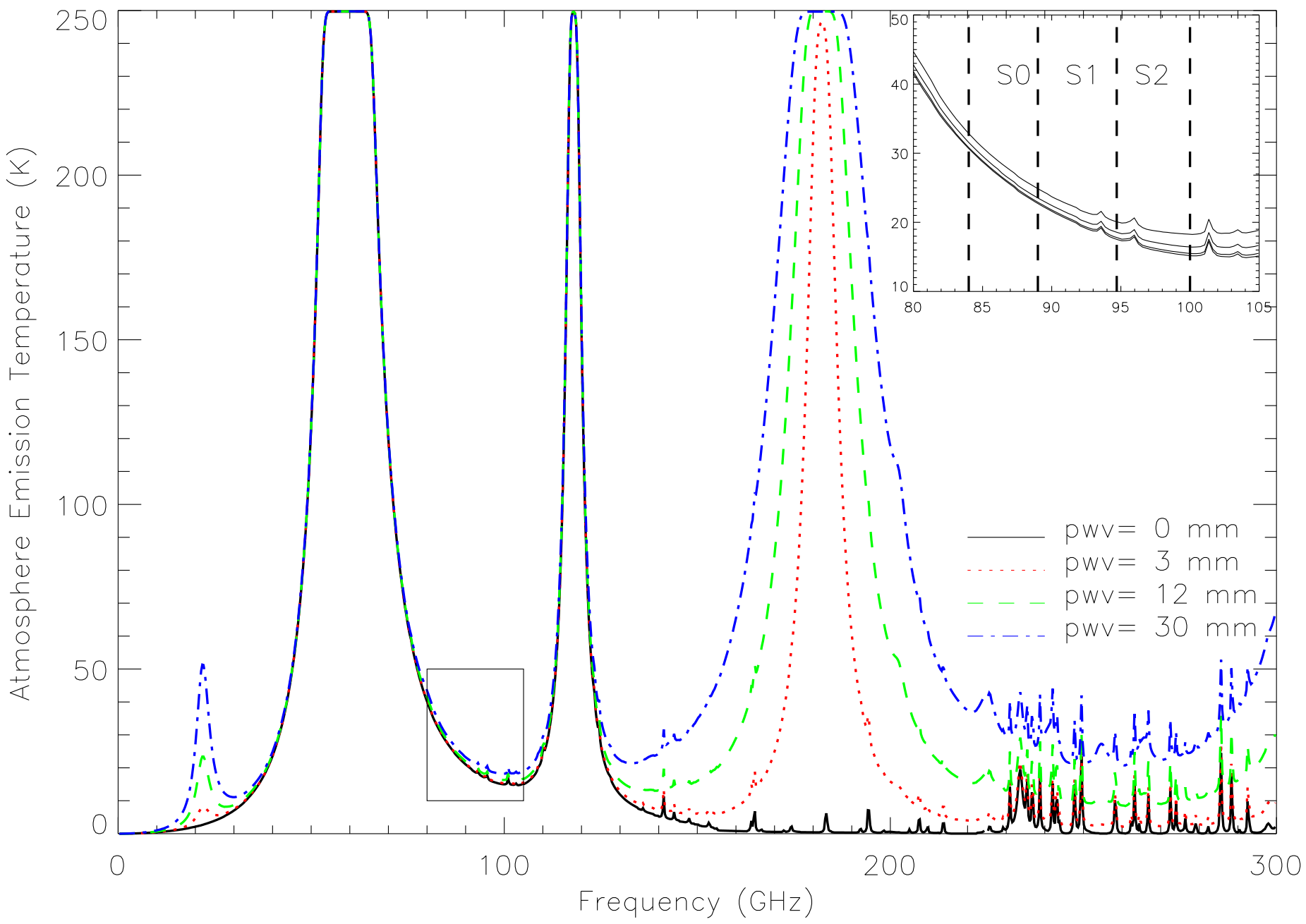

Figure 5.17: Atmospheric zenith emission temperature vs frequency for different values of PWV. Inset is a zoom on the three CAPMAP frequency bands. Note that although S0 is more affected by the wing of the 60 GHz $O_2$ line than S1 and S2, the variation in PWV does not change the emission temperature in our bands. The three values of PWV (3mm, 12mm, and 30mm) are representative of the best, average and worst observing conditions. The line-shapes were generated using E. Grossman AT software [46], with the following parameters: full Lorentzian line profile, elevation=100 m, latitude = 40°, zenith angle = 0°, physical temperature of the atmosphere = 250K. The AT software produces the atmospheric transmission, $t$, which was converted into a zenith sky temperature as $T_{atm} = 250(1 - t)$

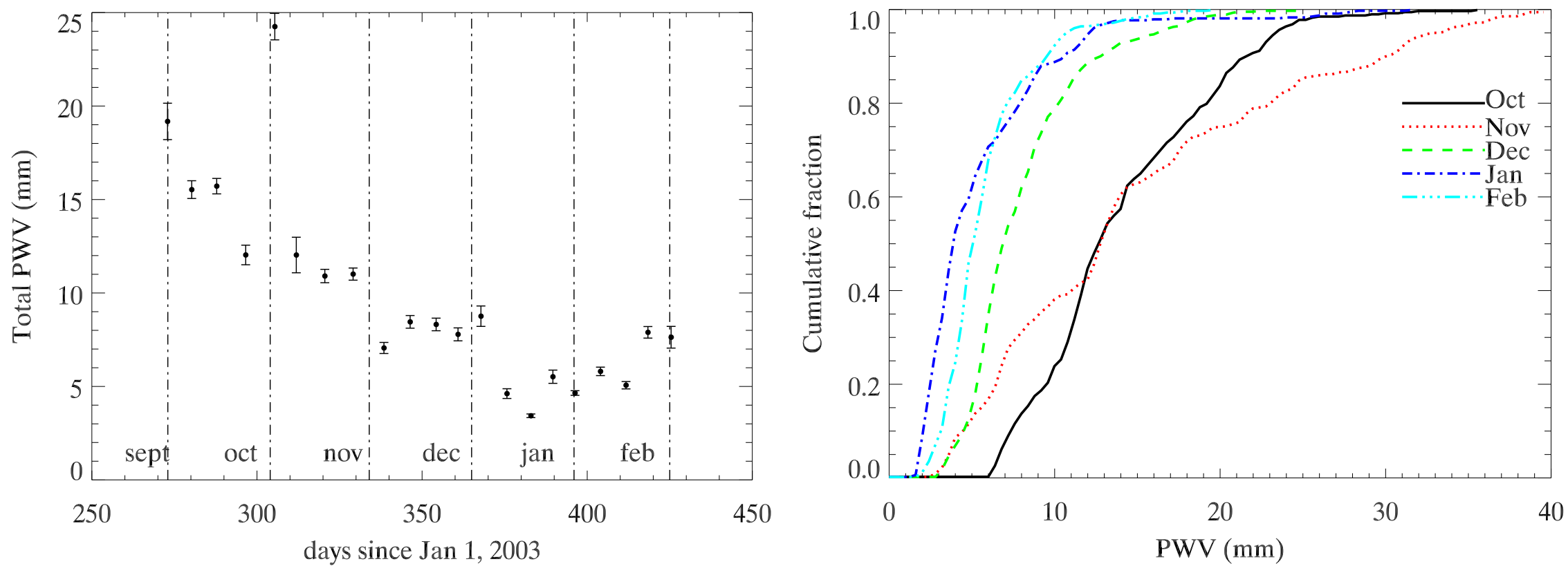

(a) (b)

Figure 5.18: (a) Time series of the total precipitable water vapor (PWV) during the winter 2004 observing season, rebinned in 10 days intervals. These data are derived from the hourly GOES satellite archives. The PWV is the average from a 100 by 100 km square centered on Crawford Hill. This clearly indicates that the level and stability of the atmospheric water vapor content is lowest during the months of January and February.
(b) Cumulative histogram of the winter months of 2004. Solid line is Oct, dotted Nov, dashed Dec, dash dotted January, and dash double dotted February. The 50% quartile for these 5 months respectively are 12.8, 12.8, 6.8, 4, 4.8 mm. Comparative numbers for dedicated microwave observing sites are listed in Figure 5.19.

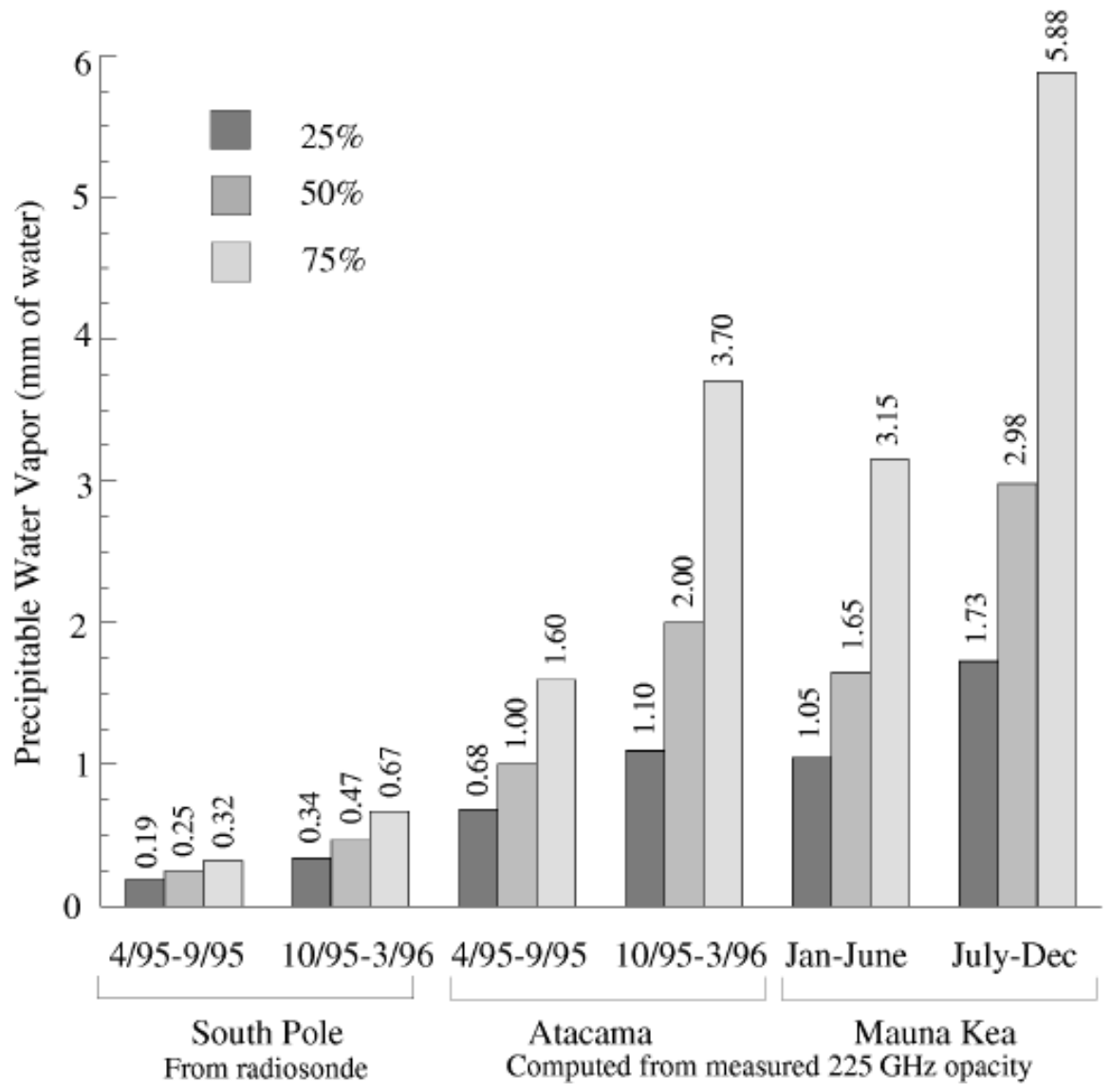

Figure 5.19: South Pole, Atacama Chile, and Mauna Kea PWV quartiles. Figure from `http://cfa-www.harvard.edu/ aas/tenmeter/pwv.htm`

# Chapter 6

# Observatory Calibration

## 6.1 Introduction

While most of the season is spent on CMB observations, a few minutes every day are dedicated to observing various celestial sources. These observations are performed to characterize the beam shape, the pointing solution, and the receiver's gains, which are necessary for the CMB polarization data analysis.

For a polarized receiver at 90 GHz with a $4'$ beam, choices of calibrator sources are limited. An ideal calibrator source should have the following properties:

- It should be intense enough to be detected with a S/N $\geq$ 3-5 in each receiver map with a few minutes integration time.
- It should be unresolved by the $4'$ beam, to allow derivation of the telescope's beam shape and pointing information. Observing a point source also simplifies the calibration of the receiver's gain.
- Its total-power and polarized emission should already be well characterized.
- A small set of such sources should provide an even coverage over the whole sky. This requirement is necessary for the pointing calibration.

No single source has all these properties, so a variety of sources were used to satisfy the calibration requirements. Jupiter was used to determine the beam shape, pointing calibration, and total power gains, because it is the brightest source in the sky at 90 GHz (after the Sun and the Moon) and remains unresolved. However, its emission is unpolarized. Tau A[1] was chosen as the primary polarized calibrator. Tau A emits strongly in the microwave (205 Jy at W-band [93]) and is relatively highly polarized (see Section 6.5.2).

[1] The Crab nebulae is a supernova remnant located at $\alpha_{2000}$ = 05h34m32.0s, $\delta_{2000}$ = +22d00m52s

The downsides of using Tau A are 1) the source is slightly larger than the CAPMAP beam and might require a complicated deconvolution to cleanly derive a polarization gain, and 2) its polarized fraction is only measured to ~15% . The chopper plate test was therefore used twice during the season as a second polarization calibrator (see Section 3.3.2). Finally, Jupiter by itself poorly constrains the pointing solution of the telescope because it is only observed along the ecliptic plane. We therefore tried to observe other pointing targets, particularly Cas A because it is circumpolar.

## 6.2 Data Reduction for Source Observations

### 6.2.1 Source Scan Strategy

During the CAPMAP03 observing season (Feb 2 to April 6, 2003), a total of 32 scans of Jupiter (of which 15 good were enough to be analyzed and 5 were deep scans (Section 6.3.3)), 45 scans of Tau A (of which 13 were analyzed), 16 scans of Cas A, two of Venus, one of Saturn, and six of the Moon, were acquired. Until March 9, the "gsca" macro [151] was used to command the telescope motion during the source scans. This macro was unreliable and crashed the telescope motion half of the time, which explains why many of the early observations are not analyzed. The other sources were observed with the newly created "ra" command [151] which is more reliable. Other unprocessed observations were corrupted by clouds which produce large $1/f$ structure in the time series and prevent a good baseline removal.

During a standard source observation, the telescope tracks the source. Superimposed on the tracking motion, the telescope scans back and forth in azimuth (az), at discrete steps in elevation (el). The az range, el steps, and offset from the source in az and el are specified in the scanning command "ra." The az rastering is done at the maximum telescope velocity. Figure 6.1 shows a typical scan, where a $0.8^\circ \times 0.8^\circ$ region centered on the source is observed. The scan starts below the source, steps in elevation at $0.01^\circ$ intervals and then covers the region again from top to bottom. This strategy ensures that the source is observed with each receiver, with a typical integration time of 2 seconds per beam-sized pixel.

A strict sources observation schedule was followed so that the resulting gaps in CMB data were uniformly spread across local sidereal time (LST) bins. The observations produce a time stream of data, encoder position, and time (see Appendix G) similar to the standard CMB time stream. The following section describes how these time streams are reduced to physical measurements. All routines are written in IDL and are available from the author [5].

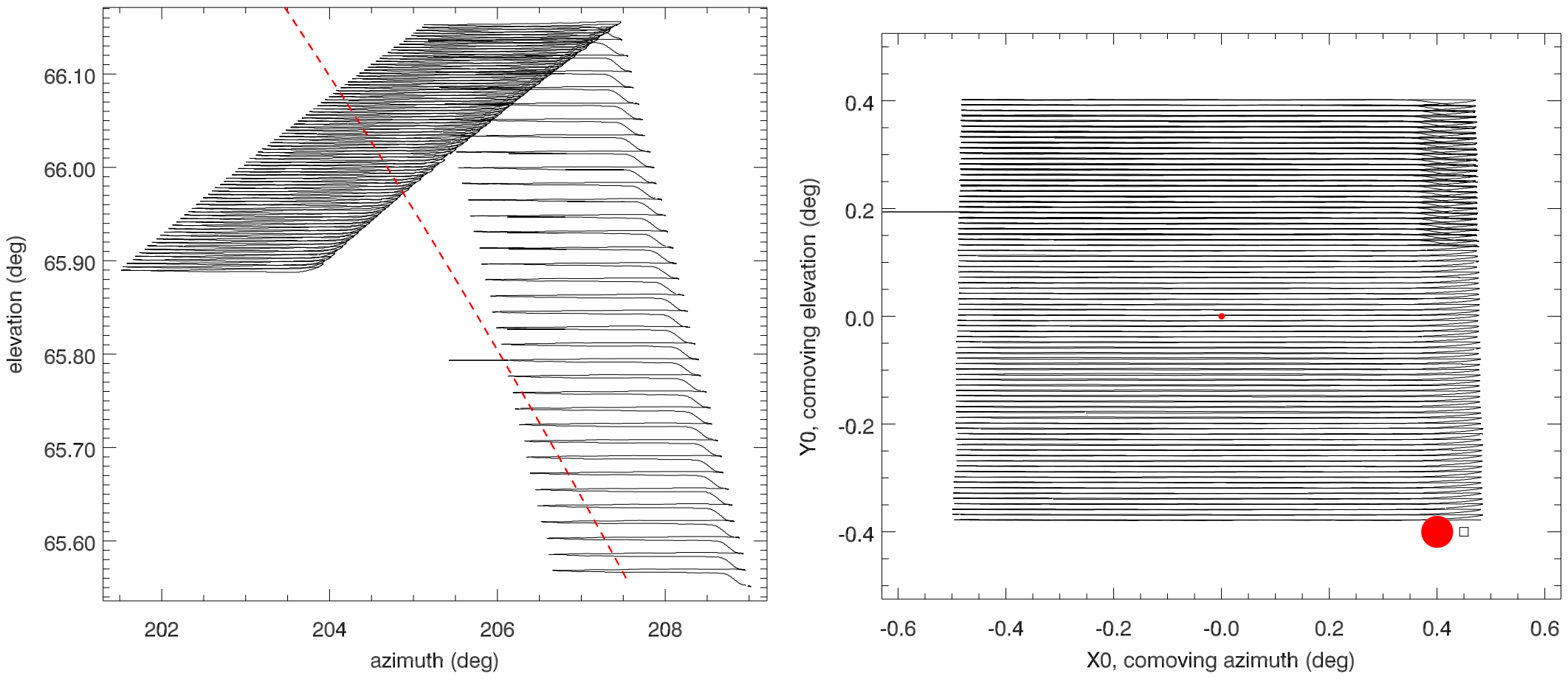

Figure 6.1: *left*: Telescope azimuth and elevation coordinates (solid line) for a typical Jupiter scan. Jupiter is setting and its position is plotted as the red dashed line. *right*: Same scan where the telescope azimuth and elevation are converted into co-moving coordinates ($X_0$ and $Y_0$) by removing the tracking motion of the telescope. The resulting scan is a square region, centered on Jupiter (the 40″ red dot in the center). The circle in the bottom right shows the FWHM of the beam and the square next to it shows the typical pixel size used during the mapping.

### 6.2.2 Observations

**1) Clean.** Once the data stream is read into memory, the encoder and radiometer data are deglitched to remove large spikes. Data points which exceed $6\sigma$, where the standard deviation is calculated on a window of 100 samples, are replaced by the average of the surrounding samples. For sources with a large S/N ratio in the time stream, these glitches are removed manually to prevent cutting the signal. Additional small sections of time are cut out from the beginning and end of the file to remove data when the telescope was not yet scanning.

**2) Ephemeris.** In the coordinate system where the telescope is tracking the source, the scan pattern is a simple rectangular region centered on the source (see Figure 6.1(b)). The telescope az and el translate into co-moving az, $X_0$, and co-moving el, $Y_0$, which are source-centered coordinates. This approach is equivalent to analyzing the data in equatorial coordinates [90], but the co-moving coordinates have the advantage that the map and final results are produced in terms of the natural coordinate system of the telescope. To convert into co-moving coordinates, the ephemeris of the source is generated using the JPL planetary[2] ephemeris package [75] evaluated at one second

[2]The JPL ephemerides provide the positions and motions of the major planetary bodies in the solar system to very high precision. The ephemeris calculation includes precession, nutation, aberration, and refraction correction.

intervals, then interpolated at 100 Hz for each sample.

An older version of the source analysis routine used the Xephem [150] software package to generate the ephemeris of the source. Xephem is a stand-alone software and cannot be run within IDL. Because a new ephemeris file must be generated for each new observation, Xephem is too cumbersome to use for the automated analysis of all the source observations. It is still used to produce the ephemeris of objects other than the nine major solar system planets. Xephem and the JPL ephemerides agree to better than 0.001°.

**3) Pointing and Refraction Corrections.** As described in Section 6.4.1, a pointing correction must be applied to the az and el values read by the encoders to recover the true position of the telescope's main beam.

$$AzEl_{enc} = AzEl_{tel} + \text{Corrections} \ , \tag{6.1}$$

where $AzEl_{tel}$ is the commanded telescope position and $AzEl_{enc}$ is the position as read by the encoders. One purpose of the source analysis described here is to improve the pointing corrections. However, we apply best-guess pointing corrections (the "original2003" pointing solution) to these data during their analysis (see Section 6.4.1).

**4) Conversion to Source-centered Coordinates.** The source-centered coordinates are:

$$X_0 = (Az_{tel} - Az_{src}) \cos \ El_{tel} \tag{6.2}$$

$$Y_0 = El_{tel} - El_{src} \tag{6.3}$$

where the *tel* and *src* subscripts are the corrected encoder position and the source position from the JPL ephemeris respectively. This conversion transforms the tracking plus scanning motion in Figure 6.1(a) into the rectangular map of Figure 6.1(b).

**5) Removal of the baseline.** The polarization channels only suffer from $1/f$ noise on time scale of many minutes ($1/f$ knee between 1 and 10 mHz, see Figure 3.19). It is therefore sufficient to remove a linear or quadratic trend from the whole ∼20-minutes polarization channel time-series.

Because the time streams of the total-power channels are dominated by $1/f$ noise (Figure 6.2(c)), a map created directly from the raw data would be striped (features in the scan direction in Figure 6.4(b)) and would be unusable. Therefore, the total-power channel time streams are filtered to remove fluctuations on time scales longer than the scan period.

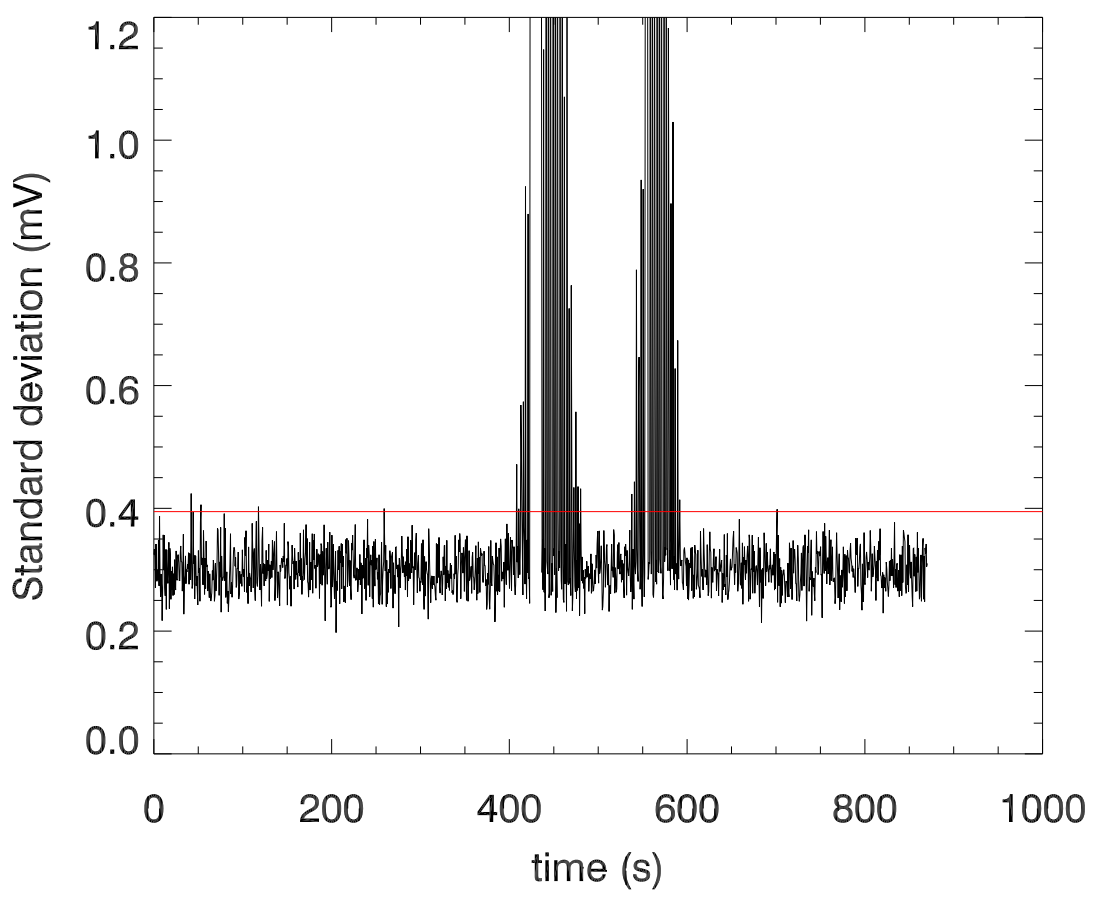

(a) Standard deviation in each 60-sample segment for the Jupiter observation versus time. The large values indicate when the source went through the beam. The horizontal line corresponds to the 3-sigma cut from the gaussian fit in sub-figure (b).

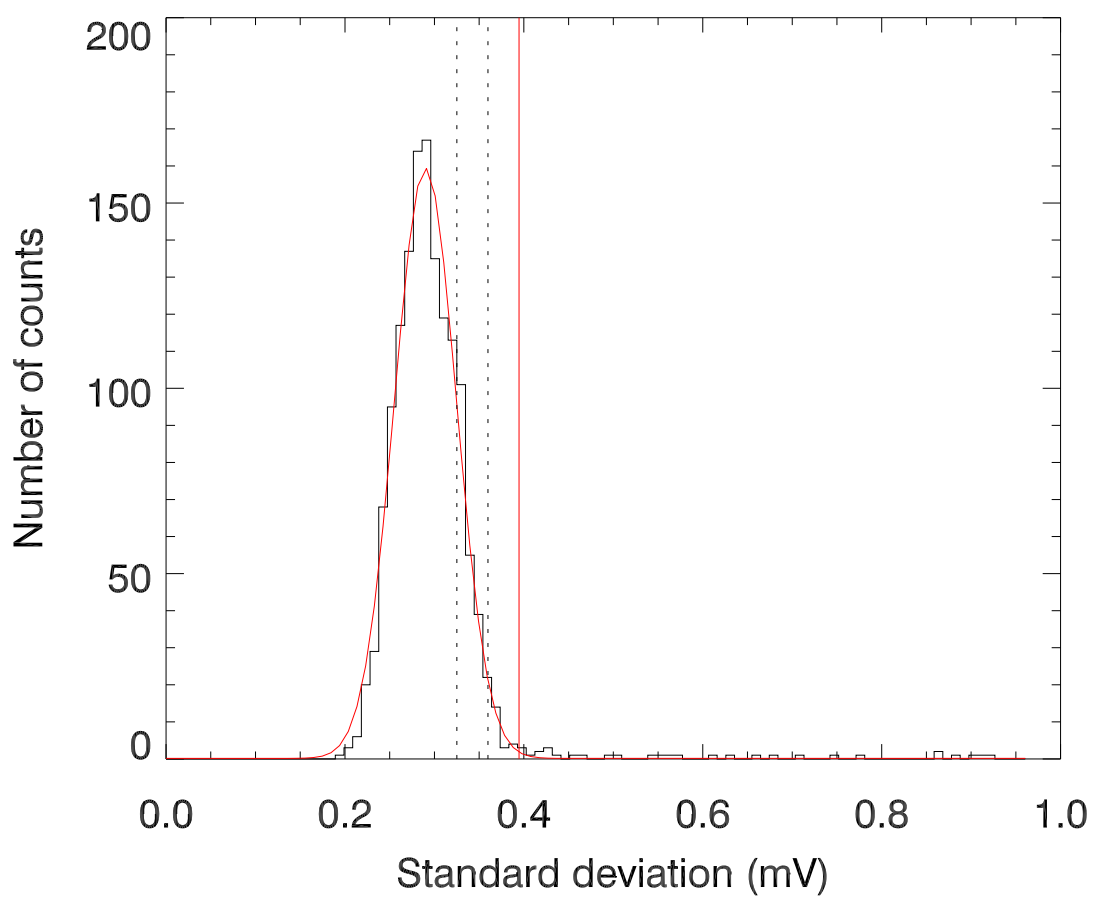

(b) Histogram of the standard deviation in sub-figure (a). The dashed lines are 1- and 2-sigma levels, determined from the gaussian fit to the histogram. The 3-sigma cut (solid line) automatically blanks out time periods when the source passes through the beam.

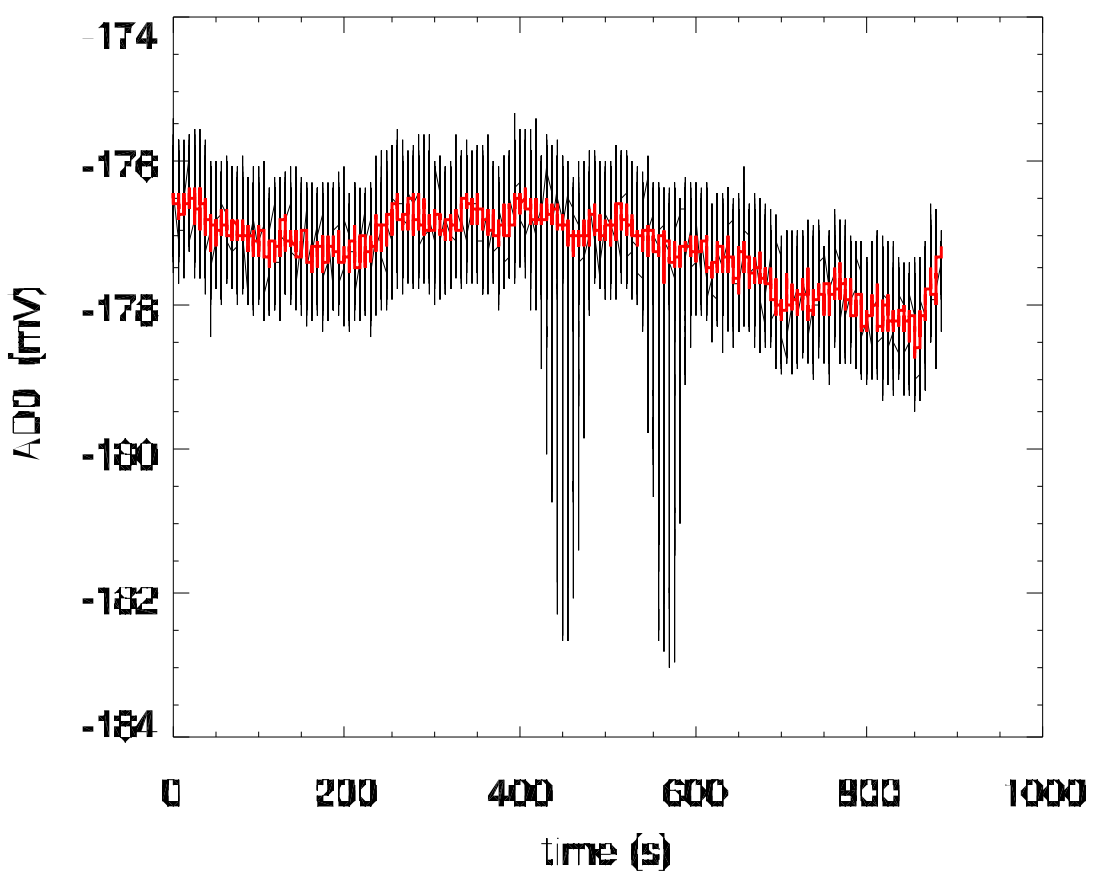

(c) Raw total power time series of channel AD0 (black) and the derived baseline (red) to be removed. Note how the baseline follows the fast variations of the raw time series without being affected by the signal from Jupiter. See Figure 6.3(b) for a blow up of this plot.

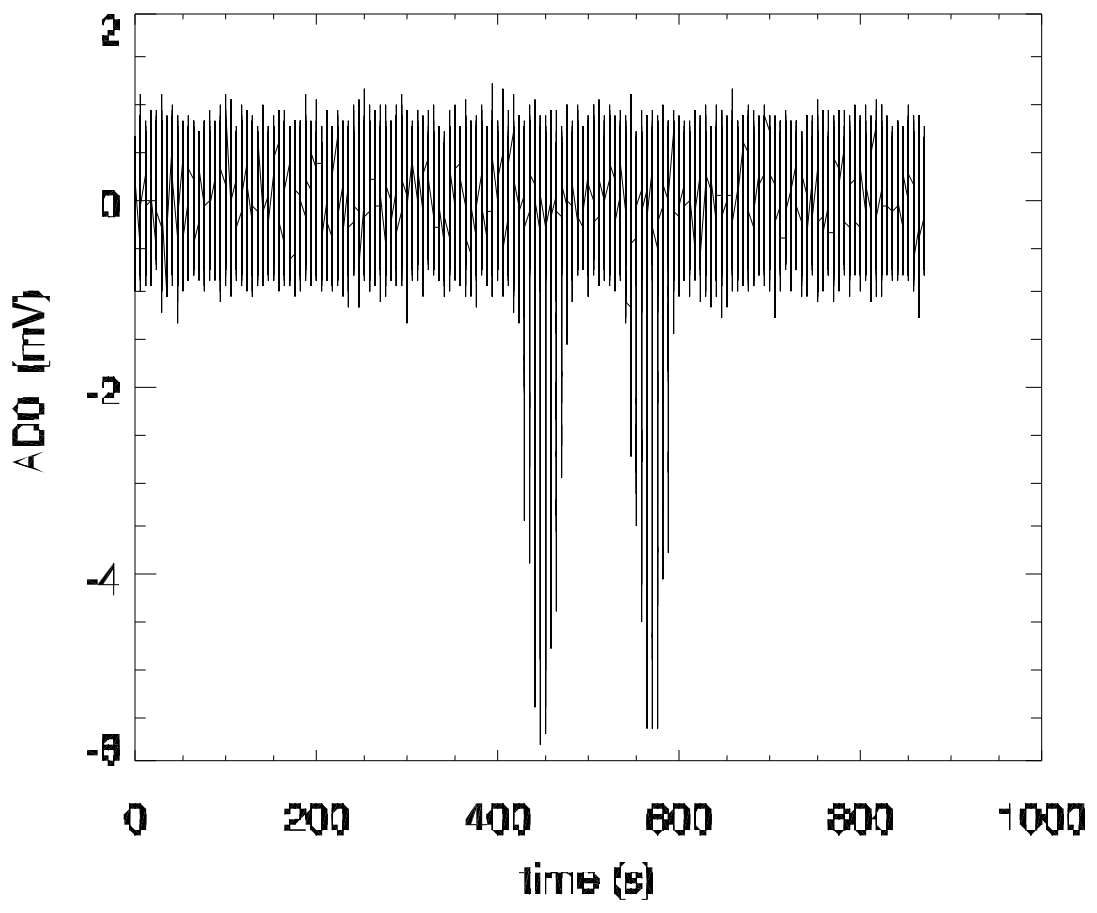

(d) Final time series with baseline removed. The noise is now white and the amplitude, width, and position of the Jupiter signal appear largely unaffected by the baseline removal. A closeup of the whitened data is shown in Figure 6.3(d).

Figure 6.2: Baseline removal algorithm.

The filtering is not done in Fourier space to avoid ringing effects. Instead, the filtering is done by an algorithm which finds and removes an averaged baseline of the time stream.

The baseline only follows the trends in the noise, not the signal from the source. This is sometimes done by blanking out the expected position of the source in the time series and interpolating through the remaining points. The problem with this method is that the exact position of the beams must be known a-priori, which makes this method sensitive to changes in the pointing solution, as well as the exact configuration of the receivers.

The algorithm presented here (Figure 6.2) is independent of the position of the source in the time series. It has been tested for both high signal-to-noise observations of Jupiter and for weaker sources such as Tau A or Cas A which are not visible in the time series but appear when binned into a map. In both regimes, the baseline removal does not seem to affect the amplitude, width, and position of the underlying signal. The algorithm proceeds as follows:

- Calculate the mean and the standard deviation for each adjacent segment of 60 samples (0.6 second) (Fig 6.2(a)). Make a histogram of the standard deviations, fit a gaussian to the histogram and blank out all segments whose standard deviation exceeds by $3\sigma$ the central standard deviation value (Fig 6.2(b)). This excludes all the blocks where the source was observed. The number of samples is the only parameter that needs to be adjusted. The optimal choice (60) is based on the fact that the baseline removal should remove as much of the low-frequency noise as possible without compromising signals at and above the scan frequency (0.3 Hz or 3 seconds). Increasing the block size results in inadequate $1/f$ removal. On the other hand, decreasing the block size can cause removal of the signal itself. A good balance is obtained empirically. Figure 6.3(a) shows how the Fourier transform of the baseline is equivalent to a low-pass filter.
- Take the means of the remaining 60-sample segments and interpolate through the missing ones to recover a time series the same size as the original data (Fig 6.2(c)) which does not contain the source signal. A closeup of the baseline is shown in Figure 6.3(b).
- Remove the baseline from the original data to obtain a whitened time stream, shown in Figure 6.2(d) and 6.3(d).

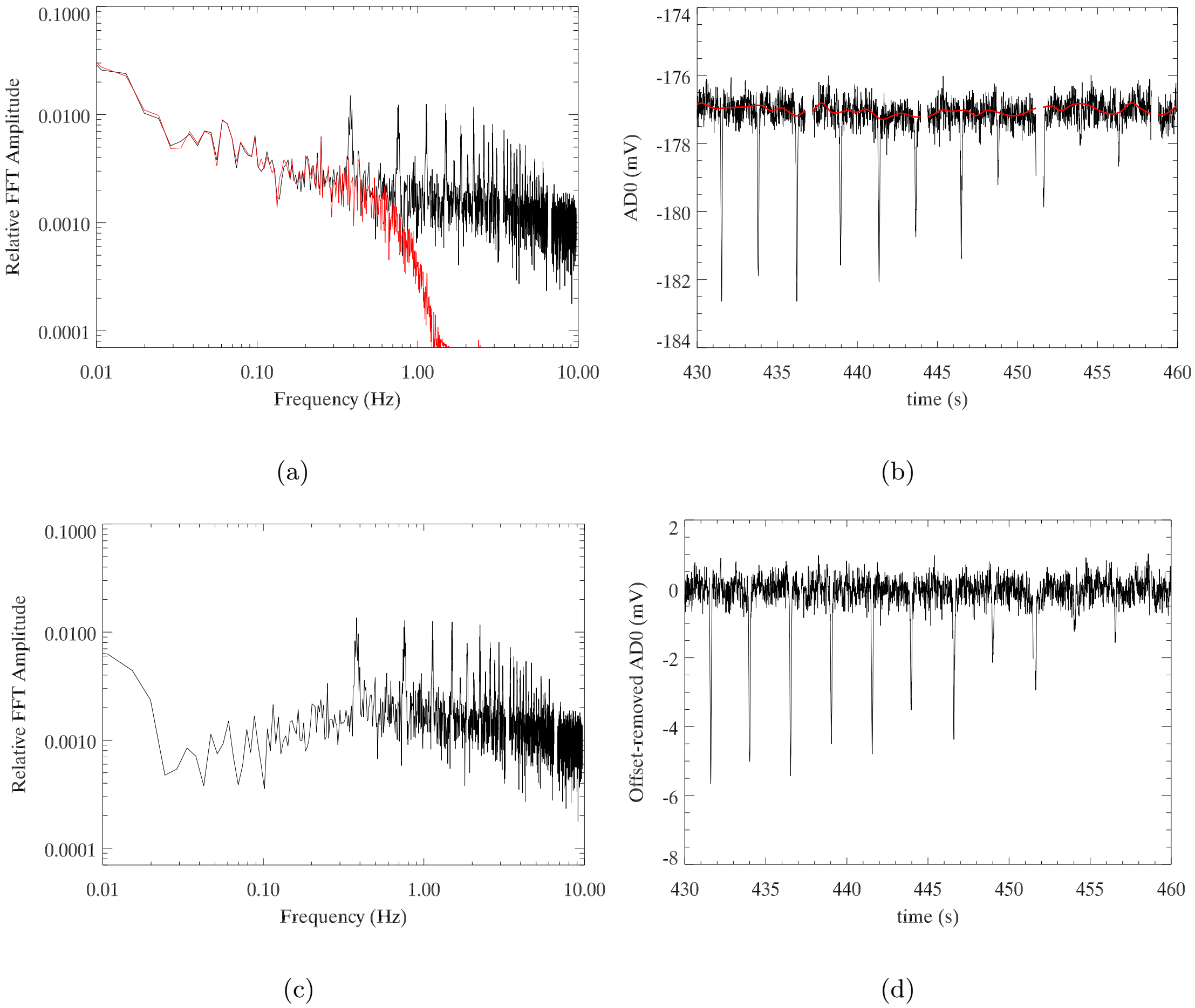

Figure 6.3: (a) Fourier transforms of the raw time-series (black) and baseline (red) of a Jupiter observation. The structure corresponds to the periodic Jupiter signal. The baseline appears as a low-pass filtered version of the original data, with the signal at the scan frequency filtered out. Subtracting the baseline therefore removes the low-frequency noise and whitens the time-series (subfigure (c)). (b) Closeup of the time-series and baseline of Figure 6.2(c). The block size of this baseline is 0.6 seconds. Note how the baseline is insensitive to the strong peaks of Jupiter. (c) Fourier transform of the whitened data. (d) Closeup of the whitened data on the same scale as the raw data.

**6) Map.** The mapping transforms the one-dimensional time stream of data into a two-dimensional array. This is done by breaking the $X_0$-$Y_0$ region into a square grid and co-adding the points that fall inside each pixel. For an elevation step size of .01°, a typical pixel size of 0.02°×0.02° (see Figure 6.1(b)) gives ∼10 samples per pixel. Observations of faint sources such as Tau A and Cas A require a larger pixel size to detect the signal, in which case beam-sized features are lost. Changing the pixel size up to 0.03° does not affect the recovered parameters (see table 6.3).

The error associated with each pixel is the standard error from the points in that pixel. This error is an underestimate for two reasons. First, the fluctuations of the points due to the atmosphere within a pixel are intrinsically smaller because these points have been filtered by the beam (when the pixel size is smaller than the beam size). Second, the fluctuations have been artificially decreased by the baseline removal. We deal with this second underestimate later, by simulating the recovered error of the fitted parameters (Section 6.2.3). The final results are not affected by a 20% variation in the baseline block size.

**7) Fit a model.** For a source smaller than the beam size, the model used to describe the resulting map is a 2D-gaussian, given by

$$G(X_m, Y_m) = A_0 + A_1\ Exp\left\{-\frac{1}{2}\left(\frac{(X_m - p_x)^2}{\sigma_x^2} + \frac{(Y_m - p_y)^2}{\sigma_y^2}\right)\right\}\ , \tag{6.4}$$

where $A_0$ is the overall map offset (usually null), $A_1$ is the amplitude of the gaussian, $\sigma_x$ ($\sigma_y$), and $p_x(p_y)$ are the widths and the centroids of the gaussian in the azimuth (elevation) axis. The gaussian fit uses a non-linear least square minimization routine (MPFIT) library developed for IDL [101]. This model is fit to the total power channels of the observations of Jupiter and to the polarization channel observations of polarized sources (Tau A, see Section 6.5.2). The six parameters, their associated error bars, and the $\chi^2$ of the fit are returned by the fitting routine and saved. The fit is performed on a 0.2°×0.2° square region around each beam because the fit does not converge for the whole map. A typical Jupiter map and gaussian fit are shown in Figure 6.4.

A second function is used to model the response of the polarization channels to an *unpolarized* source. This function was developed specifically to fit the quadrupolar pattern observed in the polarization channels response to Jupiter. The model function $M(X_m, Y_m)$, is the sum of a gaussian monopole, a gaussian dipole, and a gaussian quadrupole:

$$M(X_m, Y_m) = G(X_m, Y_m) + D(X_m, Y_m) + Q(X_m, Y_m)\ , \tag{6.5}$$

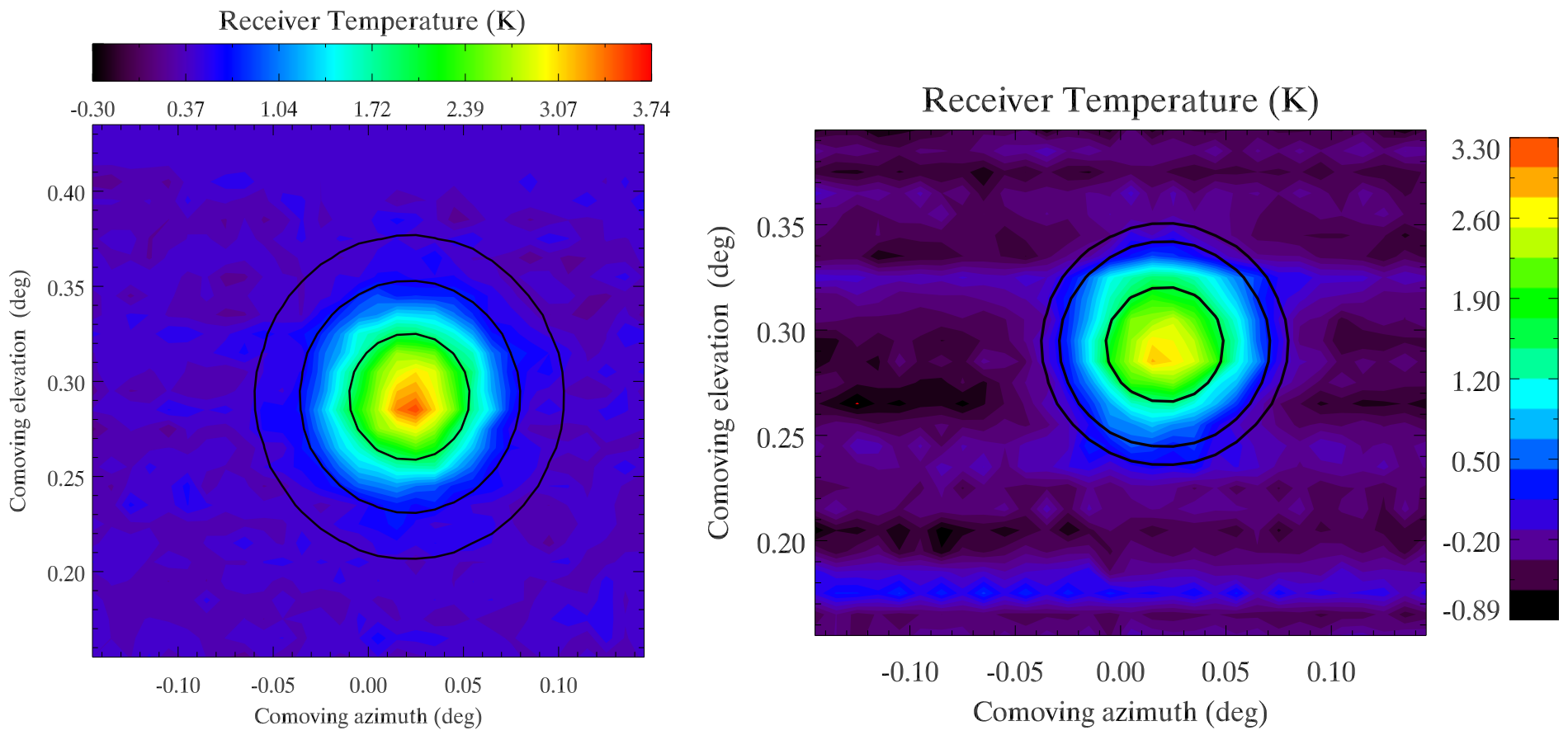

Figure 6.4: *left*: Map and gaussian fit of AD0 total-power channel for a single Jupiter observation. The contour levels are in receiver temperature. The fit contour levels are at -20 dB, -10 dB, and -3 dB from the peak level. This map is made from a 6 minute-long observation. This map has S/N *simeq* 10. *left*: Same map but with the baseline removal step omitted. Note that the striping is not very strong because receiver A has the least $1/f$ noise.

where the different terms are the partial derivatives of a gaussian with respect to one or both variables:

$$\begin{aligned} G(X_m, Y_m) &= P_0 \; E(X_m, Y_m) \; , \\ Q(X_m, Y_m) &= \frac{P_1 \; e^1 \; (X_m - p_x) \; (Y_m - p_y) \; E(X_m, Y_m)}{\sigma_x^3 \sigma_y^3} \; , \\ D(X_m, Y_m) &= \frac{P_2 \; e^{0.5} \; (Y_m - p_y) \; E(X_m, Y_m)}{\sigma_y} \; , \\ E(X_m, Y_m) &= Exp\left\{ -\frac{1}{2}\left( \frac{(X_m - p_x)^2}{\sigma_x^2} + \frac{(Y_m - p_y)^2}{\sigma_y^2} \right) \right\} . \end{aligned} \tag{6.6}$$

The widths ($\sigma_x$, $\sigma_y$) and centroids ($p_x$, $p_y$) are allowed to vary from the total-power beam widths and centroids. $P_0$, $P_1$, and $P_2$ are the amplitudes of the monopole, dipole, and quadrupole terms ( the factors of $e$ are included to make these half the peak to peak amplitude). Note that the diploe term is in the elevation direction. Because the parameters for these functions are not at all orthogonal (in fact they are very degenerate), it is necessary to start with the parameters close to the realistic values. An example map of the polarized response to Jupiter along with the best fit is shown in Figure 6.5. Further results are presented in Section 6.3.1

For each observation of Jupiter, for each channel, a set of 13 parameters (6 for the total-power channels, 7 for the polarization channels) and their associated errors are produced.

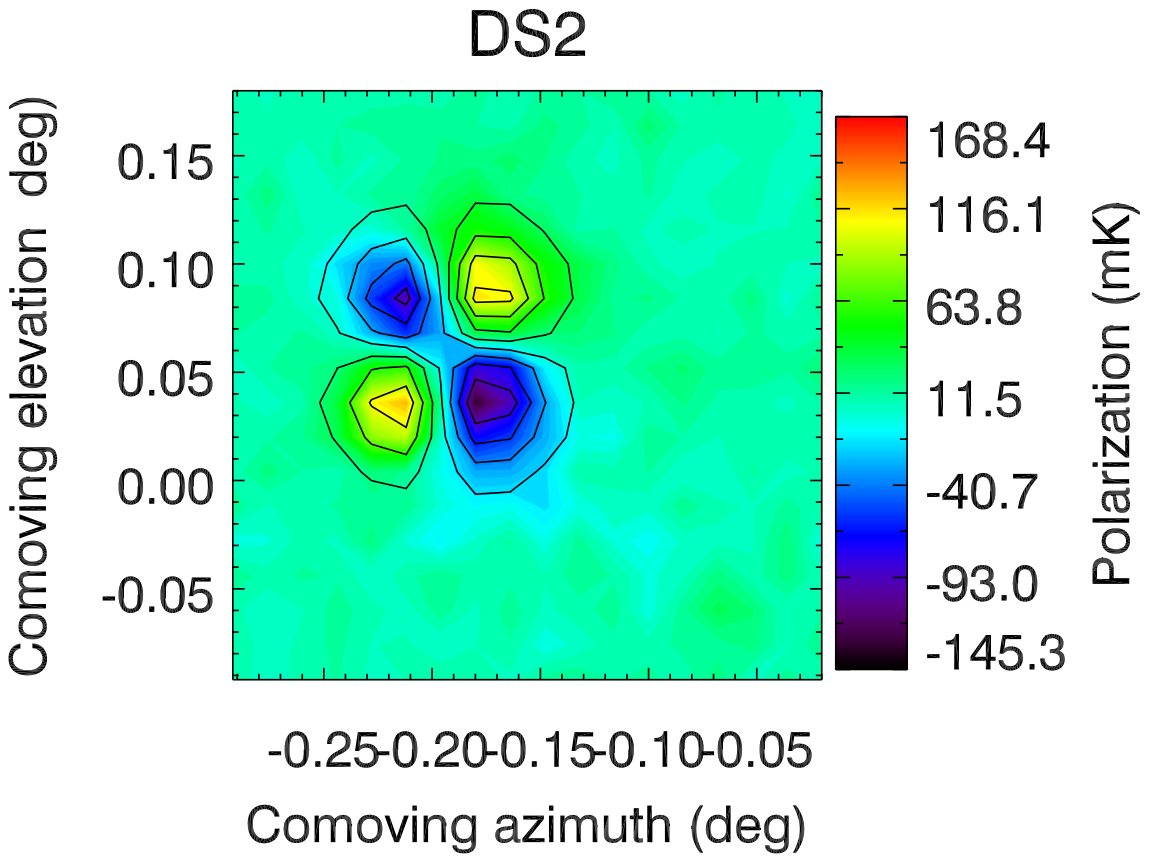

Figure 6.5: Map and fit of DS2 polarization-channel response to a single Jupiter observation. The contour levels are in mK receiver temperature evenly spaced from peak to peak. This quadrupolar feature is the cross-polar pickup of the polarization channel, caused by the lens anti-reflection grooving (see [142] for details).

The errors for the total-power channel parameters are replaced by the simulation errors presented in the next section.

### 6.2.3 Simulation of the Errors

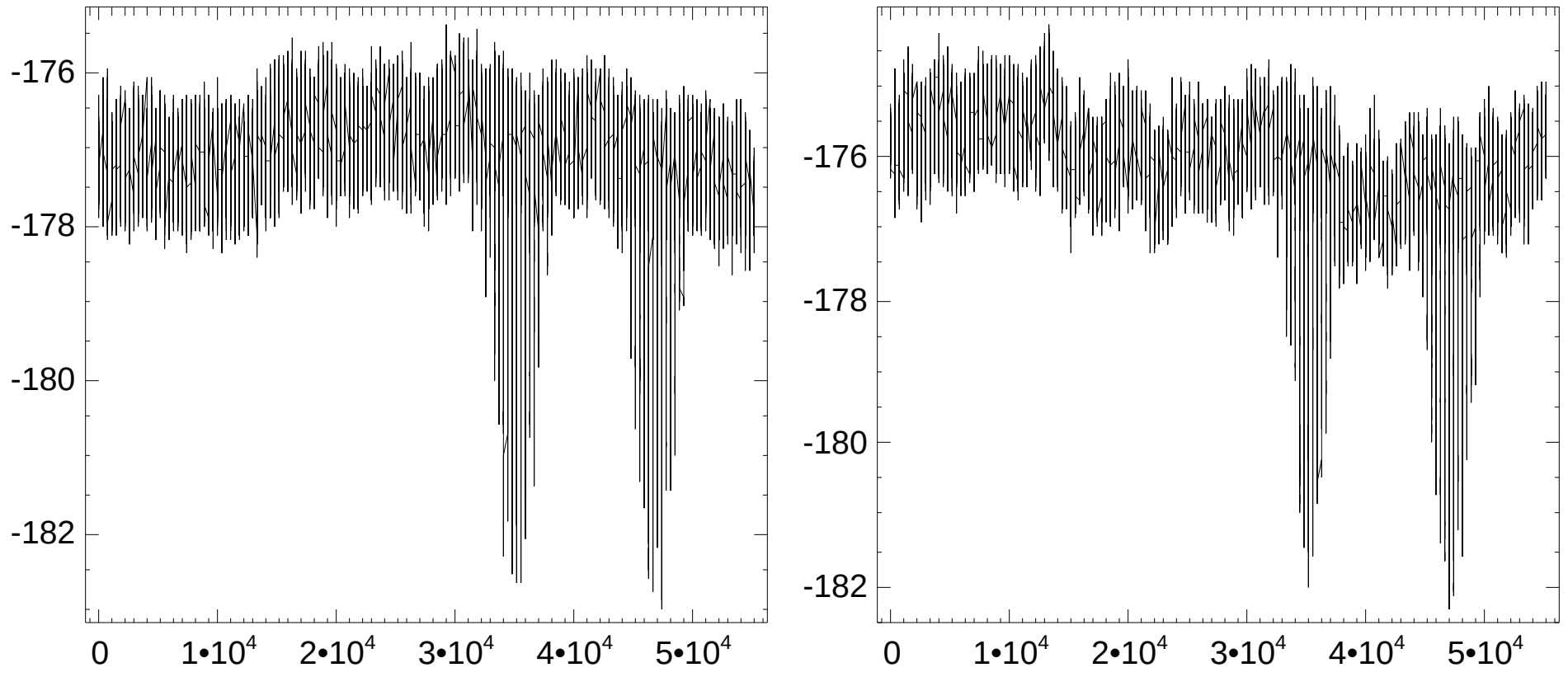

Figure 6.6: Real and simulated time-series of a Jupiter observation as a function of time (in ms). The fake time stream is generated from a noiseless Jupiter signal plus a realization of $1/f$ noise whose power spectra (slope and level), are tuned to match the average properties of the observations. Which one is the fake time stream[4]?

The parameter errors derived from the gaussian fit are underestimated because the errors in the map are too. The dominant cause of this underestimation is the baseline removal which reduces the fluctuations in each map pixel. Ideally, one could obtain a parameter error by repeating the same Jupiter observations $N$ times and averaging the resulting parameter to get a mean value and use the standard deviation as its error. Since this is not possible, we simulate many realizations of each observation. Figure 6.7 shows a flowchart of the simulations. For each observation, $\sim$300 simulations are performed . Each simulation starts with a time-series, from a fake noiseless Jupiter signal plus a random realization of $1/f$ noise. Examples of a real and fake Jupiter time series are shown in Figure 6.6. The $1/f$ noise is adjusted to have the same power spectrum as that of the observation. The simulated time-series then goes through exactly the same steps 1 through 7 as the real data. The distribution of the parameters from the 300 simulations are recorded and the resulting error on each parameter is taken to be the width of the distribution of that parameter. The parameter errors from simulations are typically 2 to 4 times larger than the errors from the fit.

[4] The right sub-figure is the fake time-series.

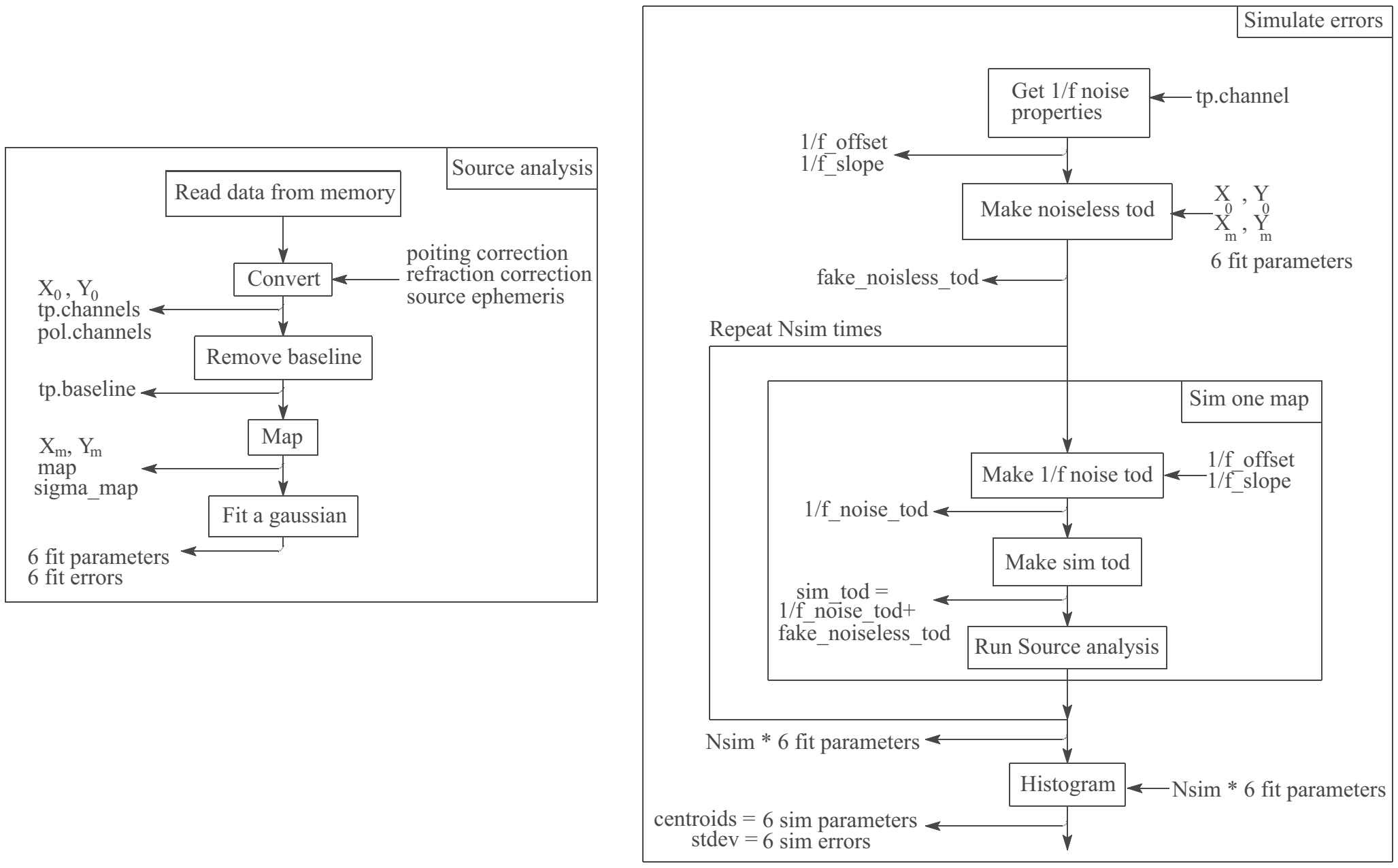

Figure 6.7: Flow Chart of the source observations analysis (left) and the error simulation analysis (right).

## 6.3 Beams

### Definitions

Summarized here are the equations used to combine all the fit parameters into final values. If the $N$ observations produce $N$ values of the parameters, $x_i$, with errors, $e_i$, the weighted mean is defined as

$$\overline{x} = \frac{\sum (x_i/e_i^2)}{\sum (1/e_i^2)} , \tag{6.7}$$

and the error on the weighted mean is

$$e_{\overline{x}} = \sqrt{\frac{1}{\sum (1/e_i^2)}} . \tag{6.8}$$

As described above, for the total-power measurements, we find $e_i$ from simulations. The RMS of the $N$ values $x_i$ is:

$$s_{\overline{x}} = \sqrt{\frac{\sum (x_i - \overline{x})^2}{N(N-1)}} . \tag{6.9}$$

The RMS is a measure of the scatter of the values, relevant for approximately equal errors. The $\chi^2$ to the weighted mean is defined as

$$\chi^2 = \sum \Big(\frac{x_i - \overline{x}}{e_i}\Big)^2 . \tag{6.10}$$

The integral probability to exceed (PTE) is:

$$\mathrm{PTE} = P_\chi(\chi^2;\nu) = \int_{\chi^2}^{\infty} P_\chi(x^2 : \nu) dx^2 , \tag{6.11}$$

which describes the probability that a random set of data points with the same number of degrees of freedom ,$\nu$, as a given set of parameter measurements will produce a value of $\chi^2$ greater than or equal to the value of $\chi^2$ calculated from Equation 6.10.

### Results

The beam widths combined over all channels and all observations are $\sigma_{\mathbf{x}} = \mathbf{0.0279 \pm .0004^\circ}$ and $\sigma_{\mathbf{y}} = \mathbf{0.0270 \pm .0003^\circ}$ (or $FWHM_x = 3.93' \pm 0.054'$ and $FWHM_y = 3.81' \pm 0.042'$ ). Given the effective frequency of the total power channels ($\nu_{eff}$= 93.5 GHz), the results agree with the polarized beam widths measured from Tau A [57] ($\sigma_{s0} = 0.0291 \pm .0009$, $\sigma_{s1} = 0.0270 \pm .0006$, and $\sigma_{s2} = 0.0254 \pm .0008$).

The individual beam widths for all observations are plotted in Figure 6.8 and summarized in Table 6.1. The PTE for 7 of the 16 averaged beam widths are less that 1%. This excess scatter in the 15 beam widths is most likely an analysis-induced effect. To check for a real effect, the beams widths were plotted against azimuth, elevation, outside temperature, humidity, and IF temperature of each observation and no statistically significant slope were visible. One possible cause is that the errors associated with each individual beam width are still underestimated, although this was the original reason for estimating the errors through simulations, which did indeed inflate the errors by a factor of 2 to 4. The parameter errors estimated from the simulation are probably correct in that they account for any correlation in the maps introduced by the baseline removal. A more likely cause for this excess scatter is imperfect baseline removal, leaving spurious structures in the map which contaminate the fits. rather than arbitrarily removing large outliers, we instead use the scatter in the parameters (RMS) as a more likely error estimate. Further work may include use of a different baseline removal algorithm (in frequency space, or by using wavelets for example [100]) to see if it eliminates the outliers. Although $\sigma_x$ and $\sigma_y$ are formally inconsistent between the eight channels and with each other, there are no visible trends with receiver position.

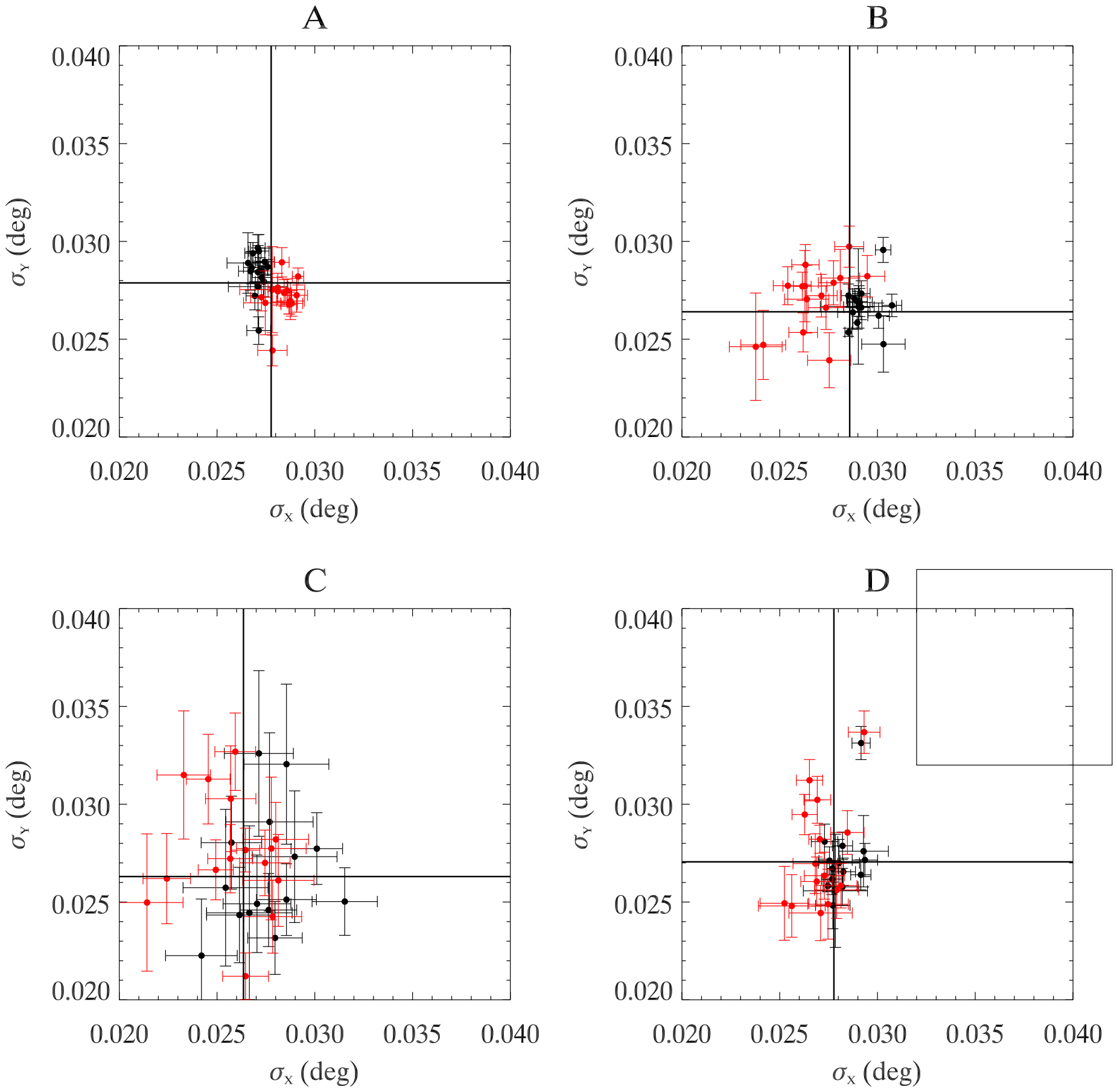

Figure 6.8: Beam widths for the 15 processed Jupiter observations of the 2003 season. Each plot has channel D0 in black and D1 in red (gray) for either receivers A, B, C, and D. The horizontal and vertical lines are the weighted means. The large square in the upper right of the D plot is the $0.01^\circ \times 0.01^\circ$ pixel size used in the mapping routine. Noisier channels such as C naturally yield noisier estimates of the beam widths.

| Radiometer | $\sigma_x$ (mdeg) | RMS | PTE | $\sigma_y$(mdeg) | RMS | PTE |
|---|---|---|---|---|---|---|
| AD0 | $27.1 \pm 0.1$ | 0.07 | 0.93 | $28.4 \pm 0.2$ | 0.3 | <.01 |
| AD1 | $28.3 \pm 0.1$ | 0.1 | 0.38 | $27.3 \pm 0.2$ | 0.2 | 0.03 |
| BD0 | $29.2 \pm 0.1$ | 0.2 | <.01 | $26.7 \pm 0.1$ | 0.3 | <.01 |
| BD1 | $26.8 \pm 0.2$ | 0.4 | <.01 | $27.6 \pm 0.3$ | 0.4 | 0.07 |
| CD0 | $28.0 \pm 0.4$ | 0.5 | 0.25 | $25.1 \pm 0.6$ | 0.8 | 0.57 |
| CD1 | $25.4 \pm 0.3$ | 0.5 | 0.02 | $26.1 \pm 0.5$ | 0.8 | <.01 |
| DD0 | $28.2 \pm 0.1$ | 0.2 | 0.04 | $26.8 \pm 0.3$ | 0.5 | <.01 |
| DD1 | $27.3 \pm 0.2$ | 0.3 | 0.22 | $27.8 \pm 0.3$ | 0.7 | <.01 |
| Combined | $27.9 \pm .05$ | 0.4 | <0.01 | $27.0 \pm 0.07$ | 0.3 | <.01 |

Table 6.1: Beam widths from all 15 Jupiter observations. The RMS and PTE statistic are described in the text. The RMS values are in the same units as the beam widths. The value for the combined beam is the weighted mean and weighted average of the eight TP-channels. The combined RMS, however, is RMS of the eight beam widths. We use the final RMS as a measure of the beam widths errors. During the observations, Jupiter's angular size varied from $45''$ to $41''$.

### 6.3.1 Cross Polarization

The amplitude of the polarized response to an unpolarized point source is derived from the fit of Equation 6.5 to each of the polarization channel maps of Jupiter. Figure 6.5 shows a typical polarized response to Jupiter. To obtain the instrumental polarization pickup, the gaussian quadrupole ($P_2$) amplitude is compared to the amplitude of the total power response to Jupiter, $A_1$, from Equation 6.4. The ratios for the 12 receiver channels are listed in Table 6.2. This quadrupolar polarization effect is dominated by the anti-reflection grooving and the large curvature of the first year CAPMAP lens [142]. The grooved lenses (B and C) are expected to display a stronger quadrupolar amplitude than the lenses only grooved on the back surface (A and D). The amplitude of the dipole and quadrupole terms are plotted against each other in Figure 6.9(a). Note that the lens shape and anti-reflection coating have been improved and the quadrupolar pickup reduced in the CAPMAP04 version of the lens (see [106]). The dipole term is thought to be associated with the telescope optics.

Although this instrumental polarized pickup may seem large (-10 to -15 dB), it only couples total power from angular scales smaller than the beam itself, as shown in Figure 6.9(b). At those angular scales ($\ell \geq 3000$), the CMB power spectrum is already very damped and can only account for a few $\mu$K rms. Thus, the polarization channels would only pickup up at most 10% of a few $\mu$K. A more careful estimate of the exact contamination from CMB anisotropy is underway.

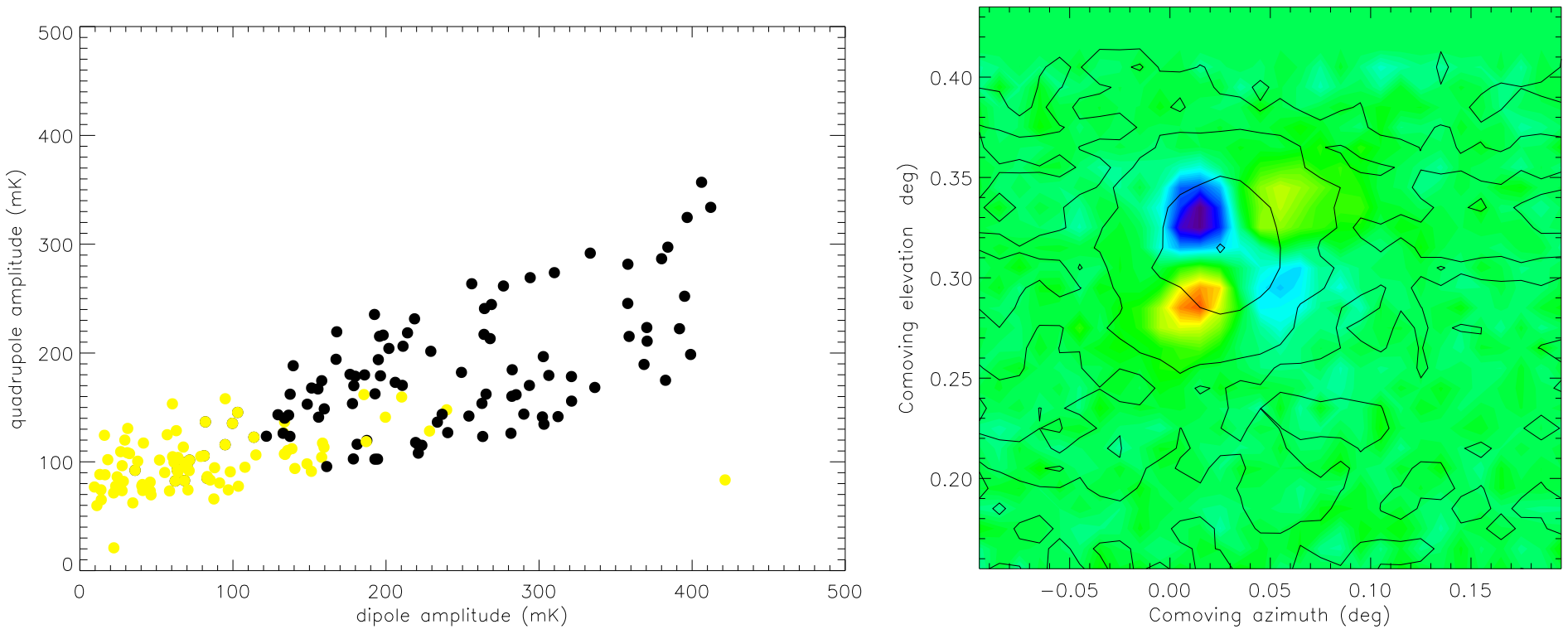

(a) (b)

Figure 6.9: (a) Amplitude of the dipole versus the amplitude of the quadrupole in the fit to the polarized response to Jupiter. Not only are the two components highly correlated, but also, the amplitudes for the half-grooved lenses (A and D in yellow) are consistently smaller than those for the fully grooved lenses (B and C in black). (b) Beam map of the polarization response to Jupiter, overlaid with the normalized total power contours. The TP contours are at 0.9, 0.5, 0.1, and 0.01 times the peak. The characteristic size of each of the four lobes is 3/4 of the TP beam FWHM. Note also the asymmetry in the quadrupole pattern due to the non-zero dipole term in the y-direction.

| Sub-channel | S0 | S1 | S2 |
|---|---|---|---|
| Receiver A | -14.0 $\pm$ 0.6 dB | -13.6 $\pm$ 0.4 dB | -13.1 $\pm$ 0.4 dB |
| Receiver B | -12.7 $\pm$ 0.6 dB | -12.0 $\pm$ 0.6 dB | -11.0 $\pm$ 0.7 dB |
| Receiver C | -11.7 $\pm$ 0.5 dB | -10.8 $\pm$ 0.5 dB | -10.0 $\pm$ 0.6 dB |
| Receiver D | -14.4 $\pm$ 0.6 dB | -14.1 $\pm$ 0.6 dB | -13.8 $\pm$ 0.6 dB |

Table 6.2: Ratio of total-power amplitude to polarization quadrupole amplitude from Jupiter observations. The total-power amplitude is the mean of the gaussian amplitude of the two total-power channels while the quadrupole amplitude is $P_2$ from the fit of Equation 6.6. Each number and its error result from averaging the ratio of each of the 15 Jupiter observations. The typical TP gaussian amplitude is $\sim$3 K and quadrupole amplitude is $\sim$100 mK in antenna temperature. The slight frequency dependence in the ratio has not been corrected for the main beam's frequency dependance.

### 6.3.2 Unexpected Encoder Two-sample Shift

Early in the season, a strange feature appeared immediately in the beam widths. The 2D-gaussian fit yielded a significantly larger beam width in the az (x) than in the el (y) direction ($\widetilde{\sigma_x} = .031$, $\sigma_y = .027$). At first we thought this was a feature of the mapping routine because a 1D-gaussian fit to each Jupiter profile in the time series consistently produced only one beam width. However, if the mapping routine were the culprit, varying the pixel size would affect the gaussian beam width, which was not the case, as shown in Table 6.3.

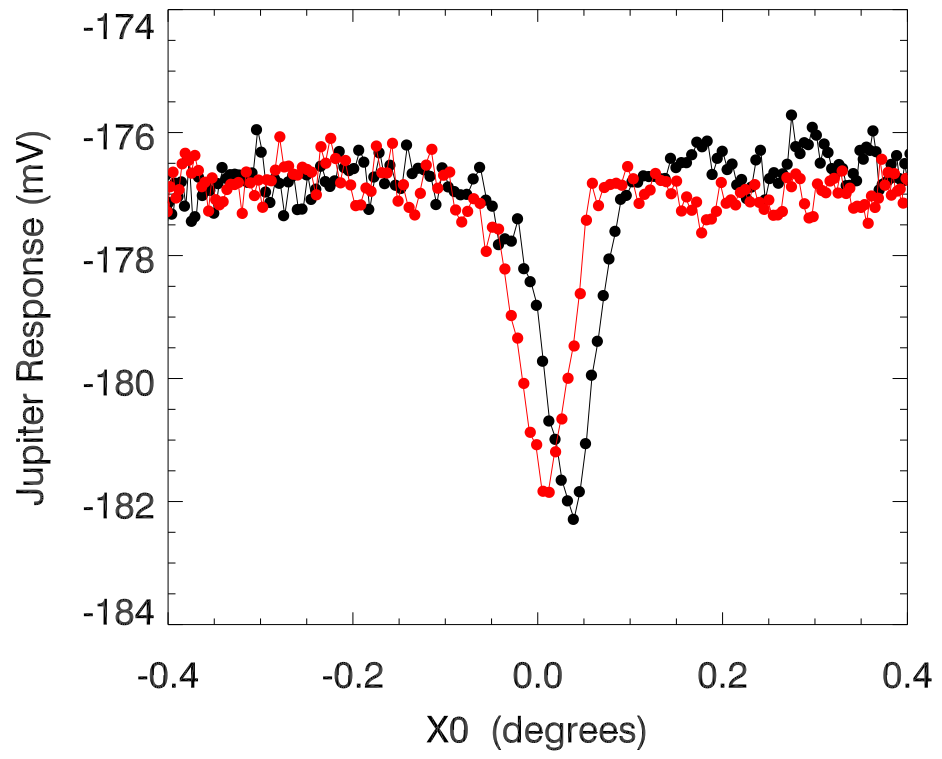

Figure 6.10: Right-going(red) and left-going(black) scan of Jupiter superimposed vs co-moving azimuth $X_0$ in degrees. The two profiles are shifted by 0.02°. This cannot be a real effect as Jupiter must be in one place. It is caused by a shift of the encoder relative to the data stream. A physical explanation for this behavior is still unclear. It may be due to a buffer in the serial port of the DAQ which always contains two encoder samples.

| $X_0$ bin size | $Y_0$ bin size | $\sigma_x$ from fit | $\sigma_y$ from fit |
|---|---|---|---|
| 0.01 | 0.01 | 0.030 | 0.026 |
| 0.02 | 0.01 | 0.030 | 0.026 |
| 0.03 | 0.01 | 0.030 | 0.026 |
| 0.007 | 0.01 | 0.029 | 0.023 |
| 0.01 | 0.02 | 0.030 | 0.026 |
| 0.01 | 0.03 | 0.030 | 0.026 |

Table 6.3: Best fit beam width for DD0 TP channel from file jupi_0212030003 for different map pixel size. The beam width are unaffected by small changes in the pixel size. The timing correction described in the text results in a smaller $\sigma_x$, consistent with $\sigma_y$ at the 3% level.

Further investigating of the 1D-gaussian fits to the time series lead to the discovery that the centroid of a Jupiter profile from left-going scans was shifted by 0.02° from the centroid of a right-going scan (see Figure 6.10). This shift thus increased the beam width in the scan direction without affecting it in the el direction. Such an observation is however unphysical: Jupiter cannot be seen in one place in one scan and in another in the next. The only explanation for this behavior is a relative shift between the encoder time stream and the data time stream. If the encoder sample is acquired at an earlier time than its associated data sample, Jupiter appears early during each scan. To fix the problem, we shift the encoder time-series two samples back with respect to the data and this exactly overlaps the left and right going Jupiter profiles from Figure 6.10. This two-sample encoder shift is independent of the telescope motion and remains true for all Jupiter observations throughout the season.

### 6.3.3 Deep Beam Maps

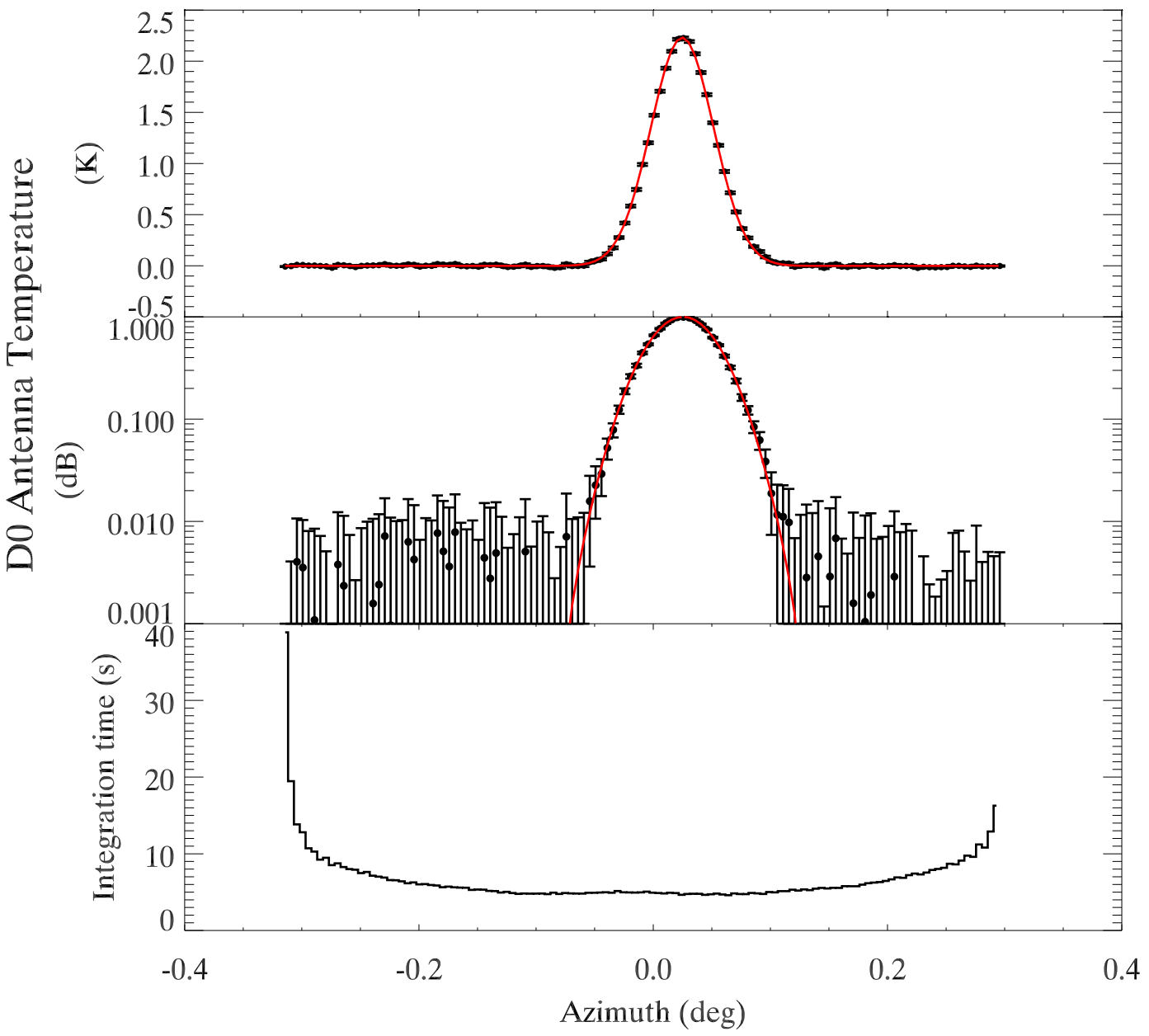

Figure 6.11: Deep beam map on receiver A using Jupiter as the source. The top and center plots are the beam maps in linear and dB scale. The bottom plot shows the integration time per bin. No sidelobes are visible down to the noise level of the map, at ∼25 dB below the peak.

On April 28, at the end of the season, a deep observation of Jupiter on each receiver was performed to try to detect the first sidelobe of the beam. The observing strategy differed from the normal scan in that there was no stepping in elevation. The scan stayed at the same elevation with respect to the source to integrate enough signal to probe the beam at the expected level of the first sidelobe. Assuming the first sidelobe appears ∼30 dB below the peak, an estimate of the power available at the level of the first sidelobe of Jupiter gives 3 mK. We performed one scan for each receiver, scanning a 0.6° region centered on Jupiter for approximately 20 min. Binning the scan in 0.01° bins, each bin gets ∼20 seconds. The beam map (Figure 6.11) shows no sidelobe down to ∼25 dB below the peak. A brighter source such as the limb of the moon would help probe the first sidelobe.

## 6.4 Pointing

An antenna is commanded to point in astronomical coordinates but responds in instrumental coordinates. Imperfections in the structure of the 7-meter telescope require pointing corrections, $\Delta Az$ and $\Delta El$ to be added to the azimuth and the elevation to match the astronomical and instrumental frames. The pointing errors are defined as

$$\begin{aligned} \Delta Az(az, el) &= Az_{obs} - Az_{cmd} \ , \\ \Delta El(az, el) &= El_{obs} - El_{cmd} \ , \end{aligned} \tag{6.12}$$

where *obs* refers to the coordinates where the source is observed and *cmd* refers to the coordinates computed by an error-free antenna control system. The pointing corrections can be parameterized and the functional form of these corrections is called the "pointing solution" of the telescope.

For an optical telescope, the pointing errors are measured by using a set of bright stars with low proper motion which span the whole sky. For a radio telescope, however, there are only a few objects at 90 GHz bright enough to be used as pointing targets, so a set of a few objects (planets, supernovæ remnants) is observed at different times in their transits to cover the whole sky. This process is slow and tedious. The star measurement is technically simpler, quicker, and usually more accurate because the stars are by nature point-like and provide a better sampling of the pointing errors over the hemisphere. However, the radio and optical measurement do not necessarily yield the same pointing errors since they do not use the same parts of the antenna. A small optical telescope attached next to the main radio beam of the main 7-meter antenna allowed us to perform both the optical and radio measurements, and compare the results (see Section 6.4.1).

Chronologically, a detailed pointing solution of the antenna had been established in 1998 by Greg Wright [147]. Before starting the 2002-2003 observing season, we performed a new star pointing measurement. This measurement was not analyzed until after the season so we operated the telescope with the old pointing solution from Greg Wright, and later compared it with the new pointing solution presented here. Because the CMB observations always point close to NCP, only the pointing solutions around the NCP region is relevant for CMB analysis.

### 6.4.1 Telescope Pointing Solution

**The telescope pointing parameters**

The 7m antenna pointing solution is based on a model initially developed for the Haystack 120-ft radio telescope [109]. This section follows closely the discussion in [109].

We choose to parameterize the pointing solution of a telescope using only terms which describe geometrical aspects of the antenna structure (as opposed to a polynomial series [138] for example). This approach has the advantage that it must yield physically realistic results. We consider the following six telescope pointing errors: 1) deviation of the azimuth-axis direction from astronomical zenith (tilt), 2) non-orthogonality of the azimuth and elevation axes (skew), 3) non-orthogonality of the antenna beam and the elevation axis (collimation), 4) gravitational deflection of the antenna structure (sag), 5) higher-order terms in the azimuth axis tilt (wobble), 6) atmospheric refraction. Each effect can induce a pointing error in azimuth, elevation or both.

**1) Azimuth-axis tilt:** The azimuth axis of an alt-az mount telescope should point towards astronomical zenith. A deviation from zenith produces a pointing error in azimuth and in elevation. The two parameters that describe this rotation are $\phi$, the amplitude of the tilt in degrees measured southward from zenith and $\kappa$, the azimuth of the axis through which the mount is rotated counter-clockwise, measured eastward from North (see Figure 6.16).

$$\Delta Az_1 = -\phi \cos(\kappa - az) \tan(el) \ , \tag{6.13}$$

$$\Delta El_1 = -\phi \sin(\kappa - az) \ . \tag{6.14}$$

This is the dominant term in the pointing solution but also the most stable one so the newly derived tilt parameters should be similar to those from previous pointing solution.

**2) Elevation axis skew:** A non-orthogonality of the azimuth and elevation axis causes an azimuth error of the form

$$\Delta Az_2 = -\epsilon \ \tan(el) \ . \tag{6.15}$$

**3) Collimation:** This term is not used in the final pointing solution but is necessary to account for the possible non-orthogonality of the optical or radio beam from the elevation axis.

$$\Delta Az_3 = -\delta \ \sec(el) \ . \tag{6.16}$$

**4) Gravitational sag:** This term is due to a deflection of the secondary support arm that causes an elevation error.

$$\Delta El_2 = \beta(el) \ .$$

**5) Wobble:** This term is specific to the 7m-antenna model. It was added to model higher order variations in the azimuth tilt. This term is kept fixed in most of this fitting because it is smaller than previous terms and requires a larger sample of measured pointing errors to yield significant parameters.

$$\begin{aligned} \Delta El_3 \ &= \ C_{w1}\ \cos(az) + C_{w2}\ \cos(2az) + C_{w3}\ \cos(3az) + \\ & \quad S_{w1}\ \sin(az) + S_{w2}\ \sin(2az) + S_{w3}\ \sin(3az) \ . \end{aligned} \tag{6.17}$$

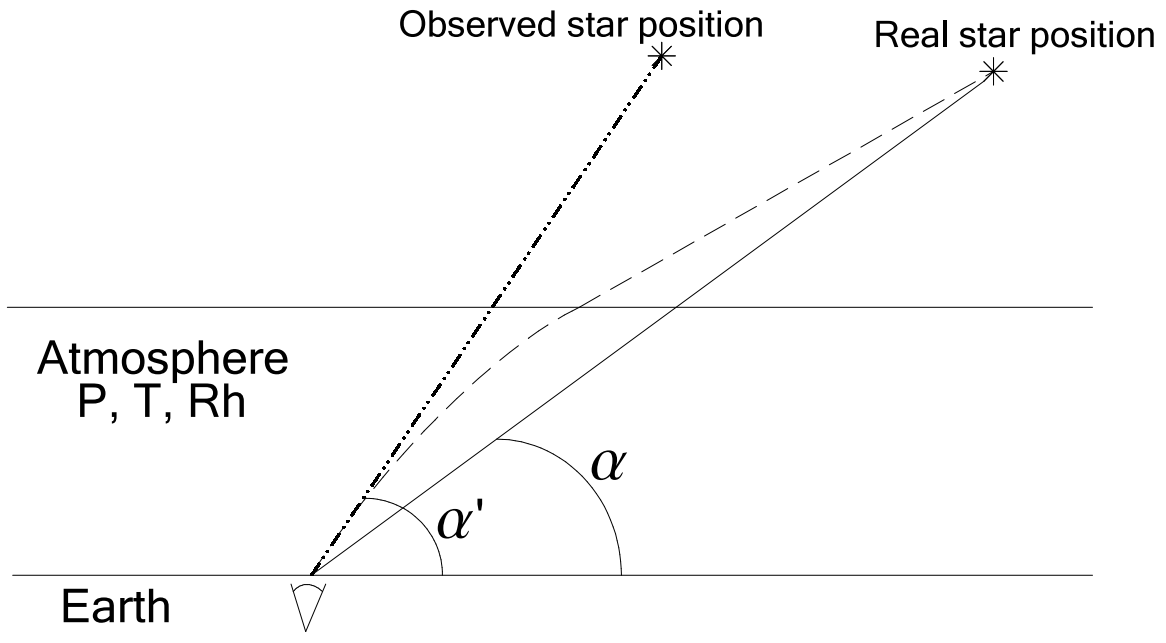

Figure 6.12: Refraction in the Earth's atmosphere bends a light ray towards vertical. Thus, the "apparent elevation" at which a star is observed, $\alpha'$, is always greater than the "true elevation", $\alpha$, in the absence of atmosphere. As a guidance, the refraction corrections for a standard atmosphere (P = 1000 mb, T = 273° K, Rh = 0%) at 20°, 40°, and 70° are 0.044°, 0.019°, and 0.006° respectively. Note that the refraction correction of the atmosphere in the microwave spectrum does not depend on frequency.

**6) Atmospheric Refraction:** Although this term is not a true instrumental effect, it behaves like one by causing an object to be observed at a higher elevation than it really is (Figure 6.12). The refraction correction is:

$$\Delta El_4 = \frac{\mathcal{R}}{\tan(el)} \ , \tag{6.18}$$

$$\begin{aligned} \mathcal{R} \ &= \ \frac{77.6\ \frac{10^{-6}}{2\pi}\ P}{T} + \left(\frac{3.75\ 10^5\ \frac{10^{-6}}{2\pi}}{T^2} - \frac{5.6\ \frac{10^{-6}}{2\pi}}{T}\right) P_{H_20} \ , \\ P_{H_20} \ &= \ R_h \left(\frac{T}{247.1}\right)^{18.36} \ . \end{aligned} \tag{6.19}$$

Here $\mathcal{R}$ is in fractional degrees (i.e. multiply by 360 to get degrees), P (the atmospheric pressure) and $P_{H_20}$ (the water vapor partial pressure) in millibar, T (the atmospheric temperature) in Kelvin, and $R_h$, the relative humidity in %. Equation 6.19 is empirically determined from the refractive indices of dry air and water vapor [11]. During the season, the refraction correction was fixed to $\mathcal{R}$(P= 1013 mb, T= 273 K, $R_h$ = 0%) = 0.01628° because small changes in the atmospheric conditions (variation of 10 mb, 10°, or 10% Rh) cause undetectable changes in the refraction correction (less than 1% of a beam width).

### Observations and Reduction of the Pointing Errors

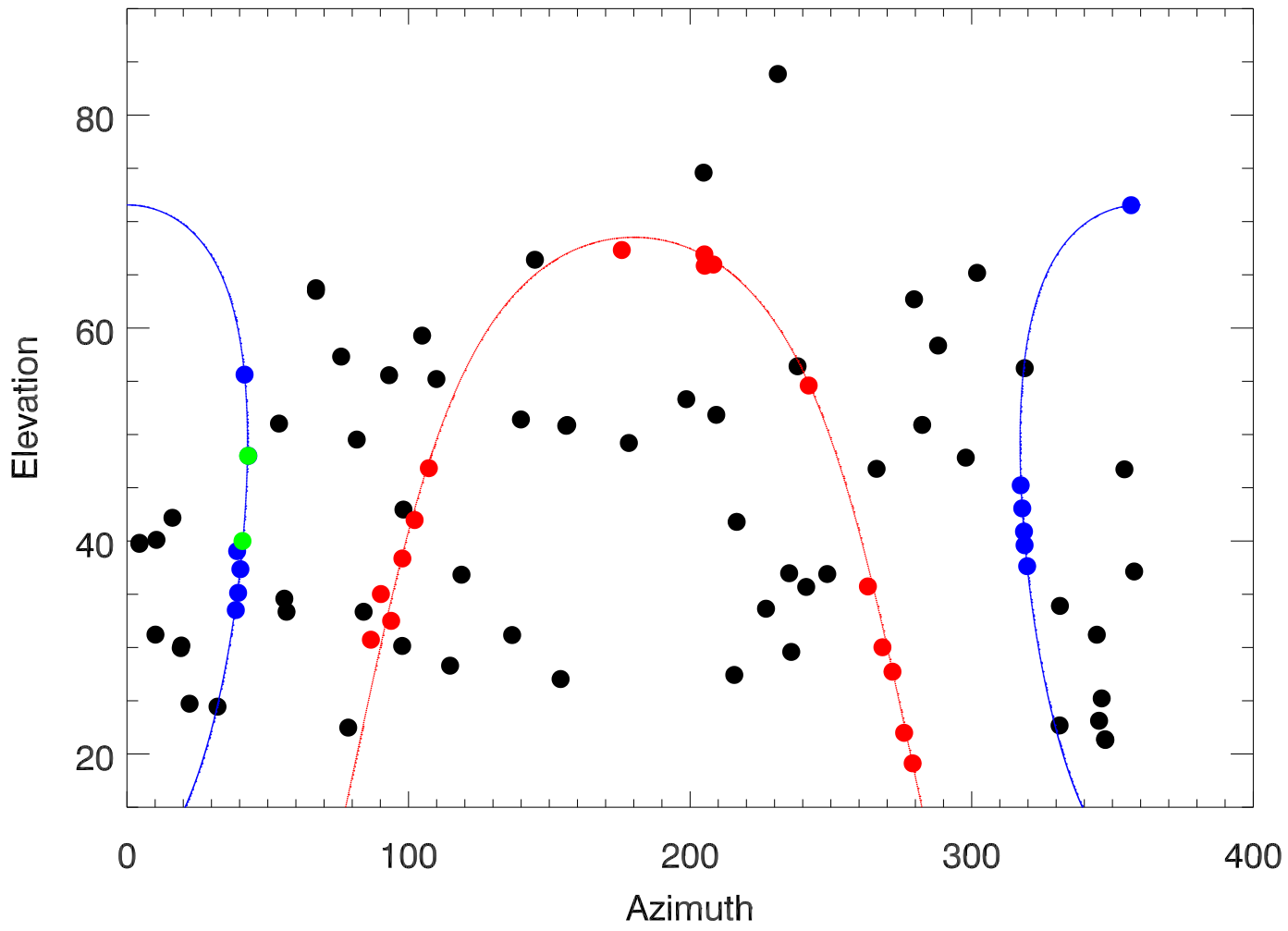

Figure 6.13: Positions in an az-el map of the sources observed. In black, the 60 star pointing measurements, in red the 15 analyzed Jupiter observations, in blue the 20 Cas A observations, of which only the green ones were detected. No source can be observed below an elevation of $\tilde{2}0°$ because of the trees surrounding the antenna.

The pointing solution of the 7m telescope was supposed to be derived from the scheduled radio observations of Jupiter and Cas A[5]. However, a satisfactory fit of the pointing solution could not be obtained because the 20 observations of Cas A only yielded two very weak detections and the Jupiter observations are all clustered in the ecliptic plane (see Figure 6.13). Therefore, the radio observations only serve to verify that the radio and optical beam axes coincide (see Section 6.4.1).

The main pointing calibration is done with a sample of 60 stars evenly distributed (see Figure 6.13), using a small optical camera meant to be aligned with the radio beam. The stars were observed on Jan 23, 2003 following the procedure described in Appendix A. Each star is centered on a cross hair of the CCD image, and the offset between the commanded position and the observed position was recorded as the pointing delta ($\Delta Az_i$, $\Delta El_i$). The main source of systematic error during this process is the repeatability in centering a faint star on the cross hair. The CCD screen being 0.1°×0.1°, we estimate the error on each measurement to be 0.003°.

[5]Cassiopeia A located at $\alpha_{2000}$ = 23h23m25.4s, $\delta_{2000}$ = +58d48m38s

The seven model parameters ($\phi$, $\kappa$, $\epsilon$, $\delta$, $\beta$, $\mathcal{R}$, and $Wobble$) are determined by fitting the pair of two-dimensional functions (Equation 6.21) to the 60 measured pointing delta doublets $\Delta Az_i$ and $\Delta El_i$:

$$\Delta Az(az, el) = C_{az} + \Delta Az_1 + \Delta Az_2 + \Delta Az_3 \quad (6.20)$$

$$\Delta El(az, el) = C_{el} + \Delta El_1 + \Delta El_2 + \Delta El_3 + \Delta El_4 \ , \quad (6.21)$$

where $\Delta Az_1$, $\Delta Az_2$, $\Delta Az_3$, $\Delta El_1$, $\Delta El_2$, $\Delta El_3$, and $\Delta El_4$ are the pointing error terms described previously. The best fit parameters are obtained using two independent codes, one written by Greg Wright in C, and another by the author in IDL. Both perform a least-square minimization. The results are presented in Tables 6.4 and 6.5 and in Figure 6.14.

Two problems were encountered during the fitting. First, the atmospheric refraction parameter $\mathcal{R}$ was accidentally set to zero during the observing of the 60 stars. Usually, $\mathcal{R}$ is known and can be fixed in the fit. Instead, leaving $\mathcal{R}$ as a free parameter, adds to the uncertainty of the other parameters and can yield unphysical solutions. Second, a set of 60 pointing errors is not enough to constrain the $Wobble$ parameters. According to Greg Wright [147], $Wobble$ is a small addition to the simple tilt model and requires a minimum of 300 pointing errors to be correctly constrained. As a result, many fits have $\mathcal{R}$ and the $Wobble$ parameters fixed to their most likely values of $\mathcal{R}$(P=1000 mb, T=273° K, Rh=0%) = 0.01627°and $Wobble = \{0\}$.

| ΔAz | $\phi$(tilt) | $\kappa$(tilt) | $\epsilon$(skew) | $\delta$(collimation) | RMS[e] | ΔAz at NCP[f] |
|---|---|---|---|---|---|---|
| original2003[a] | 0.009 | 175.93 | -0.0087 | | | 0.00023 |
| star2003[b] | 0.01 | 170 | -0.0072 | 0.325 | 0.0011 | 0.0021 |
| star2003[c] | 0.011 | 168 | -0.0088 | 0.327 | 0.0015 | 0.0016 |
| star2003[d] | 0.012 | 171 | -0.0106 | 0.330 | 0.0016 | 0.0010 |

Table 6.4: Best fit parameters from different fits to the azimuth errors. All units are degrees
[a]The original pointing parameters established 5 years ago and used during this observing season.
[b]New parameters from a fit to the 2004 star measurements using the C code and letting all the parameters free.
[c] Same as (b) but with the IDL code.
[d] Same as (c) but with the tilt parameters fixed to the directly measured values.
[e]The RMS column is the RMS of the residual pointing error in the 60 star measurements after the specific pointing solution of a given row is accounted for.
[f]ΔAz and ΔEl at NCP columns are the actual value of the pointing solution at (Az, El)=(0°,40.392°). These are the only two numbers that need to be applied to the CMB observations to convert from the encoder readout values to the actual position of the central beam (Eq 6.1).

Table 6.4 shows that:

| ΔEl | $\phi$ | $\kappa$ | $\beta$ | $\mathcal{R}$ | $Wobble^e(\times 10^{-3})$ | RMS | ΔEl @ NCP |
|---|---|---|---|---|---|---|---|
| original2003[a] | 0.009 | 175.93 | -0.0002 | 0.01627 | [7.3, -0.26,0.,-0.82,-0.3,0.] | | 0.008 |
| star2003[b] | 0.035 | 162 | 0.0002 | 0.035 | [-20., 0.8, -10., 2., -3., 2.] | 0.0011 | .04 |
| star2003[c] | 0.054 | 170 | -0.0004 | 0.01627 | {0} | 0.014 | -0.0059 |
| star2003[d] | 0.012 | 171 | -0.0002 | 0.01627 | {0} | 0.033 | 0.009 |

Table 6.5: Best fit parameters from different fits to the elevation errors. All units are degrees
[a] The original pointing parameters established 5 years ago and used during this observing season
[b] New parameters from a fit to the star measurement using the IDL code, letting all the parameters free.
[c] Same as (b) but with the refraction and wobble parameters fixed.
[d] Same as (c) but with the tilt parameters fixed to the directly measured values.
[e] the six Wobble parameters defined as $[S_{w1}, S_{w2}, S_{w3}, C_{w1}, C_{w2}, C_{w3}]$

- Both codes give similar results.
- The fit to the ΔAz errors with all the parameters left free is consistent with the original pointing solution.
- With the tilt parameters fixed to the directly measured values (see Section 6.4.1), the fit at NCP remains consistent with the original solution.
- In all cases, the scatter of the residual az pointing errors after applying the derived pointing solution remains small ($\leq \frac{1}{50}$ FWHM) and the correction at NCP insignificant.

Table 6.5 shows that:

- The fit to the ΔEl errors with the $Wobble$ and $\mathcal{R}$ parameters unconstrained produces results which are inconsistent with the original solution. However, the fit value of $\mathcal{R} = 0.035$ is unphysical so we disregard this fit.
- Once $Wobble$ and $\mathcal{R}$ are fixed, the sag parameter agrees well with the original value but the tilt parameter remains inconsistent with both the original value and the tilt parameter determined from the fit to the ΔAz errors. This led us to obtain a direct measurement of the tilt parameters to clear up this discrepancy (see Section 6.4.1).
- With the tilt parameters fixed to the directly measured values, the sag becomes consistent with the original solution.

The final parameters in the last row of Table 6.4 and 6.5 provide a correction at NCP which is within 0.001° of the original pointing solution. Only the large RMS for the residuals of the elevation pointing errors indicates a discrepancy between the fit and the data points. Trying different combinations of fixed parameters doesn't improve the fit. The RMS of the residual el pointing errors remain closes to 0.03° when the tilt is fixed to $\phi = 0.012$ and $\kappa = 171$. Although this is not ideal, only the RMS for the residuals of the radio pointing errors gives the best measure of the accuracy of the radio pointing solution.

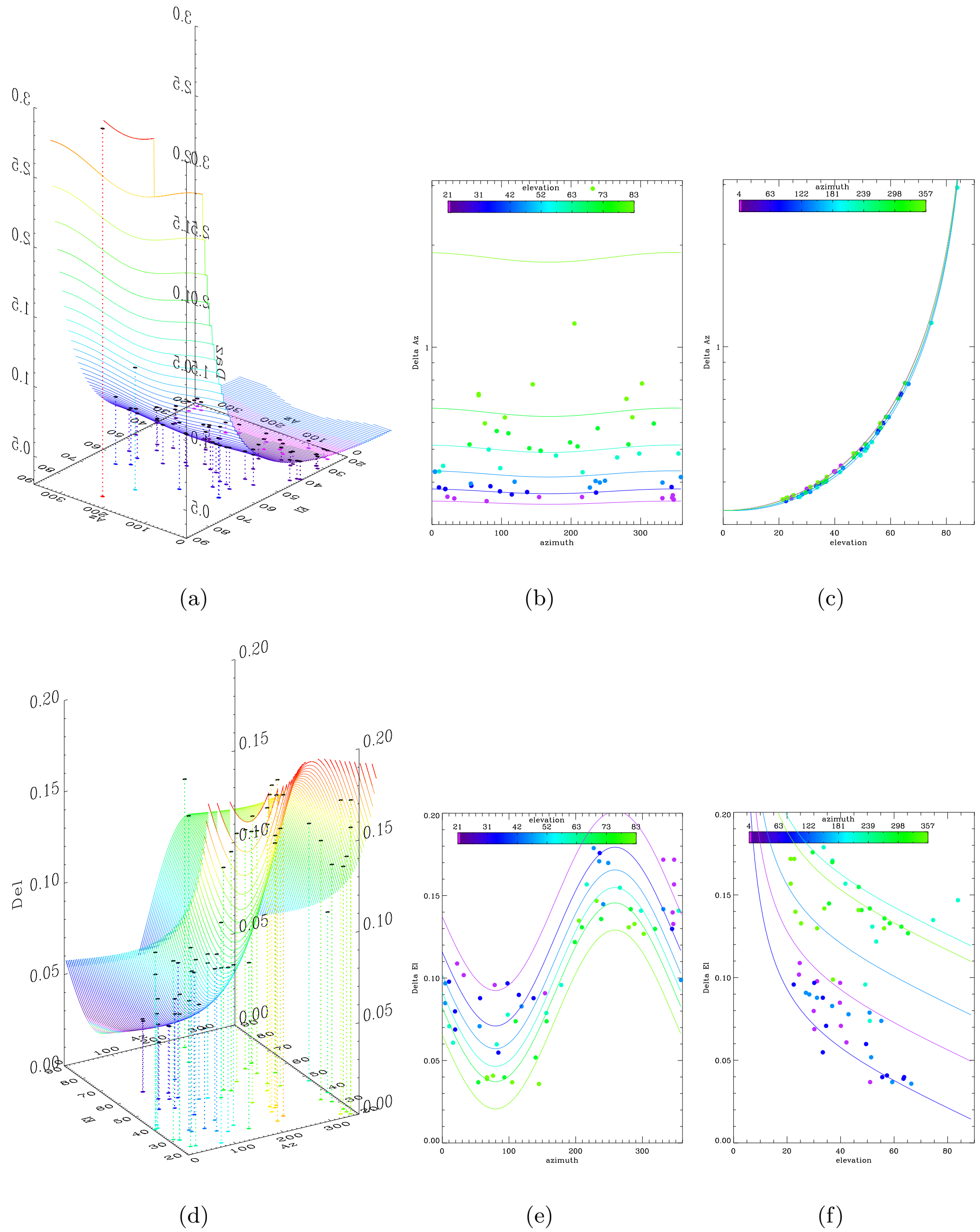

Figure 6.14: (a) $\Delta Az$ pointing deltas for the 60 stars observed vs Az and El, along with the best fit pointing solutions. The projections on the Az and El planes are shown in (b) and (c). (d)$\Delta El$ pointing errors for the 60 stars observed vs Az and El, with the best fit model. Projections on the Az and El planes are shown in (e) and (f). The parameters for the best fit models are in the last rows of Table 6.4 and 6.5. Note that CMB observations are made at az $=0°$ and el$=40°$.

### Direct measurement of azimuth axis tilt

In order to resolve the disagreement in the best-fit azimuth axis tilt parameter between the azimuth and elevation data, the tilt is measured directly with a precision level[6]. The tiltmeter has a resolution of $1''$, and the angle to measure is on the order of $0.01°$ ($= 36''$), so the tiltmeter's resolution should be sufficient to unambiguously determine the tilt. The tiltmeter is mounted in the vertex cab, as close as possible to the top of the azimuth axis. The two axes of the tiltmeter are aligned such that its x-axis is parallel to a horizontal line from the focus to the secondary and the y-axis is parallel to the elevation axis. The sign convention is such that when the x-axis angle is negative the secondary is high. The telescope is fixed in elevation near $0°$ and the x- and y-axis angle are recorded around the whole azimuth range at $10°$ intervals. The telescope is stopped for about a minute at each position to let the bubble in the tiltmeter relax and give a stable reading.

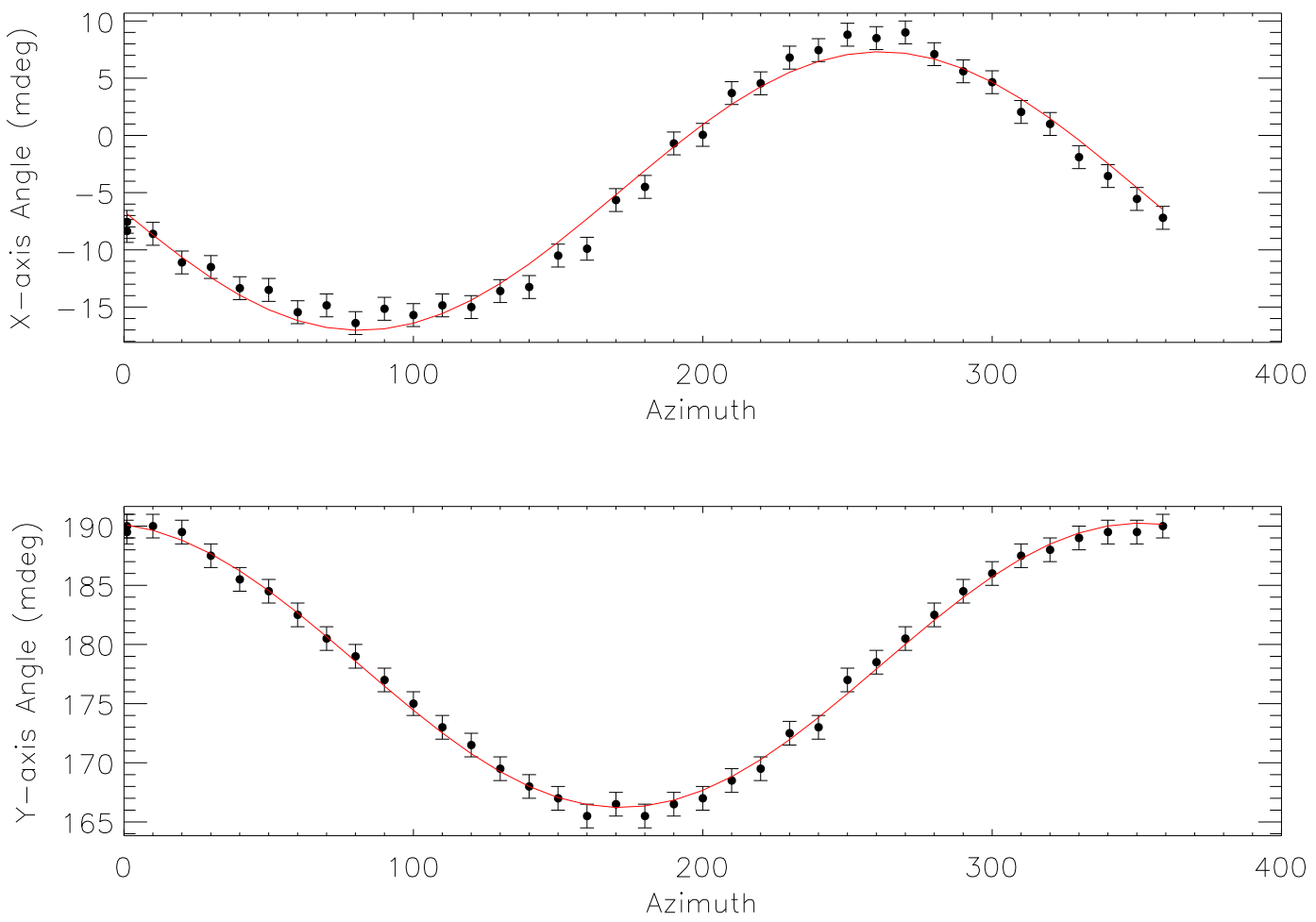

Figure 6.15: Measured tilt of the alt-az mount of the telescope. *top*: Angle of the axis parallel to the secondary arm and best fit $-4.8 + 12.5\cos(261 - az)$ (in millidegrees). *bottom*: Angle of axis parallel to the elevation axis and best fit $178 + 12.0\cos(351 - az)$. As expected the two angles are $90°$ out of phase. The amplitude of the tilt ($\phi$) is simple to deduce from the plots but the direction in which the azimuth axis is tilted takes a few steps. In the top plot, the x-axis is most positive at an azimuth of $81°$, which means, from the sign convention that the azimuth axis is actually tilted towards $261°$. Therefore, the level axis (the axis through which one needs to rotate the horizontal plane to obtain the tilt of the azimuth axis) is directed towards $171°$ (see Figure 6.16). Note that we could have also chosen the diametrically opposite axis for the level axis (az = -9°) and reversed the sign of the tilt amplitude.

The results are plotted in Figure 6.15 along with a sinusoidal fit to the two angles. The fit yields $\phi = (12° \pm 0.2°) \times 10^{-3}$ and $\kappa = 171 \pm 1°$ for the tilt parameters. These results are in excellent agreement with the values determined from the azimuth errors as well as

[6] with a tiltmeter, model 711-2 from Applied Geomechanics [137].

the original 1998 values.

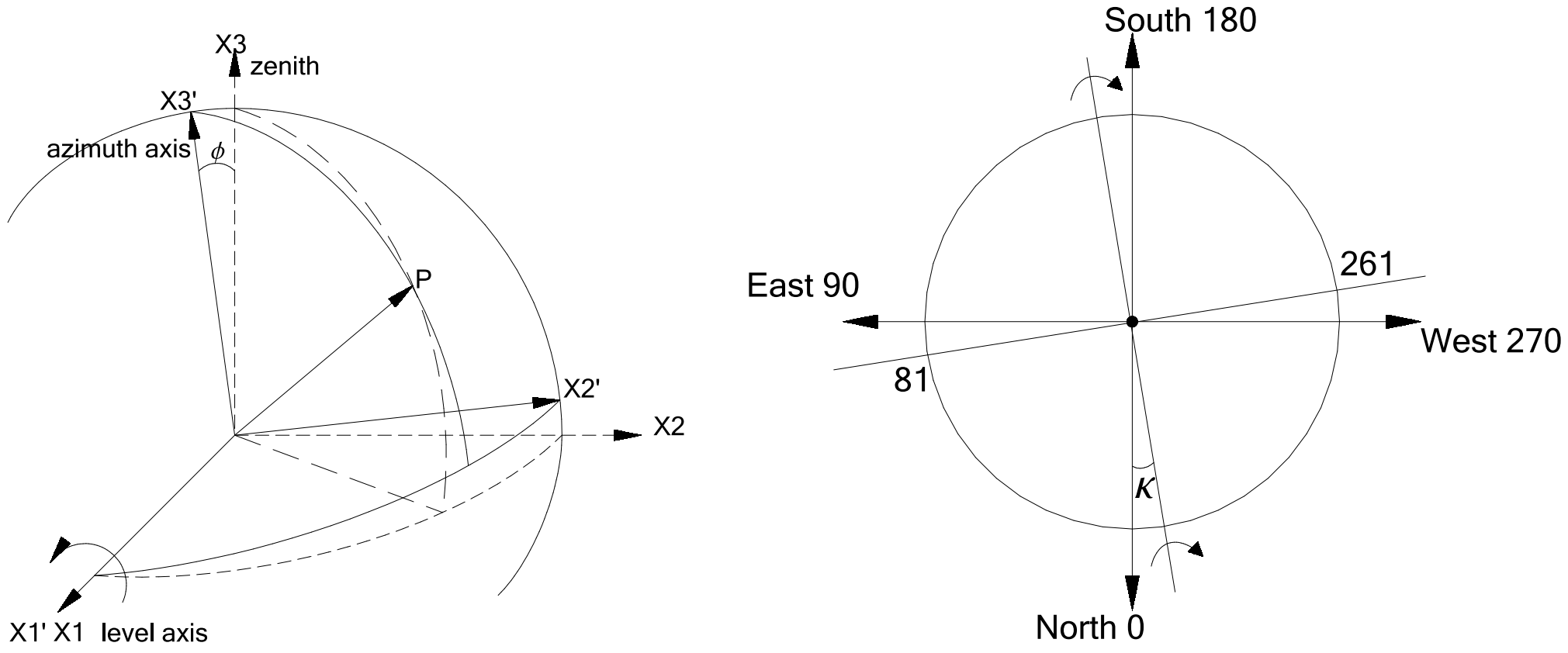

Figure 6.16: Illustration of the telescope azimuth-axis tilt. The primed coordinate system is rotated counter-clockwise through an angle $\phi$ around the $X_1$ axis (the level axis). For a counter-clockwise rotation, the azimuth of the level axis is $\kappa$, measured eastward from North, and $180 - \kappa$ for a clockwise rotation. We find $180 - \kappa = 171°$

### Comparison of the radio and optical pointing

As described earlier, the pointing errors from the Jupiter observations cannot be fit to a pointing solution with all the parameters free because of the poor sky coverage of the Jupiter observations. However, the Jupiter pointing errors can be checked against the best fit parameters from the star pointing errors.

| TP channel | $C_{az}$ | $\delta$ | Az RMS | $C_{el}$ | El RMS |
|---|---|---|---|---|---|
| AD0 | -0.02 | .0009 | 0.0014 | 0.28 | 0.013 |
| AD1 | -0.02 | .0004 | 0.0013 | 0.28 | 0.013 |
| BD0 | -0.02 | 0.22 | 0.0014 | 0.069 | 0.013 |
| BD1 | -0.02 | 0.22 | 0.016 | 0.068 | 0.013 |
| CD0 | -0.02 | 0.004 | 0.032 | -0.015 | 0.014 |
| CD1 | -0.02 | 0.004 | 0.0018 | -0.015 | 0.014 |
| DD0 | -0.02 | -0.22 | 0.0017 | 0.06 | 0.013 |
| DD1 | -0.02 | -0.22 | 0.002 | 0.06 | 0.012 |

Table 6.6: Comparison of the star and Jupiter pointing parameters. All entries are in degrees. All pointing parameters apart from the collimation $\delta$, and the constants are fixed. The fact that the AZ and El RMS remain small even with the pointing solution derived from the star observations means that the pointing solution is a good fit.

The observed X- and Y-centroids ($p_x$ and $p_y$) of the 15 analyzed Jupiter observations, uncorrected for refraction and pointing (skipping step 3 of Section 6.2.2), are fit to the same functions used for the star pointing errors (Equation 6.21) with only the collimation and the two constants $(C)az$ and $C_{el}$) as free parameters. The other parameters are fixed to their best values ($\phi = 0.012$, $\kappa = 170$, $\epsilon = -0.01$, $\beta = -0.0002$, $\mathcal{R} = 0.01627$ and $Wobble = \{0\}$). The results are summarized in Table 6.6. The Az RMS is much smaller that the El RMS, just as in the star pointing errors. This probably means that the pointing solution for the elevation errors could be improved by adding another term to the pointing solution. The final Az RMS is **0.0014**° (= 5″) and the final El RMS is **0.013**° (= 0.7') The pointing error is sufficiently small (less than 1/5 FWHM) to be a minor effect on the main CMB analysis.

### 6.4.2 Relative Pointing of the Array

The X- and Y-centroids of the 15 Jupiter observations are mainly used to determine the relative position of the beams with respect to the telescope central ray. Again the pointing solution is relevant because each observation is made at different az and el, so the pointing corrections are different. Applying the pointing solution reduced the spread of the distribution of centroids (see Figure 6.17(right)). The expected position of the beams with respect to the telescope central ray in shown in Figure 6.17(left). The whole array is pointed ∼0.06° low, and ∼0.02° to the left of NCP. This offset converts to a dewar positioned 1" too high and 0.3" offset from the nominal on-focus position. The array pointing offset could have been zeroed for CMB observation at NCP by offsetting the telescope central ray by the opposite amount. A negative sign furtively introduced itself in the calculation which resulted in a scan pattern offset by twice the array offset, as shown in Figure 6.17(left). This mis-pointing unfortunately destroys the useful redundancy of the scan strategy described in Section 2.4, but improves the Q-U coverage. The mean centroids for each total-power channel are summarized in Table 6.7. Note that the values in Table 6.7 for each receiver position are in comoving coordinates, while the absolute telescope pointing corrections in Table 6.4 and 6.5 are in encoder coordinates.

| TP channel | $p_x$ (deg) | RMS | PTE | $p_y$ (deg) | RMS | PTE |
|---|---|---|---|---|---|---|
| AD0 | $-.0223 \pm 0.0001$ | 0.0002 | $< .01$ | $-.286 \pm 0.0001$ | 0.0013 | $< .01$ |
| AD1 | $-.0217 \pm 0.0001$ | 0.0002 | $< .01$ | $-.286 \pm 0.0001$ | 0.0013 | $< .01$ |
| BD0 | $-.2456 \pm 0.0001$ | 0.0002 | $< .01$ | $-.066 \pm 0.0001$ | 0.0013 | $< .01$ |
| BD1 | $-.2477 \pm 0.0002$ | 0.0002 | $< .01$ | $-.065 \pm 0.0002$ | 0.0015 | $< .01$ |
| CD0 | $-.0252 \pm 0.0005$ | 0.0004 | $< .01$ | $0.161 \pm 0.0004$ | 0.0015 | $< .01$ |
| CD1 | $-.0255 \pm 0.0001$ | 0.0003 | $< .01$ | $0.163 \pm 0.0002$ | 0.0014 | $< .01$ |
| DD0 | $0.1999 \pm 0.0001$ | 0.0002 | $< .01$ | $-.059 \pm 0.0001$ | 0.0014 | $< .01$ |
| DD1 | $0.1992 \pm 0.0002$ | 0.0002 | $< .01$ | $-.058 \pm 0.0002$ | 0.0013 | $< .01$ |

Table 6.7: Beam centroids relative to the central ray. Each row is the statistic averaged over all 15 Jupiter observations. The standard error and RMS are after the best pointing solution has been removed. The poor value of the PTE indicate that there remains a small pointing effect, and/or that the errors are still underestimated (see Section 6.2.3).

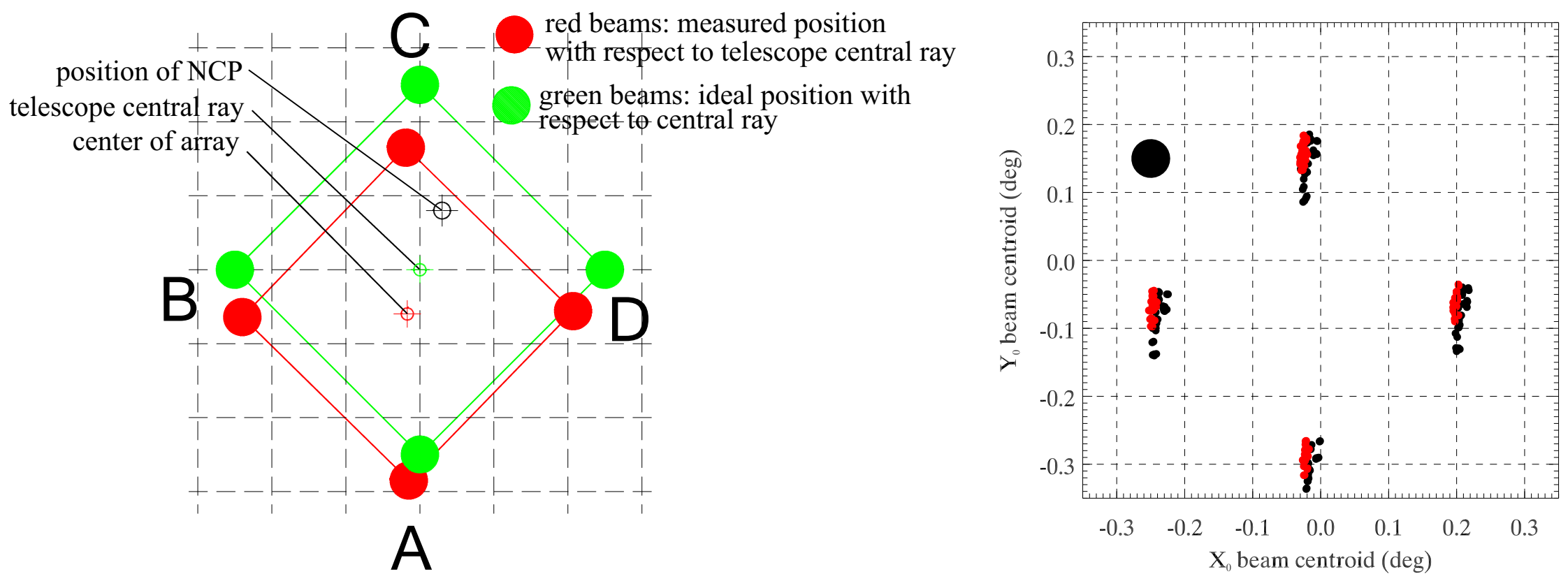

Figure 6.17: *left*: Relative position of each of the four beams in the array with respect to the central ray and NCP during the azimuth scans. The underlying grid spacing is 0.1°. The ideal configuration would have the NCP at the center of the array but the inadvertent misplacement improved our QU coverage. *right*: Beam centroids from the 15 Jupiter observations with the central ray at the origin. In black, the beam centroids without pointing correction applied. In red, accounting for pointing and refraction correction. The pointing correction improves the RMS in Az and El but the El RMS is still notably worse than the Az. A 0.06° FWHM beam is shown for scale.

## 6.5 Calibration

### 6.5.1 Total Power Gains

The amplitudes of the 2D-gaussian fits to the Jupiter maps provide an independent way to measure the gains of the eight total power channels, using the recent measurement of Jupiter's temperature at W-band by the WMAP satellite [115]. The TP gains are also derived from sky-dips tests, performed throughout the season. The Jupiter observations also provide a measure of the stability of the gains. The TP gains are useful for monitoring the receiver's noise temperature and the sky noise temperature, which in turn gives us the sky opacity during the CMB observations.

During the Jupiter observations, the contribution to the receiver's system temperature can be decomposed into the following terms:

$$T_{sys} = \overbrace{T_{rec} + T_{cmb} + T_{sky}}^{T_{off}} + T_{jup} \ , \tag{6.22}$$

where $T_{off}$ is the offset temperature while Jupiter is not in the beam. $T_{off}$ has contribution from the receiver system temperature, $T_{rec}$, and the sky temperature, $T_{sky}$. The fraction of the radiation coupled from the sky is implicitly assumed to be close to 1. Because of its finite emissivity, the atmosphere absorbs part of an external signal and replaces it with noise emitted at the physical temperature of the atmosphere[7]. Assuming that the physical temperature of the atmosphere is 250 K, Equation 6.22 can be re-written as

$$T_{off} = G\ V_{off} = T_{rec} + 1.1\ e^{-\tau} + 250(1 - e^{-\tau})\ , \tag{6.23}$$

and

$$T_{jup} = G\ V_{jup} = 171\ (\Omega_J/\Omega_A)\ e^{-\tau}\ , \tag{6.24}$$

where the numerical factors for the Raleigh-Jeans temperature of Jupiter (171 K) and the CMB (1.1 K) have been inserted from [115]. The term $\Omega_J/\Omega_A$ accounts for the beam dilution, where $\Omega_J = \frac{\pi}{4} D_e D_p$ is the solid angle subtended by Jupiter, and $\Omega_A = 2\pi\sigma_x\sigma_y$ is the beam solid angle. For this gain calculation, we assume $\sigma_x = \sigma_y = 0.027°$ and retrieve $D_e$ and $D_p$, Jupiter's equatorial and polar diameter, from the JPL ephemeris [75]. We also use the values of the eight receiver temperatures derived from sky dips [58]. For each observation and for each channel, the amplitude of the Jupiter signal $V_{jup}$, and the mean offset $V_{off}$ (in mV) are measured from the 2D-gaussian fit. We then solve Equation 6.23 and 6.24 for the gain $G$ (in mV/K), and the loss $e^{-\tau}$. The results are given in Table 6.8, which includes for comparison the receiver noise temperature and gains derived from the sky dip tests. Figure 6.19 shows the measured atmospheric loss, $e^{-\tau}$, for the eight TP channels.

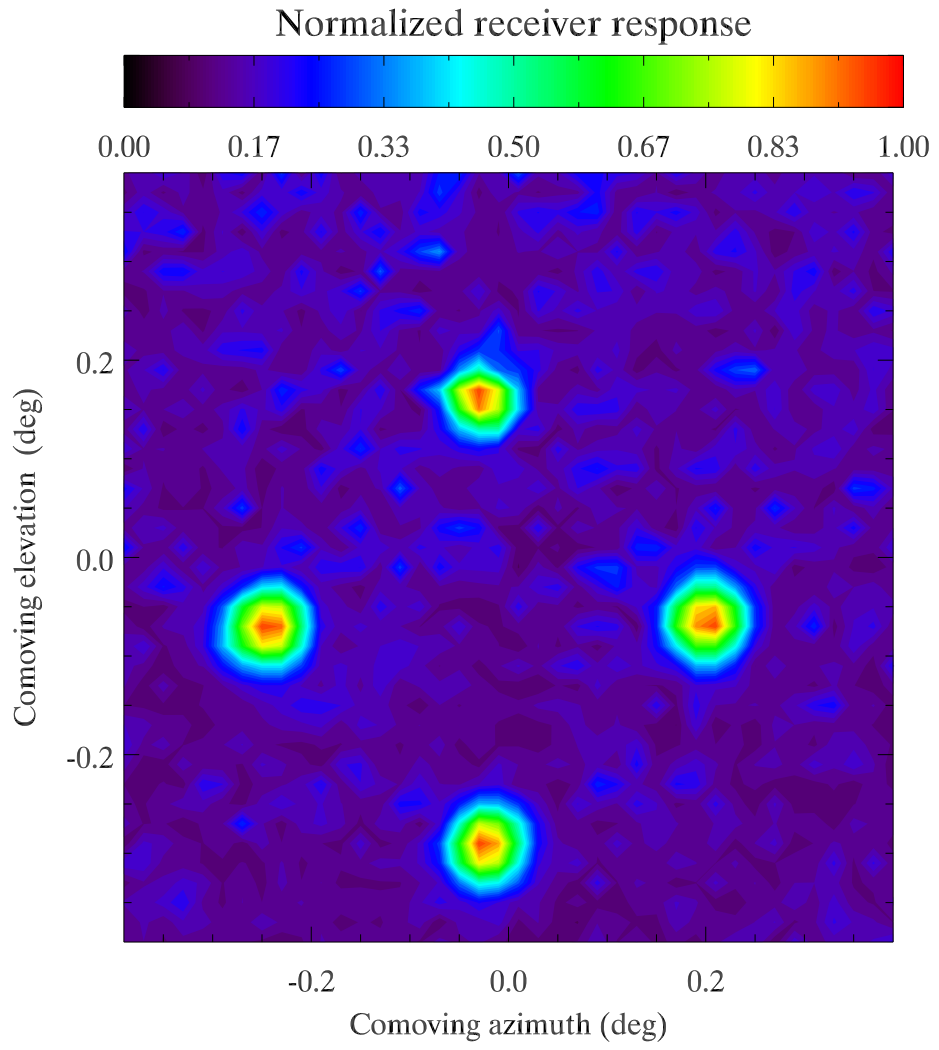

Figure 6.18: Mosaic of the measured beams. Different noise levels in each receiver map lead to apparent artefacts away from the beam centers. Receiver C which is notably noisy has been binned with three times the bin size, leading to the apparently smaller beam size. Only 6 min of observation go into this mosaic.

[7]For reminders, the effective observed temperature of a source of temperature $T_s$ viewed through a lossy medium of loss $L$ at a temperature $T_m$ is $T_o = LT_s + (1 - L)T_m$. The optical depth, $\tau$, describes the opacity of the atmosphere: $e^{-\tau}$ is then the transmission and $\epsilon = (1 - e^{-\tau})$ is the absorption. The optical depth is elevation dependant and goes as $\tau_0/\cos(el)$, where $\tau_0$ is the zenith optical depth.

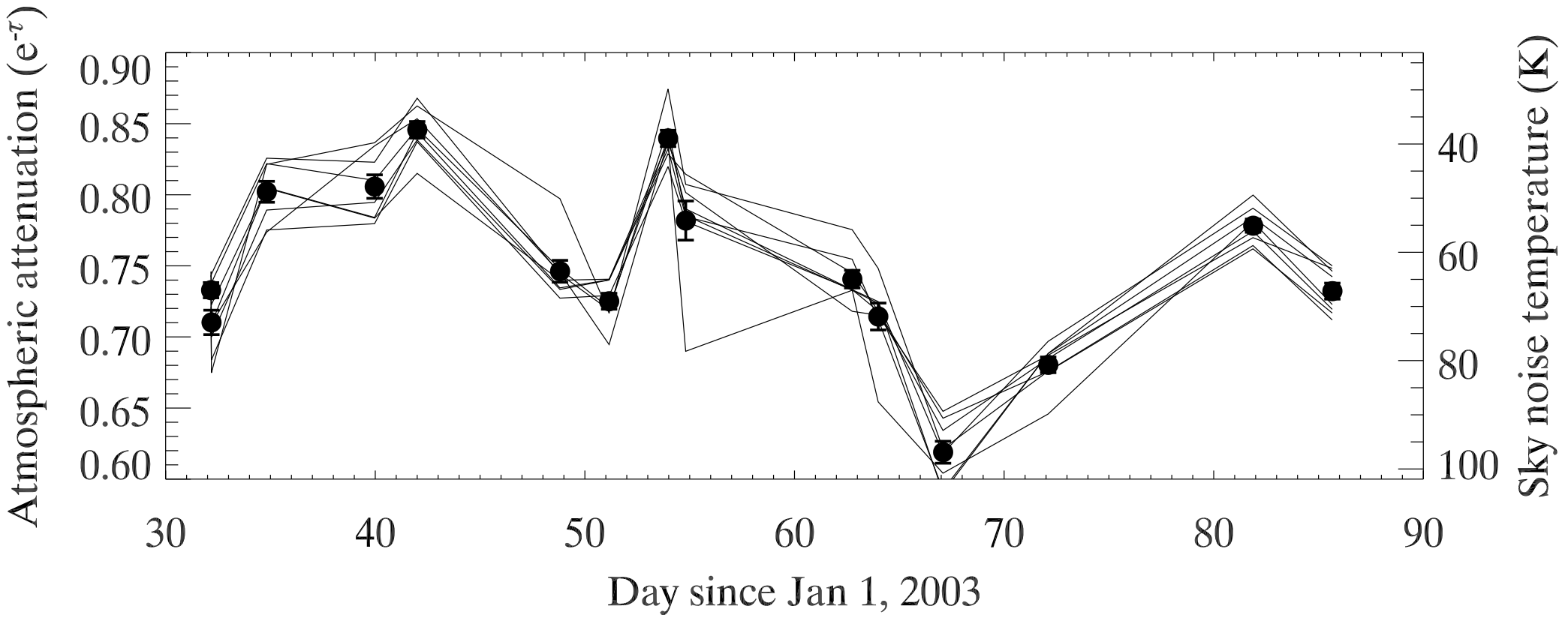

Figure 6.19: Atmospheric attenuation for the eight TP channels during the 15 Jupiter observations. As expected, the attenuations for the eight TP channels are identical. The attenuations are plotted at the zenith angle of the observation (from 20° to 70° elevation), so the zenith noise temperature is actually smaller. See text for a derivation of the formula for the loss. An attenuation of 0.8 at 40° is typical.

### IF temperature gain dependence

| | AD0 | AD1 | BD0 | BD1 | CD0 | CD1 | DD0 | DD1 |
|---|---|---|---|---|---|---|---|---|
| | Results from the **Jupiter observations** | | | | | | | |
| Gain (mV/K) | -1.49 | -2.66 | -2.24 | -1.61 | -2.74 | -1.47 | -4.75 | -4.01 |
| Gain stability (%) | 2.0 | 2.1 | 2.5 | 2.3 | 4.6 | 4.2 | 2.7 | 4.2 |
| $dG/dT_{IF}$ (%/K) | 0.72 | 0.69 | 0.35 | 0.42 | 0.73 | 0.53 | 0.70 | 0.76 |
| | Results from the **Sky-dip test** | | | | | | | |
| Noise temperature (K) | 79 | 107 | 64 | 62 | 114 | 78 | 70 | 65 |
| Gain (mV/K) | -1.56 | -2.95 | -2.43 | -1.66 | -2.93 | -1.49 | -5.07 | -4.3 |
| $dG/dT_{IF}$ (%/K) | 0.69 | 0.49 | 0.09 | 0.14 | 0.43 | 0.31 | 0.39 | 0.38 |

Table 6.8: *top*: Gain and IF temperature correction factor ($dG/dT_{IF}$) of the eight total power channels derived from **Jupiter observations**. The gains are normalized to $T_{IF} \simeq 33K$. The gain stability is the fractional gain change (RMS) for the 15 measurements.
*bottom*: Gain and receiver temperature of the eight total power channels derived from **sky-dips test**. The gains are normalized to $T_{IF} \simeq 33K$. The noise temperature is that of the telescope which includes of course the receiver's noise temperature as well as any added noise from the various windows and optical elements. Sky-dip data analyzed by M. Hedman [58].

During the first five and the last two Jupiter observations, the IF temperature was not regulated to its nominal value of 33° C. Because the most temperature sensitive devices are the IF amplifiers, there is a clear correlation between the IF temperature and the uncorrected total power gains. In order to derive the temperature-dependant gain correction, we first solve Equation 6.23 and 6.24 assuming no IF correction factor to find $G$. The IF temperature correction factor, $\alpha_{IF}$, is then derived by minimizing the $\chi^2$ to a mean of the quantity $\Big(G\ (1+\alpha_{IF})(T_{IF}-33)\Big)^2$ by varying $\alpha_{IF}$. The resulting fractional gain factors are given in Table 6.8.

### 6.5.2 Polarized Gains

Observations of Tau A are used as a secondary polarization-channel calibration method. This discussion follows closely the analysis by Matt Hedman [57]. The expected polarized flux from Tau A is derived from a combination of the WMAP data set and published literature results. The Tau A intensity, measured from the WMAP publicly available maps [93], peaks at 41.7 mK at W-band (at an effective frequency of 93.5 GHz [115]). The linearly polarized fraction is derived from various published values from 10 to 90 GHz, summarized in Figure 6.20.

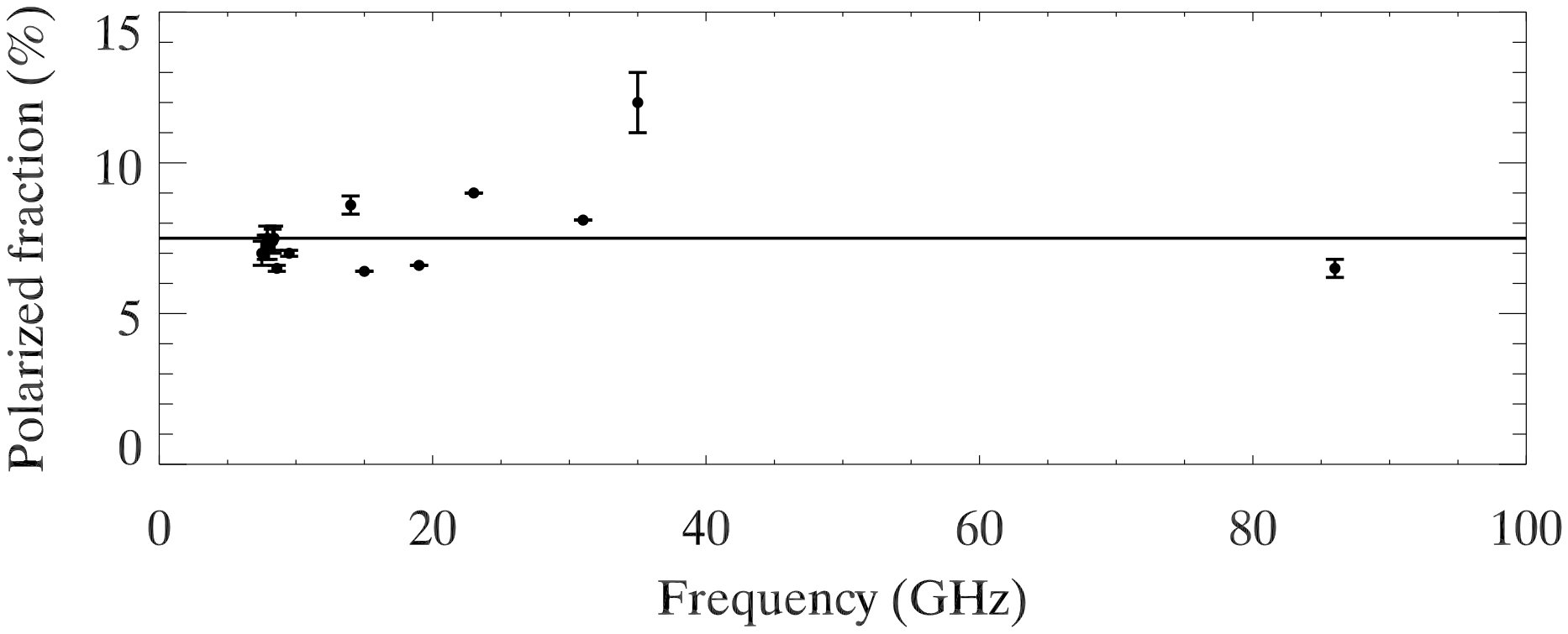

Figure 6.20: Published values of the average polarized fraction of Tau A as a function of frequency. The sources in increasing frequency order are [63, 64, 105, 17, 45, 104, 149, 74, 76, 103]. We adopt the mean of all the values, 7.5 ± 1%, for the polarized fraction at 90 GHz as there does not seem to be any strong frequency dependency.

For lack of better measurements, we assume the mean value that the polarized fraction is 7.5% ± 1%, which corresponds to the leading source of uncertainty (13%) in the polarized signal from Tau A. The predicted maximum polarized signal from Tau A for WMAP W-band channel is thus $3.1 \pm 0.4$mK

Tau A's spectral index is measured to be $-0.3$ [2]. To scale the apparent temperature (in thermodynamic CMB units) with the frequency of each receiver sub band, we apply a frequency correction factor, derived from the following relation:

$$\delta T_{CMB} = \frac{\mathrm{d}T}{\mathrm{d}I}\Big|_{CMB} \delta I = \delta I \Big(\frac{\mathrm{d}I}{\mathrm{d}T}\Big|_{CMB}\Big)^{-1} \propto \nu^{-2+\alpha}\frac{(e^x-1)^2}{e^x x^2}\ , \tag{6.25}$$

where $\alpha$ is the spectral index of the source (ie. $\delta I \propto \nu^{\alpha}$), and $x = h\nu/kT_{CMB}$. Using 87,

91.5, and 97 GHz as the effective frequencies of S0, S1, S2 yields the frequency correction factors, 1.15, 1.04, and 0.93 respectively. Changes in the exact frequency of each receiver sub-band only affect these factors at a sub-percent level.

The expected signal from Tau A also scales as the beam size. The beam sizes of the three sub-bands are $\sigma_{S0} = 0.0291°$, $\sigma_{S1} = 0.0270°$, and $\sigma_{S1} = 0.0254°$, measured using the same fitting algorithm as the one used to derive the beam widths of the total power channel using Jupiter (see Section 6.7). From the beam solid angles calculated using the formula $\Omega = 2\pi\sigma^2$, and the WMAP W-band beam solid angle of $0.206\times10^{-4}$ steradian [115], the beam size correction factors are 12.45, 14.44, and 16.42 for, S0, S1, and S2, respectively. The expected polarized signals from Tau A for CAPMAP's three polarization channels are therefore 44.8, 47.0, 47.8 mK, with a 15% uncertainty due to the polarized intensity and a 6% uncertainty due to the beam sizes.

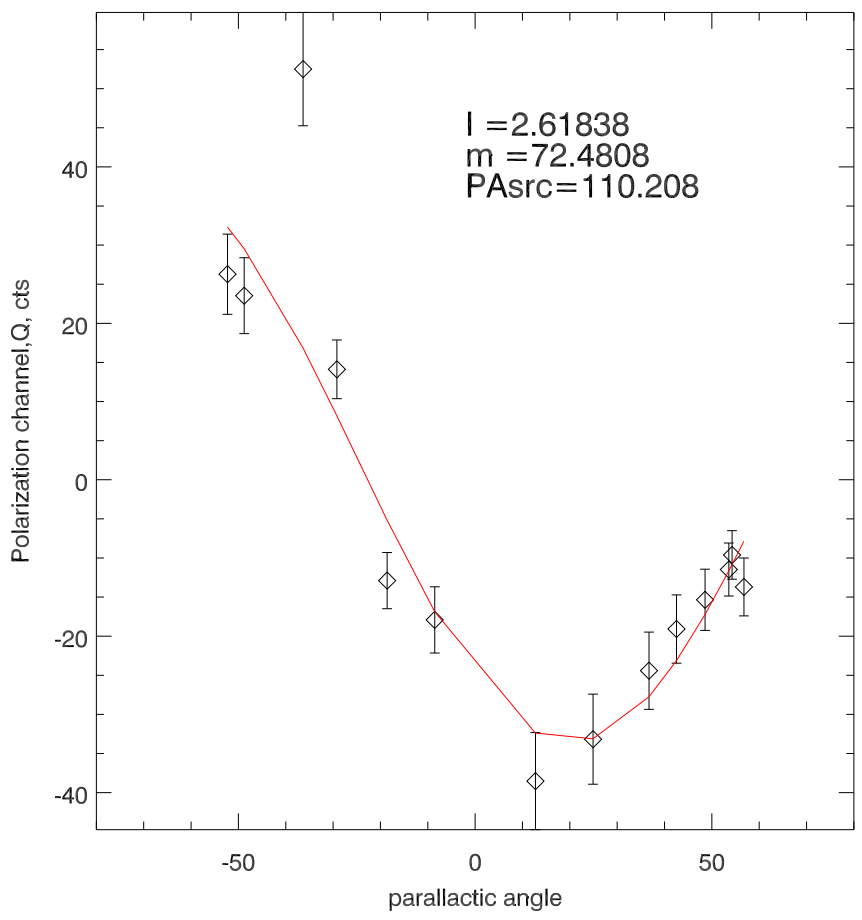

Figure 6.21: Measurement of the polarized signal from Tau A (in counts) as a function of parallactic angle and the best fit curve. The fit model is $I + \frac{m}{2}\cos((p - \phi_{source})/2)$, with the best parameters labeled on the plot. This measurement was made using the PIQUE receiver [56] installed on the Crawford Hill antenna. The measured position angle of Tau A agrees well with the published value of 155° [10] (see text). Given the PIQUE S0 channel gain of 1.8 counts/mK, the measured polarized amplitude of 40 mK also agrees with the expected signal derived from the WMAP numbers (see text).

In order to compare each measured Tau A signal with a correct estimate of the expected signal, an atmospheric correction factor is applied, to account for the variable loss though the atmosphere. The atmospheric correction factor is calculated from the total-power channels, using the total power receiver noise and gain calculated in Section 6.5.1. Finally, a correction factor is applied to account for the orientation of Tau A polarized direction relative to the receiver detection axis. The apparent variation of the signal from Tau A is proportional to $\cos((p - \phi_{source})/2)$ where $p$ is the parallactic angle[8] of the observation defined as $\sin p = \cos(lat)\sin(az)/\cos(dec)$ [108] and $\phi_{source}$ is the position angle of the source. The position angle of Tau A is fairly well measured to be 155° [10]. Given that the receiver's detection axis is rotated 45° from vertical, the polarized signal will be maximum when the parallactic angle is $155° - 45° = 110°$, so $\phi_{source} = 110°$ (see Figure 6.21).

For each of the analyzed observations (13 out of 45 total), the same processing is performed on the raw data as that described in Section 6.2.2. The amplitude of each measured signal is derived from the 2D-gaussian fit to the map of Tau A. The amplitudes are then adjusted for the clock phase and divided by the expected signal for each observation. The resulting estimates of the polarized gains are given in Table 6.9, along with the gains from the chopper plate tests from Section 3.3.2.

| Gain (mV/K) | As0 | As1 | As2 | Bs0 | Bs1 | Bs2 | Cs0 | Cs1 | Cs2 | Ds0 | Ds1 | Ds2 |
|---|---|---|---|---|---|---|---|---|---|---|---|---|
| Tau A | 27.0 | 16.7 | 15.4 | 14.1 | 5.6 | 8.5 | 19.7 | 9.0 | 10.7 | -24.5 | -22.6 | -15.2 |
| | ±1.4 | ±0.7 | ±0.9 | ± 1.2 | ±0.2 | ±0.5 | ±1.1 | ±0.4 | ±0.9 | ±1.6 | ±1.2 | ±1.0 |
| Chopper | 21.0 | 15.9 | 14.1 | 14.2 | 5.6 | 9.9 | 18.2 | 9.1 | 10.6 | -25.3 | -23.4 | -19.1 |
| plate | ±1.5 | ±1.0 | ±1.0 | ± 1.4 | ±0.4 | ±0.5 | ±1.5 | ±0.8 | ±0.6 | ±1.2 | ±1.6 | ±0.6 |

Table 6.9: Derived polarized gains (in mV/K) from the Tau A observations during the CAPMAP03 season. The values in the table are in thermodynamic units. Note that these gains are with the IF box at 40° C, and the LO at 18° C. Analysis by M Hedman.

[8] The parallactic angle is the angle on the sphere between the arc from the source to NCP and the source to zenith.

Chapter 7

# CMB Data and Conclusions

This chapter describes the processing of the primary CMB polarization data. The steps to convert the raw data to binned data and to maps are outlined. At this time, the data analysis is still underway so the details of the final likelihood analysis of the maps will be published in [8]. Some preliminary results concerning the polarized scan synchronous signals are detailed. All the data reduction presented here is done in IDL and the routines are available at [5].

## 7.1 Data Processing Outline

### 7.1.1 From Raw to Binned Data

Binning is done for each file of the scaz scan strategy (see Section 5.1.1). Short files (<10 seconds), as well as files where the telescope motion stopped or became erratic are discarded. This leaves 1359 30-minutes files for the nominal CAPMAP03 season (Feb 2 to April 6, 2003) (see Chapter 5). The steps to producing the binned data follow.

**Clean.** The data from one file is read into a structure (see Appendix G). The encoder time-series is shifted two samples forward with respect to the data time-series (Section 6.3.2). The samples where an error flag is on or where the DAQ stopped acquiring are blanked, out along with 100 neighboring samples. Large spikes in the polarization channels (due to an insect on the window, a plane, or a bird in the beam) are blanked with 10 neighboring samples. These glitches happen evenly throughout the season, and remove less than 1% of the total data.

**Get the turnaround points.** Find the extremum points of the azimuth motion and keep those indices in memory.

**Convert in sky coordinates.** Convert the azimuth and elevation of the central ray into sky coordinates of each receiver by

$$\begin{aligned} X_0 &= (Az_{tel} - 360)\cos(40.3149) + p_x \\ Y_0 &= El_{tel} - 40.3149 + p_y \end{aligned} \tag{7.1}$$

where $Az_{tel}$ and $El_{tel}$ are the position of the central ray (where the pointing corrections at NCP from Tables 6.4 and 6.5 have been accounted for) and $p_x$ and $p_y$ are the relative positions of each receiver with respect to the central ray (from Table 6.7).

**Housekeeping channels.** Convert all the housekeeping channels into physical units (see Appendix G.

**Rebin** The data are then binned as follows. Each scan (either left-going or right-going) is gridded into 24 azimuth bins (0.05° wide which yields ~25 samples per azimuth bin). The samples within each bin are averaged and the 24 averages are saved along with the time of the center of the scan, the number of elements in each bin, the mean, slope, and quadratic fit vs azimuth to the data in that scan, and a left/right-going flag. This binning is done on each of the ~450 scans per file and is saved as the az-binned data. This step is repeated four times to record the az-binned data with either nothing removed, or with the offset, the slope or the quadratic removed. The offsets, slopes, and quadratics are calculated from the unbinned data in each scan. A plot of the offsets, slopes, and quadratics throughout the season is shown in Figure 7.2. A time-binned data is produced in parallel which contains the averages of the data for each scan: the time, all the data channels, the working housekeeping channels, and the interpolated weather-station data.

Once the binning is performed on all the 1359 files, 21 arrays are saved to disk. One time-binned data array, 69 columns by 593996 rows, where 593996 is the total number of scans (160 Mb of memory), and 20 az-binned data arrays (one for each channel) 57 columns by 593996 rows (132 Mb of memory each). These arrays are saved as IDL ".sav" files and can be loaded in memory in a few seconds.

### 7.1.2 From binned data to a map

**Corrections.** Three corrections are applied to each channel: the clock gain correction factor (Section 5.2.2), the IF temperature correction factor (Section 6.5.1), and a factor to account for the variable loss through the atmosphere (Section 6.5.1). These correction factors are applied on each scan.

**Calibrate.** Convert the data from mV to thermodynamic temperature units using the total power and polarimetry channel gains (see Section 6.5).

**Selection and data quality.** The criteria to select the astronomically interesting data while discarding the data affected by atmospheric fluctuations is still being fine-tuned. The selection criteria are based on the total-power channels in order to be unbiased with respect to the primary polarization data. The selection cut discards between 30 and 50% of the data, leaving between 270 and 370 hours of polarization data which go into the final analysis. This is in agreement with the expected amount of data acquired for the CAPMAP03 season (Chapter 2).

**Deployment periods.** The whole season is broken up into shorter segments. Historically, these were periods separated by a tarping and untarping of the PIQUE telescope during bad weather. Here, the season is broken into 19 deployment periods (see Table 7.1), separated by periods of obvious bad weather.

**Collapse data.** This step takes the az-binned data and collapses every contiguous 100 scans of 24 bins into a single row of 24 means, 24 standard errors, and 24 hits. This forms the collapsed data, a 73 columns by 5939 rows array. The function of this step is to further compress the data but most importantly to provide a correct estimate of the errors. The collapsing action can be done on any number of scans. Collapsing a given channel over all the scans provides a measurement of any remaining "azimuthal structure".

**Map.** The mapping starts from the collapsed data. The Healpix [42] pixelisation scheme and routines[1] are used here, but different pixelisation methods are used by other CAPMAP team members. Note that this step is still preliminary. First convert the time, and co-moving coordinates ($X_0$, $Y_0$) of each collapsed data sample into a right ascension (Ra), declination (Dec), and parallactic angle (Pa), for the particular receiver of interest. Given a Healpix pixelisation with Nside = 2048 (which yields a $1.7'$ pixel on a side), Healpix converts the list of Ra and Dec from each sample to a list of pixel indices for the samples. The simplest way to form a map is to co-add all the samples contained in a given pixel. At this point, the Pa is accounted for by simply forming East or West maps (when the receiver is either East of West of NCP). An example of the maps produced is shown in Figure 7.1.

[1] `http://www.eso.org/science/healpix/`

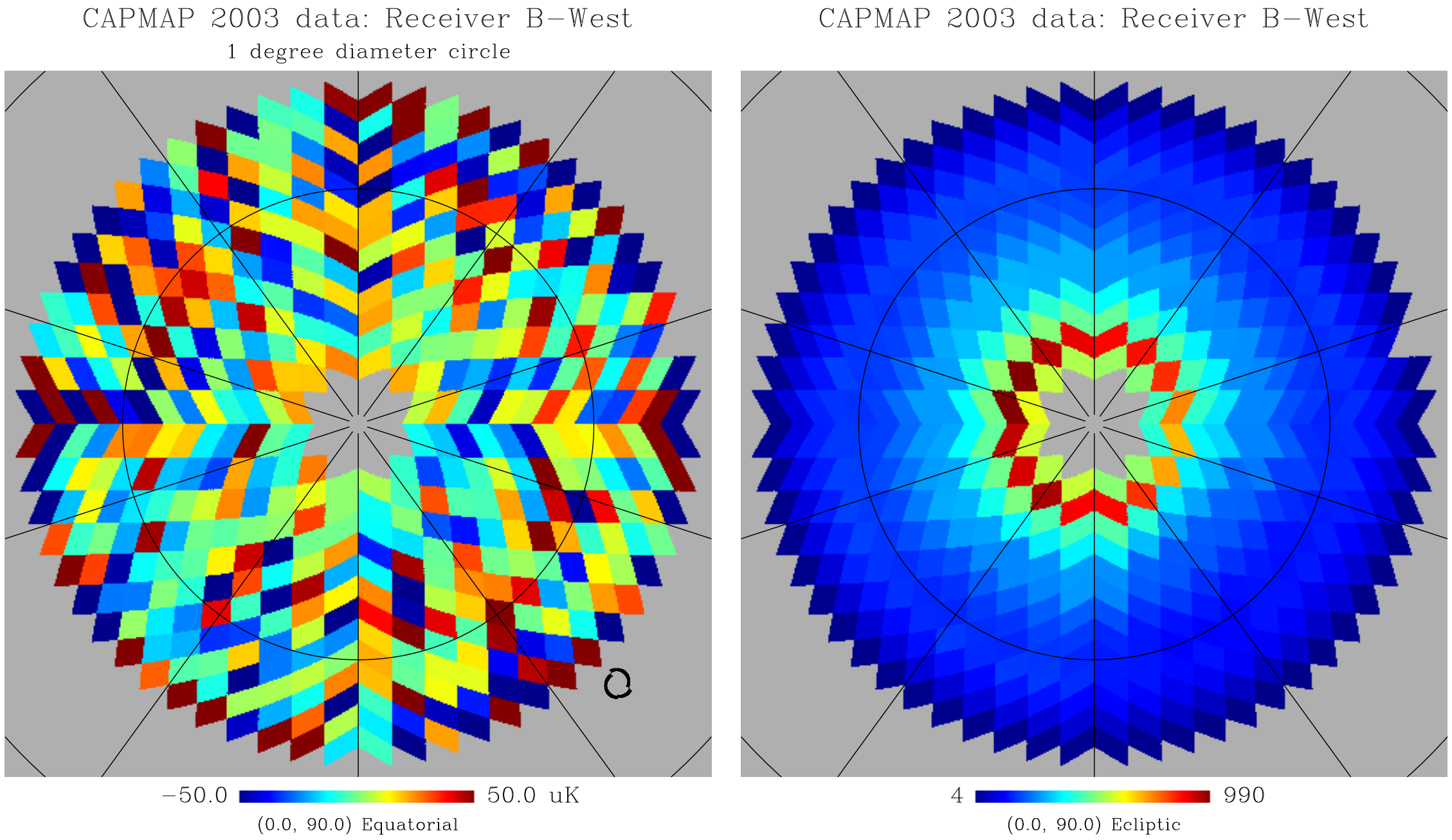

Figure 7.1: Example map produced with the Healpix pixelisation (Nside = 1024) using ∼450 hours of all the data with receiver B, when it is located west of the NCP. Note that this corresponds approximately to 1/8 of the full CAPMAP03 data set. Gnomic view centered on the NCP in equatorial coordinates. The inner circle is 0.5° radius. This pixelisation makes beam-sized pixels. The small circle in the lower corner represents the beam. The left map shows the data while the right map gives the number of 100-scan collapsed-data- sample hits per pixel (one hit therefore equals $100 \times 1/24 \times 4$ seconds = 16 seconds). In the left map, the offset, slope, quadratic and universal structure have been removed for each scan. For these maps , the error per pixel is between 20 $\mu$Kand 50 $\mu$K, in agreement with the predicted CAPMAP03 $\sigma_{pix}$ (in Table 2.1 $\sigma_{pix} = 19.1$ $\mu$Kwith all the receivers).

## 7.2 Scan Synchronous Structure

As is already known, the absolute output of the polarimeter is not stable enough on long time scales (periods greater than a few minutes) which is why the telescope is scanned across the sky. In order to discard the polarimeter's finite sensitivity to total-power fluctuations (see Section 3.3.3), the offset is removed for every telescope azimuthal scan. However, we find that it is also necessary to remove a slope and a quadratic after every azimuth scan to ensure that the noise across the scan is white. The offset, slopes, and quadratics removed and averaged over a deployment period are plotted in Figure 7.2. Once the offset, slope, and quadratics are removed, there is a small remaining azimuthal structure (<<

1 mK), which we call the "universal structure" (US). Because it is fixed in azimuth, it cannot be attributed to a signal on the sky, and is most likely caused by a sidelobe of the secondary mirror missing the primary and hitting the ground. Further details on the universal structure will be published in [8].

## 7.3 Conclusions

The results on the polarization of the CMB for the CAPMAP03 season will be presented in [8]. The final sensitivity after preliminary cuts yields an experimental weight (senfac) of $\sigma_E = 0.7\ \mu$K. Given this sensitivity, we should expect a detection of the CMB polarization E-mode, as predicted in Section 2.1.

At the same time, the full CAPMAP instrument, with 16 receivers, including four Q-band receivers, is being fielded. Observations with 12 receivers were made during the winter 2003-2004, and all 16 receivers will observe in 2004-2005. The additional data will be combined with the current data in a future analysis.

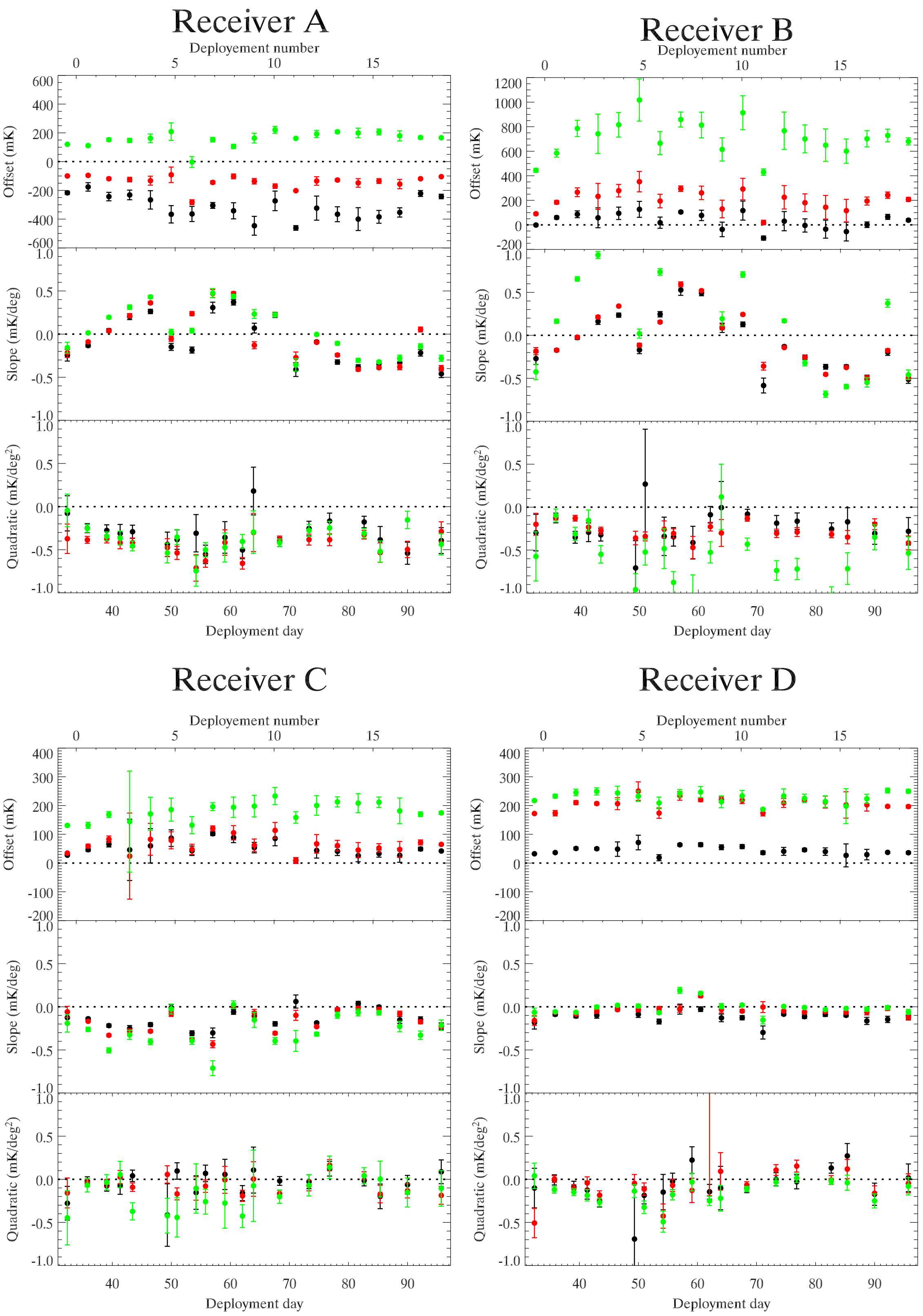

Figure 7.2: Plot of the average offset, slope, and quadratic for each of the 19 deployment periods. S0 is black; S1 is red; S2 is green. The values are summarized in Table 7.1. The offset are expected to be non-zero. A weak selection cut on 6$\sigma$ of the points with slopes > 6$\sigma$ of the average slope has been applied (removing less that 1% of the data.)

| Deployment index | Date | Hours | Offset (mK) | | | Slope (mK/°) | | | Quadratic (mK/$°^2$) | | |
|---|---|---|---|---|---|---|---|---|---|---|---|
| | | | AS0 | AS1 | AS2 | AS0 | AS1 | AS2 | AS0 | AS1 | AS2 |
| 0 | 32.4 | 2.1 | -217.1 | -99.9 | 120.2 | -0.25 | -0.22 | -0.16 | -0.08 | -0.37 | -0.37 |
| 1 | 35.8 | 32.6 | -176.1 | -95.9 | 111.2 | -0.13 | -0.09 | 0.01 | -0.25 | -0.39 | -0.39 |
| 2 | 39.1 | 36.2 | -243.4 | -119.0 | 152.3 | 0.04 | 0.04 | 0.19 | -0.28 | -0.38 | -0.38 |
| 3 | 41.3 | 13.2 | -231.8 | -125.3 | 147.2 | 0.20 | 0.21 | 0.31 | -0.31 | -0.42 | -0.42 |
| 4 | 43.4 | 39.2 | -266.3 | -132.8 | 162.1 | 0.26 | 0.36 | 0.43 | -0.29 | -0.42 | -0.42 |
| 5 | 49.3 | 21.3 | -366.6 | -91.8 | 208.4 | -0.15 | -0.05 | 0.02 | -0.44 | -0.48 | -0.48 |
| 6 | 51.0 | 32.3 | -364.3 | -282.3 | -3.1 | -0.18 | 0.24 | 0.04 | -0.38 | -0.54 | -0.54 |
| 7 | 54.2 | 7.4 | -304.9 | -144.8 | 152.8 | 0.31 | 0.47 | 0.48 | -0.31 | -0.71 | -0.71 |
| 8 | 55.8 | 35.4 | -342.2 | -102.8 | 105.3 | 0.37 | 0.47 | 0.44 | -0.55 | -0.63 | -0.63 |
| 9 | 59.1 | 16.5 | -446.3 | -136.9 | 163.4 | 0.07 | -0.13 | 0.23 | -0.36 | -0.42 | -0.42 |
| 10 | 62.0 | 45.2 | -273.1 | -170.7 | 220.8 | 0.22 | 0.23 | 0.22 | -0.50 | -0.66 | -0.66 |
| 11 | 63.9 | 5.2 | -461.2 | -202.3 | 161.4 | -0.41 | -0.27 | -0.36 | 0.18 | -0.30 | -0.30 |
| 12 | 68.3 | 133.6 | -323.6 | -134.8 | 191.7 | -0.09 | -0.10 | -0.00 | -0.41 | -0.38 | -0.38 |
| 13 | 73.3 | 57.6 | -366.2 | -128.2 | 207.3 | -0.32 | -0.24 | -0.11 | -0.25 | -0.38 | -0.38 |
| 14 | 76.8 | 50.6 | -400.0 | -148.2 | 199.5 | -0.38 | -0.41 | -0.30 | -0.17 | -0.38 | -0.38 |
| 15 | 82.6 | 91.0 | -384.2 | -135.2 | 206.0 | -0.34 | -0.39 | -0.32 | -0.18 | -0.33 | -0.33 |
| 16 | 85.4 | 14.6 | -353.5 | -156.7 | 179.3 | -0.33 | -0.38 | -0.28 | -0.39 | -0.53 | -0.53 |
| 17 | 90.0 | 14.1 | -221.8 | -119.0 | 167.5 | -0.22 | 0.05 | -0.14 | -0.54 | -0.50 | -0.50 |
| 18 | 95.7 | 9.9 | -242.4 | -104.7 | 166.0 | -0.46 | -0.40 | -0.28 | -0.39 | -0.29 | -0.29 |
| | | | BS0 | BS1 | BS2 | BS0 | BS1 | BS2 | BS0 | BS1 | BS2 |
| 0 | | | -1.5 | 90.2 | 444.2 | -0.27 | -0.18 | -0.43 | -0.30 | -0.20 | -0.20 |
| 1 | | | 59.7 | 183.7 | 584.7 | -0.17 | -0.17 | 0.16 | -0.13 | -0.12 | -0.12 |
| 2 | | | 86.4 | 266.0 | 785.5 | -0.03 | -0.02 | 0.66 | -0.35 | -0.13 | -0.13 |
| 3 | | | 57.7 | 232.0 | 741.9 | 0.16 | 0.21 | 0.94 | -0.29 | -0.23 | -0.23 |
| 4 | | | 94.3 | 278.6 | 815.5 | 0.23 | 0.34 | 1.28 | -0.32 | -0.28 | -0.28 |
| 5 | | | 126.4 | 352.1 | 1017.3 | -0.17 | -0.11 | 0.02 | -0.71 | -0.36 | -0.36 |
| 6 | | | 17.9 | 194.1 | 665.5 | 0.25 | 0.15 | 0.74 | 0.27 | -0.34 | -0.34 |
| 7 | | | 104.8 | 294.7 | 858.1 | 0.53 | 0.59 | 1.63 | -0.34 | -0.26 | -0.26 |
| 8 | | | 77.0 | 260.6 | 812.9 | 0.49 | 0.52 | 1.62 | -0.35 | -0.31 | -0.31 |
| 9 | | | -36.8 | 129.4 | 614.4 | 0.09 | 0.10 | 0.19 | -0.41 | -0.47 | -0.47 |
| 10 | | | 116.3 | 291.8 | 913.9 | 0.13 | 0.24 | 0.71 | -0.09 | -0.23 | -0.23 |
| 11 | | | -107.7 | 19.2 | 429.6 | -0.58 | -0.36 | -1.12 | -0.01 | -0.30 | -0.30 |
| 12 | | | 30.1 | 224.7 | 766.4 | -0.13 | -0.14 | 0.17 | -0.08 | -0.13 | -0.13 |
| 13 | | | -4.7 | 180.2 | 700.2 | -0.26 | -0.25 | -0.32 | -0.19 | -0.31 | -0.31 |
| 14 | | | -35.0 | 143.5 | 649.2 | -0.37 | -0.45 | -0.69 | -0.16 | -0.29 | -0.29 |
| 15 | | | -54.9 | 115.9 | 600.8 | -0.36 | -0.37 | -0.60 | -0.25 | -0.32 | -0.32 |
| 16 | | | 1.3 | 194.1 | 701.6 | -0.51 | -0.49 | -0.55 | -0.17 | -0.35 | -0.35 |
| 17 | | | 64.9 | 241.8 | 728.2 | -0.20 | -0.18 | 0.37 | -0.31 | -0.20 | -0.20 |
| 18 | | | 38.2 | 207.1 | 680.2 | -0.51 | -0.49 | -0.46 | -0.28 | -0.42 | -0.42 |

Table 7.1: Summary of the average offset, slope, and quadratic for each deployment period. The errors for each parameters are visible in Figure 7.2. On average the error on the offset is 20 mK, on the slope 0.1 mK and on the quadratic 0.1 mK. It is likely that these uncertainties are dominated by real variations during the period, rather than by random scatter on the values.

| Deployment | | | Offset (mK) | | | Slope (mK/°) | | | Quadratic (mK/°$^2$) | | |
|---|---|---|---|---|---|---|---|---|---|---|---|
| | | | CS0 | CS1 | CS2 | CS0 | CS1 | CS2 | CS0 | CS1 | CS2 |
| 0 | | | 27.4 | 34.5 | 130.8 | -0.13 | -0.06 | -0.19 | -0.28 | -0.16 | -0.16 |
| 1 | | | 46.3 | 58.9 | 131.3 | -0.14 | -0.17 | -0.26 | -0.02 | -0.04 | -0.04 |
| 2 | | | 65.2 | 82.8 | 168.7 | -0.22 | -0.33 | -0.51 | -0.08 | -0.04 | -0.04 |
| 3 | | | 45.8 | 24.1 | 144.1 | -0.25 | -0.28 | -0.33 | -0.07 | 0.03 | 0.03 |
| 4 | | | 59.7 | 82.6 | 171.1 | -0.21 | -0.28 | -0.41 | 0.04 | -0.09 | -0.09 |
| 5 | | | 86.5 | 79.0 | 185.7 | -0.01 | -0.08 | -0.03 | -0.41 | 0.06 | 0.06 |
| 6 | | | 42.7 | 47.5 | 131.7 | -0.31 | -0.38 | -0.39 | 0.10 | -0.17 | -0.17 |
| 7 | | | 102.0 | 120.3 | 196.0 | -0.30 | -0.43 | -0.71 | -0.15 | -0.11 | -0.11 |
| 8 | | | 88.6 | 105.3 | 194.0 | -0.06 | 0.01 | 0.02 | 0.07 | -0.08 | -0.08 |
| 9 | | | 53.6 | 61.8 | 197.8 | -0.11 | -0.07 | -0.15 | 0.06 | -0.01 | -0.01 |
| 10 | | | 85.3 | 113.6 | 232.9 | -0.20 | -0.31 | -0.40 | -0.16 | -0.19 | -0.19 |
| 11 | | | 9.2 | 8.5 | 158.3 | 0.06 | -0.10 | -0.40 | 0.11 | 0.00 | 0.00 |
| 12 | | | 43.7 | 66.7 | 200.2 | -0.19 | -0.23 | -0.32 | -0.02 | -0.17 | -0.17 |
| 13 | | | 40.8 | 60.5 | 212.8 | -0.04 | -0.03 | -0.10 | -0.03 | -0.07 | -0.07 |
| 14 | | | 25.7 | 45.0 | 208.2 | 0.04 | -0.02 | -0.07 | 0.12 | 0.16 | 0.16 |
| 15 | | | 33.0 | 52.6 | 211.4 | -0.00 | -0.03 | -0.07 | -0.01 | 0.04 | 0.04 |
| 16 | | | 26.7 | 47.2 | 181.4 | -0.15 | -0.08 | -0.23 | -0.20 | -0.17 | -0.17 |
| 17 | | | 48.9 | 71.7 | 169.8 | -0.15 | -0.17 | -0.33 | -0.06 | -0.13 | -0.13 |
| 18 | | | 41.9 | 64.9 | 174.3 | -0.23 | -0.22 | -0.21 | 0.09 | -0.19 | -0.19 |
| | | | DS0 | DS1 | DS2 | DS0 | DS1 | DS2 | DS0 | DS1 | DS2 |
| 0 | | | 32.2 | 172.2 | 217.1 | -0.18 | -0.16 | -0.06 | -0.10 | -0.51 | -0.51 |
| 1 | | | 36.4 | 173.3 | 232.5 | -0.09 | -0.06 | -0.06 | -0.01 | 0.00 | 0.00 |
| 2 | | | 50.8 | 210.2 | 245.3 | -0.11 | -0.09 | -0.06 | -0.09 | -0.11 | -0.11 |
| 3 | | | 49.8 | 206.9 | 249.2 | -0.10 | -0.05 | -0.00 | -0.13 | -0.04 | -0.04 |
| 4 | | | 48.5 | 206.5 | 243.8 | -0.02 | -0.03 | 0.02 | -0.25 | -0.18 | -0.18 |
| 5 | | | 71.4 | 250.2 | 232.4 | -0.09 | -0.03 | 0.01 | -0.69 | -0.05 | -0.05 |
| 6 | | | 18.6 | 173.7 | 208.9 | -0.17 | -0.02 | -0.07 | -0.19 | -0.11 | -0.11 |
| 7 | | | 63.5 | 233.8 | 242.3 | -0.03 | -0.02 | 0.19 | -0.15 | -0.43 | -0.43 |
| 8 | | | 63.8 | 220.3 | 247.0 | -0.03 | 0.13 | 0.16 | -0.02 | -0.07 | -0.07 |
| 9 | | | 54.9 | 220.4 | 213.1 | -0.13 | -0.03 | 0.01 | 0.22 | -0.13 | -0.13 |
| 10 | | | 57.3 | 218.0 | 234.0 | -0.13 | -0.05 | 0.02 | -0.14 | 16.26 | 16.26 |
| 11 | | | 36.0 | 173.5 | 186.8 | -0.30 | -0.00 | -0.15 | -0.10 | 0.09 | 0.09 |
| 12 | | | 40.6 | 209.0 | 233.3 | -0.09 | -0.05 | 0.00 | -0.11 | -0.07 | -0.07 |
| 13 | | | 45.9 | 220.0 | 224.8 | -0.11 | -0.06 | -0.01 | 0.00 | 0.11 | 0.11 |
| 14 | | | 39.7 | 212.1 | 213.9 | -0.09 | -0.06 | -0.03 | -0.03 | 0.15 | 0.15 |
| 15 | | | 26.6 | 202.0 | 197.9 | -0.10 | -0.07 | -0.02 | 0.13 | -0.01 | -0.01 |
| 16 | | | 29.3 | 202.4 | 223.6 | -0.16 | -0.06 | -0.03 | 0.27 | 0.12 | 0.12 |
| 17 | | | 37.1 | 197.1 | 252.1 | -0.15 | -0.02 | -0.01 | -0.17 | -0.16 | -0.16 |
| 18 | | | 35.9 | 196.5 | 249.7 | -0.10 | -0.12 | -0.06 | 0.02 | -0.05 | -0.05 |

Table 7.2: Average offset, slope and quadratic for each deployment period. The errors for each parameters are plotted in Figure 7.2. On average the errors are 20 mK, 0.1 mK, and 0.1 mK for the offset, the slope and the quadratic respectively. These uncertainties are dominated by real variations during the period, rather than by random scatter on the values.

# Appendix A

# Antenna Pointing Guide

## A.1 Optical Pointing Observation Procedure

Listed here are the steps necessary to obtain measurements of the optical pointing errors using stars. This optical pointing procedure should only be used as a secondary pointing method, when radio sources are not available.

1. Choose a large set of stars, bright enough to be easily detected in your optical camera, with low proper motion, and well distributed around the whole sky. The simple fact that higher elevations span less solid angle tends to bias the star's distribution towards lower elevations, which tends to degrade the fit to the elevation pointing errors. Note the 88% of the stars observed for this seasons' optical pointing had elevation less that 60°. Such a catalog was compiled by Greg Wright and is located in the OBS computer in the directory */obs/obs/lib*.

2. Verify that the OBS computer time (displayed in AMON) is within one second of the GPS time on the two large red displays and on the "Odetics" receiver). If not, you will not be able to point the telescope to any star. It has also happened that the "Odetics" GPS receiver itself stopped tracking time, which does not cause a difference between OBS time and GPS time. You can verify that the GPS time is synched with a network time server such as `http://tycho.usno.navy.mil/what.html`

3. Open the optical telescope's window and verify it is clean and not covered in dew. Turn on the power to the CCD camera, the TV screen, and the digital crosshair in the control room.

4. Set the current telescope pointing correction to zero. Type `-> PNTZ`

5. Check that the atmospheric refraction values are correctly set (in AMON). Set to the current pressure, temperature, and relative humidity for an optical measurement or to the standard value (P=1000 mb, T=0°C, Rh=0%) for a radio measurement. The current atmospheric values can be retrieved on `http://www.weatherunderground.com`. Use the command `-> re` to change the refraction values.

6. Type the command `-> pd nf:"filename"` to open a new file called "filename" to record the star pointing data.

7. Move the telescope to a star in the catalog. The star should be visible in the TV screen of the CCD camera. The field of view is approximately 0.1° × 0.1°. Type the command `-> cp n:"133" cn:"stars'"` to point the telescope to star number 133 in the "stars" catalog. Use the commands `-> a w d x` to nudge the telescope up, right, down, or left respectively by 0.001°.

8. After having found the star in the CCD screen and centered it on the digital crosshair, type `-> pd rec:` to record this star's pointing errors. This records the telescope azimuth, elevation, and azimuth and elevation errors, as defined in Equation 6.12.

9. Repeat steps 7 and 8 until enough stars pointing errors have been recorded.

## A.2 Recommendations for improved source observation

Here are set of suggestions to improve the absolute antenna pointing for a future observing season.

To improve the radio beam calibration (for total power and polarization calibration, array and telescope pointing, and beam size), general rules are:

- The accuracy of the pointing solution –much more than the gain or beam size measurement– is very dependent on the sky coverage of sources.

- Sources observed on cloudy days are useless for analysis, and will most likely be discarded.

- Sources with a map S/N less than 3 are useless and have been discarded in the past for analysis. For example, the best total power last year had a 0.5K scatter on one minute timescale. A standard scan on Cas A (`->ra casa: azr:0.3,0.3,0.01`) gave a 2 second integration on beam sized pixel but made no detection of Cas A at W-band.

Only a few scans (3 times the above scan with half the elevation step size) with 20 seconds integration per pixel produced a weak detection.

- Unless observed with a high S/N (see suggestions below), extended sources such as Cas A or Tau A are useless for pointing information .

Here is a tentative specific calibration plan for the CAPMAP04 (second observing season):

1. After the deployment of each successive dewar, assuming all systems (IF and LO temperature control, dewar temperature, encoder and housekeeping channels) are running nominally, perform one day of "full sky beam mapping" on Jupiter. That is map a square region 0.8° ×0.8° with 0.01° elevation step size centered on that array at 20 mn intervals, throughout the full range where Jupiter is visible (`->ra ss:'ju' azr:0.4,0.4,0.01`). Each observation should last approximately 10 mn which should give around 30 observations. This will provide a good basis for the pointing solution and the beams. A similar full sky observation of Tau A can also be done at the same time, staggered with Jupiter. Because this creates an even coverage of Jupiter in the sky, the time and sky position of further observations are not as important for the pointing solution.

2. After this first set of observation with each dewar, only a "crude" observation of Jupiter with the whole 16 receiver array is necessary every 3 to 4 days or on order of 15 total useable observations for the season. This cruder observation will be used mainly for total power gain calibration. If the furthest receiver is placed 18" away from the focal point, it should point 0.75° away form the array center in the sky. The whole array could then be mapped by observing a 1.6° × 1.6° region with a 0.02° elevation step size (`->ra ss:'ju' azr:0.8,0.8,0.02`). Such a map would take approximately 20 mn. It will be necessary to verify that the edge receivers are not truncated in which case the mapped region can be increased.

3. In order to fill in the gaps where pointing information has not been determined with Jupiter, I suggest observing another source approximately 30 times at different times of day to fill in the sky coverage. However, this source should only be observed with a single receiver. First because the relative position of the receivers in the array is already well known and does not change, and second because this source is usually weaker and requires deeper integration to provide useable pointing information. This secondary source should therefore be observed with only the best receiver. Cas A or Orion are appropriate targets and should both be investigated early in the season.

Based on observations from the 2003 season, a clear detection of these sources require a double pass on a 0.26° ×0.26° square region with a 0.01° elevation step size centered on the given receiver (`->ra ss:'ju' azr:0.13,0.13,0.005`). This should only take a few minutes. It will be necessary to verify that the source was indeed detected to tune the necessary integration time.

4. To ensure an even sky coverage for pointing, I suggest generating a hard copy plot of the daily path of Jupiter and Cas A and having each successful observation of the source at a given azimuth and elevation be noted on the plot. This will make poorly covered region immediately evident.

5. Finally, a "crude" map of Tau A every 3 days can be performed on the whole 16 receiver array to calibrate the polarization channels (`->ra taua:  azr:0.8,0.8,0.02`)

6. Because of the initial first day "full sky beam map", large time gaps in the observation are not as critical and would only affect the polarization gain calibration.

7. For the "crude" observation, I strongly advise against a separate measurement of each sub array or each receiver individually, as varying atmospheric condition between each observation would introduce systematic differences between each measurement. Although some time is wasted in the center of the array where there is no receiver, a single longer, wider scan is preferable.

8. If one is interested in looking for new sources or scanning "interesting" regions of the sky, I strongly advise making a smaller map with the single best receiver rather than trying to map the whole array.

# Appendix B

# MMIC HEMT body preparation

This appendix describes the procedure for building the bodies of the MMIC HEMT amplifiers used in the CAPMAP experiment. The part numbers, company names, contact information, and specific specific steps are provided.

### 1) Machining the bodies

A MMIC LNA body consists of three pieces. Two brass pieces which enclose the microwave amplifier and provide the waveguide for input and output, and 1 aluminum cap. Bill Dix, in the Princeton machine shop, has been building these bodies and has the drawings for these pieces. He is aware of the special care that must be taken during machining (ie. go slowly around the waveguide bends to prevent deformation, keep inner and flange surfaces as smooth as possible). The bodies are machined in groups of six.

### 2) Gathering additional parts

Various other pieces are needed to finish the body. Some of these items have long lead times and should be ordered early.

**Pins** These bodies require 2 types of alignment pins.

- Large inner alignment pins (2 per body): These pins attach to the 2 inner surfaces. They are cylinders 3/32" in diameter and 0.3" long. Bill Dix machines these. Currently, they have tapered ends, but we think hemispherical ends would be preferable to ensure a smooth fitting when putting the 2 pieces together.
- Small outer alignment pins (4 per body): These pins attach to the diagonal holes of the waveguide mating flange. We order them from Aerowave (P/N: 00-0230, 30$ for 100, or free sample if you talk to Leon, see Aerowave catalog, page 36)

**Pin insertion tool** Bill made a special tool to insert each of these pins correctly into the bodies. Get it.

**Connectors** There are two types of connectors associated with the bodies. We currently order these from Glenair (`http://www.glenair.com`, phone: (818)-247-6000)

- Inner socket connector (1 per body): This is the socket connector that brings the power to the MMIC amplifier. We let JPL do the attachment because it requires bonding, (P/N: MWSL-5P-5C4-.250, 5.81$ each, 6 to 8 weeks delivery.)
- Outer pin connector (1 per body): This connector is not absolutely necessary for building the bodies themselves. It is the connector that plugs into the previous one. It is useful however to just order it at the same time. It comes with 18" of color-coded copper wires, (P/N: MWSL-5S-6E5-18, 8.49$ each, 6 to 8 weeks delivery.)

### 3) Numbering the bodies

Both parts of each body must be numbered. It is best to scratch a serial number on a non-critical surface such as the one opposite of the power bias hole.

### 4) Deburing the bodies

Because brass is a pretty soft metal, it must be debured in order to remove excess material which was partly removed from the body. These usually hang on holes and sharp corners. It is best to actually see this task done once before performing this process on the fragile bodies. Deburring bodies is a meticulous task which can take approximately 30 to 40 minutes per body.

- Open the body into the 2 halves.
- With a compressed air duster, blow air through all holes and around channels to get rid of any residual metal shavings and grease.
- Get the small Bausch and Lomb microscope (in Suzanne or Lyman's lab) and shine a strong light from above it to illuminate the body.
- Cut the end of a wooden cotton tip into a thin wedge. This is your deburring tool. You can press as hard as you want without scratching the metal. **Do not use a metal tool**.

- Check inside surface with medium amplification. Check especially around the holes and next to the power bias hole.
- Check inside both waveguide channels with maximum amplification. Look at the walls and the bottom of the channels by tilting the body while adjusting the focus.
- Check the transistor chamber.
- Turn it on its side and check both outside surfaces and the pin holes.
- Repeat for the other half.

## 5) Cleaning

Once the bodies are debured, it is good to give them a quick clean before the next step. This can be done in Joe Horvath's lab. It simply consists of putting the bodies in water with a small amount of Citranox (an acid cleaning detergent to remove metal oxides) and putting this in the ultrasonic bath for ∼20 seconds. Then remove all the water with the Nitrogen blow dry on the left side of the sink.

## 6) Lapping

The inner and the flange surfaces must be very flat to prevent radiation from leaking out of the waveguide. To flatten both of those surface, we lap the bodies. We use a dry lapping process, described to us by the JPL machine shop guys (Jim Crosby and Pete Bruneau at (818) 354-4412). This process consists of polishing the surface with finer and finer polishing paper.

**Order lapping material.** For this dry lapping process, you will need the following material ordered from Allied High Tech (phone: (800)-675-1118): a large flat glass plate to support the lapping film (P/N: 69-10030), a rubber squeegee to fix the film on the glass plate (P/N: 50-05518), and an assortment of different grit size aluminium oxide lapping films (P/N: 50-20070 or see `http://www.alliedhightech.com` for other films).

**Attach film to glass plate.** Squirt water on glass plate and apply the shinny side of lapping film of glass plate. Squeegee all air bubbles away from below the film. The film is now firmly attached to glass.

**Lap the inside surface.** Starting from the roughest grit (roughly 30$\mu$m), rub the inner surface of the body against the film, keeping the body at 45° from the rubbing motion. Rotate the body by 90° every 10 to 20 laps. You can see when you are done when you cannot see anymore surface marks from the previous lapping or machining stage. Switch to a finer grit until desired polish is attained. For the MMIC bodies, we lapped with 30, 12,and 1$\mu$m paper.

**Verify the surface accuracy.** Once you think the surface is flat enough, it is useful to have a method to check the surface accuracy. The JPL machinists who taught us this dry lapping method use either an optically flat plate and the interference pattern of a single frequency light on the surface of interest, or an electronic indicator. Since we have neither of these, we use a third method. We bring the two flat surfaces under test together (ie. we bolt the 2 sides of the MMIC body) and shine a bright light behind it. If light is visible though the interface, then the surface needs more polishing.

**Lap the outside.** Once, the inside surfaces are deemed flat enough, we can insert the 2 inner pins (see section B), and go on to lap the two outer waveguide flange surfaces.

## 7) Gold Platting

The MMIC bodies are gold plated with 50$\mu$inches of hard gold to prevent oxidation and to obtain the best conductivity. We send out the MMIC bodies to be gold plated by a company called Specialized Plating Inc (phone: (978)-373-8030 , approximately 5$ each.) It is a good idea to clean the bodies again before sending them out for gold plating, although the company recleans them before plating.

## 8) Inserting the pins

Once the bodies come back from gold plating, we use a special tool designed by Bill to insert both the inner and outer pins.

# Appendix C

# A-priori Phase-Matching Procedure

This section provides a more detailed description of the procedure used to arrive to the final lengths RF and LO sections.

The lengths of the five custom waveguides in the LO section (A, B, C, D, and E, see Figure C.1) are determined by equating the distance from the top of the OMT to the center of the hybrid tee in the main arm and side arm. This ensures that the two arms indeed mechanically join at the OMT and at the hybrid tee. This equality is carried out separately in each of three axes and gives the following three equations (in inches):

**Equation for x-axis**

$$\begin{aligned} 1.15 &= 0.625 + 0.6 + 0.6 + 1.0 - 0.11 - 1.0 - 1.0 - D - 1.0 - 0.75 \\ D &= 0.315 \end{aligned}$$

**Equation for y-axis**

$$\begin{aligned} 1.15 &= 1.0 + 0.24 + 0.9 + A + 1.0 - 1.0 - C - 1.0 \\ A - C &= -0.01 \end{aligned}$$

**Equation for z-axis**

$$1.0{+}0.6{+}0.6{+}SS1{+}1.0{-}0.11{+}1.0{+}\,B{+}1.0$$

$$\begin{aligned} &= 0.5{+}1.0{+}SS2{+}0.24{+}0.9{+}E{+}1.0{+}1.0{+}0.28{+}\,1.0 \\ B &= E + 0.83 \\ E &= 1.0 \\ B &= 1.83 \end{aligned}$$

An additional constraint comes from the phase matching requirement. Both radiometer arms contain the same components and waveguide bends up to the OMT, phase-switch and five custom LO sections. To ensure phase matched arms, the length of the custom waveguide in each arms in equated to give the last necessary equation:

**Equation for phase matching**

$$
\begin{aligned}
E + D + 0.89 &= A + B + C \\
A + C &= 0.375 \\
A &= 0.182 \\
C &= 0.192
\end{aligned}
$$

The OMT and phase-switch are a special case because they are unpaired in the radiometer. The length difference between the main and the side arm of the OMT was measured with the VNA across the RF band. The equivalent waveguide length of the side arm was, on average for the six OMTs tested, 1.75 mm greater than the main arm with a maximum deviation of any OMT of 0.15 mm. The 3" stainless steel section, SS1 and SS2, were therefore machined with the same asymmetry to account for the OMT path length difference.

Similarly, the phase switch comes only in the side arm of the receivers. Because it is an active component, its equivalent path length must be determined and balanced with a similar length in the main arm. A set of ten phase-switches was swept on the VNA to measure their equivalent waveguide path lengths at 82 GHz. We found a path length of 22.6 mm (0.89") with the largest deviation of 2.5 mm.

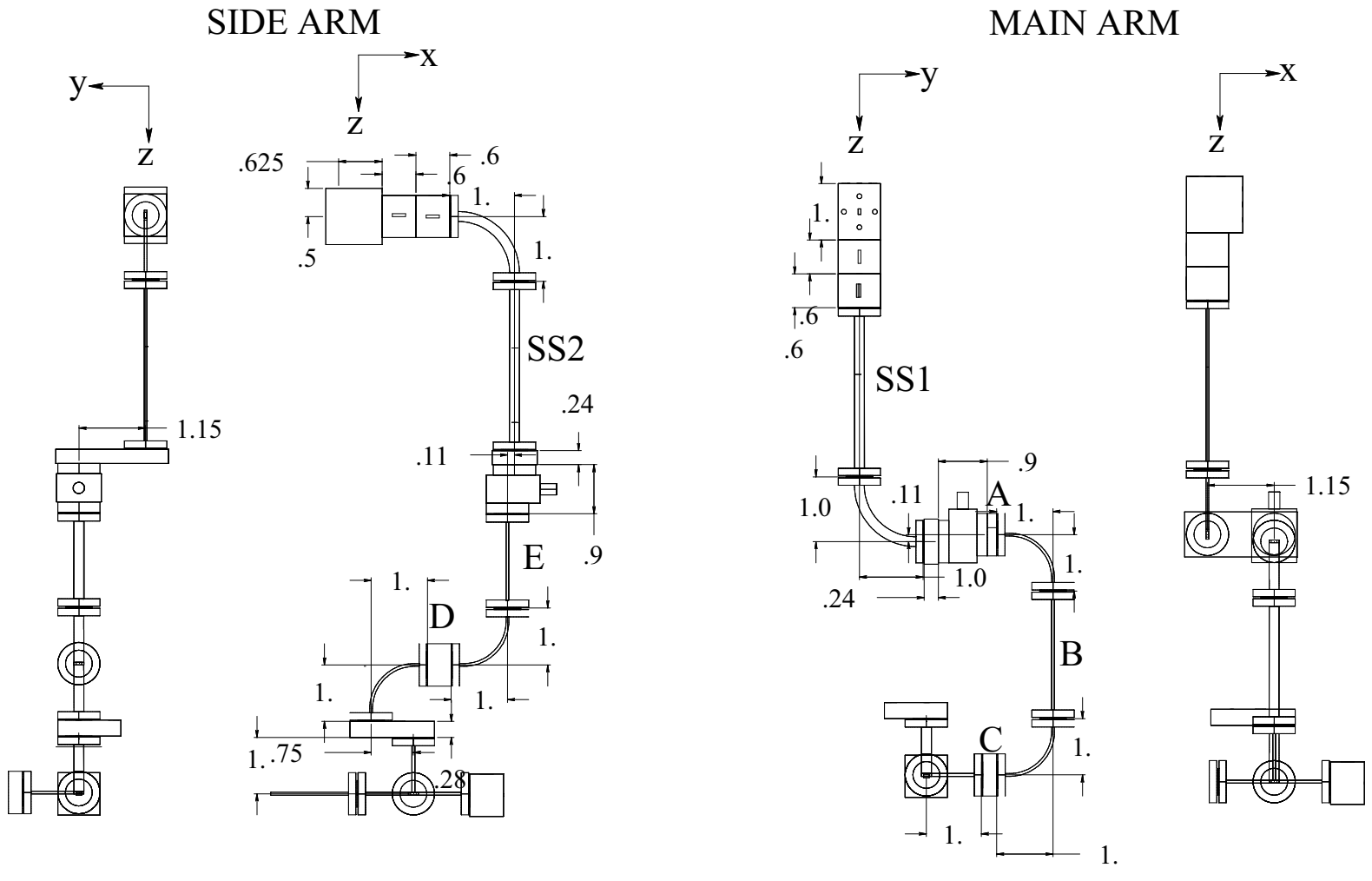

Figure C.1: Length determination for initial phase-matching of RF and LO section of receivers. All numbers are in inches.

# Appendix D

# Radiometer Component List

This is a list of all the components present in a a single receiver. Devices such as standard waveguide, custom built waveguide, connectors, wiring, feed-though, and SMA coaxial cables are not included.

| Component | Part Number | Manufacturer | Price |
|---|---|---|---|
| | RF components | | |
| OMT | Vertex RSI | 111590-1 | 5675$ |
| Feed Horn | Custom Microwave | Custom | 3000$ |
| MMIC LNA | JPL | Custom | |
| Bandpass Filter | Microwave Resources | FLWS-94 | 782$ |
| Mixer (matched pair) | Spacek Labs | 2M94-019LN | 6500$ |
| Local Oscillator | Spacek Labs | GW-820 | 3000$ |
| Phase Switch | Pacific Millimeter Products | 75MS-82 | 2000$ |
| Hybrid Tee | Millitech | HBT-10-R-0000 | 258$ |
| Power amplifier | JPL | Custom | |
| | IF components | | |
| Warm IF amplifier | Miteq | AFS44-02001800-25-KCR-S-44 | 2750$ |
| 0° 3 dB splitter | Mac technology | P8248-2 | 254$ |
| 90° hybrid splitter | Sage Labs | 2375-9 | 490$ |
| Triplexer | ES Microwave, LLC | 3SM-7/12.7-10PM | 3800$ |
| In-Line phase shifter | Weinshel | 917-12 | 276$ |
| Attenuator | Weinshel | - | 40$ |
| Detector diode | Agilent Technologies | HP8472B w/ 002 option | 400$ |
| Multiplier | Miteq | DB0218LW2 | 280$ |
| E- (H-) bends | Aerowave | 10-1890(101990) | 150$ |

Table D.1: Microwave Component List for a single CAPMAP receiver.

## Wiring

The bias wiring for MMICs and thermometers in and out of the dewar is done through 26-pin and 32-pin Bendix hermetic connectors. On the inside on the dewar, the Bendix connectors's solder cups are soldered to the insulated wires of an MDM (25 or 37 pin) connector mounted on an L bracket. From that point, the bias wiring goes through the Tekdata cables. Tekdata cables are made of 0.1 mm (4 mils) Manganin wires (32 wires or 24 wires). The wires are twisted two by two and weaved in nomex weaving fibre to give the braid some strength. Both ends of the cable are terminated with an MDM connector. The back of the MDM connectors is potted in Stycast 2850 to provide strain relief.

# Appendix E

# Antenna Control Hardware

The antenna control software is exhaustively described in Michelle Yeh's OBS software Guide [151]. Parts of the hardware (including the 24-bits digital I/O card, the remote-enable circuit, the 16-bin D/A card, and the GPS receiver timing card) related to the OBS computer is also described in that guide. A comprehensive antenna maintenance guide [7] contains some additional information about some antenna hardware.

## E.1 Micro controller and BCD output

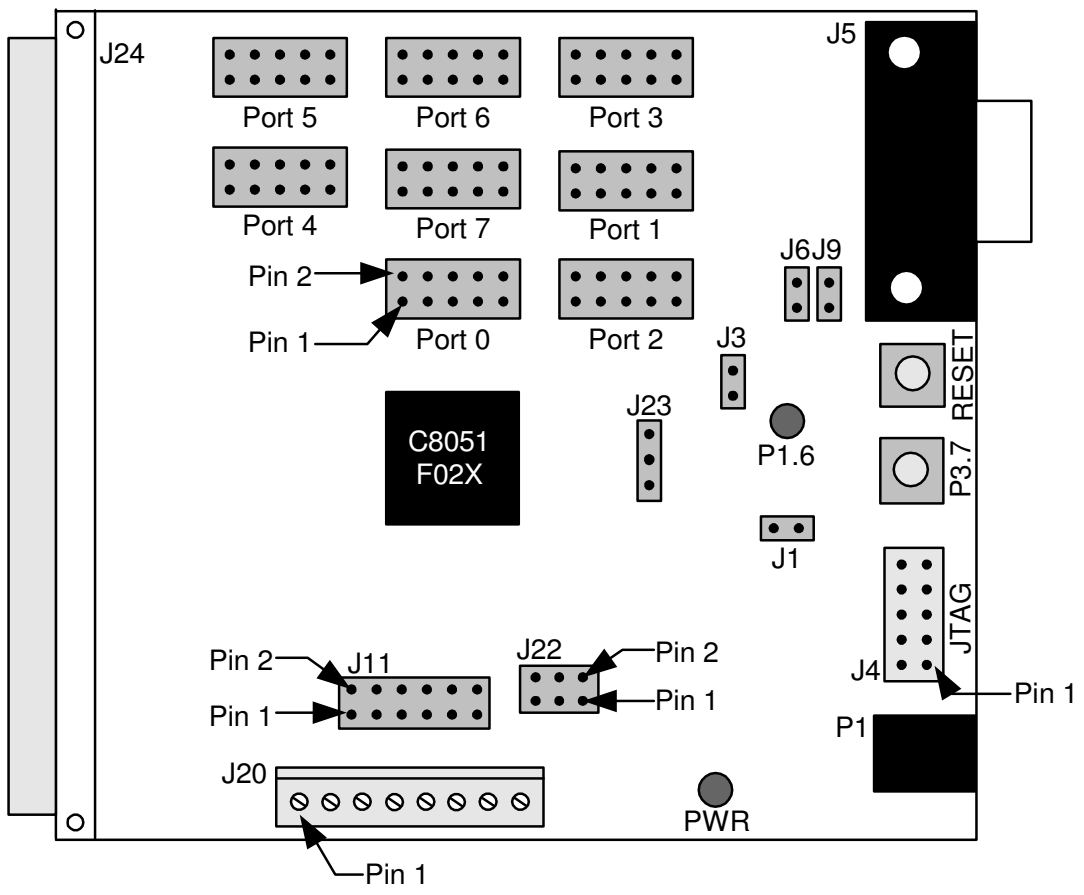

Figure E.1: Schematic of the C8051F020 $\mu$controller board, used to send the encoder information to the DAQ.

The wiring pinout for the position indicator BCD output as well as the pinout to the $\mu$controller is listed in Table E.1.

| BCD value | Bendix Pin | wire color | Az pins | El pins |
|---|---|---|---|---|
| sign | A | white | P4.1 | P4.5 |
| 100 | B | grey | P4.2 | x |
| 200 | C | purple | P4.3 | x |
| 10 | D | blue | P4.4 | P5.1 |
| 20 | E | green | P1.2 | P5.2 |
| 40 | F | yellow | P1.3 | P5.3 |
| 80 | G | orange | P1.4 | P5.4 |
| 1 | H | red | P1.5 | P5.5 |
| 2 | J | brown | P1.6 | P5.6 |
| 4 | K | black | P1.7 | P5.7 |
| 8 | L | white | P1.8 | P5.8 |
| .1 | M | grey | P2.1 | P6.1 |
| .2 | N | purple | P2.2 | P6.2 |
| .4 | P | blue | P2.3 | P6.3 |
| .8 | R | green | P2.4 | P6.4 |
| .01 | S | yellow | P2.5 | P6.5 |
| .02 | T | orange | P2.6 | P6.6 |
| .04 | U | red | P2.7 | P6.7 |
| .08 | V | brown | P2.8 | P6.8 |
| .001 | W | black | P3.1 | P7.1 |
| .002 | X | white | P3.2 | P7.2 |
| .004 | Y | grey | P3.3 | P7.3 |
| .008 | Z | purple | P3.4 | P7.4 |
| .0001 | a | blue | P3.5 | P7.5 |
| .0002 | b | green | P3.6 | P7.6 |
| .0004 | c | yellow | P3.7 | P7.7 |
| .0008 | d | orange | P3.8 | P7.8 |
| conversion complete | e | - | - | - |
| digital common | f | - | - | - |

Table E.1: BCD $\mu$controller pinout and functions. Pn.n refers to a pin on one of the 8 ports on the C8051F020 $\mu$controller. All Pn.9 pins are +3.3V and Pn.10 GND. Pin f of both bendix connector is attached to all Pn.10. In addition, the az and el conversion complete signals are wired to P0.7 and P0.8 respectively, and the 100Hz strobe signal from the DAQ computer is wired to P0.5

```
#include <c8051f020.h>
#include <stdio.h>

//-----------------------------//
// Global CONSTANTS //
//----------------------------//

#define BAUDRATE   115200    //Baud rate of UART in bps
#define SYSCLK     22118400  // SYSCLK frequency in Hz

//-------------------------------//
// Function PROTOTYPES //
//-------------------------------//

void SYSCLK_Init (void);
void PORT_Init (void);
void UART0_Init (void);

//-------------------//
// MAIN Routine //
//------------------//
void main (void) {
char strobe;
char p0start, p0end, p1, p2, p3, p4, p4low, p4high, p5, p6, p7;

WDTCN = 0xde;  // disable watchdog timer
WDTCN = 0xad;
SYSCLK_Init (); // initialize oscillator
PORT_Init ();  // initialize crossbar and GPIO
UART0_Init (); // initialize UART0

/* super-loop */
while (1) {
p0start=P0 & 0xd0;  //latch everything and P0 twice
p3 = P3;
p7 = P7;
p2 = P2;
p6 = P6;
p1 = P1;
p5 = P5;
p4 = P4;
p0end=P0 & 0xd0 ;

//if (conversion complete)
if ((p0start==p0end) && (!(p0end & 0x80)) && (!(p0end &
0x40))) {
if (p0end & 0x10) {   // if clock is high
p4low = p4 & 0x0f;      //mask to include only low nibble
p4high = p4 & 0xf0;    //mask to include only high nibble
p4high = p4high >>=4; //shift all bits to low end
putchar(0xbb); //start byte        //send data to serial port
// *** send elevation BCD *** /
putchar(p4high);
putchar(p5);
putchar(p6);
putchar(p7);
// *** send azimuth BCD *** /
putchar(p4low);
putchar(p1);
putchar(p2);
putchar(p3);
putchar(0xaa); //end byte
/*wait for clock to fall back */
while (1) {
p0end = P0;
strobe = p0end & 0x10;
if (!strobe) break;
}  //end while
//end if clock high
} //end if conv complete
}  //end while
}  //end main


//-------------------------------//
// Initialization Subroutines //
//-------------------------------//

//------------------//
// SYSCLK_Init //
//-----------------//
// This routine initializes the system clock to use an 22.1184MHz
crystal as its clock source, necesary for UART functioning.

void SYSCLK_Init (void)
{
int i;          // delay counter
OSCXCN = 0x67; //start ext. osc. with 22.1184MHz crystal

for (i=0; i < 256; i++); //XTLVLD blank. interval (>1ms)
while (!(OSCXCN & 0x80)); //Wait for crystal osc. to settle
OSCICN = 0x88; // select external oscillator as SYSCLK
source and enable missing clock
}

//--------------//
// PORT_Init //
//--------------//
// Configure the Crossbar and GPIO ports

void PORT_Init (void)
{
XBR0    = 0x04;  // Enable UART0
XBR1    = 0x00;
XBR2    = 0x40;  // Enable crossbar and weak pull-ups

/* setup ports for digital input */
P0MDOUT = 0x00;
P1MDOUT = 0x00;
P2MDOUT = 0x00;
P3MDOUT = 0x00;
P74OUT = 0x00;

  /* initialize port latches to all 1s for digital input open drain */
P0 = 0xff;
P1 = 0xff;
P2 = 0xff;
P3 = 0xff;
P4 = 0xff;
P5 = 0xff;
P6 = 0xff;
P7 = 0xff;
}

//-----------------//
// UART0_Init//
//----------------//
// Configure the UART0 using Timer1, for  and 8-N-1.

void UART0_Init (void)
{
SCON0   = 0x40; //SCON0: mode 1, 8-bit UART, disable RX
TMOD    = 0x20; // TMOD: timer 1, mode 2, 8-bit reload
TH1    = -(SYSCLK/BAUDRATE/16);    // set Timer1 reload
value for baudrate
TR1    = 1;        // start Timer1
CKCON |= 0x10;     // Timer1 uses SYSCLK as time base
PCON  |= 0x80;    // SMOD00 = 1, double UART baud rate
TI0    = 1;       // Indicate TX0 ready
}
```

Figure E.2: C code to program the $\mu$controller.

## E.2 SCR

### Description

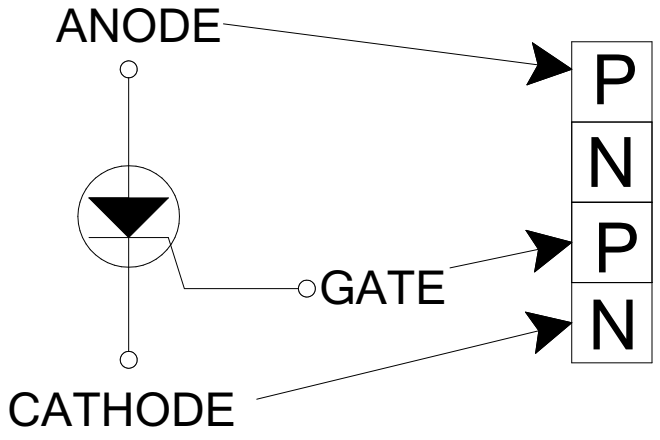

Figure E.3: Schematic diagram of an SCR and its semiconductor junction equivalence

A Silicon Controlled Rectifier (SCR) is a conventional rectifier (like a diode) controlled by a gate signal. However, unlike a diode, the application of a forward voltage is not enough for conduction. A gate signal controls the conduction. An SCR is a four-part semiconductor component (either PNPN or NPNP) schematically represented in Figure E.3. It can be modelled as two standard bipolar junction transistors connected as shown in Figure E.4.

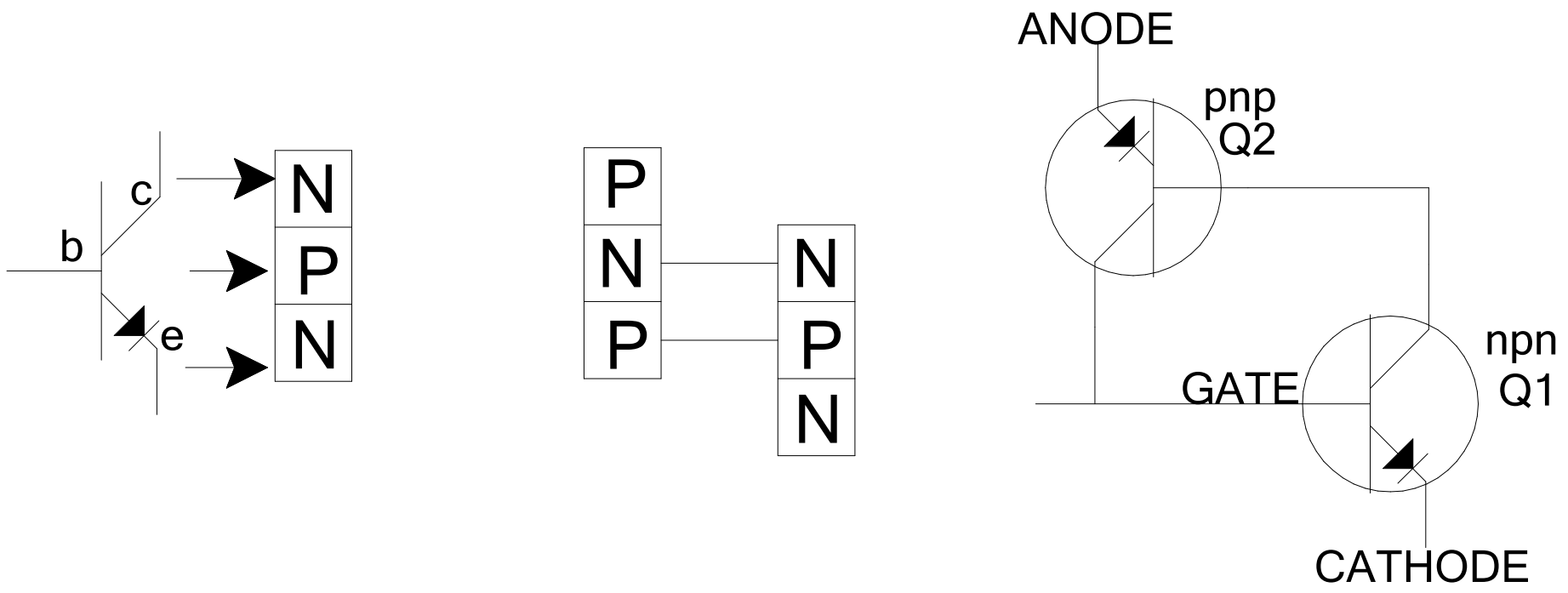

Figure E.4: Model of an SCR with two bipolar junction transistor connected. A single BJT is shown on the left. An SCR is similar in operation as 2 BJTs where the collector of Q1 to the base of Q2 and the collector of Q2 to the base of Q1. The model functions as follows: A small negative voltage (or 0V) will bias Q1 off and the SCR is stably off (no current flows between the anode and the cathode). A positive voltage applied to the gate will bias Q1 into conduction, causing its collector current to rise, which will bias Q2 into conduction, which in turn will tend to increase the gate voltage. Very fast, this positive feedback loop puts both transistors in saturation. The positive voltage initially applied to the gate to turn the SCR on is no longer necessary. The SCR will remain on until the Anode-Cathode voltage is turned off.

## Operation

The SCR is controlled from the off state (high anode-cathode resistance) to the on state ( low resistance) by applying a signal to the gate. Once turned on, it remains on, even after removal of the gate signal, until the anode cathode voltage drops to zero. One purpose of such a device is that it can supply large amounts of current using a very small trigger signal at the gate. This is comparable to a mechanical switch except that an SCR takes only a few microseconds to switch ON or OFF.

In the case of the telescope control, the SCR play the role of a fast, high current switch. Figure E.6 shows the wiring between the SCR logic circuit board, the SCR, and the the motors. Because each SCR can only deliver current in one direction, and because the 240 V AC power is on two separate lines 180° apart, there are four SCR's in parallel with each motor. One for each of the 4 cases: from top to bottom in Figure E.6 positive current-positive line 1 phase, positive current-positive line 2 phase, negative current-negative line 1 phase, negative current-negative line 2 phase. Each of the four SCR, one at any given time, receives an individual trigger from the SCR logic circuit board in Figure E.7.

There are 4 SCR logic circuit board, one for each motor. The circuit is designed to decide which of the SCR will be triggered and when. In such a system, the amount of power delivered by the SCR to the motor is regulated by the amount of time the SCR stays on. Because the SCR is automatically turned off when the Anode-Cathode voltage drops to zero, every half cycle of the 60 Hz AC power, the amount of power delivered to the motor is controlled by relative phase at which the SCR is turned on (Figure E.5).

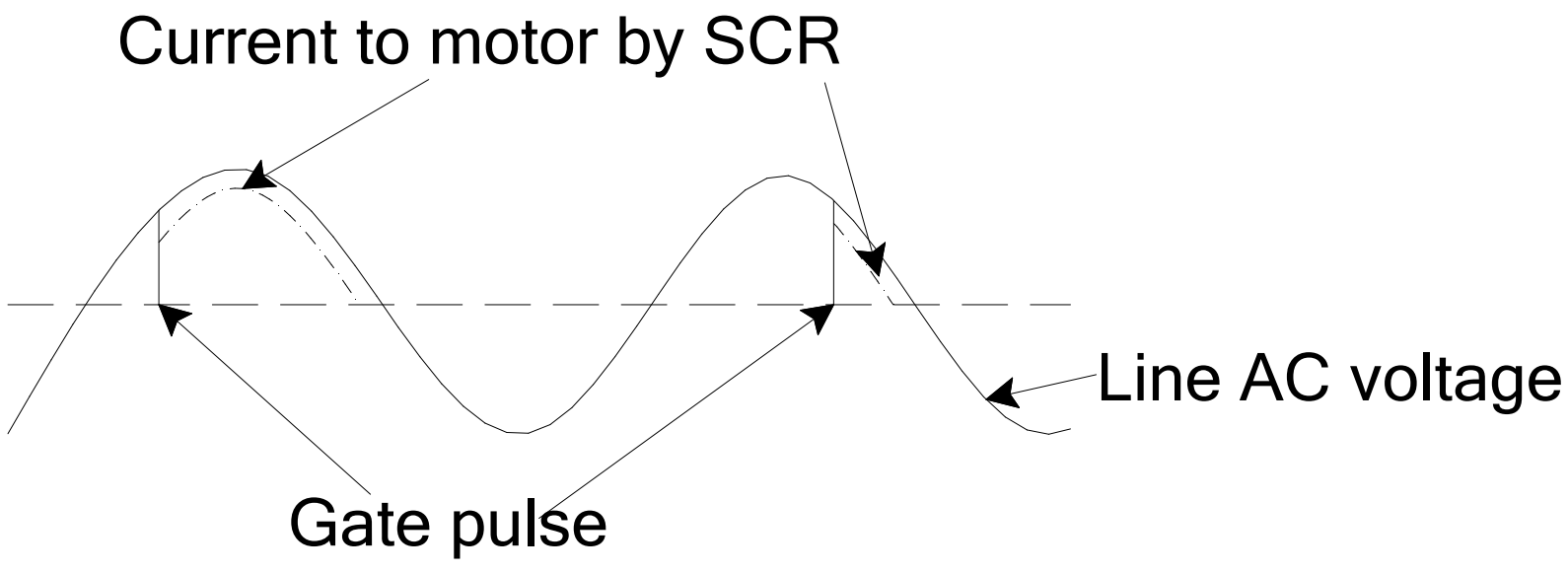

Figure E.5: Example of an SCR timing with respect to the phase of the AC line.

The SCR logic circuit board functions as a high gain feedback loop. The velocity request (from the computer or the slew knobs) is turned into a torque command input in the wire-wrap board. The SCR logic circuit board (see Figure E.7) starts by taking a weighted sum of this torque command input and Δ tachometer ( the difference of one axis' two motor's tachometer reading). In order to keep the two motors in an opposing state a low torque, a symmetric bias voltage is added. This triple sum, which can be monitored on test point (TP) 18, represents the first input to the feedback loop. The second input in simply the current at the given time in that particular motor (visible on TP11 after the differential amplifier). Note that the motor current can range from positive to negative 10 Amps. These two inputs are differenced in an integrating summing amplifier (B6). The purpose of the rest of the circuit behind this part is to do whatever it can to null the negative input of B6. The output of B6 (TP6) is a thus a voltage whose amplitude will determine at what phase to trigger one of the SCR associated with this motor. This is done by comparing the output of B6 to a pair of 0 to 10V triangle wave in phase with the two quadrature 60 Hz AC power line signals. The comparators therefore turn on when the triangle wave reaches the commanded voltage. A set of logic then takes that trigger signal and chooses depending on the phase of the 60Hz line and the requested motor direction, which of the four NAND gates should trigger. Only one NAND gate should trigger at a time. The trigger is then passed on to the SCR which is turned on until the next zero crossing of the appropriate 60Hz line.

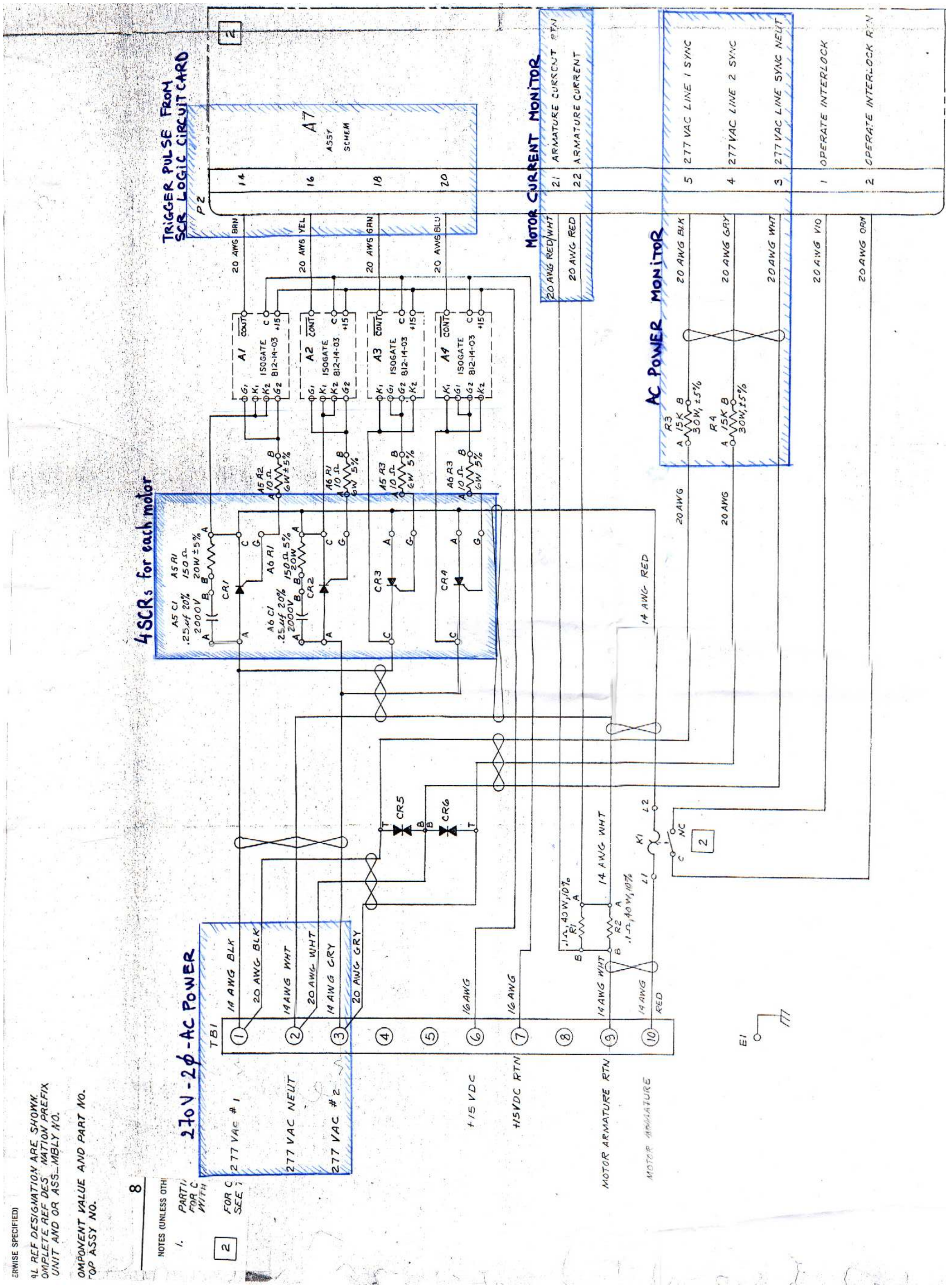

Figure E.6: SCR power amplifier wiring diagram. The main elements of the circuits are surrounded in blue.

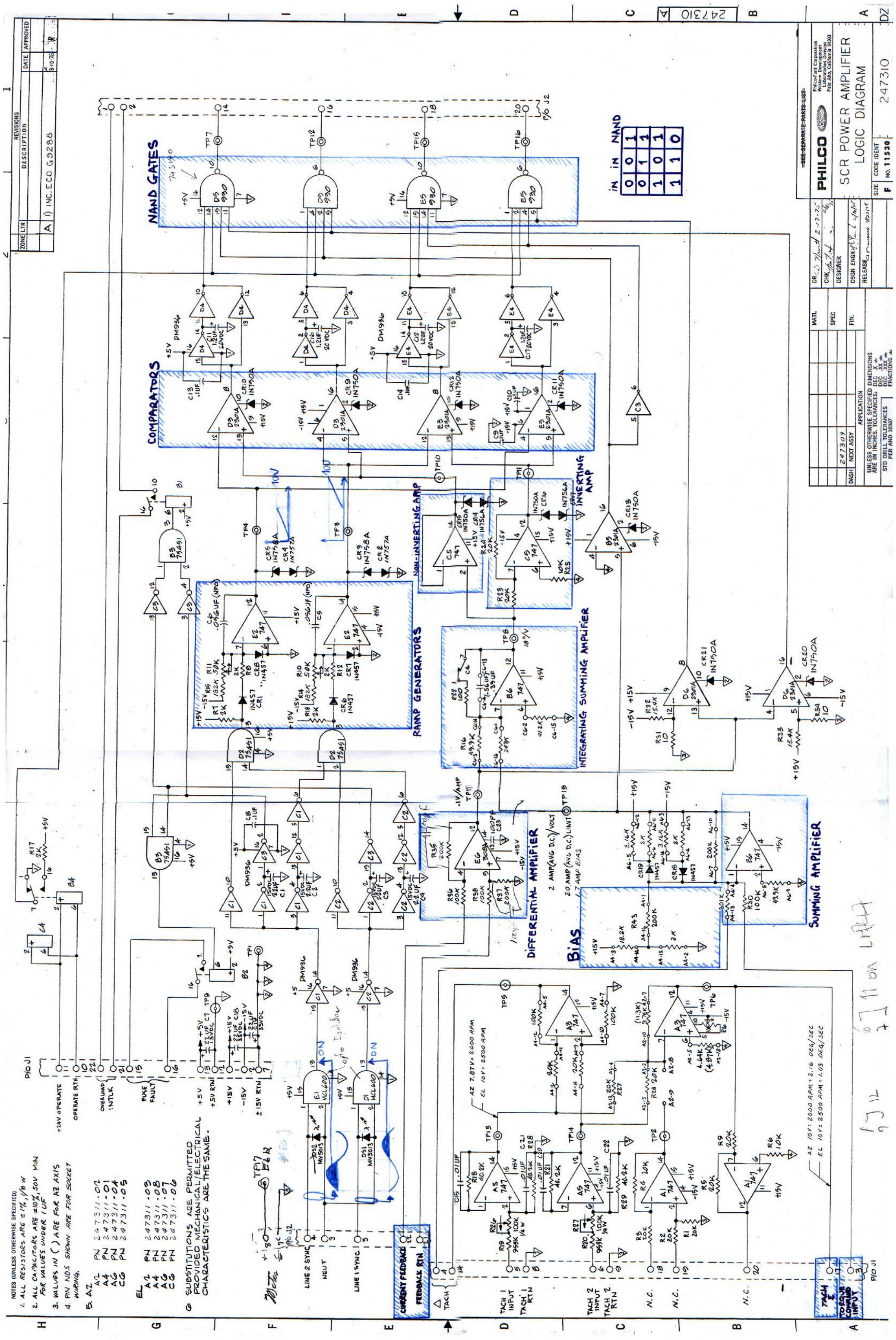

Figure E.7: SCR power amplifier logic diagram.The main components are surrounded in blue. Refer to the text for an explanation of the circuit's function. This diagram along with the diagram in Figure E.6 are maintained at the telescope stack of documentation

# Appendix F

# Off-axis Conic section

## F.1 Conics Equation

The geometry of an off-axis ellipsoid, used to design the tertiary mirror in Section 4.2.1 are recorded here. The ellipsoid is specified in terms of an incoming and outgoing radius of curvature,$R_1$ and $R_2$, and the half angle though which the beam is turned, $\theta_i$. The shape of the ellipsoid is typically written in the (x,y,z) coordinate system as $\frac{x^2}{b^2} + \frac{z^2}{a^2} = 1$, but for the shape $z'(x', y')$in terms of the $(x', y', z')$ coordinate system is more useful to machine the mirror.$z'$ is the normal to the mirror center and the $x'$-$y'$ plane is the symmetry plane of the conic surface. Referring to Figure F.1, the vertex to near focus distance, $f_0$, and eccentricity $e$ are

$$f_0 = \frac{R_1 + R_2 - \sqrt{R_1^2 + R_2^2 - 2R_1R_2\cos 2\theta_i}}{2} \tag{F.1}$$

$$e = 1 - \frac{2f_0}{R_1 + R2} \tag{F.2}$$

and the semi-axis lengths are

$$a = \frac{f_0}{1-e} \tag{F.3}$$

$$b = f_0\sqrt{\frac{1+e}{1-e}}\ . \tag{F.4}$$

The angle between the major axis and the ray connecting the near focus to the mirror center is $\theta_p$:

$$\theta_p = \pi - \cos^{-1}\left[\frac{R_1^2 + A_0^2 - R_2^2}{2R_1A_0}\right] \tag{F.5}$$

$$A_0 = \sqrt{R_1^2 + R_2^2 - 2R_1R_2\cos 2\theta_i} \tag{F.6}$$

and

$$\psi = \theta_p - \theta_i\ . \tag{F.7}$$

The equation for the mirror surface, centered on $(x', y') = (0, 0)$ is

$$z' = \frac{-B - \sqrt{B^2 - 4AC}}{2A} \tag{F.8}$$

where

$$A = a^2 \sin^2 \psi + b^2 \cos^2 \psi \tag{F.9}$$

$$B = 2[b^2 x' \sin\psi \cos\psi - a^2 x' \sin\psi \cos\psi - b^2 \cos\psi(\sqrt{a^2 - b^2} + R_1 \cos\theta_p) - a^2 R_1 \sin\psi \sin\theta_p] \tag{F.10}$$

$$C = b^2[x' \sin\psi - (\sqrt{a^2 - b^2} + R_1 \cos\theta_p)]^2 + a^2(x' \cos\psi + R_1 \sin\theta_p)^2 - a^2 b^2 - a^2 y'^2 \tag{F.11}$$

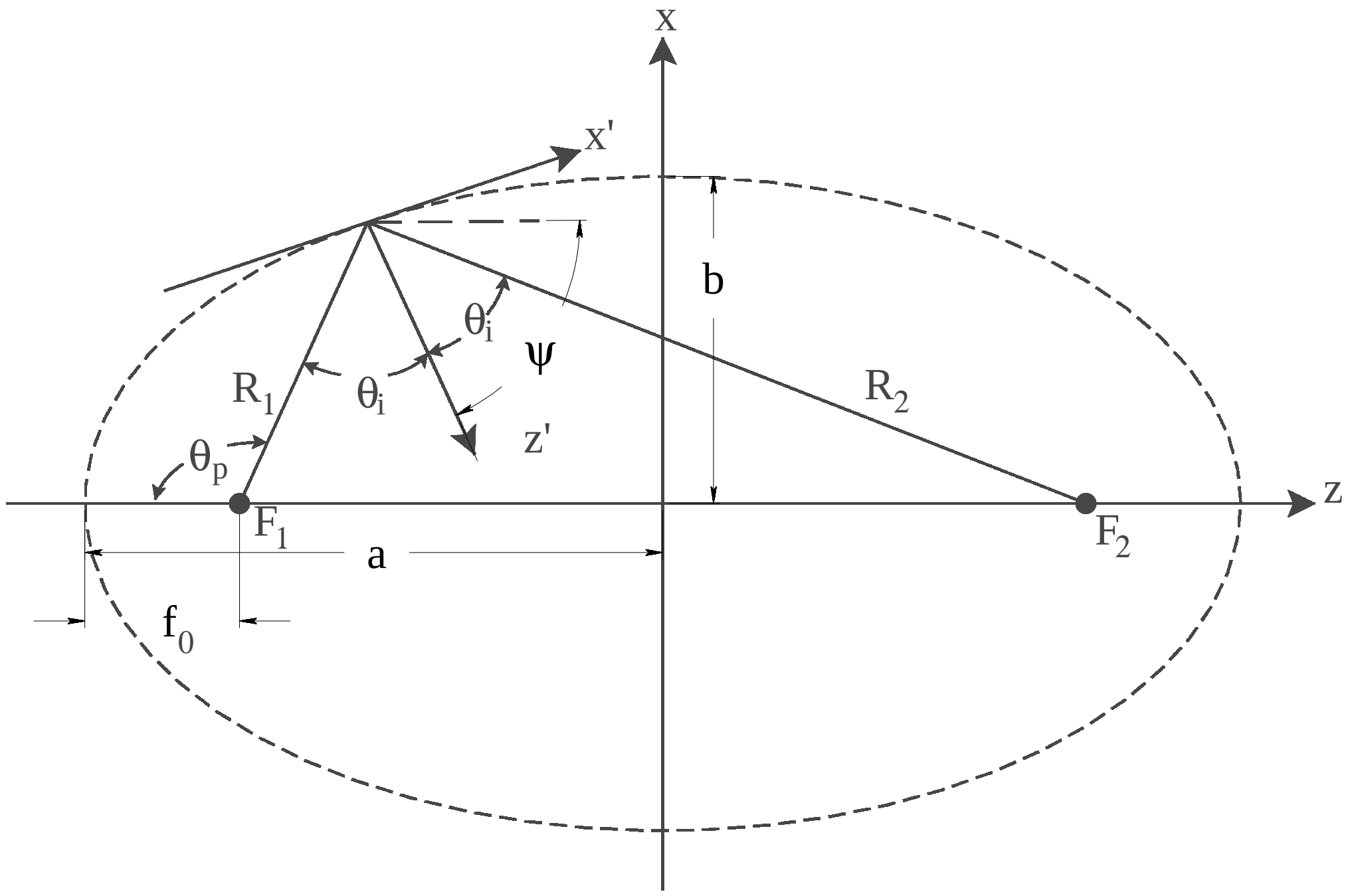

Figure F.1: Geometrical relations for an off-axis ellipsoid. The y-axis comes out of the page.

## F.2 CNC mirror milling code

```
#1
F1=265.415                                          ;SEMI-MAJOR AXIS
F2=87.309                                           ;SEMI-MINER AXIS
R4=.5                                               ;RADIUS OF BALLMILL
B4=9.0                                              ;STARTING RADIUS
E4=17.75                                            ;ENDING RADIUS
R1=B4
W1=8.0                                              ;Y WIDTH OF MIRROR
Y1=-(W1/2.0)-R4                                     ;Y-CLERANCE
Y2=(W1/2.0)+R4                                      ;OTHER SIDE Y-CLEARANCE
Z4= -F1*SQR(1-E4*E4/F2/F2)+F1+1.0                   ;ZED CLERANCE

L1

A1=ATN(F1*R1/F2/F2/SQR(1-R1*R1/F2/F2))              ;CALCULATE RADIUS & HEIGHT OF CUT
Z1=-F1*SQR(1-R1*R1/F2/F2)+F1-R4+R4*COS(A1)+.200     ;FINISH CUT DEPTH
R2=R1-R4*SIN(A1)
X1=SQR(R2*R2-Y1*Y1)
G0X1Y1Z4                                            ;MOVE ABOVE METAL TO START OF CUT
G0X1Y1Z1                                            ;LOWER TO CUT HEIGHT
G3X1Y2I0.0J0.0F20.                                  ;CUT CIRCLE
G0X1Y2Z4                                            ;RAISE TO CLEAR HEIGHT
R1=R1+.1                                            ;SET STEP SIZE FOR ROUGH CUT (100 MILS)
?R1<LE>E4                                           ;REPEAT
=L1!
$

#2
F1=265.415
F2=87.309
R4=.5
B4=9.0
E4=17.75
R1=B4
W1=8.0
Y1=-(W1/2.0)-R4
Y2=(W1/2.0)+R4
Z4= -F1*SQR(1-E4*E4/F2/F2)+F1+1.0

L2
A1=ATN(F1*R1/F2/F2/SQR(1-R1*R1/F2/F2))
Z1=-F1*SQR(1-R1*R1/F2/F2)+F1-R4+R4*COS(A1)+0.02     ;DEPTH OF FINISH CUT
R2=R1-R4*SIN(A1)
X1=SQR(R2*R2-Y1*Y1)
G0X1Y1Z4
G0X1Y1Z1
G3X1Y2I0.0J0.0 F10.
G0X1Y2Z4
R1=R1+.04                                           ;STEP OF ROUGH CUT (40 MILS)
?R1<LE>E4
=L2!
$

#3
F1=265.415
F2=87.309
R4=.5
B4=9.0
E4=17.75
R1=B4
W1=8.0
Z4= -F1*SQR(1-E4*E4/F2/F2)+F1+1.0
Y1=-(W1/2.0)-R4/5
Y2=(W1/2.0)+R4/5

L3

A1= ATN(F1*R1/F2/F2/SQR(1-R1*R1/F2/F2))
Z1=-F1*SQR(1-R1*R1/F2/F2)+F1-R4+R4*COS(A1)          ;DEPTH OF FINISH CUT
R2=R1-R4*SIN(A1)
X1=SQR(R2*R2-Y1*Y1)
G0X1Y1Z4
GOX1Y1Z1
G3X1Y2I0.0J0.0F20.
G0X1Y2Z4
R1=R1+.004                                          ;SET FINER STEPS 4 MILS
?R1<LE>E4
=L3!
$

G0G70G75G90G94T7M6
S3000M3
=#1
=#2
=#3
M22
```

Figure F.2: Code for machining the tertiary mirror on CNC mill.

# Appendix G

# 2003 Data Format

## G.1 DATA Structure

During the 2003 CAPMAP observing season (Jan to June 2003), data were sampled at 100 Hz and saved to file every 30mn. The demodulated data files are named in the following manner: `scan_type_mmddyyhhmn_av_000.dat`, where `scan_type` is a 4 letter word which described the current scan, `mmddyy` is the month, day, year , `hhmm` the EST hour and minute, and the three last digits are incremented for each new file of a given run. A similar file without the `_av_` subscript is saved for the raw data.

The rest of this description refers to only the demodulated data files. The files are saved in binary format with the following structure. One frame (~1 second) contains 1 housekeeping block and 100 data blocks. This structure then repeats for 2000 frames before a new file is opened. One frame thus adds up to 10294 bytes, and one file to ~ 20Mb. In logical order, the blocks are:

**Houskeeping** 47 words 2 bytes . The breakout of the housekeeping channels is given in G.2.

**Error flag** 1 word of 2 bytes.

**GPS time** 2 words of 4 bytes (number of seconds since Jan, 1 1970, microseconds since last second)

**Encoder** 2 words of 4 bytes (azimuth, elevation). The encoder values saved are $10^4$ times the real values.

**Radiometer data** 21 words of 4 bytes each (clock, AS0, AD0, AS1, AD1, AS2, BS0, BD0, BS1, BD1, BS2, CS0, CD0, CS1, CD1, CS2, DS0, DD0, DS1, DD1, DS2)

## G.2 Houskeeping Channels

Table G.1 lists the function of the 47 housekeeping channels.

| Channel # | Function | Comment |
|---|---|---|
| 1. | Sky monitor X | not deployed |
| 2. | Sky monitor Y | not deployed |
| 3. | Cold Cathode Pressure | working |
| 4. | Pirani Pressure | working |
| 5. | MMIC Bias monitor | not implemented |
| 6. | LO-PA Bias monitor | not implemented |
| 7. | MMIC TC Current | not implemented |
| 8. | MMIC TC temperature | not implemented |
| 9. | IF TC Current | working |
| 10. | IF TC temperature | working |
| 11. | LO TC Current | working |
| 12. | LO TC temperature | working |
| 13. | Shroud B temperature | working |
| 14. | Horn B temperature | not functioning |
| 15. | A MMIC temperature | not functioning |
| 16. | B MMIC temperature | working |
| 17. | C MMIC temperature | working |
| 18. | D MMIC temperature | working |
| 19. | Filter B temperature | Broken |
| 20. | Power Amp A temperature | not functioning |
| 21. | LO temperature | working |
| 22. | Power Amp B temperature | working |
| 23. | Shroud D temperature | working |
| 24. | Power Amp D temperature | not functioning |
| 25. | 80 K TC temperature | working |
| 26. | 80 K TC Current | working |
| 27. | 20 K TC temperature | flaky |
| 28. | 20 K TC Current | working |
| 29. | Tilt Meter: El. | working, 10 degrees/Volt |
| 30. | Tilt Meter: Az. | working, 10 degrees/Volt |
| 31-38 | Mirror temperature | working |
| 39-46 | Room temperature | working |
| 47 | Monitoring addressing | not implemented |

Table G.1: Housekeeping channel description